\documentclass{optica-article}

\journal{opticajournal} % for journals or Optica Open

\articletype{Research Article}

\usepackage{lineno}
\usepackage{bm}
\usepackage{ dsfont }
\usepackage[T1]{fontenc}
\usepackage[utf8]{inputenc}
\usepackage{bbold}

\usepackage{fancyhdr}
\fancypagestyle{plain}{ % also use fancy style for the first page
  \fancyhf{}
  \cfoot{\thepage}
}

\newenvironment{biography}[2][]
  {\begingroup
   \leftskip=0pt plus 1fil
   \rightskip=0pt plus -1fil
   \parfillskip=0pt plus 2fil
   \vspace{1.5\baselineskip}
   \noindent\textbf{#2} \par
   \small#1\par}
  {\par\endgroup}
\def\_#1{{\bf #1}}
\def\l#1{\label{eq:#1}}
\def\r#1{(\ref{eq:#1})}
\def\_#1{{\bf #1\mit}}
\def\=#1{\overline{\overline #1}}
\newcommand{\ve}[1]{\mathbf{#1}}
\def\w{\omega}
\def\omegaf{\omega_{\rm F}}

\def\atone{\frac{{\rm d} A_1(t)}{{\rm d} t}}
\def\attwo{\frac{{\rm d} A_2(t)}{{\rm d} t}}

\def\aattone{\frac{{\rm d}^2 A_1(t)}{{\rm d} t^2}}
\def\aatttwo{\frac{{\rm d}^2 A_2(t)}{{\rm d} t^2}}

\def\atconjone{\frac{{\rm d} A_1^*(t)}{{\rm d} t}}
\def\atconjtwo{\frac{{\rm d} A_2^*(t)}{{\rm d} t}}

\def\aattconjone{\frac{{\rm d}^2 A_1^*(t)}{{\rm d} t^2}}
\def\aattconjtwo{\frac{{\rm d}^2 A_2^*(t)}{{\rm d} t^2}}

\begin{document} 

\title{Theory and applications of photonic time crystals: a tutorial}

\author{
Mohammad M. Asgari,\authormark{1,*} 
Puneet Garg,\authormark{2,*} 
Xuchen Wang,\authormark{3,*} 
Mohammad S. Mirmoosa,\authormark{4,*} 
% Aristeidis G. Lamprianidis,\authormark{2} 
Carsten Rockstuhl,\authormark{2,3,\dag} 
and Viktar Asadchy\authormark{1,\dag}}
[*]{These authors contributed equally to this work.}

[\dag]{Corresponding authors: viktar.asadchy@aalto.fi, carsten.rockstuhl@kit.edu}
% \author{
% Mohammad M. Asgari,\authormark{1} 
% Puneet Garg,\authormark{2} 
% Xuchen Wang,\authormark{2} 
% Mohammad S. Mirmoosa,\authormark{3} 
% Aristeidis G. Lamprianidis,\authormark{2} 
% Carsten Rockstuhl,\authormark{2,4} 
% and Viktar Asadchy\authormark{1,*}}

\vspace{2mm}
%\email{\red{\uppercase{The order of authors is preliminary}}}
\address{
\authormark{1}Department of Electronics and Nanoengineering, Aalto University, Maarintie 8, 02150 Espoo, Finland\\
\authormark{2}Institute of Theoretical Solid State Physics, Karlsruhe Institute of Technology, Kaiserstr. 12, 76131 Karlsruhe, Germany\\
\authormark{3}Institute of Nanotechnology, Karlsruhe Institute of Technology, Kaiserstr. 12, 76131 Karlsruhe, Germany\\
\authormark{4}Department of Physics and Mathematics, University of Eastern Finland, Yliopistokatu 7,  80130 Joensuu, Finland
}

%\pagenumbering{arabic}

% \email{\authormark{*}viktar.asadchy@aalto.fi} %% email address is required; see note below about the corresponding author designation
% use {asbstract*} to suppress the copyright line. Copyright information will be added in production

\begin{abstract*} 
This tutorial offers a comprehensive overview of photonic time crystals  -- artificial materials whose electromagnetic properties are periodically modulated in time at scales comparable to the oscillation period of light while remaining spatially uniform. Being the temporal analogs to traditional photonic crystals, photonic time crystals differ in that they exhibit momentum bandgaps instead of energy bandgaps. The energy is not conserved within momentum bandgaps, and eigenmodes with exponentially growing amplitudes exist in the momentum bandgap. Such properties make photonic time crystals a fascinating novel class of artificial materials from a basic science and applied perspective.  
This tutorial overviews the fundamental electromagnetic equations governing photonic time crystals and explores the groundbreaking physical phenomena they support. Based on these properties, we also oversee a diverse range of applications they unlock. Different material platforms suitable for creating photonic time crystals are discussed and compared. Furthermore, we elaborate on the connections between wave amplification in photonic time crystals and parametric amplification mechanisms in electrical circuits and nonlinear optics. Numerical codes for calculating the band structures of photonic time crystals using two approaches, the plane wave expansion method and the transfer matrix method, are provided.
The tutorial will be helpful for readers with physics or engineering backgrounds. It is designed to serve as an introductory guide for beginners and to establish a reference baseline reflecting the current understanding for researchers in the field. 

% \red{The planned total page number is at minimum 54 pages.  } 

\end{abstract*}

%%%%%%%%%%%%%%%%%%%%%%%%%%  body  %%%%%%%%%%%%%%%%%%%%%%%%%%
\tableofcontents

\section{Introduction} 

Materials constitute the cornerstone for most technological and many societal developments. It is by no means a surprise that periods of humanity are named according to the materials that shaped them. While we had to consider those materials given to us by nature for a long time, we got increasingly used to the fact that tailor-made materials with properties on demand come in reach by combining intrinsic materials with a suitable geometry or structure. In the context of the tutorial at hand, the material properties at stake are those that govern the interaction of electromagnetic fields or light with matter, and we are interested in electromagnetic or optical materials. In particular, the ever-evolving demands of contemporary society call for artificial materials that can control light propagation in a way inaccessible to natural materials. 

Over the past several decades, a diverse spectrum of artificial composites and material systems has emerged. This includes but is not limited to photonic crystals~\cite{joannopoulos_photonic_2008}, metamaterials~\cite{simovski_introduction_2020}, metasurfaces~\cite{achouri_electromagnetic_2021}, nanocolloids~\cite{sanchez-dominguez_nanocolloids_2016}, and two-dimensional materials~\cite{novoselov_2d_2016}. The materials explored in these endeavors have subsequently found extensive utility in various industrial applications. A unifying feature of these advanced materials is their inherent \textit{spatial} inhomogeneity, which accounts for their complex electromagnetic properties and sharply distinguishes them from naturally occurring, spatially uniform materials. 

\begin{figure}[t]
\centerline{\includegraphics[width= 1\columnwidth]{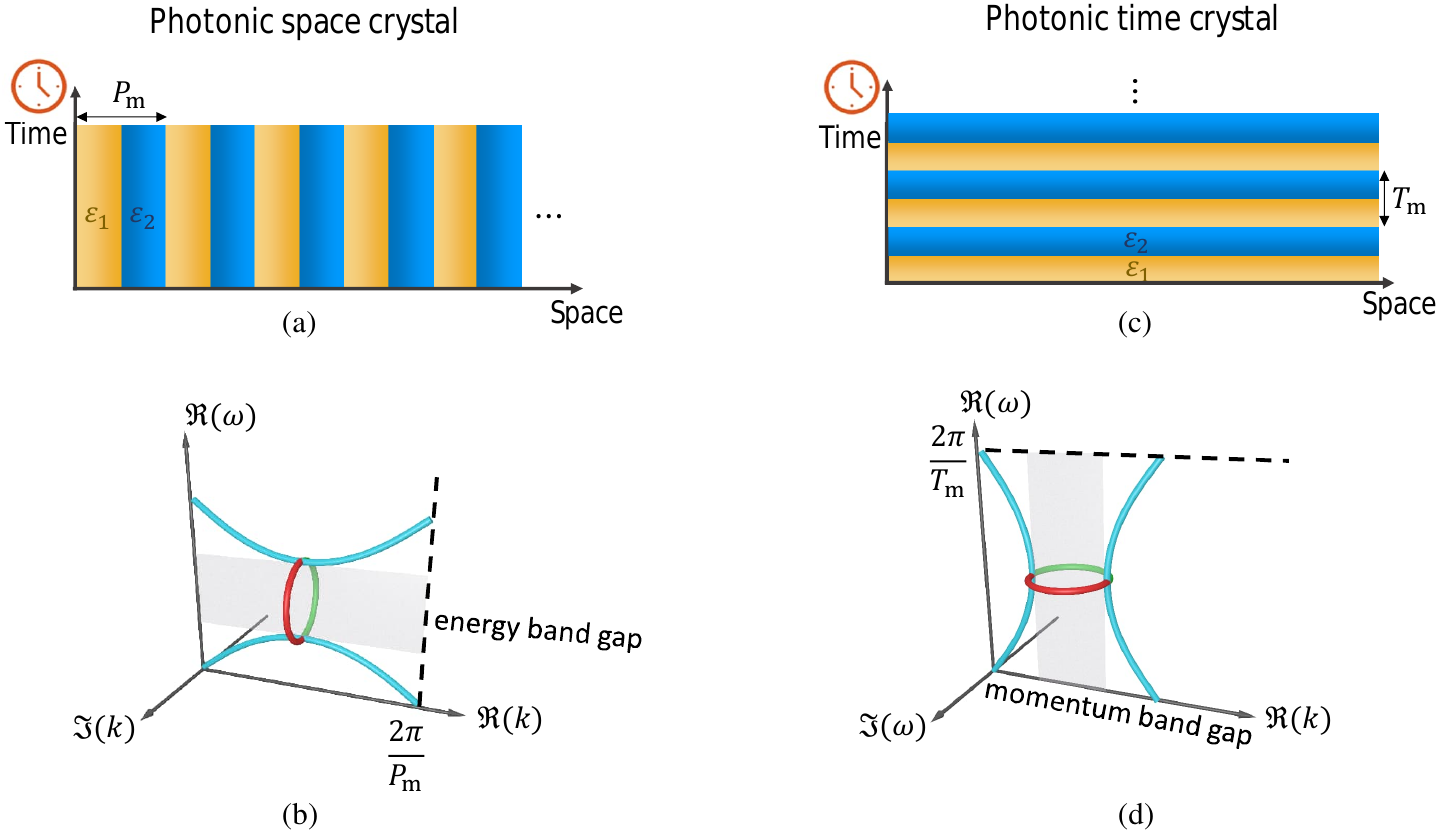}}
\caption{(a) Schematics of a conventional one-dimensional photonic (space) crystal where permittivity $\varepsilon$ is modulated along one spatial coordinate while it is constant in time. Moreover, the crystal is uniform along the other two spatial directions. 
(b) Characteristic band diagram of a one-dimensional photonic crystal. Here, the dispersion relation is shown, expressing the functional dependency of the frequency on the real part of the wavenumber $\Re (k)$ that expresses the phase variation of the eigenmode along the direction of the periodic material modulation. The third coordinate axis denotes the imaginary part of wavenumber $\Im (k)$. For simplicity, only the two lowest bands (shown in blue) are depicted inside the Brillouin zone (whose edge is indicated with the dashed line). There is a frequency domain, the energy bandgap (shown in light gray), in which no propagating solutions to Maxwell's equations exist. Inside the energy bandgap, two eigenmodes are also supported (shown in red and green), but they decay in space along the positive and negative spatial directions, respectively. (c) Schematics of a photonic time crystal where the permittivity is a periodic function of time only. Otherwise, the spatial distribution of the permittivity is uniform in all three directions. (d) Characteristic band diagram of a photonic time crystal. The third coordinate axis denotes the imaginary part of frequency $\Im (\omega)$. Inside the momentum bandgap, two eigenmodes are supported: one exponentially growing (shown in red) and one exponentially decaying (shown in green) in time.
}
\label{Fig:photonic_space_time}
\end{figure}

Among the categories mentioned above, photonic crystals are particularly noteworthy, as they are arguably the most thoroughly investigated material platform, well-established, and application-ready. In the most basic one-dimensional geometry, photonic crystals constitute a multilayer structure with a periodicity $P_{\rm m}$ comparable to the light wavelength. In its simplest form, each period comprises alternating layers of two distinct materials characterized by permittivities 
$\varepsilon_1$ and $\varepsilon_2$ (see illustration in Fig.~\ref{Fig:photonic_space_time}(a)). When discussing the propagation of light in such a system, we usually ask: what do the eigenmodes look like, and what is the dispersion relation of the associated eigenvalues? Here, the eigenmodes are the elementary solutions to Maxwell's equations in such a medium without external sources. Once they are known, an arbitrary solution to Maxwell's equation can be written as a superposition of such eigenmodes weighted with suitable amplitudes. In extension, the dispersion relation is a governing equation that relates the parameters that characterize the eigenmodes. For a plane wave, being the eigenmode of the homogeneous space, the parameters are the frequency and the wavevector components.       

Now, a fundamental principle of physics, the Noether theorem, says that for every continuous symmetry of a physical system, there exists a conserved quantity. Owing to their continuous time-translational symmetry, the frequency of the eigenmode of photonic crystals remains conserved. However, because of only the discrete space-translational symmetries, their wavenumbers $k$ are conserved only up to the addition of a multiple of the reciprocal lattice vector $2\pi N/P_{\rm m}$~\cite[p.~35]{joannopoulos_photonic_2008}, with $N$ being an integer. Consequently, just as the photonic crystal is periodic in space, expressed as $\varepsilon(z+ P_{\rm m}) = \varepsilon(z)$ for the one-dimensional version, the dispersion relation exhibits a periodicity in the reciprocal space, expressed as $\omega(k+ 2\pi/P_{\rm m}) = \omega(k)$. For a finite permittivity contrast, i.e., $\varepsilon_1 \neq \varepsilon_2$, eigenmodes that differ by $2\pi/P_{\rm m}$ in wavenumber (momentum) have lifted frequency degeneracy, resulting in a frequency (energy) bandgap (see illustration in Fig.~\ref{Fig:photonic_space_time}(b)). 

Remarkably, these photonic bandgaps have been known since 1887~\cite{rayleigh_xvii_1887}.  
Most importantly, light with frequencies within these photonic bandgaps cannot propagate inside the photonic crystal. Indeed, within such a bandgap, two eigenmodes exist but with complexwavenumbers (see Fig.~\ref{Fig:photonic_space_time}(b)). These eigenmodes are evanescent and decay exponentially in space in either the positive or negative spatial directions along the crystal's periodicity. Those eigenmodes have a non-vanishing amplitude after excitation only close to the photonic crystal interface, and the requirement on an exponential decay explains which of the two eigenmodes is excited. 

The significance of frequency bandgaps cannot be overstated, as they fundamentally drive the practical applications of two- and three-dimensional photonic crystals. Examples of such applications include optical fibers, light-emitting diodes, solar cells, and biosensors, among others~\cite{knight_allsilica_1996,wierer_iiinitride_2009,liu_advance_2019,fenzl_photonic_2014,munzberg2018superconducting,bielawny2009intermediate}. However, it is not just the bandgap that draws attention. The promise to tailor the dispersion relation and, with that, the isofrequency surfaces, holds the key to controlling the light propagation comprehensively.  

In special relativity and electrodynamics, the interwoven relationship between time and space coordinates naturally invites exploration into the behavior of light in materials with \textit{temporal} inhomogeneity. These sophisticated materials, variously termed ``temporal," ``dynamic," or ``time-varying" in the scientific literature, exhibit properties that vary over time on scales comparable to the oscillation period of the incident electromagnetic wave~\cite{richards_analysis_2012,kalluri_electromagnetics_2018,caloz2019spacetime,Caloz2020spacetime2,galiffi2022photonics,engheta2023four}. It is crucial to differentiate these rapidly modulated time-varying systems from adiabatically slowly-varying tunable and reconfigurable materials~\cite{he_tunable_2019a}. The latter always operate in the steady-state regimes, possibly switching between several states. In sharp contrast, time-varying systems, as considered here, always operate in the transient regime due to the ultrafast modulation, which enables novel, unique phenomena in light-matter interactions.

Considering the possibility of rapidly modulating material properties, the concept of photonic crystals can be extended to a temporal context. We will consider a spatially uniform material but with material parameters, such as the permittivity, that oscillate periodically in time with a period $T_{\rm m}$ (as illustrated in Fig.~\ref{Fig:photonic_space_time}(c)). 
Such materials, which possess continuous space-translational and discrete time-translational symmetries, have been recently categorized as ``photonic time crystals", ``temporal photonic crystals", ``pure-time crystals", etc.~\cite{zurita2009reflection,lustig2018topological,caloz2019spacetime}. For the purposes of this tutorial, we will use the term ``photonic time crystals" (PTCs), and it is these PTCs that we will focus on here. 

Please note the subtle but important choice of terminology of a {\bf photonic} time crystals. That is done to distinguish PTCs from the newly introduced ``time crystals", which are quantum many-body systems that spontaneously violate time-translation symmetry in the presence of a drive that is uncorrelated with the system's periodicity~\cite{wilczek2012quantum}. As we often encountered at conferences or more general discussion questions to which extent PTCs are related to these ``time crystals", we wanted to emphasize upfront that there is no relation between them.

Intriguingly, the band structure of PTCs is analogous to that of spatial photonic crystals but rotated by 90 degrees in the $\omega -  k$ plane~\cite{biancalana2007dynamics} (see Fig.~\ref{Fig:photonic_space_time}(d)). In the following chapters, we will explain in more detail how to derive this dispersion relation and elaborate on its physical meaning. 
But for the sake of an introduction, we wanted to highlight that in PTCs, the wavenumbers of eigenmodes are conserved thanks to their spatial uniformity. In contrast, their frequencies are conserved only up to the addition of a multiple of $2\pi N /T_{\rm m}$. Unlike conventional photonic crystals, which feature energy bandgaps, PTCs exhibit ``momentum bandgaps". Notably, these momentum bandgaps host two eigenmodes with complex frequencies (see Fig.~\ref{Fig:photonic_space_time}(d)): one of these eigenmodes decays over time while the other undergoes amplification. Such mode amplification is allowable in PTCs since they are non-energy-conserving systems, i.e., non-Hermitian, that can extract energy from the modulation source to permit such an amplification. The phenomenon of momentum bandgaps represents one of the principal drivers for the recent rapid growth of scientific interest in PTCs. 
However, despite most research on PTCs has been emerging only in the past five years, examining and recognizing the foundational historical developments that have led to this new field remains crucial.

The seminal work on what is now known today as PTCs was first published in 1958 by F.~Morgenthaler~\cite{morgenthaler1958velocity}. He investigated wave propagation in a bulk spatially uniform material with time-varying permittivity and permeability. While groundbreaking, his work was limited to solving special cases of temporal modulation. 
Around the same period, a separate line of research began to examine wave propagation in systems characterized by a spatiotemporal modulation. This research opened up possibilities for parametric amplification in diverse systems such as transmission lines~\cite{cullen_travellingwave_1958,tien_travelingwave_1958}, guided-mode systems~\cite{simon1960action,averkov_wave_1959,oliner1961wave,ostrovskii_nonresonance_1971}, and nonlinear optical materials~\cite{armstrong_interactions_1962,kingston_parametric_1962,kroll_parametric_1962,akhmanov_concerning_1962}.
Another pivotal contribution to exploring PTCs was made by D.~Holberg and K.~Kunz~\cite{holberg1966parametric}. They focused on materials with a distinct type of permittivity modulation, resulting in eigenmodes that could be analytically described using Mathieu functions. Importantly, they were among the first to demonstrate theoretically eigenmode amplification within a momentum bandgap.
However, even earlier discussions of momentum bandgaps and the complex-frequency modes they support can be attributed to a 1958 publication by P.~Sturrock~\cite{sturrock_kinematics_1958}. Sturrock employed general kinematic principles to analyze wave functions in propagating systems. Subsequent studies further explored dispersion relations with momentum bandgaps in systems with space-time modulations~\cite{cassedy_temporal_1962,cassedy1963dispersion,cassedy1967dispersion2}.
Other impactful early works in the area of PTCs addressed various aspects. Examples are the excitation problem of the crystal~\cite{felsen1970wave}, momentum bandgaps at time-varying impedance boundaries~\cite{cassedy1965wave}, finite spatial extents of PTCs~\cite{fante_transmission_1971,harfoush_scattering_1991}, and self-modulated crystals based on electron plasmas~\cite{yablonovitch_selfphase_1974}.
In the early 21st century, the field witnessed significant expansion with the introduction of a generalized framework by the group of E. O'Reilly~\cite{biancalana2007dynamics}. This framework encompassed wave propagation in bulk materials subject to a spatiotemporal rectangular-shaped modulation, and it posited that energy and momentum bandgaps are specific examples within a broader category of forbidden bandgaps.
Finally, the group of P. Halevi explored different aspects of wave propagation and parametric resonances in PTCs, including those with temporal modulation of a general form~\cite{zurita2009reflection,martinez2016temporal,martinez2018parametric}, as well as performed the first experimental observation of the momentum bandgap~\cite{Reyes-Ayona2015Observation, reyes2016electromagnetic}.

Over the past decade, the field of PTCs has witnessed significant expansion, garnering substantial interest from the research community. This growth is attributed to the following two primary factors:
\begin{itemize}
\item The distinct and intrinsically significant effects PTCs exhibit on light-matter interactions have been recognized. Analogous to how traditional photonic crystals offer remarkable spatial light concentration and suppress the spontaneous emission of quantum emitters within the crystal~\cite{bykov_spontaneous_1972a,bykov_spontaneous_1975b,yablonovitch_inhibited_1987a,john_strong_1987a}, PTCs, when operating in optical domains proximal to electronic transitions in solids, have the potential to intensify interaction with matter strongly. This has been evidenced by the ability of PTCs to amplify the spontaneous emission from excited atoms~\cite{lyubarov2022amplified}, support subluminal Cherenkov radiation~\cite{dikopoltsev2022light,gao2023free}, facilitate superluminal momentum-gap solitons~\cite{pan2023superluminal}, and host a temporal counterpart of Anderson localization~\cite{carminati_universal_2021,sharabi2021disordered,apffel2021time}, among other phenomena.

\item Current advancements in material science reveal great potential for fabricating the first PTCs that operate within the optical domain~\cite{saha2023photonic}. The realization of such PTCs necessitates temporal permittivity modulations at exceptionally high frequencies, typically twice that of the light probing the response, and with sufficiently large amplitudes~\cite{zurita2009reflection,lustig2018topological,hayran2022ℏomega}. Experimental findings suggest that transparent conductive oxides, specifically indium tin oxide (ITO) and aluminum-doped zinc oxide (AZO), could meet both criteria when operated in the epsilon-near-zero domain~\cite{alam2016large,bohn_spatiotemporal_2021,zhou2020broadband,caspani2016enhanced,lustig2023time, tirole2024second}. 
\end{itemize}

However, to reach the successful synthesis of PTCs from these materials, two primary challenges remain to be addressed: (i) the high pumping power requirements that could induce thermal degradation of the modulated material~\cite{hayran2022ℏomega} and (ii) the inadvertent excitation of a ``dynamic-grating" nonlinear effect, potentially hindering the observation of the PTC state~\cite{khurgin2021fast}. Notwithstanding these obstacles, ongoing research endeavors in this domain persist, and alternative promising methodologies for constructing optical PTCs continue to emerge~\cite{wang2023unleashing,narimanov_ultrafast_2023,dong2024non,zhang2024longitudinal}. 

% space-time crystals \cite{Complete optical isolation created by indirect interband photonic transitions, Unusual electromagnetic modes in space-time-modulated dispersion-engineered
% media, Uniform-velocity spacetime crystals, Spatiotemporal plane wave expansion method for
% arbitrary space–time periodi
% Early work on time modulation in materials. recently experimented with high photoexcitation in GaN; as a result of electron-hole recombination, the initial free-electron density decayed from 1019 to 3 10 18cm in a time about 800 ps.
% \cite{Luminescence decay in highly excited GaN grown by hydride vapor-phase epitaxy}

In this tutorial, we present the governing physics and potential applications of PTCs in an educational manner, accessible to readers without a solid background in this field. We start with the description of the eigenmodes in PTCs and the properties of the momentum bandgaps (Chapter~\ref{secEigen}). This is followed by considerations of practical aspects of realistic PTCs, such as effects of frequency dispersion, topology, finite spatial and temporal extent of the crystal, etc. (Chapter~\ref{secAspects}). Next, Chapter~\ref{secRelations} investigates the relations between PTCs and other related concepts, including parametric amplification in distributed systems, nonlinear optics, and quantum time crystals. Material candidates for implementation of PTCs are reviewed in Chapter~\ref{secMaterials}. Chapter~\ref{secApplications} describes the implications and potential applications of PTCs. Finally, in Chapter~\ref{secSpacetime}, we explore the additional opportunities in PTCs provided by introducing spatial modulation. We will finish this tutorial with an outlook and concluding remarks in the two last chapters. For the convenience of the reader, we attached as Supplementary Material two numerical codes for calculating the band structures of photonic time crystals
using the plane wave expansion method (Code~1~\cite{code1}) and the transfer matrix method (Code~2~\cite{code2}).

Although this is the first in-depth tutorial article focusing on PTCs, it is important to acknowledge other related reviews and perspectives. The theory and applications of general time-modulated materials and systems were reviewed in~\cite{shvartsburg2005optics,yuan2021synthetic,galiffi2022photonics,Hayran_using_2023,ptitcyn2023tutorial,ortega-gomez_tutorial_2023,yin_scattering_2023}. 
Future visions and challenges for PTCs were overviewed in perspectives~\cite{lustig2023photonic,won_it_2023,boltasseva2024photonic}. Discussion on potential material platforms for PTCs can be found in~\cite{saha2023photonic,lobet2023new}. Physics and applications of space-time metamaterials and metasurfaces were described in~\cite{shaltout_2019_spatiotemporal,caloz2019spacetime,Caloz2020spacetime2,taravati2022microwave,engheta_metamaterials_2021,engheta2023four,galiffi2022photonics}.

Finally, it is important to note that throughout this article,
time-harmonic oscillations in the form $e^{j \omega t}$ are assumed
according to the conventional electrical engineering
notation \cite{pozar_microwave_2012,cheng_field_1983}.
Moreover, all the \textit{field quantities} in the frequency domain are marked with the tilde symbol ``$\sim$'' on top to differentiate them from the same quantities in the time domain. For example, the electric fields in the two domains are denoted as $\mathbf{E}(t)$ and $\tilde{\mathbf{E}}(\omega)$. Material parameters, in contrast, are identified with their specific domain through explicit notation in the argument brackets. We strive to be consistent and do not wish to drop any of the arguments, as the arguments tell us something about the space in which they live. For example, $\varepsilon(t)$ describes a material that is non-dispersive but has time-varying properties. In contrast, $\varepsilon(\omega)$ would describe a stationary but dispersive material, etc.

\section{Eigenmodes in PTCs }\label{secEigen} 

% Make a small introduction to the topic of this section. Mention that werExplain the importance of this section. 
This chapter introduces the basic properties of PTCs from first principles. It is structured into two sections. In the first section, we derive the general solution of Maxwell's equations in a PTC (see the illustration of the geometry in Fig.~\ref{Fig:photonic_space_time}(c)). We stress upfront that two different approaches are chosen. One of them solves Maxwell's equations fully in the Fourier domain, while the other exploits a transfer matrix technique.
By discussing the details of the solutions, we overview the concepts of the momentum bandgap and other electromagnetic effects inside the momentum bandgap in the second section of this chapter.
The rigorous mathematical and physical frameworks developed here are pivotal in their own right and serve as prerequisites for comprehending the complex phenomena discussed in subsequent chapters of this tutorial.

%%%%%%%%%%%%%%%%%%%%%%%%%%%%%%%%%%%%%%%%%%%%%%%%%
%%%%%%%%%%%%%%%%%%%%%%%%%%%%%%%%%%%%%&&&&&&&&&&&&
%%%%%%%%%%%%%%%%%%%%%%%%%%%%%%%%%%%%%%%%%%%%%%%%% 

\subsection{Governing equations}
This section considers the basic concepts and concisely describes the time- and frequency-domain constitutive relations associated with a linear time-varying medium. Although we outline constitutive relations applicable for the general temporally nonlocal (dispersive) medium, we focus here, in Chapter~\ref{secEigen}, on the instantaneous (dispersionless) response when deriving the time-domain wave equation and plotting the band structure of an exemplary medium. We stress that this assumption of an instantaneous response in this chapter is only made to permit a more transparent discussion of the effects linked to the time modulation. The band structure analysis of time-varying media in the presence of frequency dispersion is explored in Chapter~
\ref{secAspects}, when more realistic aspects of PTCs are discussed.

In the following, we also outline two different methods for calculating the band structure of a periodically time-varying medium based on the Floquet theorem. The Floquet theorem is sometimes also referred to as the Bloch theorem. Both theorems reveal that in periodic systems, wave functions can be expressed as a product of a periodic function and a simpler, well-understood function (like a plane wave).
While the Bloch theorem is historically applied for spatially periodic systems (like electronic or photonic crystals), the Floquet theorem is typically applied for temporally periodic systems (like certain quantum or classical wave problems, including PTCs)~\cite{blakemore1985solid}. One of the methods for calculating the band structure of a periodically time-varying medium is based on the plane-wave expansion~\cite{zurita2009reflection}. The other method exploits a transfer matrix formalism, sometimes called the ABCD-matrix formalism~\cite{lustig2018topological}. The former method is suitable for PTCs with a continuous and smooth variation of the time-dependent permittivity, i.e., ideally in a sinusoidal fashion. In contrast, the latter method is convenient when a stepwise change in the permittivity is considered. The results should be independent of the chosen method, but the numerical convenience strongly differs depending on the modulation type. Also, traditional numerical techniques to solve Maxwell's equations can be used to obtain the band structure, of course \cite{ozer2020batio3}.

%%%%%%%%%%%%%%%%%%%%%%%%%%%%%%%%%%%%%%%%%%%%%%%%%%%%%%%%%%%%%%%%%%%%%%%%%%%%%
%%%%%%%%%%%%%%%%%%%%%%%%%%%%%%%%%%%%%%%%%%%%%%%%%%%%%%%%%%%%%%%%%%%%%%%%%%%%%
%%%%%%%%%%%%%%%%%%%%%%%%%%%%%%%%%%%%%%%%%%%%%%%%%%%%%%%%%%%%%%%%%%%%%%%%%%%%% 

%(\red{Sajjad, could you add reference \cite{PhysRevA.108.043504} somewhere in this section? this paper discusses the wavenumber-domain analysis for the response function, instead of time domain)}

\subsubsection{Description of a linear, isotropic, spatially homogeneous, time-invariant, and dispersive medium} 
To set the stage, we remind the reader of the constitutive relation for most natural materials. These constitutive relations must supplement the Maxwell equations to make them solvable.

In nature, we encounter a temporal nonlocal response as an inherent characteristic of any material that stems from inertia. Specifically, it implies that a response, such as the induced polarization density $\_P(\_r,t)$ (here $\_r$ is the spatial dependence), always experiences a delay to an action, such as the applied electric field $\_E(\_r,t)$ or magnetic field $\_H(\_r,t)$. Combined with causality, we assert that the response depends on the action only in the past, and the delay time between the response and the action is always positive. Moreover, for a material whose properties do not depend on time, i.e., it is a time-invariant medium, the absolute times do not matter but only the time difference between action and response. In electromagnetic theory, this fundamental principle defines the conventional constitutive relation. In the time domain, this constitutive relation must be written as a convolution integral, in which the polarization density is found as the convolution of the impulse response function $R(\_r,t^{\prime})$ with the electric field. For simplicity, we assume an isotropic medium. Anisotropic materials will matter in Section~\ref{Anisotropicmedium}. Under all these assumptions, the polarization is expressed as 
\begin{equation}
\_P(\_r,t)=\varepsilon_0\int_{0}^{+\infty}R(\_r,t^{\prime})\_E(\_r,t-t^{\prime})\mathrm{d}t'\,.
\l{equ1}
\end{equation} 
Here, $t^{\prime}$ is the delay time between the action and the response, i.e., the electric field and the polarization density. The second variable $t$ is the observation time. The response function expresses the induced polarization density for a punctual excitation in time. It is the response of the medium, and the total polarization is just the sum, or in a continuous manner, the integral, of the response due to all the excitations from the past.  We assumed that the medium is not bianisotropic. A bianisotropy would have caused an electric response in the polarization density from the magnetic field.

Indeed, the above convolution integral properly models a causal and nonlocal response in time (see, e.g., Refs.~\cite{mirmoosa2022dipole}~and~\cite[p.~330]{jackson_classical_1999}). It should be noted that throughout the paper, we assume a spatially local material response. It implies that the induced polarization only depends on the electric field at the same spatial location. However, the theory could be further generalized to spatially nonlocal materials. For a spatially homogeneous medium, as assumed here for the moment, there is also no explicit space dependency of the response function, i.e., $R(\_r, t^{\prime})=R(t^{\prime})$.

Expressing constitutive relations in the time domain on the base of such a convolution is to some extent inconvenient. Therefore, we go to a reciprocal space for linear systems by Fourier transforming all involved quantities. The electric field, for example, can be written as 
\begin{equation}
\_E(\_r,t)={1\over2\pi}\int_{-\infty}^{+\infty}\tilde{\_E}(\_r,\omega)\exp(j\omega t)\mathrm{d}\omega\,.
\l{foru}
\end{equation}
Then, we can write the constitutive relation as a product in the frequency domain, which reads as 
\begin{equation}
\tilde{\_P}(\_r,\omega)=\varepsilon_0\chi(\omega)\tilde{\_E}(\_r,\omega)\,,   
\end{equation} 
where the susceptibility $\chi(\omega)$ is the Fourier transform of the response function $R(t^{\prime})$. Again, for an isotropic medium, this is a scalar function. For an anisotropic medium, it would be a tensor. 

It remains to be mentioned that the electric flux density, also called the electric displacement field, in the time domain is the sum of the electric field multiplied by the vacuum permittivity and the polarization density, 
\begin{equation}
\_D(\_r,t)=\varepsilon_0\_E(\_r,t)+\_P(\_r,t)\,.
\end{equation} 
In the frequency domain, it reads as
\begin{equation}
\tilde{\_D}(\_r,\omega)=\varepsilon_0\tilde{\_E}(\_r,\omega)+\tilde{\_P}(\_r,\omega)\,.
\end{equation} 

\begin{figure*}[t!]
\centerline{\includegraphics[width= 0.4\columnwidth]{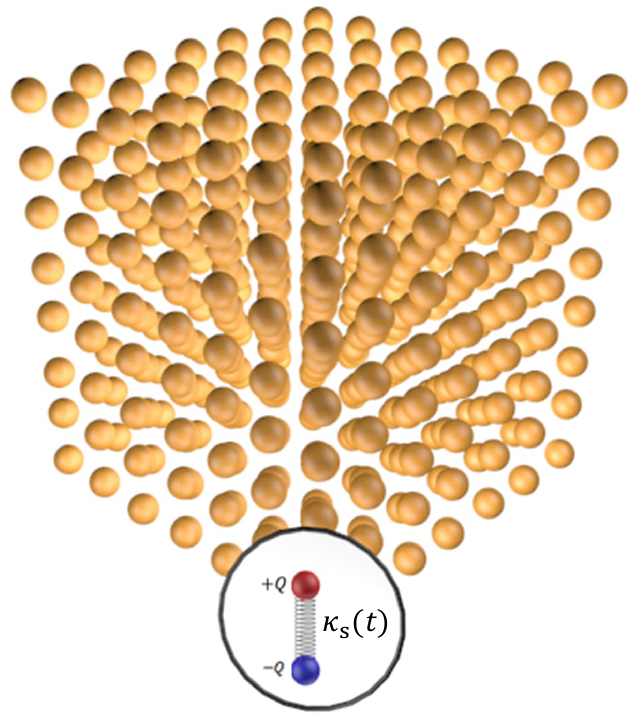}}
\caption{A conceptional representation of a linear time-varying artificial medium. Here, we consider the medium to be made from periodically arranged meta-atoms that change, in this specific example, their properties in time. For that, the meta-atoms are described as harmonic oscillators with a time-varying spring constant $\kappa_{\rm s}(t)$.  
}
\label{Fig:3.1GEWEABCD}
\end{figure*}

So far, so conventional. An intriguing question arises when considering a linear medium that is not temporally invariant. For example, the number of atoms per unit volume, the damping coefficient, or the resonance frequency of atoms' response varies over time (see, e.g.,~Fig.~\ref{Fig:3.1GEWEABCD}). In this case, the invariance under time translation is broken, and the medium becomes nonstationary (in other words, if the cause shifts in time, the macroscopic response does not shift by the same amount of time). Foundational questions arise: How does the constitutive relation change in the frequency domain? And how does this change affect the dispersion curves for plane wave solutions of Maxwell's equations? In the following, we are going to answer these questions. 

%%%%%%%%%%%%%%%%%%%%%%%%%%%%%%%%%%%%%%%%%%%%%%%%%%%%%%%%%%%%%%%%%%
%%%%%%%%%%%%%%%%%%%%%%%%%%%%%%%%%%%%%%%%%%%%%%%%%%%%%%%%%%%%%%%%%%
%%%%%%%%%%%%%%%%%%%%%%%%%%%%%%%%%%%%%%%%%%%%%%%%%%%%%%%%%%%%%%%%%%

\subsubsection{Extension to a time-varying medium}

For a homogeneous, time-varying dielectric medium that is linear and causal, we write the general nonlocal constitutive relation between the electric flux density $\_D(\_r,t)$ and the electric field $\_E(\_r,t)$ as
\begin{equation}
\_D(\_r,t)=\varepsilon_0\_E(\_r,t)+\varepsilon_0\int_0^{+\infty}R(t^{\prime},t)\_E(\_r,t-t^{\prime})
\mathrm{d}t^{\prime}\,.
\label{eq:DEX}
\end{equation} 
% \begin{equation}
% \_D(\_r,t)=\varepsilon_0\_E(\_r,t)+\varepsilon_0\int_0^{+\infty}\bigg[\int_V\hat\chi(\_r',\_r,\gamma,t)\_E(\_r-\_r',t-\gamma)d\_r'\bigg]d\gamma,
% \label{eq:tnsndd}
% \end{equation} 
Compared to Eq.~\r{equ1}, here, the response function depends on the two time-variables $t^{\prime}$ and $t$. The former expresses the memory of the material, and the latter expresses the explicit time dependency of the material properties.  

% Regarding the spatial characteristic, the above equation means that 1) the medium is spatially inhomogeneous and 2) the electric flux density at each point in space depends on the electric field at other locations. However, for avoiding difficulty, let us assume that the response is local in space, i.e.,~neglect spatial dispersion, and assume that $\hat\chi(\_r',\_r,\gamma,t)=\delta^3(\_r')\hat\chi(\_r,\gamma,t)$, in which $\delta^3(\_r)$ is the three-dimensional Dirac delta function. Under this assumption,  Eq.~\eqref{eq:tnsndd} is simplified as 

In Eq.~\eqref{eq:DEX}, the electric field at each point in space can be an arbitrary function of time. Therefore, we can write the electric field again as the inverse Fourier transform of $\tilde{\_E}(\_r,\omega)$, and we replace the electric field in the constitutive relation (see Eq.~\eqref{eq:DEX}) by the above expression. Consequently, we achieve an important expression for the instantaneous electric flux density, which reads as
\begin{equation}
\_D(\_r,t)={\varepsilon_0\over2\pi}\int_{-\infty}^{+\infty}\varepsilon_{\rm{T}}(\omega',t)\tilde{\_E}(\_r,\omega')\exp(j\omega' t)\mathrm{d}\omega'\,.
\label{eq:DtepEw}
\end{equation}
Here, we introduced the temporal complex dielectric function $\varepsilon_{\rm{T}}(\omega',t)=1+\chi_{\rm{T}}(\omega',t)$, and the temporal complex susceptibility is defined as 
\begin{equation}
\chi_{\rm{T}}(\omega',t)=\int_0^{+\infty}R(t^{\prime},t)\exp(-j\omega' t^{\prime})\mathrm{d} t^{\prime} \,.
\end{equation} 
The subscript 'T' shall remind the reader that this quantity lives partially in time and partially in the frequency domain. It can be seen that the complex susceptibility $\chi_{\rm{T}}(\omega',t)$ is merely the Fourier transform of the response function $R(t^{\prime},t)$ with respect to the delay time $t^{\prime}$. These definitions and expressions (discussed in \cite{mirmoosa2022dipole,ptitcyn2023tutorial} and recently reviewed in \cite{koutserimpas2024time}) closely mirror those employed in the study of linear variable networks, originally introduced in 1950~\cite{1701220}.  

According to Eq.~\eqref{eq:DtepEw}, we explicitly observe that by taking the second Fourier transform, we fully leave the time domain and can investigate light-matter interactions utterly in the frequency domain. However, this second Fourier transform should be done concerning the time variable $t$. Therefore, we must apply a different angular frequency notation rather than $\omega'$. Here, we use the letter $\omega$ (that we employed before for the time-invariant case) for this purpose. Knowing that $\exp(j\omega' t)$ gives rise to a shift in the frequency domain, we conclude that~\cite{mirmoosa2022dipole} 
\begin{equation}
\tilde{\_D}(\_r,\omega)={\varepsilon_0\over2\pi}\int_{-\infty}^{+\infty}\varepsilon(\omega',\omega-\omega')\tilde{\_E}(\_r,\omega')\mathrm{d}\omega'\,.
\label{eq:FDPEREL}
\end{equation}  
Consequently, in this equation, there are two angular frequencies. One of them ($\omega'$) refers to the temporal nonlocality or dispersion, and the presence of the other one ($\omega$) is due to the time-variance of the material and Fourier transform concerning the observation time.   

%%%%%%%%%%%%%%%%%%%%%%%%%%%%%%%%%%%%%%%%%%%%%%%%%%%%%%%%%%%%%%%%%%%%%%
%%%%%%%%%%%%%%%%%%%%%%%%%%%%%%%%%%%%%%%%%%%%%%%%%%%%%%%%%%%%%%%%%%%%%%
%%%%%%%%%%%%%%%%%%%%%%%%%%%%%%%%%%%%%%%%%%%%%%%%%%%%%%%%%%%%%%%%%%%%%%

\subsubsection{Assumption of an instantaneous response}
To simplify the discussion in this chapter, we assume that the response of the material is instantaneous. In that case, we say that the medium is dispersionless, i.e., it shows the same response at every frequency.

A medium has an instantaneous (dispersionless) response if the induced electric flux density depends only on the electric field at the observation time $t$. From this point of view, there is no temporal nonlocality, and the response function must be represented by a Dirac delta distribution, allowing a temporal local response. For that, the argument of the Dirac delta distribution should be only the time variable $t^{\prime}$. However, since the system is time-varying, the response function continues to include another term depending on the time variable $t$. Accordingly, we eventually write the response function as ${R}(t^{\prime},t)=\delta(t^{\prime})\chi(t)$. By having this expression and applying Eq.~\eqref{eq:DEX}, in the time domain, we see that $\_D(\_r,t)=\varepsilon_0\varepsilon(t)\_E(\_r,t)$, where $\varepsilon(t)=1+\chi(t)$. 
% Here, $\varepsilon(t)$ is an arbitrary function, and there is no restriction for its behavior in time. 

Since in a PTC, $\varepsilon(t)$ is a periodic function with a temporal period $T_{\rm m}$ (see Fig.~\ref{Fig:photonic_space_time}(c)), we notice that $\varepsilon(t+T_{\rm{m}})=\varepsilon(t)$. Due to this property, we expand $\varepsilon(t)$ into the Fourier series: 
\begin{equation}
\varepsilon(t)=\sum_p \epsilon_p e^{jp\omega_{\rm{m}} t}\,,  
\label{eq: eps}
\end{equation}
in which $\epsilon_p$ are the Fourier coefficients, and $\omega_{\rm{m}}=2\pi/T_{\rm{m}}$ denotes the angular modulation frequency. 

In the end, to briefly mention another feature of the instantaneous response for the interested reader, we refer to the modification of Eq.~\eqref{eq:FDPEREL}. In fact, due to the presence of the Dirac delta distribution that we explained above, the relative permittivity $\varepsilon(\omega',\omega)$ does not include the independent variable $\omega'$, and it is only a function of $\omega$. In this scenario, Eq.~\eqref{eq:FDPEREL} becomes a convolution integral, and the electric flux density $\tilde{\_D}(\_r,
\omega)$ is proportional to the convolution of the relative permittivity $\varepsilon(\omega-\omega')$ and the electric field $\tilde{\_E}(\_r,\omega')$ in the frequency domain.     

%%%%%%%%%%%%%%%%%%%%%%%%%%%%%%%%%%%%%%%%%%%%%%%%%%%%%%%%%%%%%%%%%%%%%%%%
%%%%%%%%%%%%%%%%%%%%%%%%%%%%%%%%%%%%%%%%%%%%%%%%%%%%%%%%%%%%%%%%%%%%%%%%
%%%%%%%%%%%%%%%%%%%%%%%%%%%%%%%%%%%%%%%%%%%%%%%%%%%%%%%%%%%%%%%%%%%%%%%%

\subsubsection{Band structure analysis based on the plane wave expansion} \label{sec: planewave expansion}
To derive the band structure of periodically time-varying media with an instantaneous response, we use in this subsection the plane wave expansion\cite{zurita2009reflection}. According to the Floquet theorem~\cite[Sec.~4.2]{ptitcyn2023tutorial}, the electric field can be written as a product of the exponential function $e^{j\omega_{\rm F} t}$ and a function periodic in time, where the periodicity is the same as that of the time-varying material properties. Here, $\omega_{\rm F}$ is called the Floquet angular frequency. Using the Fourier series, we can express the electric field for an eigenwave propagating along the $+z$ direction inside a PTC as  
\begin{equation}
\_E(\_r, t)=\sum_n E_n (\omegaf) e^{j\omega_nt}e^{-jk z}\_a_x\,,   
\label{eq:BLOCHT}
\end{equation}
in which $\omega_n=\omega_{\rm F}+n\omega_{\rm m}$ is the $n$-th frequency harmonic, $E_n(\omegaf)$ is the amplitude of the frequency harmonic $\omega_n$ at the Floquet angular frequency $\omegaf$, and $k$ represents the phase constant (wavenumber). Notice that the phase constant is fixed for all these harmonics and equals $k$. That is because the material modulation occurs in time, while the susceptibility and permittivity are uniform in space. Hence, the phase constant is a conserved quantity. The polarization of the field was arbitrarily assumed to be in the $x$-direction. Generally, in the spatially homogeneous medium, the eigenmodes are elliptically polarized. Moreover, for convenience, we have chosen here a complex notation. However, it should be explicitly said that the experimentally observable field corresponds only to the real part of that quantity. 

In addition to the Floquet theorem, we also need to infer the wave equation, and, certainly, we need to start from Maxwell's equations in the absence of sources, which are written for our study as 
\begin{subequations}
\begin{equation}
\nabla \times \mathbf{E}(\_r, t)=-\mu_0\frac{\partial \mathbf{H}(\_r, t)}{\partial t}\,,\quad\nabla \cdot \mathbf{E}(\_r, t)=0\,,\label{eq: max1}
\end{equation}
\begin{equation}
\nabla \times \mathbf{H}(\_r, t)=\varepsilon_0\frac{\partial}{\partial t}\Big[\varepsilon(t)\mathbf{E}(\_r, t)\Big]\,,\quad\nabla \cdot \mathbf{H}(\_r, t)=0\,.\label{eq: max2}
\end{equation}
\end{subequations}
Regarding Eq.~\eqref{eq: max1}, we apply the curl operator to both sides of the equation, and, subsequently, we use the information of Eq.~\eqref{eq: max2} about the curl of the magnetic field as well as the information about the divergences of the fields that are zero. 
After traditional algebraic manipulations, we achieve the wave equation  
\begin{equation}
\nabla^2\_E(\_r, t)-\frac{1}{c^2}\frac{\partial^2}{\partial t^2}\Big(\varepsilon(t)\mathbf{E}(\_r, t)\Big)=0\,. 
\label{Eq: governing}
\end{equation}
Here, $c=1/\sqrt{ \varepsilon_0 \mu_0}$ represents the vacuum speed of light. 
% This partial differential equation is the wave equation of such a time-varying system.
Substituting Eqs.~\eqref{eq: eps} and \eqref{eq:BLOCHT} into this equation results in the master equation for finding the band structure of the PTC. In Eq.~\eqref{Eq: governing}, the Laplacian operator is easily replaced by the square of the phase constant because this operator is about space variations (i.e.,~$\nabla\rightarrow-jk$ and $\nabla^2\rightarrow-k^2$). The master equation is derived as follows:
\begin{equation}
\sum_n \sum_p \frac{[\omega_{\rm F}+(n+p)\omega_{\rm m}]^2}{c^2}\epsilon_p E_n e^{j[\omega_{\rm F}+(n+p)\omega_{\rm m}]t}=k^2\sum_n E_ne^{j\omega_n t}\,. \label{eq: masters equation}
\end{equation}
On the left-hand side of Eq.~(\ref{eq: masters equation}), we shift the index $n$  to $n-p$. Consequently, Eq.~(\ref{eq: masters equation}) can be simplified to
\begin{equation}
    \sum_{n} \sum_p \frac{(\omega_{\rm F}+n\omega_{\rm m})^2}{c^2}\epsilon_p E_{n-p} e^{j\omega_{n}t}=k^2\sum_n E_ne^{j\omega_n t}\,. \label{eq: simply masters equation}
\end{equation}
We can see from Eq.~(\ref{eq: simply masters equation}) that both sides share the same basis $e^{j\omega_n t}$. Therefore, the summation in terms of $n$ can be removed. By shifting now the index $p$ to $n-p$, we infer that
% Therefore, by introducing the Kronecker delta function $\delta_{nm}$ as the result of having two indices of summation $n$ and $m$, we derive that 
\begin{equation}
\sum_p{(\omega_{\rm F}+n\omega_{\rm{m}})^2\over c^2}\epsilon_{n-p}E_{p}-k^2E_n\delta_{pn}=0, 
\label{eq:summatrixBS}
\end{equation} 
where $\delta_{pn}$ is the Kronecker delta function, which is valid for each index $n$~\cite{zurita2009reflection}. Equation~\eqref{eq:summatrixBS} could also be written in a matrix form where a square matrix of infinite size $\={U}$ is multiplied by a vector $\overline{V}$ describing the field amplitudes of harmonics for a given Floquet frequency $\omega_{\rm F}$. 
% It is worth mentioning that due to the conservation of energy and finite strength of modulation, we are allowed to ponder a square matrix consisting of only $2N+1$ rows and columns. 
 We can write Eq.~\eqref{eq:summatrixBS} in the following form: 
\begin{equation}
\underbrace{
\begin{bmatrix}
\ddots & \ddots & \ddots & \ddots & \ddots \\
\ddots & \zeta_{-1}\epsilon_0-k^2 & \zeta_{-1}\epsilon_{-1} & \zeta_{-1}\epsilon_{-2} & \ddots \\
\ddots & \zeta_{0}\epsilon_{1} & \zeta_{0}\epsilon_0-k^2 & \zeta_{0}\epsilon_{-1} & \ddots \\
\ddots & \zeta_{1}\epsilon_2 & \zeta_{1}\epsilon_1 & \zeta_{1}\epsilon_0-k^2 & \ddots \\
\ddots & \ddots & \ddots & \ddots & \ddots 
\end{bmatrix}
}_{\={U}(\omega_{\rm F}, k)} \cdot
\underbrace{\begin{bmatrix}
\vdots \\ E_{-1} \\ E_0 \\ E_{1}\\ \vdots
\end{bmatrix}}_{\overline{V}}=0\,, 
\label{eq:MATRIXFVEC}
\end{equation} 
in which $\zeta_n (\omega_{\rm F})=(\omega_{\rm F}+n\omega_{\rm{m}})^2/c^2$.

%\begin{equation}
%\underbrace{\begin{bmatrix}
%\ddots & \ddots & \ddots & & & \\
% & \zeta_{-1}\epsilon_1 & \zeta_{-1}\epsilon_0-k^2 & \zeta_{-1}\epsilon_{-1} & & \\
% & &  \zeta_{0}\epsilon_1 & \zeta_{0}\epsilon_0-k^2 & \zeta_{0}\epsilon_{-1}& \\
%& & & \zeta_{1}\epsilon_1 & \zeta_{1}\epsilon_0-k^2 & \zeta_{1}\epsilon_{-1} & \\
%& & & & \ddots & \ddots & \ddots \\
%\end{bmatrix}}_{\={U}} \cdot
%\underbrace{\begin{bmatrix}
%\vdots \\ E_{-1} \\ E_0 \\ E_{1}\\ \vdots
%\end{bmatrix}}_{\overline{V}}=0, 
%\label{eq:MATRIXFVEC}
%\end{equation}

We observe that the square of the phase constant $k^2$ is accompanied by the Fourier coefficient $\epsilon_0$ (not to be confused with vacuum permittivity $\varepsilon_0$), and both appear only in the diagonal elements of the matrix. 
In most PTCs, amplitudes of high-order harmonics decay rapidly with the harmonic number. Therefore, it is possible to truncate the square matrix to that of a finite size, i.e.,  $(2N+1) \times (2N+1)$, where $N$ means considering the harmonics from $-N\leq n \leq +N$. 

Equation~(\ref{eq:MATRIXFVEC}) defines an eigenvalue problem and has a nontrivial solution
if the determinant of the square matrix vanishes, i.e., $\det\left[\={U}(\omega_{\rm F}, k)\right]=0$. 
% Equation~\eqref{eq:MATRIXFVEC} indicates that for having a nonzero eigenvector, the determinant of the square matrix must be zero. 
% On the other hand, as also shown implicitly by Eq.~\eqref{eq:MATRIXFVEC}, this determinant is dependent on the Floquet angular frequency $\omega$ and the phase constant $\beta$. 
It allows us to calculate the band structure, demonstrating the relation between $\omega_{\rm F}$ and $k$.
From Eq.~(\ref{eq:MATRIXFVEC}), one can observe an important property of the band structure, its periodicity. Indeed, one can replace all $ \zeta_{n} (\omega_{\rm F})$ in the determinant of the matrix with equivalent terms $\zeta_{n-1} (\omega_{\rm F}+ \omega_{\rm m})$. The resulting determinant will be the same as the initial one (due to the infinite size of the matrix) with the only change that all parameters depend on $\omega_{\rm F} +\omega_{\rm m}$. Thus, the eigenvalue solution of the new determinant $\det\left[\={U}(\omega_{\rm F}+\omega_{\rm m},k)\right]$  must be equal to that of the initial determinant  $\det\left[\={U}(\omega_{\rm F},k)\right]$, which leads to:
\begin{equation}
k(\omega_{\rm F} +\omega_{\rm m}) = k(\omega_{\rm F})\,.
     \label{Eq:dets}
\end{equation}
Relation~(\ref{Eq:dets}) brings us to the important implication: the
band structure of a PTC is periodic in $\omega_{\rm F}$ with the period being the modulation frequency $\omega_{\rm m}$. Moreover, from Eq.~(\ref{eq:MATRIXFVEC}) one can find that eigenmode solutions for two different eigenfrequencies $\omegaf$ and $\omegaf+q\omega_{\rm m}$ ($q \in \mathds{Z}$) are related to each another through the relation~\cite{zurita2009reflection}
\begin{equation}
E_n(\omegaf+ q\omega_{\rm m}) = E_{n-q}(\omegaf)\,.
     \l{pereigen}
\end{equation}
Therefore, without loss of information, it is sufficient to consider only the first period of the band structure $\omega_{\rm F} \in [-\omega_{\rm m}/2, \omega_{\rm m}/2]$, known as the first Brillouin zone of the PTC. 

\begin{figure}[t]
    \centering
    \includegraphics[width=0.9\columnwidth]{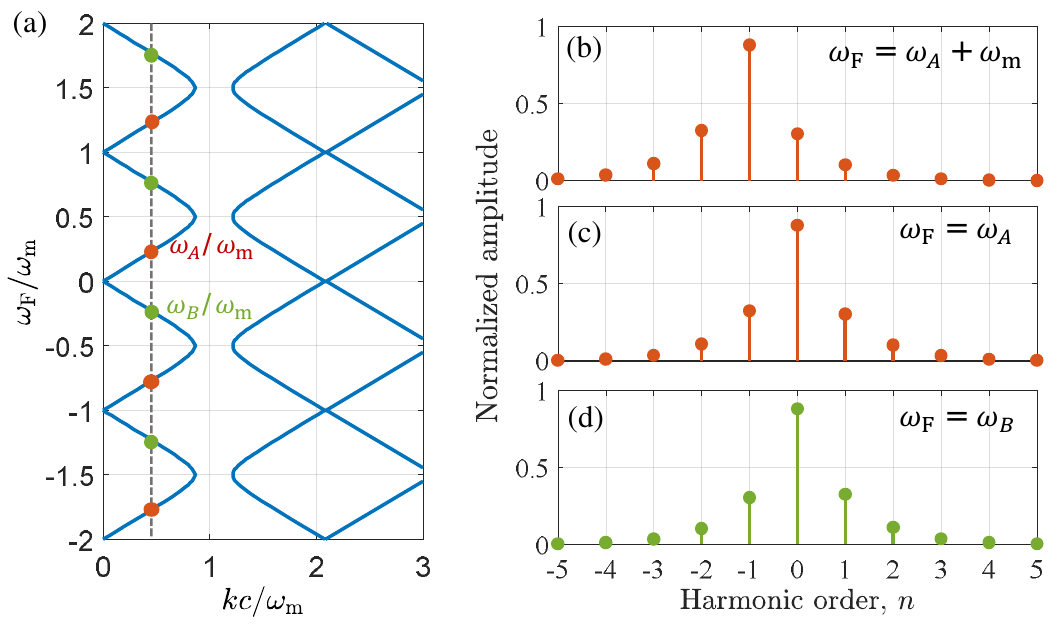}
    \caption{ (a) Band structure of the PTC calculated using Eq.~(\ref{eq:MATRIXFVEC}). The permittivity of the PTC is given by $\varepsilon(t) = \varepsilon_{\mathrm{av}}(1+m_\varepsilon\cos(\omega_{\rm m}t))$, where $\varepsilon_{\mathrm{av}} = 5$ and $m_\varepsilon= 0.6$. Distribution of the normalized harmonic magnitudes for different eigenmodes in the PTC. The eigenfrequencies of these modes are: (b) $\omega_{\rm F}=\omega_{A}+\omega_{\rm m}$,  (c)  $\omega_{\rm F}=\omega_{A}$, and (d)  $\omega_{\rm F}=\omega_{B}$. Note that in (b), (c), and (d), the harmonic distribution is asymmetric with respect to the fundamental harmonic. This asymmetry is caused by the asymmetry of the matrix $\overline{\overline{U}}$ in Eq.~(\ref{eq:MATRIXFVEC}) since $\omega_{\rm F}\neq \omega_{\rm m}/2$ ($k$ is outside the momentum bandgap).
    }
    \label{fig:bandstructure_harmonic}
\end{figure}

As the simplest but illustrative example, consider a sinusoidal permittivity modulation defined as $\varepsilon(t) = \varepsilon_{\mathrm{av}}(1+m_\varepsilon \cos(\omega_{\rm m}t))$, where $\varepsilon_{\mathrm{av}}$ denotes the time-averaged permittivity, and $m_\varepsilon$ is the relative amplitude of the permittivity modulation ranging from 0 to 1. The band structure, depicted in Fig.~\ref{fig:bandstructure_harmonic}(a), is computed using Eq.~(\ref{eq:MATRIXFVEC}). The exact numerical parameters are indicated in the figure legend, and we plot the band structure for a frequency interval $\omega_{\rm F}\in [-2\omega_{\rm m}, 2\omega_{\rm m}]$. From Fig.~\ref{fig:bandstructure_harmonic}(a), it is evident that the band structure is periodic in the frequency domain.

The most important feature of PTCs is the bandgap that appears in the momentum domain, as shown in Fig.~\ref{fig:bandstructure_harmonic}(a). What we see here is the hallmark of a PTC. We see that for a certain momentum interval, Maxwell's equations do not have a propagating solution at any possible real frequency.
We will elaborate on that momentum bandgap in much more detail below. Afterward, we discuss the eigenmodes and return to Figs.~\ref{fig:bandstructure_harmonic}(b)-(d).

\begin{figure}[t]
\centerline{\includegraphics[width= 0.7\columnwidth]{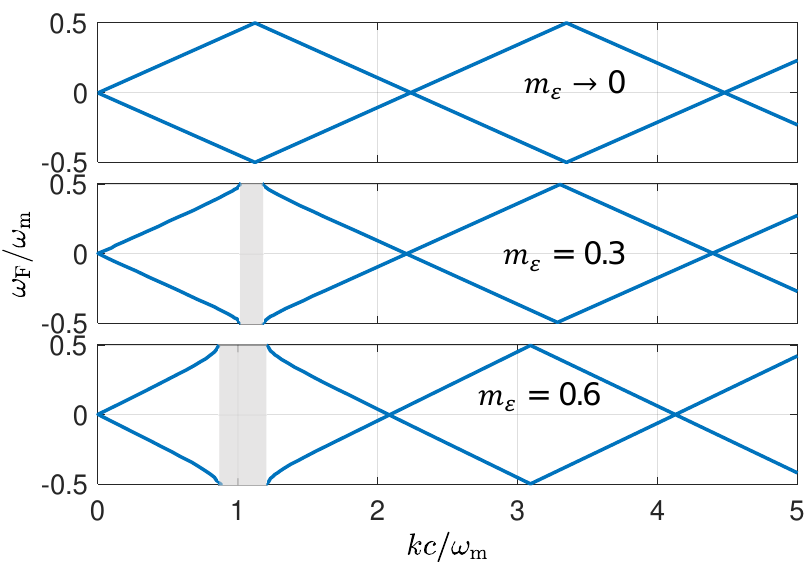}}
\caption{Band structure of a PTC calculated for three different values of the modulation depth. The permittivity of the PTC is given by $\varepsilon(t)=\varepsilon_{\mathrm{av}}(1+m_\varepsilon\cos(\omega_{\mathrm{m}}t))$ where $\varepsilon_{\rm av}=5$. The shaded regions depict the momentum domain in which a bandgap exists. }
\label{Fig:bandstructrue_3cases}
\end{figure}

The size of the aforementioned momentum bandgap is affected by the modulation depth $m_\varepsilon$. Figure~\ref{Fig:bandstructrue_3cases} shows the band structure in the first Brillouin zone for three different modulation amplitudes of the same PTC as discussed before. The plots were created using Code~1 attached as Supplementary Material~\cite{code1}.
% This figure presents the band structure within the first  Brillouin Zone $-0.5<\omega/\omega_{\mathrm{m}}<0.5$, and it's important to note that the actual band structure repeats periodically in the frequency domain, spaced by $\omega_{\mathrm{m}}$.}
Within the so-called empty-lattice approximation, i.e., when $m_\varepsilon \rightarrow 0$ (upper plot in the figure), the dispersion relation of the PTC can be obtained from the corresponding dispersion relation of the same material without temporal modulations upon ``folding'' the dispersion relation into the first Brillouin zone. This folding of the bands occurs due to the periodicity of the band structure given by  Eq.~(\ref{Eq:dets}). Once a band reaches one zone edge (e.g., at $\omega_{\rm m}/2$), through the translation by $-\omega_{\rm m}$,  it emerges and continues from the opposite edge $-\omega_{\rm m}/2$. 
Therefore, due to the periodicity, different branches of the dispersion relation cross at the edge of the Brillouin zone, i.e., when $\omega= \pm\omega_\mathrm{m}/2$ (see upper plot in Fig.~\ref{Fig:bandstructrue_3cases}). 
In the same plot, in the region of small values of wavenumber $k$, one can see a linear dispersion emerging from $\omega_{\rm F}=0$. This is the effective-medium regime, where the modulation frequency of the material properties is much higher than the frequency of the considered light. The light would effectively see a homogeneous medium, and the slope of the dispersion relation reflects the effective properties. In this specific case, the permittivity of the effective medium corresponds to the average permittivity relative to which a modulation occurs. 
% However, because of the time modulation, the dispersion relation is periodic in the reciprocal space ($\omega$ domain) and displaced by a multiple of the modulation frequencies. The periodicity in reciprocal space explains the different branches. 

When the modulation amplitude $m_\varepsilon$ is noticeable (see the two bottom plots in Fig.~\ref{Fig:bandstructrue_3cases}), an avoided crossing occurs at the Brillouin zone edge, and a bandgap in the momentum domain opens (shown with grey shaded region). The curvature of the bands near the bandgap is not linear. In this region of $k$, the light starts to probe the temporal modulation of the material properties of the PTC. 
Such a momentum bandgap is the hallmark of PTCs. With the increase in the modulation amplitude, the bandgap size increases. 
Interestingly, for a sinusoidal modulation with only one frequency, only one momentum bandgap is formed, that is, between the first and second bands. If higher order harmonics are added to the modulation function $\varepsilon(t)$, more bandgaps will be open at higher momenta. In particular, when the modulation function is stepwise, i.e., it has an infinite number of non-zero Fourier coefficients in its spectrum given by Eq.~(\ref{eq: eps}), infinitely many momentum bandgaps open, as discussed in the next subsection. 
% Inside the momentum bandgap, the eigenfrequency becomes complex values. 
% The real part of the eigenfrequency corresponds to the edge of the Brillouin zone ($\omega=\omega_{\rm m}/2$), and the imaginary part is non-zero. 
The wave phenomena occurring inside the bandgap will be explained in Section~\ref{momband}.

Looking at Fig.~\ref{Fig:bandstructrue_3cases}, one can observe that the band structure is mirror-symmetric with respect to the $\omega_{\rm F}=0$ axis. The reason can be directly observed from Eq.~(\ref{eq:MATRIXFVEC}). If $\omega_{\rm F}$ is an eigenfrequency, i.e., $\det\left[\={U}(\omega_{\rm F}, k)\right]=0$, then $\det\left[\={U}(-\omega_{\rm F}, k)\right]=0$ must be satisfied, since $\zeta_n(\omega_{\rm F}) = \zeta_{-n}(-\omega_{\rm F})$.  Therefore, we have $k(\omega_{\rm F}) = k(-\omega_{\rm F})$. It should be mentioned, however, that this degeneracy does not hold in a general case of PTC, which could consist of, e.g., magneto-optical materials and possess an asymmetric band structure. For example, the emergence of asymmetric band structures in spatial photonic crystals is a well-explored topic~\cite{figotin2001nonreciprocal,yu2007one}.
% \begin{equation}
%     k(\omega) = k(-\omega). \label{eq:folding}
% \end{equation}
% Eq.~(\ref{eq:folding}) indicates that the band structure is symmetric with respect to $\omega = 0$, which implies that the band structure is folded with respect to the $k$-axis.

Next, we discuss the eigenmodes of Eq.~(\ref{eq:MATRIXFVEC}). For a \textit{given} $k$ (see the dashed line in Fig.~\ref{fig:bandstructure_harmonic}(a)), there are infinitely many eigenfrequencies that could be grouped into two sets: $\omega_{A, n} = \omega_A + n \omega_{\rm m}$ and $\omega_{B, n}  = \omega_B  + n \omega_{\rm m}$,
 where $n \in \mathds{Z}$. 
% two eigenmodes in the considered PTC each of which consists of an infinite array of frequency harmonics spaced by $\omega_{\rm m}$, see Eq.~(\ref{eq:BLOCHT}). 
These two arrays of harmonics are marked by red and green dots in Fig.~\ref{fig:bandstructure_harmonic}(a), respectively. 
For each eigenfrequency belonging to $\omega_{A, n}$, one can derive from Eq.~(\ref{eq:MATRIXFVEC}) a formally different eigenmode characterized by complex amplitudes of each harmonic $E_{n}(\omegaf)$ (see Eq.~(\ref{eq:BLOCHT})). However, as it is clear from Eq.~\r{pereigen}, the spectral weights of different frequency harmonics $E_n(\omegaf)$ remain the same for all these eigenmodes with the only change in relabeling the harmonic numbers~\cite[p.~47]{joannopoulos_photonic_2008}. This is depicted in Figs.~\ref{fig:bandstructure_harmonic}(b) and (c), where the amplitudes of the different harmonics contributing to the eigenmodes are compared for two chosen eigenfrequencies $\omega_A$ and $\omega_A+\omega_{\rm m}$. Indeed, the harmonic spectra for $\omega_A+\omega_{\rm m}$ are the same as for $\omega_A$ if shifted by one order to the left. 
In Fig.~\ref{fig:bandstructure_harmonic}(c), the harmonic with order $n=0$ corresponds to the Floquet frequency $\omega = \omega_A$ in the PTC. 
The corresponding harmonic amplitude distribution $E_n$ for the eigenfrequencies $\omega_{B,n}$ is depicted in Fig.~\ref{fig:bandstructure_harmonic}(d). In the considered example of the PTC, it is equivalent to that shown in Fig.~\ref{fig:bandstructure_harmonic}(c) under the sign flip of the harmonic number $n$. However, in the general case, they could be different. 
Thus, for a given wavenumber $k$, the electric field inside a PTC can be viewed as a superposition of eigenmodes~(\ref{eq:BLOCHT}) with known amplitudes $E_n(\omegaf)$. This solution can be used in problems where the wavenumber $k$ in the PTC is fixed to a single value, e.g., the problem of time interface between free space and PTC (see Section~\ref{tfptcs} for more details). However, in the general case, this solution is not complete, as discussed below.

From Fig.~\ref{Fig:bandstructrue_3cases}, one can observe that for a given Floquet frequency $\omegaf$, there is a discrete but infinite number of solutions for wavenumber $k$. We denote them by $k_p$, where $p \in \mathds{Z}$. Therefore, $k_p$ is generally a function of $\omegaf$, and we will write it as $k_p(\omegaf)$.
Index $p$ can be used to indicate the number of the band in the photonic band structure. 
Thus, the general solution for the electric field inside a PTC can be written in the following form~\cite{zurita2009reflection}
\begin{equation}
\_E(\_r, t)=\sum_{p=1}^\infty \sum_{n=-\infty}^\infty {E}_{p,n} (\omegaf) e^{j\omega_nt}e^{-jk_p(\omegaf) z}\_a_x\,,   
\l{generalzurita}
\end{equation}
where $E_{p,n} (\omegaf)$ are the eigenmodes complex amplitudes. This solution for the electric field should be used when the wavenumber of the eigenmodes inside the PTC is not fixed, as it is in the scenario of spatially finite PTCs (see Section~\ref{sfptcs} for more details).

Finally, we wish to stress that the band structure with such a technique can also be analyzed while simultaneously considering a time-varying permeability \cite{Mart2017Standing}.

\subsubsection{{Band structure analysis based  on the transfer matrix method}}\label{sec: transfer matrix}

% Eigenmodes and band structures have been extensively studied with respect to PTCs. We will describe how they are calculated and also mention the novel emerging effects such as momentum bandgap (MBG), parametric amplification, and parametric absorption.\\

In the previous subsection, we discussed how to obtain the dispersion relation of the eigenmodes in a PTC directly from the wave equation. The quantities that entered that description are the Fourier coefficients describing the time-dependent and periodic modulation of the permittivity. Naturally, such an approach is convenient for a smoothly varying periodic modulation. Ideally, a sinusoidal modulation is considered so that only three Fourier coefficients are sufficient to describe the permittivity modulation. In contrast, stepwise modulations are more demanding to describe because of the many Fourier coefficients that must be considered. For such problems, an alternative approach is much more suitable. It is usually referred to as the transfer-matrix method or the ABCD-matrix method. It allows us to propagate the fields through one unit cell in time by slicing the medium into time segments with constant material properties and connecting the solutions in the different time segments using suitable interface conditions. By imposing additional Floquet-periodic boundary conditions onto time segments corresponding to one period, we can solve for the dispersion relation \cite{gaxiola2021temporal,Chegnizadeh2018general}. The method is well known when solving light propagation in spatially stratified media \cite{yeh_yariv_hong_1977}, and its dual version was adapted recently for PTCs~\cite{lustig2018topological,ramaccia2021temporal,sadhukhan2023defect,shaltout2016photonic,koutserimpas2018electromagnetic}.    

Before developing the calculation method for the band structure, it is important to introduce the concept of temporal interface conditions.
The corresponding expressions for a spatial interface are well known. As a reminder, let us consider a spatial interface between two different materials. We assume that each material occupies one half-space along the $x$-coordinate, and the interface shall be at $x=x_0$. Provided that there are no electric and magnetic surface currents flowing at the interface, the tangential electric and magnetic fields must be continuous when transiting through the interface~\cite[Chapter~7.3.6]{griffiths2021introduction}. Also, the tangential component of the wavevector is a preserved quantity, as the interface is assumed to be translational invariant in the corresponding spatial directions. To satisfy those interface conditions, reflection and refraction must happen. 

Analogous to a spatial interface, its temporal counterpart is an interface occurring in time between two materials with different properties. It signifies an abrupt transition at time moment $t=t_0$ from one material property to another. The materials continue to be spatially uniform, i.e., the properties are assumed to be identical at each spatial coordinate inside the media. 
Integrating the Maxwell curl equations around the switching time, i.e., from $t_0^-=t_0-s$ to $t_0^+=t_0+s$ and taking the limit of $s\rightarrow 0$, we have
\begin{subequations}
  \begin{equation}
        \lim_{s\rightarrow 0} \int_{t_0-s}^{t_0+s} \frac{\partial \mathbf{B}(\_r,t)}{\partial t}~{\rm d}t= -\lim_{s\rightarrow 0} \int_{t_0-s}^{t_0+s} \nabla\times \mathbf{E}(\_r,t)~{\rm d}t\,, \label{eq: limit B}
  \end{equation} 
  \begin{equation}
        \lim_{s\rightarrow 0} \int_{t_0-s}^{t_0+s} \frac{\partial \mathbf{D}(\_r,t)}{\partial t}~{\rm d}t= \lim_{s\rightarrow 0} \int_{t_0-s}^{t_0+s} (-\mathbf{J}(\_r,t)+\nabla\times \mathbf{H}(\_r,t))~{\rm d}t\,. \label{eq: limit D}
  \end{equation}  \label{eq: limit }
\end{subequations}
Since the field and source ($\mathbf{E}(\_r,t)$, $\mathbf{H}(\_r,t)$, and $\mathbf{J}(\_r,t)$) are finite, the integration on the right-hand side of Eq.~(\ref{eq: limit B}) and Eq.~(\ref{eq: limit D}) is zero~\cite{galiffi2022photonics}. Therefore, their left-hand sides of the equations are also zero, indicating the continuity of field fluxes across the switching time moment $t_0$. This continuity corresponds to the interface condition, and it reads as
\begin{subequations}
\begin{equation}
    \mathbf{D}(\_r,t=t_0^+)=\mathbf{D}(\_r,t=t_0^-)\,,
\end{equation}
\begin{equation}
     \mathbf{B}(\_r,t=t_0^+)=\mathbf{B}(\_r,t=t_0^-)\,.
\end{equation} \label{eq: temporal boundary condition }
\end{subequations}
To ensure the temporal interface conditions in Eq.~\eqref{eq: temporal boundary condition } at the switching moment, waves undergo phenomena similar to spatial reflection and refraction, referred to as temporal reflection and refraction \cite{sivan_time_2011,Xiao:14,wang2023controlling}. Reflected and refracted waves at spatial interfaces are separated in space since they propagate in opposite directions and in different half-spaces. However, temporal reflected and refracted waves are not spatially separated. %Time can only evolve in the forward direction. Therefore, both of these waves propagate in the entire media but in opposite directions.  
Time refraction has been recently experimentally observed at optical \cite{lustig2023time, zhou2020broadband, shaltout2016doppler, tirole2024second}, mid-infrared~\cite{bohn_spatiotemporal_2021}, and terahertz frequencies~\cite{deng_frequencyselective_2024}, but time reflection proves more challenging due to the difficulty in altering optical properties strongly and instantaneously. Analogously to the spatial interface case, generating strong time reflections requires a strong impedance mismatch at the temporal boundary, which translates into the necessity to have the strong amplitude of the temporal modulations (currently very challenging at optical frequencies)~\cite{galiffi2022photonics,bar2023time}.
Therefore, time reflection has been observed only at microwave frequencies for electromagnetic waves~\cite{jones2023time, moussa2023observation,galiffi2023broadband}, as well as for water wave \cite{peng2020time}, cold atoms~\cite{dong2023quantum}, and synthetic dimensions~\cite{long2023time}.
% \textcolor{red}{Should we cite Nader's paper about time-jumps?}

When periodically repeating spatial interfaces, a photonic crystal is created, and the energy bandgaps are formed through reflections at consecutive spatial interfaces. The impossibility of traveling waves within a frequency bandgap inside a photonic crystal is usually explained as an interference phenomenon. Consider a plane wave illuminating from outside a one-dimensional photonic crystal. Then, light will experience a sequence of reflections at the spatial interfaces. The optical thickness of the layers is now suitably adjusted so that reflected fields in the backward direction constructively interfere while they destructively interfere in the forward direction, forbidding light propagation inside the photonic crystal. 
% Light for the frequencies in the bandgap cannot propagate through the photonic crystal.   

Similarly, when temporal interfaces are repeated periodically in time by jumping between two values of the permittivity, a PTC is created, and the momentum bandgap emerges. The momentum bandgaps are formed due to multiple instances of temporal reflection/refraction at these temporal interfaces \cite{galiffi2022photonics}. Thus, the band structure of PTCs can be ascertained through these temporal interface conditions \cite{lustig2016topology,sadhukhan2023defect,salem2015temporal}. The derivation of the band structure can parallel the approach for spatial photonic crystals~\cite{yeh_yariv_hong_1977}. 

Recall that in calculating the band structure for photonic crystals, we essentially establish a relation that expresses how the (continuous) tangential components of the field evolve from one interface to the next interface. The field is written between the interfaces as a superposition of a forward and a backward propagating field. 
A $2\times2$ matrix can express the evolution of the field from one interface to the following interface. Multiplication of a sequence of matrices allows us to express how the field evolves across one unit cell. Finally, imposing Bloch-periodic boundary conditions, i.e., requiring that the amplitude is the same and only the phase varies by the predefined Bloch phase, allows us to disclose the dispersion relation. A similar procedure will be put in place in case of a PTC.     

\begin{figure}[t]
\centerline{\includegraphics[width= 0.7\columnwidth]{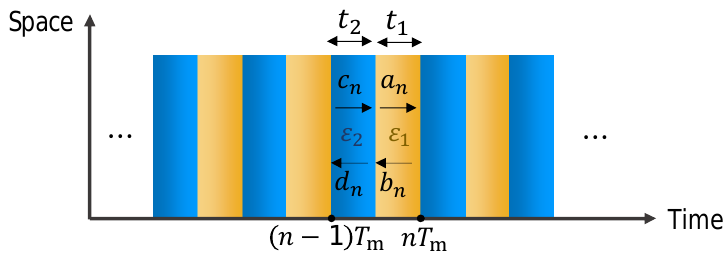}}
\caption{Schematic of a PTC made from a sequence of temporal slabs with constant permittivities. Such a structure is perfectly suitable for applying the ABCD-matrix method in the calculation of the band structure. Arrows denote the propagation directions of plane waves denoted by corresponding complex amplitudes $a_n$, $b_n$, $c_n$, and $d_n$.}
\label{Fig:ABCDschematic}
\end{figure}

For that, let us consider a homogeneous, nonmagnetic, isotropic, bulk material.
% At the moment of $t=t_0$, the material undergoes a sudden change in its properties. 
For definiteness, we assume an $x$-polarized plane wave propagating along the $z$-direction. As shown in Fig.~\ref{Fig:ABCDschematic}, the medium experiences a periodic stepwise temporal permittivity modulation with a period of $T_{\rm m}$, i.e., $\varepsilon(t+T_{\rm m})=\varepsilon(t)$.
% In one period, the permittivity is $\varepsilon(t)=\varepsilon_1$ for $0<t<t_1$ and $\varepsilon(t)=\varepsilon_2$ for $t_1<t<T_{\rm m}$ with a duration of $t_2=T_{\rm m}-t_1$. 
Let us consider the $n$-th period of the time evolution, where $(n-1)T_{\rm m}<t<nT_{\rm m}$. The permittivity variation in this period is divided into two temporal segments or slabs, with duration $t_1$ and $t_2$, where $t_1+t_2=T_{\rm m}$:
\begin{equation}
\varepsilon(t) = 
\begin{cases}
\varepsilon_2, & (n-1)T_{\rm m}<t<(n-1)T_{\rm m}+t_2\, \\
\varepsilon_1, & nT_{\rm m}-t_1<t<nT_{\rm m}\, 
\end{cases}
\end{equation}
In the time segment filled with yellow color in Fig.~\ref{Fig:ABCDschematic}, the permittivity is constant and equal to $\varepsilon_1$ for a duration of $t_1$. The forward and backward complex amplitudes of the plane waves in this temporal segment can be denoted as $a_n$ and $b_n$. The frequency of the wave is $\omega_1$.
Therefore, we choose an ansatz for the electric flux in the media as a sum of forward and backward propagating waves in time, which can be written as
    \begin{equation}
       D_x(z,t)= \left[a_n e^{j\omega_1(t-nT_{\rm m})}+b_n e^{-j\omega_1(t-nT_{\rm m})}\right] e^{-jkz}, \quad  nT_{\rm m}-t_1<t<nT_{\rm m}. \label{eq: Dx_a}
    \end{equation} 
The corresponding magnetic flux is the time derivative of the electric flux, which can be calculated by $\nabla \times \mathbf{B}(\_r,t)={\mu_0}\frac{\partial \mathbf{D}(\_r,t)}{\partial t}$. The magnetic flux can be calculated as,
  \begin{equation}
    B_y(z,t)=\frac{\omega_1\mu_0}{k} \left[a_n e^{j\omega_1 (t-nT_{\rm m})}-b_n e^{-j\omega_1 (t-nT_{\rm m})}\right] e^{-jkz},  \quad nT_{\rm m}-t_1<t<nT_{\rm m}\,.\label{eq: by_a}
\end{equation}

In the time segment filled with blue color in Fig.~\ref{Fig:ABCDschematic}, the permittivity is constant and equal to $\varepsilon_2$, and the forward and backward complex amplitudes of the propagating plane waves in this temporal segment are denoted as $c_n$ and $d_n$. The frequency of the wave in this second segment is $\omega_2$. A frequency change ($\omega_1 \neq \omega_2$) occurs because the spatial homogeneity requires the conservation of the momentum \cite{ortega-gomez_tutorial_2023}. Still, the dispersion relation of each material needs to be satisfied, suggesting that frequency and momentum are related by 
\begin{equation}
    k^2=\frac{\omega_{1,2}^2}{c^2}\varepsilon_{1,2}\, .
    \label{eqdisp}
\end{equation}
Therefore, we have the basic relation between the two frequencies $\omega_1\sqrt{\varepsilon_1}=\omega_2\sqrt{\varepsilon_2}$. 
The electric and magnetic fluxes in that time segment (filled with blue color) can be likewise written as a superposition of forward and backward waves,
  \begin{equation}
       D_x(z,t)= \left[c_n e^{j\omega_2(t-nT_{\rm m})}+d_n e^{-j\omega_2(t-nT_{\rm m})}\right] e^{-jkz}, \quad   (n-1)T_{\rm m}<t<(n-1)T_{\rm m}+t_2\, \label{eq: Dx_c}
    \end{equation} 
    and
  \begin{equation}
    B_y(z,t)=\frac{\omega_2 \mu_0}{k} \left[c_n e^{j\omega_2 (t-nT_{\rm m})}-d_n e^{-j\omega_2 (t-nT_{\rm m})}\right] e^{-jkz},  \quad  (n-1)T_{\rm m}<t<(n-1)T_{\rm m}+t_2\,.\label{eq: by_c}
\end{equation}
Having spelled out these ansatz for the fields, we need to connect them by imposing the interface conditions given by Eq.~\eqref{eq: temporal boundary condition }. 
First, we consider the field continuity at the switch moment of $t=(n-1)T_{\rm m}$. According to Eq.~(\ref{eq: Dx_a}) and Eq.~(\ref{eq: Dx_c}), the continuity of $D_x$ can be written as the following equation,
\begin{equation}
    a_{n-1}+b_{n-1}=c_{n} e^{-j\omega_2 T_{\rm m}}+d_{n} e^{j\omega_2 T_{\rm m}}\,. \label{eq: cn+dn}
\end{equation}
Combining Eq.~(\ref{eq: by_a}) and Eq.~(\ref{eq: by_c}), the continuity of magnetic flux at $t=(n-1)T_{\rm m}$ requires that,
\begin{equation}
a_{n-1}-b_{n-1}=\frac{\omega_2}{\omega_1}c_{n}e^{-j\omega_2 T_{\rm m}}-\frac{\omega_2}{\omega_1}d_{n}e^{j\omega_2 T_{\rm m}}\,. \label{eq: cn-dn}
\end{equation}
Equations~\eqref{eq: cn+dn} and \eqref{eq: cn-dn} can be arranged as a matrix operation,
\begin{equation} 
\begin{bmatrix}
       1& 1\\
       1& -1
   \end{bmatrix} \cdot
   \begin{bmatrix}
       a_{n-1}\\b_{n-1}
   \end{bmatrix}=
   \begin{bmatrix}
       e^{-j\omega_2 T_{\rm m}}&e^{j\omega_2 T_{\rm m}}\\
       \frac{\omega_2}{\omega_1}e^{-j\omega_2 T_{\rm m}}& -\frac{\omega_2}{\omega_1} e^{j\omega_2 T_{\rm m}}
   \end{bmatrix} \cdot
   \begin{bmatrix}
       c_{n}\\d_{n}
   \end{bmatrix}\,. \label{eq: matrix1}
\end{equation}
Similarly, utilizing the flux continuity at the time moment of $t=nT_{\rm m}-t_1$, we obtain another matrix equation,
\begin{equation}
\begin{bmatrix}
       e^{-j\omega_1 t_1}&e^{j\omega_1 t_1}\\
      \frac{\omega_1}{\omega_2}  e^{-j\omega_1 t_1}& -\frac{\omega_1}{\omega_2}e^{j\omega_1 t_1}
   \end{bmatrix} \cdot
   \begin{bmatrix}
       a_{n}\\b_{n}
   \end{bmatrix}=
    \begin{bmatrix}
        e^{-j\omega_2 t_1}&  e^{j\omega_2 t_1}\\
        e^{-j\omega_2 t_1}& - e^{j\omega_2 t_1}
   \end{bmatrix} \cdot
   \begin{bmatrix}
      c_n\\d_n
   \end{bmatrix}\,.\label{eq: matrix2}
\end{equation}
Combining Eqs.~(\ref{eq: matrix1}) and (\ref{eq: matrix2}), we obtain the fields at two consecutive periods (temporal unit cells) to be connected as
\begin{equation}
     \begin{bmatrix}
       a_{n-1}\\b_{n-1}
   \end{bmatrix}=
   \begin{bmatrix}
       A&B\\
       C&D\\
   \end{bmatrix}\cdot
   \begin{bmatrix}
       a_{n}\\b_{n}
   \end{bmatrix}\,, \label{eq: abcd}
\end{equation}
where the matrix elements are given by
\begin{subequations}
\begin{equation}
    A=e^{-j\omega_1 t_1}\left[ \cos\omega_2 t_2-\frac{j}{2}\left(\frac{\omega_2}{\omega_1} +\frac{\omega_1}{\omega_2}\right)\sin\omega_2 t_2\right]\,,
\end{equation}

   \begin{equation}
    B= - \frac{j}{2}e^{j\omega_1 t_1}
    \left(\frac{\omega_2}{\omega_1} - \frac{\omega_1}{\omega_2}\right)
   \sin\omega_2t_2\,,
\end{equation}

   \begin{equation}
    C=B^{*}= \frac{j}{2}e^{-j\omega_1 t_1}
    \left(\frac{\omega_2}{\omega_1} - \frac{\omega_1}{\omega_2}\right)
   \sin\omega_2t_2\,,
   \end{equation}
   \begin{equation}
   D=A^{*}=e^{j\omega_1 t_1}\left[ \cos\omega_2 t_2+\frac{j}{2}\left(\frac{\omega_2}{\omega_1} +\frac{\omega_1}{\omega_2}\right)\sin\omega_2 t_2\right]\,.
\end{equation}\label{eq:ABCD}
\end{subequations}
One can see that Eq.~(\ref{eq:ABCD}) and Eq.~(\ref{eq: abcd}) possess the exact dual resemblance to those derived for the spatial case in Ref.~\cite{yeh_yariv_hong_1977}. The duality is apparent under the replacement of $\omega \leftrightarrow k$ and $t \leftrightarrow z$. Before we continue, we draw the reader's attention to one point. From Eq.~(\ref{eq:ABCD}), we see that if $\omega_2t_2$ is identical with $p\pi$ ($p=\pm1,\,\pm2,...)$, the matrix elements $B$ and $C$ vanish, and the elements $A$ and $D$ are described by $e^{\pm j\omega_1 t_1}\cos(p\pi)$. In other words, the ABCD-matrix is diagonalized with elements whose absolute value is unity. Hence, it adds only an additional phase to the wave propagating between the corresponding time slab.

Thus, the ABCD-matrix connects the amplitudes of the fields at two neighboring temporal unit cells of the PTC. The ABCD-matrix can be explicitly expressed once the permittivities in the two temporal segments $\varepsilon_1$ and $\varepsilon_2$ and their durations $t_1$ and $t_2$ are given. Naturally, more complicated unit cells consisting of more than two temporal segments can be considered using this ABCD method. In essence, it would boil down to a multiplication of an increasing number of matrices. While closed-form expressions are challenging to obtain, a numerical evaluation is particularly easy. Moreover, even continuous profiles can be handled. It would require slicing the temporal profile into very short intervals within which the permittivity can be assumed constant. Independent of these details, we can always obtain a $2\times2$ ABCD-matrix that links the amplitudes of the same waves in two consecutive periods.      

To obtain the dispersion relation, we must impose that the amplitudes of the waves from one unit cell to the next unit cell are preserved, and the only permissible change is an additional phase accumulation. According to the Floquet theorem, the fields in the neighboring temporal unit cells have a phase difference (same for all the frequency harmonics) of $e^{j (\omega_{\rm F}+ p \omega_{\rm m}) T_{\rm m}} = e^{j\omega T_{\rm m}}$ (here, $\omega_{\rm F}$ is the Floquet frequency and $p$ is an arbitrary integer), given as 
\begin{equation}
     \begin{bmatrix}
       a_{n}\\b_{n}
   \end{bmatrix}=e^{j\omega_{\rm F} T_{\rm m}}\begin{bmatrix}
       a_{n-1}\\b_{n-1}
   \end{bmatrix}\,. \label{eq: an+1 an}
\end{equation}
Combining Eq.~(\ref{eq: abcd}) and Eq.~(\ref{eq: an+1 an}), we obtain the equation 
\begin{equation}
\underbrace{ \begin{bmatrix}
       A&B\\
       C&D\\
   \end{bmatrix}}
     _{\={M}(k)} \cdot
   \begin{bmatrix}
       a_{n}\\b_{n}
   \end{bmatrix}=e^{-j\omega_{\rm F} T_{\rm m}} \begin{bmatrix}
       a_{n}\\b_{n}
   \end{bmatrix}\,. \l{igen}
\end{equation}
Note that in this equation, the matrix elements $A$, $B$, $C$, and $D$ are functions of $\omega_1$ and $\omega_2$ which, in turn, are functions of momentum $k$ (see Eq.~\eqref{eqdisp}). Therefore, Eq.~\r{igen} is an eigenequation relating Floquet frequencies $\omega_{\rm F}$ to wavenumber $k$. 
The above eigenequation can be solved by requiring that 
\begin{equation}\label{eq:band diagram}
    \det\left[\={M}(k)-e^{-j\omega_{\rm F} T_{\rm m}}\={I}\right]=0\,,
\end{equation}
where $\={M}(k)$ is the ABCD-matrix. To satisfy the above equation, we have
\begin{equation}
    e^{-j\omega_{\rm F} T_{\rm m}}=\frac{A+D}{2}\pm \sqrt{\frac{(A+D)^2}{4}-1}\,. \label{eq: dispersion}
\end{equation}
Note that Eq.~(\ref{eq: dispersion}) is derived using the fact of $AD-BC=1$, which can be verified by Eqs.~(\ref{eq:ABCD}).
Since $D=A^{*}$, $A+D$ is a real number. 
% Equating the real part of Eq.~(\ref{eq: dispersion}), we have
% \begin{equation}
%     \omega(k)=\frac{1}{T_{\rm m}} \cos^{-1}\frac{A+D}{2}\,. \label{eq: dispersionrelation}
% \end{equation}
%
The ranges of wavenumber $k$ where $|A+D| < 2$ correspond to eigenmodes with purely real Floquet frequencies (allowed bands):
\begin{equation}
    \omega_{\rm F}(k)= \mp \frac{1}{T_{\rm m}} \left[\cos^{-1}\left(\frac{A+D}{2}\right)-2p\pi\right], \quad p \in \mathds{Z}\,. \label{eq: dispersionrelation}
\end{equation}
These eigenmodes are propagating waves. 

In contrast, for wavenumbers $k$ where $|A+D| > 2$, we obtain from Eq.~(\ref{eq: dispersion}) that the eigenmodes have complex Floquet frequencies. For wavenumbers $k$ that satisfy $A+D<-2$, 
\begin{equation} 
\begin{array}{cc} \displaystyle
    \Re[\omega_{\rm F}(k)]=\frac{\omega_{\rm m}(2p+1)}{2},  \quad p \in \mathds{Z}  \vspace{2mm} 
 \\ \displaystyle
    \Im[\omega_{\rm F}(k)]= \frac{1}{T_{\rm m}} \ln \left[  \left( -\frac{A+D}{2}\mp \sqrt{\frac{(A+D)^2}{4}-1} \right) \right]\, .
\end{array}
\label{eq: complex}
\end{equation}
For $A+D>2$
\begin{equation} 
\begin{array}{cc} \displaystyle
    \Re[\omega_{\rm F}(k)]=\omega_{\rm m}p,  \quad p \in \mathds{Z}  \vspace{2mm} 
 \\ \displaystyle
    \Im[\omega_{\rm F}(k)]= \frac{1}{T_{\rm m}} \ln \left[  \left( \frac{A+D}{2}\pm \sqrt{\frac{(A+D)^2}{4}-1} \right) \right]\, .
\end{array}
\label{eq: complex1}
\end{equation}
The $\pm$ signs determine the sign of the imaginary part of the eigenfrequency.
These solutions correspond to the forbidden bands (momentum bandgaps) of the PTC. The band edges occur when $|A+D| = 2$.

\begin{figure*}[t!]
\centerline{\includegraphics[width= 0.7\columnwidth]{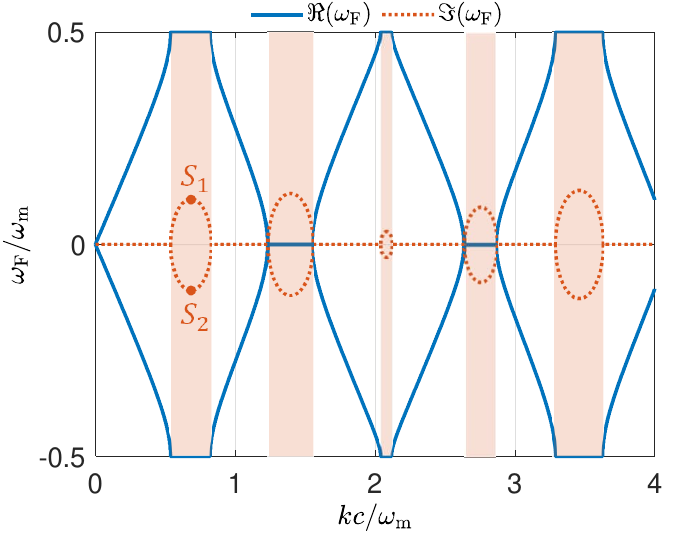}}
\caption{Band structure of PTCs with periodic stepwise modulation where $\varepsilon_1=5$ and $\varepsilon_2=1$. For real values of wavenumber $k$, complex-valued eigenfrequencies are plotted. The momentum bandgaps are shown with shaded regions.}
\label{Fig:ABCDsbandstructure}
\end{figure*}

To plot the complex solutions of the band structure, we choose $\varepsilon_1=5$ and $\varepsilon_2=1$, $t_1=t_2=T_{\rm m}/2$. For a given real $k$, the eigenfrequencies $\omega$ are generally complex. The real and imaginary parts of $\omega$ are calculated using Eqs.~(\ref{eq: dispersionrelation})-(\ref{eq: complex1}) and are shown in Fig.~\ref{Fig:ABCDsbandstructure}. The plot was created using Code~2 attached as Supplementary Material~\cite{code2}.
% For $k$ that obtains complex eigenfrequencies, it is inside the momentum bandgap. 
Unlike sinusoidal modulation, where only one bandgap appears, stepwise modulation generates a series of bandgaps. The next section discusses the physical meaning of the complex eigenfrequency.

Using the transfer matrix method, it is also possible to extract the amplitudes of each harmonic for a given eigenmode, that is $E_{p,n} (\omega_{\rm F})$ in Eq.~\r{generalzurita}.  
In the plane-wave expansion method, the calculation of the harmonics spectrum is straightforward, as the eigenmode can be directly obtained by finding the eigenvector of Eq.~(\ref{eq:MATRIXFVEC}).
In contrast, when the ABCD-matrix technique is used, calculating the harmonic spectrum becomes not so straightforward, but can be computed by Fourier transforming the eigenfield in one temporal unit cell $T_{\rm m}$ of the PTC. Next, we briefly introduce the method to calculate the eigenmode. 

For a specific wavenumber $k$, by solving the eigenvalue equation Eq.~\r{igen}, the Floquet frequencies $\omega_{\rm F}$ can be obtained. Note that there are multiple discrete solutions of $\omega_{\rm F}$. Choosing one solution of $\omega_{\rm F}$ with a spacing of $p\omega_{\rm m}$, we can obtain the corresponding eigenvector $[a_n, b_n]^{\rm T}$ of Eq.~\r{igen}, which corresponds to the amplitudes of forward and backward waves in the time section when $\varepsilon(t)=\varepsilon_1$ as shown in Fig.~\ref{Fig:ABCDschematic}. Using Eq.~(\ref{eq: matrix2}), the amplitudes of forward and backward waves in the next time segment $[c_n, d_n]^{\rm T}$ can be calculated. Knowing $a_n, b_n, c_n$, and $d_n$, the total field in one temporal unit can be obtained from Eqs.~(\ref{eq: Dx_a}) and (\ref{eq: Dx_c}). For a purely real $\omega_{\rm F}$, the total field has a constant amplitude in one temporal unit cell. Therefore, one can directly take the Fourier transform of this time-domain field and obtain the harmonics spectrum corresponding to a chosen $\omega_{\rm F}$. The calculated spectrum is the same as that determined by solving the eigenvectors from Eq.~(\ref{eq:MATRIXFVEC}).
For a complex $\omega_{\rm F}$ (when $k$ is located inside the bandgap), the total field is exponentially growing or decaying in time with a factor of $e^{\Im(\omega_{\rm F})t}$. In this case, by normalizing the field by this exponential factor before taking the Fourier transform, we can obtain the harmonics amplitudes of the eigenmode. Therefore, the methods based on transfer matrix and plane-wave expansion provide the same information. Which of the two methods is more suitable depends on the specific temporal modulation profile of $\varepsilon(t)$.
%%%%%%%%%%%%%%%%%%%%%%%%%%%%%%%%%%%%%%%%%%%%%%%%%%
%%%%%%%%%%%%%%%%%%%%%%%%%%%%%%%%%%%%%%%%%%%%%%%%%%
%%%%%%%%%%%%%%%%%%%%%%%%%%%%%%%%%%%%%%%%%%%%%%%%%%

{\subsection{Electromagnetic effects inside the momentum bandgap}\label{momband}}
% \red{Include Xuchen's rule here. Point out that it will be later used in Sec. 6.3}

 In the previous section, we discussed the band structure and mentioned the presence of the bandgaps along the momentum axis. In this section, we elaborate on one enticing feature associated with such bandgaps, which results in important consequences, at least from the application perspective. An important question is: What happens to an electromagnetic wave in a PTC whose wavenumber is located inside the momentum bandgap?

To quantitatively understand the wave effects within the momentum bandgap, we examine Fig.~\ref{Fig:ABCDsbandstructure} in further detail.
We select the wavenumber $k$ inside the first momentum bandgap (the first shaded region) shown in the figure. The eigenfrequencies, in this case, are complex-valued. The real parts of the eigenfrequencies remain constant, $\Re(\omega_{\rm F})=\frac{(2p+1)\omega_{\rm m}}{2}$, where $p \in \mathds{Z}$, while the imaginary part delineates an ellipse in the band structure, as shown in Fig.~\ref{Fig:ABCDsbandstructure}. 
For a given $k$ inside the momentum bandgap, there are both positive and negative solutions for the imaginary part, $\Im(\omega_{\rm F})=\pm \omega_{\rm im}$ with $\omega_{\rm im}>0$, denoted as $S_1$ and $S_2$ respectively in the figure. 
The solution with $\Im(\omega_{\rm F})=- \omega_{\rm im}$ signifies fields that are exponentially growing, with the rate of $e^{+\omega_{\rm im}t}$~\cite{gaxiola2023growing}. This is the salient property of PTCs, which allows them to amplify light in time. Remember that the system we describe here is not passive and the external energy supply comes from the mechanism responsible for the temporal modulation of the PTC. Therefore, the energy conservation principle is not violated since we deal with an open and active system.
In Section~\ref{tvcapacitor}, we discuss a simple physical picture of how the energy from the temporal modulations is transferred to the signal wave to support its growing amplitude.  

The second solution for the eigenfrequencies with $\Im(\omega_{\rm F})=+ \omega_{\rm im}$ corresponds to waves that are exponentially decaying over time at the rate of $e^{- \omega_{\rm im}t}$. It is difficult to probe these modes experimentally as they decay rapidly, over the time scale of the wave oscillations. Therefore, inside the momentum bandgap of the PTC, typically only the dominant (growing in time) modes are considered. Nevertheless, under proper conditions, the PTC can be made to operate in a phase-sensitive regime, allowing excitation of only the decaying mode~\cite{hayran2023beyond}. This can be accomplished, e.g., by making PTC subwavelength and/or by adding a back reflector to ensure the standing pattern of the signal wave.% (see the relevant discussion in Subsection~\ref{phasesens}). 
% Decaying this solution over time while the wavenumber is real is equivalent to the idea that the medium virtually absorbs the corresponding field, or there is a loss in the system, although the medium is lossless. To precisely understand the energy exchange mechanism between the medium and the device providing the energy, one needs to calculate the energy flow using Poynting's theorem by considering both solutions and considering the excitation before the medium permittivity starts to change.

As was discussed in the introduction, there is a beautiful duality symmetry between PTCs and conventional photonic crystals (see Fig.~\ref{Fig:photonic_space_time}). Photonic crystals also host two types of eigenmodes within the energy bandgap: one decaying in the positive and one decaying in the negative spatial direction. However, if the photonic crystal is semi-infinite (consider, e.g., the interface between free space and the photonic crystal), only one mode is physical because the other mode corresponds to an exponentially growing field in the direction away from the interface, which is impossible due to energy conservation. In contrast, in PTCs, both eigenmodes are always physical because PTCs are not bound by the energy conservation law.  
% It is worth noting that there is a fundamental difference compared to conventional spatial photonic
% crystals.  In such a spatial photonic crystal, for a real-frequency excitation, we cannot find a mode with a pure real-valued wavenumber in some frequency domains, this domain is called an energy bandgap. The field cannot propagate with a constant amplitude inside the medium if it temporally oscillates at a frequency within the energy bandgap. In fact, within the energy bandgap, there are two solutions for the momentum or wavenumber. These two solutions are complex. 
% One of them results in an exponentially growing field amplitude in space. However, this solution is physically not feasible because, unlike PTCs, the system is closed and passive. Therefore, there is no reason for growing the amplitude of the field to infinity in space. The excited mode corresponds to the exponentially decaying solution.
% And something very similar happens for PTCs. When we fix the wavenumber, 
% we cannot find a mode with a pure real-valued frequency in some wavenumber domains. The field cannot propagate with a constant amplitude inside the medium if it spatially oscillates at a wavenumber within the momentum bandgap. It exponentially grows or decays in time at a rate given by the imaginary part of the eigenfrequency (see Fig.~\ref{Fig:ABCDsbandstructure}). 
\begin{figure}[t]
\centerline{\includegraphics[width= 1\columnwidth]{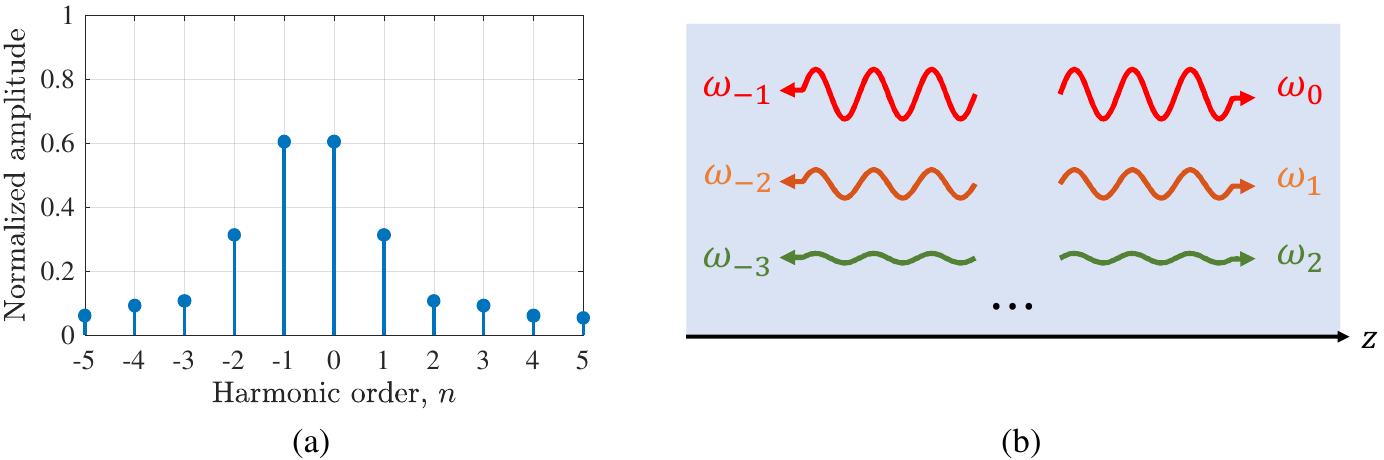}}
\caption{(a) Spectrum of harmonics amplitudes of the eigenmode whose $k$ is located in the center of the momentum bandgap and the eigenfrequency $\omega_{\rm F}$ is complex-valued with $\Re(\omega_{\rm F})=\frac{\omega_{\rm m}}{2}$. The harmonics amplitudes are normalized such that $\sum_n|E_n (\omega_{\rm F})|^2=1$. The parameters of the periodic stepwise modulation are the same as those mentioned in the caption of Fig.~\ref{Fig:ABCDsbandstructure}. (b) Illustration depicting propagation directions and relative amplitudes of the harmonics shown in (a). Note that although frequencies of different harmonics are different, their wavelengths are the same inside the PTC since they share the same wavenumber. }
\label{Fig:standingwave}
\end{figure}

Now, let us analyze the harmonics' weights $E_n (\omega_{\rm F})$ of the eigenmodes inside the bandgap (see Eq.~(\ref{eq:BLOCHT})). 
% Until this stage, we understand the physical meaning of complex eigenfrequency in the momentum bandgap of PTCs. Next, we discuss what really happens when the wave falls inside the momentum bandgap. 
% To know the details of the eigenmode, we need to decompose the eigenmode. 
Similarly to the analysis performed for plotting Fig.~\ref{fig:bandstructure_harmonic}(a), we find for each harmonic of a given eigenmode its magnitude and phase. Figure~\ref{Fig:standingwave}(a) illustrates the harmonics amplitude distribution of the eigenmode whose wavenumber is selected at the center of the momentum bandgap $k_{\rm bg}$ and the eigenfrequency has the real part equal to $\Re(\omega_{\rm F})=\frac{\omega_{\rm m}}{2}$. 
% We define the harmonics by $\omega_n=\omega_{\rm F}+n\omega_{\rm m}$. 
The frequencies of all the harmonics $\omega_n$ share the same imaginary part $\pm \omega_{\rm im}= \pm0.108 \omega_{\rm m}$. Note that the distribution of $E_n (\omega_{\rm F})$ is the same for decaying and growing eigenmodes. 
From Fig.~\ref{Fig:standingwave}(a), it is evident that the eigenmode encompasses harmonics with amplitudes symmetrically distributed across the spectrum. It is important to note that this symmetry property always holds when the wavenumber falls inside the momentum bandgap. This is different from the case we analyzed in Fig.~\ref{fig:bandstructure_harmonic}(c), where the wavenumber was outside the bandgap, and the harmonics amplitude distribution was asymmetric. 

Let us consider a pair of the dominant harmonics. The $-1$-st harmonic, i.e., $\Re(\omega_{-1})=-\omega_{\rm m}/2$, possesses the same amplitude as the fundamental harmonic, i.e., $\Re(\omega_0)=\omega_{\rm m}/2$. Thus, this degenerate pair of dominant harmonics has real frequencies with opposite signs. 
Since the two harmonics share the same momentum $k_{\rm bg}$,
their phase velocities are opposite  $v_{\rm ph}=\pm \frac{\omega_{\rm m}}{2 k_{\rm bg}}$. 
Therefore, this pair of dominant harmonics represents a standing wave~\cite{Mart2017Standing} whose amplitude grows or decays exponentially over time due to the nature of the complex eigenfrequency. From Eq.~(\ref{eq:BLOCHT}) we obtain:
\begin{equation}
E(z,t) = |E_0| e^{\omega_{\rm im} t} e^{-j k_{\rm bg }z } \left[ 
e^{j \phi_0} e^{j \frac{\omega_{\rm m}}{2} t} +
e^{j \phi_{-1}} e^{-j \frac{\omega_{\rm m}}{2} t}
\right]\, , 
\l{stw1}
\end{equation}  
where we neglected the decaying eigenmode and wrote only the dominant harmonics with $n=0$ and $n=-1$; $\phi_0$ and $\phi_{-1}$ are the phases of the complex amplitudes $E_0$ and $E_{-1}$, respectively. The instantaneous real-valued electric field $E_{\rm inst} (z,t) = \Re [E(z,t)]$ then reads 
\begin{equation}
E_{\rm inst} (z,t) = 2 |E_0| e^{\omega_{\rm im} t} 
\cos \left[k_{\rm bg }z - \frac{\phi_0 +\phi_{-1}}{2}  \right]
\cos \left[ \frac{\omega_{\rm m}}{2} t + \frac{\phi_0 -\phi_{-1}}{2}  \right]\, , 
\l{stw2}
\end{equation}  
The standing wave pattern of the instantaneous field in Eq.~\r{stw2} is apparent as the spatial and temporal variations of the electric field are decoupled. The nodes of the electric field are separated from one another by the integer of $\pi/k_{\rm bg }$. Nevertheless, the wave given by Eq.~\r{stw2} is not a conventional standing wave since its amplitude $2 |E_0| e^{\omega_{\rm im} t}$ is quickly growing in time. 

% Indeed, the decaying modes are evanescent in time. 
For other higher-order harmonic pairs with the same wavenumber, such as $\pm \frac{3\omega_{\rm m}}{2}$, the behavior is analogous: they form a standing wave pattern with smaller amplitudes (see Fig.~\ref{Fig:standingwave}(a)) that grow exponentially at the same rate $e^{\omega_{\rm im} t}$ as the fundamental pair. As the harmonic order increases, their amplitudes diminish, and their contributions are negligible. 

Finally, it is important to discuss the shape of the imaginary part of eigenfrequency inside the bandgap (as shown in Fig.~\ref{Fig:standingwave}(a)). 
For PTCs consisting of lossless material, inside the momentum bandgap, the imaginary part of eigenfrequency forms an ellipse which is symmetrically located with respect to the wavenumber axis ($\omega_{\rm F} =0$). When considering PTCs whose material includes dissipation, the imaginary-frequency ellipse will be shifted upward in the considered $e^{j\omega t}$ convention. Thus, for a sufficiently high amount of loss for a given modulation strength, all the solutions of the imaginary eigenfrequency become positive. Such a regime could be useful for enhancing light absorption~\cite{hayran2023beyond}. Furthermore, the shape of the imaginary eigenfrequency contours inside the momentum bandgap is not always elliptical and can be to some degree engineered, providing interesting possibilities for predefined direction-dependent amplification of plane waves~\cite{wang2023unleashing}.

{\section{Aspects of realistic PTCs \label{secAspects}}}
In the previous chapter, we introduced the fundamentals of PTCs and analyzed their eigenmodes while assuming the most basic scenario, i.e., that of a dispersionless and linear material. Moreover, the PTC was infinitely extended in space and time, and the crystal lattice was perfectly periodic in time. 

These idealistic assumptions were highly beneficial to getting a glimpse into the fundamental properties of PTCs and obtaining closed-form analytical expressions for some defining quantities. In particular, the dispersion relation could be explicitly written out, and properties such as the growth or decay rate of the amplitudes of the eigenmodes within the momentum bandgap could be obtained. However, many more features characterize realistic systems. We need to consider them for a reliable prediction, also in the light of possible experimental observations. Therefore, the following chapter is written with the purpose of considering an increasing number of aspects that cause a deviation of the idealistic towards a more realistic description of light propagation in PTCs. We admit at this point that the more realistic description naturally leads to a more complex description. This gives rise to further aspects at the technical and scientific level.

We start by elaborating on the impact of temporal dispersion, continuing with the consideration of spatially finite PTCs and finally considering temporally finite PTCs. These should be the most important aspects we must accommodate in a more realistic setting. But there were other assumptions that do not necessarily apply and which give rise to interesting effects. For example, the intrinsic material might be characterized by an anisotropy, whereas we have assumed isotropic materials so far. Furthermore, the time variation might deviate from a perfect periodic one, allowing us to introduce temporal defects. These defects will be discussed in a dedicated section, leading straight to the question of the general impact of the disorder on the properties of the considered PTCs. Finally, we also discuss a deviation from a linear response, and we will consider nonlinear PTCs at the end of this chapter.

% some basic concepts of PTCs concerning dispersion, topology, defects, and their comparison with time crystals. 

%%%%%%%%%%%%%%%%%%%%%%%%%%%%%%%%%%%%%%%%%%%%%%%%%%%%%%%%%%%%%%%%%%%%%%%%%%%

\subsection{Effects of temporal dispersion }  

In the preceding chapter, we derived the wave equation and determined the band structure for a PTC. That derivation of the dispersion relation has been based on the assumption of an instantaneous response in time that leads to a vanishing dispersion of the materials in the frequency domain. The materials were characterized by a non-dispersive susceptibility or permittivity, i.e., it has not been a function of the frequency. This assumption is generally nonphysical and only approximately valid for systems exhibiting minimal temporal variations and/or insignificant temporal dispersion. An example includes materials such as lithium niobate and silicon in their transparency frequency regions~\cite{barton2021wavefront,guo2019nonreciprocal}. To operate in the transparency region, all the relevant resonances that cause dispersion in the material must be located at frequencies far away from the operational frequency considered for the PTC. These resonances are usually electronic transitions in the material in the UV range or vibrational excitations in the infrared. 

However, the modulation depth of the material parameters in these dispersionless materials is typically very low, ranging from $10^{-4}$ to $10^{-3}$~\cite{williamson2020integrated}. That small modulation depth is detrimental to observing notable momentum bandgaps and the possible amplification of eigenmodes. There are two ways to overcome this limitation. First, we need to operate close to a resonance frequency of the material. An electron plasma is a typical example of a material where this is possible. In passing, we note that the plasma is a particular case where the resonance frequency is zero as there is no restoring force, but the argument applies as well. In perspective, operating close to a resonance frequency permits a substantial modification of the material properties. However, it also comes along with dispersion and absorption. 

Second, we can operate in a regime where even a tiny change in the material properties might greatly impact how light propagates in such a material. For example, operating in the epsilon-near-zero-regime would be one of the options, where a small absolute change in the permittivity tremendously affects the light propagation. Such operation is possible at visible or near-infrared frequencies in transparent conducting oxides \cite{kinsey2015epsilon}. Yet, these materials exhibit strong dispersion at relevant frequencies, particularly in the epsilon-near-zero region where the modulation depth is significant. Consequently, incorporating temporal dispersion (that should be considered from the equation of motion~\cite{mirmoosa2022dipole,GPMSMSTEMPRRAPS, GPTitcynEMPRA}) is crucial in analyzing PTCs \cite{mirmoosa2022dipole, EnghetaDisTVMPRB, sloan2020dispersion,sloan2022optical,sotoodehfar2022waves,horsley2023eigenpulses}. This section aims to consider the properties of PTCs in the presence of material dispersion for a more realistic description.

% materials based on which we obtained the band structure (e.g., see Eq.~\eqref{Eq: governing}).
% However, as mentioned at the beginning of the aforementioned section, frequency dispersion is an inherent feature that cannot be neglected. Therefore, in this part, we consider the realistic scenario of frequency-dispersive materials, and, accordingly, 
We start by deriving the Helmholtz equation in the frequency domain under the assumption of a dispersive medium. Then, in two dedicated subsections, we consider two specific examples of periodic temporal modulation in a dispersive PTC: a change in the plasma frequency and a change in the resonance frequency. It should be noted that the plasma frequency corresponds to a change in the number density and/or effective mass of the particles that constitute the material.

\subsubsection{Helmholtz equation in a PTC made from dispersive media}

Generally, in the absence of a time modulation, material properties can be discussed on phenomenological grounds either using a Drude model (applicable to free electrons in a metal or a plasma) or using a Lorentz model (relevant to bound resonances such as electronic or vibrational resonances in materials made from atoms or molecules). Of course, artificial materials, i.e., metamaterials, can also be considered, where artificially structured meta-atoms cause the dispersion. The dispersion of arbitrary materials can always be written as a superposition of a Drude term and a finite number of Lorentzian oscillators. Therefore, the material dispersion up to an arbitrary degree of precision can be expressed with these two different models. Consequently, even though applied and discussed here for some specific types of dispersions, the approach in this subsection can be used for all kinds of materials. Moreover, even though we consider only the case where a few selected parameters of the models are time-dependent, similar derivation and considerations can be done for all kinds of parameters.

To derive general expressions while considering a given constitutive relation in the frequency domain, we start from Maxwell's equations that have been Fourier-transformed regarding both time variables involved in the response function to obtain the Helmholtz equation. The Maxwell equations in the frequency domain read as ~\cite{ptitcyn2023floquet} 
\begin{eqnarray}
\nabla\times\tilde{\_E}(\_r,\omega) & = &-j\omega\mu_0\tilde{\_H}(\_r,\omega)\label{Thementionedfirstequation}\,,\\
\nabla\times\tilde{\_H}(\_r,\omega) & = & j\omega\tilde{\_D}(\_r,\omega)\label{Thementionedsecondequation}\,,\\
\nabla\cdot\tilde{\_D}(\_r,\omega) & = & 0\label{Thementionedthirdequation}\,,\\
\nabla\cdot\tilde{\_B}(\_r,\omega) & = & 0\,,
\label{eq:Maxwell_Frequency}
\end{eqnarray}
that is supplemented by the constitutive relation in Eq.~\eqref{eq:FDPEREL}. As a reminder, it reads as:
\begin{equation}
\tilde{\_D}(\_r,\omega)={\varepsilon_0\over2\pi}\int_{-\infty}^{+\infty}\varepsilon(\_r,\omega',\omega-\omega')\tilde{\_E}(\_r,\omega')\mathrm{d}\omega'\,.
\end{equation}  

To derive the Helmholtz equation, we apply the curl operator to both sides of the Faraday law, i.e., the first of the four Maxwell equations, i.e., Eq.~\eqref{Thementionedfirstequation}. Using the vector algebra relation $\nabla\times\nabla\times\_V(\_r)=\nabla(\nabla\cdot\_V(\_r))-\nabla^2\_V(\_r)$, in which $\_V(\_r)$ is an arbitrary differentiable vector field in space, we obtain $\nabla(\nabla\cdot\tilde{\_E}(\_r,\omega))-\nabla^2\tilde{\_E}(\_r,\omega)=-j\omega\mu_0\nabla\times\tilde{\_H}(\_r,\omega)$. 

On the other hand, the curl of $\tilde{\_H}(\_r,\omega)$ is related to the electric flux density through the Amp\`ere-Maxwell law, i.e., the second of the Maxwell equations, i.e., Eq.~\eqref{Thementionedsecondequation}. By plugging the Amp\`ere-Maxwell law and the constitutive relation into the modified Faraday law, we deduce that
\begin{equation}
\begin{split}
\nabla\Big(\nabla\cdot\tilde{\_E}(&\_r,\omega)\Big)-\nabla^2\tilde{\_E}(\_r,\omega)\cr
&-{\omega^2\mu_0\varepsilon_0\over2\pi}\int_{-\infty}^{+\infty}\varepsilon(\_r,\omega',\omega-\omega')\tilde{\_E}(\_r,\omega'){\rm d}\omega'=0\,.
\end{split}
\label{eq:HH1}
\end{equation}
However, this is not yet the final expression. In addition, Gauss law, i.e., the third of the Maxwell equations, i.e. Eq.~\eqref{Thementionedthirdequation}, states that the divergence of the electric flux density $\tilde{\_D}(\_r,\omega)$ should be zero. Let us recall that $\nabla\cdot(f(\_r)\_V(\_r))=\nabla f(\_r)\cdot\_V(\_r)+f(\_r)\nabla\cdot\_V(\_r)$. Here, $f(\_r)$ is an arbitrary differentiable scalar function. By using this algebraic rule, we find the Gauss law in the form
\begin{equation}
\int_{-\infty}^{+\infty}\varepsilon(\_r,\omega',\omega-\omega')\nabla\cdot\tilde{\_E}(\_r,\omega') {\rm d}\omega'=-\int_{-\infty}^{+\infty}\nabla\varepsilon(\_r,\omega',\omega-\omega')\cdot\tilde{\_E}(\_r,\omega'){\rm d}\omega'\,.
\label{eq:HH2}
\end{equation}
If we wish to discuss again the eigenmodes of a PTC, but now in the presence of dispersion, Eqs.~\eqref{eq:HH1} and \eqref{eq:HH2} must be solved simultaneously.

As our purpose is to calculate the eigenmodes of the Maxwell equations, we shall assume from now on that the space is filled homogeneously with the same material. The permittivity will not depend on the spatial coordinate, i.e., $\varepsilon(\_r,\omega',\omega-\omega')=\varepsilon(\omega',\omega-\omega')$, and the gradient of the permittivity vanishes. Therefore, we have 
\begin{equation}
\int_{-\infty}^{+\infty}\varepsilon(\omega',\omega-\omega')\nabla\cdot\tilde{\_E}(\_r,\omega'){\rm d}\omega'=0\,. 
\label{eq:GTV}
\end{equation} 
At this point, we use the separation of variables method~\cite{ptitcyn2023floquet}, assuming that the electric field is written as the product of two functions: $\tilde{\_E}(\_r,\omega)=\_R(\_r)G(\omega)$. One of them will only depend on the spatial coordinate. The other will only depend on frequency. As we will see later, this assumption is valid, and the general solution for the field is a superposition of all possible eigenfunctions written as the product of the two mentioned functions. 

If we substitute this ansatz into Eq.~\eqref{eq:GTV}, we observe that the term $\nabla\cdot\_R(\_r)$ can be taken out of the integral, and two possibilities arise that guarantee that the equation continues to apply. The first possibility is that $\int_{-\infty}^{+\infty}\varepsilon(\omega',\omega-\omega')G(\omega'){\rm d}\omega'=0$. However, this possibility cannot be realized, as we see from Eq.~\eqref{eq:HH1}. The second possibility requires that $\nabla\cdot\_R(\_r)=0$ meaning that $\nabla\cdot\tilde{\_E}(\_r,\omega)=0$. Accordingly, Eq.~\eqref{eq:HH1} is simplified, and we finally infer that 
\begin{equation}
\nabla^2\tilde{\_E}(\_r,\omega)+{\omega^2\mu_0\varepsilon_0\over2\pi}\int_{-\infty}^{+\infty}\varepsilon(\omega',\omega-\omega')\tilde{\_E}(\_r,\omega'){\rm d}\omega'=0\,.
\label{eq:HHEQU}
\end{equation} 
This is the Helmholtz equation for dispersive time-varying media. 

As a quick sanity check, for the conventional time-invariant case, the temporal complex permittivity corresponds to $\varepsilon_{\rm{T}}(\omega',t)=\varepsilon(\omega')$. Thus, the Fourier transform gives rise to $\varepsilon(\omega',\omega)=2\pi\varepsilon(\omega')\delta(\omega)$. Substituting this relation into Eq.~\eqref{eq:HHEQU} and knowing that $\int_{-\infty}^{+\infty}f(\omega')\delta(\omega-\omega'){\rm d}\omega'=f(\omega)$, we obtain $\nabla^2\tilde{\_E}(\_r,\omega)+k^2\tilde{\_E}(\_r,\omega)=0$ ($k^2=\omega^2\mu_0\varepsilon_0\varepsilon(\omega)$), which is the classical text-book result~\cite{cheng_field_1983}. 

Equation~\eqref{eq:HHEQU} is general and accurate for any linear time-varying causal medium. However, in the following, we concentrate on the particular scenario of periodic temporal modulation applicable to PTCs. Hence, we can use the Floquet theorem given by Eq.~\eqref{eq:BLOCHT} and introduce an alternative version for Eq.~\eqref{eq:HHEQU}. 
In fact, if we take the expression given for the electric field and apply it to Eq.~\eqref{eq:DtepEw}, the electric flux density is simplified to
\begin{equation}
\mathbf{D}(\_r,t)=\sum_n\varepsilon_0{\varepsilon}_{\rm{T}}(\omega_n,t)E_ne^{j\omega_nt}e^{-jkz}\mathbf{a}_x\,.
\label{dwt}
\end{equation}
This equation expresses that the amplitude of the electric flux density for each harmonic $D_n$ is time-dependent ($D_n=\varepsilon_0{\varepsilon}_{\rm{T}}(\omega_n,t)E_n$). As it is clear, this is because the temporal complex relative permittivity ${\varepsilon}_{\rm{T}}(\omega_n,t)$ is a function of time. By having Eq.~\eqref{dwt} and employing time-domain Maxwell's equations, one can deduce the desired equation as 
\begin{equation}
\nabla^2\mathbf{E}(\_r,t)-{1\over c^2}{\partial^2\over\partial t^2}\sum_n{\varepsilon}_{\rm{T}}(\omega_n,t)E_ne^{j\omega_nt}e^{-jkz} \mathbf{a}_x =0.
\label{weq}
\end{equation}
Indeed, regarding periodic modulation, Eq.~\eqref{weq} is the time-domain version of Eq.~\eqref{eq:HHEQU}, which was illustrated fully in the frequency domain. Both equations are valid, and we can use one of them depending on the nature of the problem under study. For example, next, we study the problem when the plasma frequency periodically changes in time, and we show that Eq.~\eqref{weq} is quite convenient for calculating the corresponding band structure. On the other hand, in the subsection that follows, we discuss a periodically modulated resonance frequency, in which we employ Eq.~\eqref{eq:HHEQU}.

%%%%%%%%%%%%%%%%%%%%%%%%%%%%%%%%%%%%%%%%%%%%%%%%%%%%%%%%%%%%%%%%%%%%%%%%%%%%%%%%%%%%%%
%%%%%%%%%%%%%%%%%%%%%%%%%%%%%%%%%%%%%%%%%%%%%%%%%%%%%%%%%%%%%%%%%%%%%%%%%%%%%%%%%%%%%%
%%%%%%%%%%%%%%%%%%%%%%%%%%%%%%%%%%%%%%%%%%%%%%%%%%%%%%%%%%%%%%%%%%%%%%%%%%%%%%%%%%%%%%

\subsubsection{Time-varying plasma frequency}

Here, we consider a PTC made of a dispersive bulk material whose plasma frequency depends on time and varies continuously. The temporal discontinuity in such media in which the plasma frequency rapidly transits once from one value to another has been carefully studied in different works such as~\cite{solis2021time}. It should be noted that changing the plasma frequency can be accomplished by changing the particle density in the system or their effective mass. As discussed above, we can distinguish a Drude model that applies to the description of free electrons. In this case, the plasma frequency corresponds to the density of free-charge carriers, such as the number of electrons per unit volume~\cite{caspani2016enhanced,alam2016large,bohn_spatiotemporal_2021,zhou2020broadband,9577104, ye2024floquet}. 

Alternatively, we can consider a Lorentz model that applies to the description of bound resonances. The permittivity of the Lorentz model in the static case reads as
\begin{equation}
\varepsilon(\omega)=1+\chi(\omega)=1+\frac{\omega^2_{\rm{p0}}}{\omega_{\rm r0}^2-\omega^2+j \gamma \omega}\,,
\label{equ_Lorentzmodel}
\end{equation}
where $\omega_{\rm r0}$ is the natural or resonance frequency of the material, and $\gamma$ represents a phenomenological damping constant. Moreover, the plasma frequency is defined as $\omega^2_{\rm{p0}}=N e^2/(\varepsilon_0 m_{\rm{e}})$ with $N$ being the number of electrons per unit volume, $e$ the electron charge, and $m_{\rm{e}}$ the electron mass. Please note that when we set $\omega_{\rm r0}=0$, we restore the phenomenological dielectric function of the Drude model. This reflects that there is no restoring force on the electrons as opposed to the bound resonance.

Instead of a natural material, we can also think of an artificial material, i.e., a metamaterial made from structured unit cells. Then, we can consider a metamaterial made from $N$ meta-atoms per unit volume, and the polarizability of the metamaterial at the effective level is described by one of the discussed phenomenological models. 

If only the number of electrons, the number of polarizable atoms or the number of meta-atoms per unit volume changes in time (i.e.,~$N_e(t)$), the interaction of the individual electrons/atoms/meta-atoms with the electric and magnetic fields is the same as the case of a static medium. Mathematically, we can state that the response function of the electron/atom/meta-atom and the corresponding polarizability are unaffected. By assuming that the electrons/atoms/meta-atoms interact weakly with each other, the time-dependent complex susceptibility is expressed as a product of a time-varying function, describing the change of density in time and the Lorentzian dispersion.

In this scenario, the frequency dispersion due to the temporal nonlocality is not affected, which means that in Eq.~\eqref{equ_Lorentzmodel}, we substitute $\omega^2_{\rm{p0}}$ by a time-varying function $\omega^2_{\rm{p}}(t)$, and, consequently, we write ${\varepsilon}_{\rm{T}}(\omega', t)=1+\omega^2_{\rm{p}}(t)/(\omega_{\rm r0}^2-\omega'^2+j \gamma \omega')$ (a detailed discussion is given by Ref.~\cite{mirmoosa2022dipole}). Notice that for a periodic temporal modulation, $\omega^2_{\rm{p}}(t)$ is a summation of the constant $\omega^2_{\rm{p0}}$ and a temporal function which fluctuates around zero. Conventionally, this fluctuation is in the form of a sinusoidal function, and the temporal complex relative permittivity is expressed as ${\varepsilon}_{\rm{T}}(\omega', t) = 1 + \chi(\omega')[1 + m_{\rm{p}}\cos(\omega_{\rm m} t)]$. Here, $m_{\rm{p}}$ is the relative modulation strength. The reason to modulate the plasma frequency ``sinusoidally'' is only for simplicity. In general, as before, one can work with any arbitrary periodic function and employ the theory of Fourier series (see Eq.~\eqref{eq: eps}). Accordingly, the expression inside the above bracket should be revised so that we have 
\begin{equation}
{\varepsilon}_{\rm{T}}(\omega', t)=1+\chi'(\omega')\sum_mf_m{\rm{e}}^{jm\omega_{\rm{m}}t}\, ,
\label{EQ:MIRSAJDT}
\end{equation} 
where $\chi'(\omega')=\chi(\omega')/\omega^2_{\rm{p0}}$, and $\omega^2_{\rm{p}}(t)=\sum_mf_m{\rm{e}}^{jm\omega_mt}$ ($f_m$ denotes the Fourier coefficients). Here, let us continue with this general form since our goal is to elucidate the effect of dispersion by obtaining an expression similar to the Eq.~\eqref{eq:summatrixBS}.  

By substituting the expression for ${\varepsilon}_{\rm{T}}(\omega', t)$ written in Eq.~\eqref{EQ:MIRSAJDT} into Eq.~\eqref{weq}, we find the following equation that connects the wavenumber $k$ to the corresponding Floquet frequency $\omega_{\rm{F}}$: 
\begin{equation}
\begin{array}{c} \displaystyle
\sum_n\sum_m{(\omega_{\rm{F}}+n\omega_{\rm{m}})^2\over c^2}\chi'(\omega_m)f_{n-m}E_m{\rm{e}}^{j\omega_nt}=\sum_n\left(k^2-{\omega_n^2\over c^2}\right)E_ne^{j\omega_nt}
\displaystyle\,.
\end{array}
\label{eig}
\end{equation}
To solve this equation, we must ensure that the coefficients corresponding to each harmonic frequency 
$\omega_n$ are equal on both the right and left sides of the equation. This requirement leads to simplifying the above equation and reducing it to 
\begin{equation}
\begin{array}{c} \displaystyle
\sum_m{(\omega_{\rm{F}}+n\omega_{\rm{m}})^2\over c^2}\chi'(\omega_{m})f_{n-m}E_m-\left(k^2-{\omega_n^2\over c^2}\right)E_n\delta_{nm}=0
\displaystyle\,.
\end{array}
\label{eqeqeqRDSF}
\end{equation}
If we compare Eq.~\eqref{eqeqeqRDSF} with Eq.~\eqref{eq:summatrixBS}, we observe that both are similar. The key difference is that in Eq.~\eqref{eqeqeqRDSF}, the field amplitudes $E_m$ are now multiplied by an additional factor $\chi'(\omega_{m})$. That factor is a constant value in the case of non-dispersive time-varying materials that we studied in the previous chapter. Indeed, in the expression above, if $\chi(\omega_m)=\omega^2_{\rm{p0}}/\omega^2_{\rm{r0}}$, which results in $\chi'(\omega_m)=1/\omega^2_{{\rm{r}}0}$, we obtain exactly Eq.~\eqref{eq:summatrixBS}. As we did before for Eq.~\eqref{eq:summatrixBS}, the left side of Eq.~\eqref{eqeqeqRDSF} can also be represented as the multiplication of a square matrix and field-amplitude vector, which is a one-column matrix (see Eq.~\eqref{eq:MATRIXFVEC}). To determine the band structure, again, the determinant of such a square matrix must be zero.     

\begin{figure}[t]
\centerline{\includegraphics[width= 0.7\columnwidth]{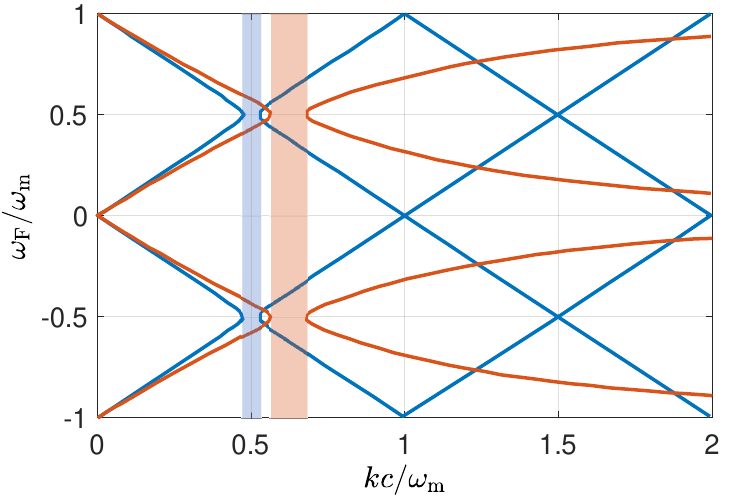}}
\caption{Band structures for a dispersive (orange curve) and a non-dispersive (blue curve) PTC. The dispersion is given by the Lorentz model (assuming that the damping coefficient is zero), and the temporal complex relative permittivity is described by $\varepsilon_{\rm{T}}(\omega',t)=1+\chi(\omega')(1+m_{\rm{p}}\cos(\omega_{\rm{m}}t))$. The non-dispersive relative permittivity is assumed as $\varepsilon(t)=\varepsilon_{\rm av}(1+m_{\varepsilon}\cos(\omega_{\rm m}t))$ where $\varepsilon_{\rm av}=5$, and $m_\varepsilon=0.2$. For non-dispersive case, we set $\omega_{\rm p0}/\omega_{\rm r0}=2$, which ensures the DC permittivity is same as the $\varepsilon_{\rm av}$. The modulation frequency is $\omega_{\rm m}=\omega_{\rm r0}$.}
\label{Fig:dispersion_non_compare}
\end{figure}

We have shown the band structure in Fig.~\ref{Fig:dispersion_non_compare} to see the dispersion effect. Accordingly, the plasma frequency (i.e., more precisely, the number of polarizable atoms) is modulated in the case of non-dispersive (blue curve) and dispersive (orange curve) dielectric media. We observe that in the dispersive case, the size of the bandgap is considerably larger which is a favorable feature for amplification of waves. Also, for the same Floquet frequency, there is a shift in the bandgap, which is due to the dispersion of the medium. The band structure of a dispersive Lorentz media with step-wise periodically modulated plasmon frequency can be calculated using the transfer matrix method \cite{10.1063/5.0187485}. 

%\red{For the rest, we use this opportunity and introduce the reader to an approximation, which is very useful for studying the band structure. This approximation is coined as the weak-modulation approximation, which works well when the modulation strength is small enough (i.e., $\delta\ll 1$)~\cite{Parametric resonances
%in a temporal photonic crystal slab,asadchy2022parametric}. According to this approximation, there are only two dominant harmonics that contribute. They are $\omega_0$ and $\omega_{-1}$ which are associated with the indices $n=0$ and $n=-1$, respectively. As a result, Eq.~\eqref{eig} is considerably simplified and reduced to the following simple form: 
%\begin{equation}
%\begin{bmatrix}
%[1+\chi(\omega_{-1})](\omega_{-1}^2/c^2) - \beta^2
%& (\delta/2c^2)\chi(\omega_{0})\omega_{-1}^2 \\
%(\delta/2c^2)\chi(\omega_{-1})\omega_{0}^2 & [1+\chi(\omega_{0})](\omega_{0}^2/c^2) - \beta^2
%\end{bmatrix}
%\begin{bmatrix}
%  E_{-1}\\
%  E_{0}  
%\end{bmatrix}=0\,. 
%\label{eq: matrix}
%\end{equation} 
%As it is crystal clear, for achieving nonzero field solutions, the determinant of the above $2\times2$ matrix must vanish, and in this way, we obtain the corresponding band structure.  
%}
%%%%%%%%%%%%%%%%%%%%%%%%%%%%%%%%%%%%%%%%%%%%%%%%%%%%%%%%%%%%%%%%%%%%%%%%%
%%%%%%%%%%%%%%%%%%%%%%%%%%%%%%%%%%%%%%%%%%%%%%%%%%%%%%%%%%%%%%%%%%%%%%%%%
%%%%%%%%%%%%%%%%%%%%%%%%%%%%%%%%%%%%%%%%%%%%%%%%%%%%%%%%%%%%%%%%%%%%%%%%%

\subsubsection{Time-varying resonance frequency} 
\label{Sec:TimeVaryingResonanceFrequency}

Next, we elucidate how to calculate the band structure of a Lorentzian dielectric material described above with a time-varying resonance frequency (here, the damping coefficient and the plasma frequency are assumed time-invariant). This scenario is significantly different from that in the previous subsection. Whereas the change of the plasma frequency requires the change of the density of the free electrons or that of the polarizable entities, the change in the resonance frequency preserves the number of particles but changes the properties of the individual oscillators. That can be implemented by various means. On the one hand, the intrinsic resonance frequency of a bulk material can be temporally modulated, for instance, by applying a strong dynamic electric bias~\cite{PhysRevB.108.035119}. However, this method can be practically challenging. On the other hand, an effective resonance frequency can be modulated using spatially structured meta-atoms, as discussed in Ref.~\cite{wang2023unleashing}. These meta-atoms can be as simple as dielectric spheres or more advanced unit cells.

Regarding such an artificial material, if the properties of the individual particle or meta-atom vary, the response function and the associated dipole polarizability are certainly modified, and, accordingly, the temporal complex susceptibility is strikingly revised (see Ref.~\cite{mirmoosa2022dipole} for a complete discussion about this subject). Therefore, we cannot simply write the Lorentzian dispersion and only change the parameters in time in the corresponding model, as we did in the previous subsection. Instead, we need to explicitly derive the response function and the corresponding polarizability for such a meta-atom. Subsequently, we can calculate the dispersion relation based on the wave equations described above. This is what we will do in this subsection.

In the following, we consider that the resonance frequency is temporally modulated as $\omega_{\rm r}(t)$. It is worth mentioning that a similar derivation could be used to obtain the band structure of a PTC with a time-varying damping coefficient. We start with the second-order differential equation that describes the polarization density of the Lorentzian material with the resonance frequency being modulated in time~\cite{mirmoosa2022dipole}:
\begin{equation}
\frac{{\rm d}^2\mathbf{P}\left(t\right)}{{\rm d}t^2}+\gamma\frac{{\rm d}\mathbf{P}\left(t\right)}{{\rm d}t}+\omega_{\rm r}^2\left(t\right){\mathbf{P}}\left(t\right)=\varepsilon_0\omega_{\rm p}^2\mathbf{E}(t)\,,
\end{equation}
in which we suppose that the resonance frequency is sinusoidally modulated as $\omega_{\rm r}^2(t)=\omega_{\rm r0}^2[1+m_{\rm r}\cos{(\omega_{\rm m} t)}]$. This modulation function is only for simplicity regarding the next steps. Accordingly, by taking the Fourier transform regarding the observation time variable $t$ from both sides, we derive that 
\begin{equation}
(\omega_{\rm r0}^2-\omega^2+j\gamma\omega)\tilde{\_P}(\omega)+\frac{m_{\rm r}\omega_{\rm r0}^2}{2}\left[\tilde{\_P}(\omega-\omega_{\rm m})+\tilde{\_P}(\omega+\omega_{\rm m})\right]=\varepsilon_0\omega_{\rm p}^2\tilde{\_E}(\omega)\,. 
\label{eq: 2}
\end{equation}
We see that if there is no modulation (i.e., $m_{\rm r}=0$), we obtain the conventional Lorentzian dispersion as $\tilde{\_P}(\omega)=\varepsilon_0\omega_{\rm p}^2/(\omega_{\rm r0}^2-\omega^2+j\gamma\omega)\tilde{\_E}(\omega)$ that was already expressed in Eq.~\eqref{equ_Lorentzmodel}. 
Equation~\eqref{eq: 2} connects the polarization density to the electric field in the frequency domain. However, on the other hand, these vectors are also related to each other through the definition of the transfer function in both time and frequency domains. Indeed, based on the definition (see Eq.~\eqref{eq:FDPEREL}), we have 
\begin{equation}
\tilde{\_P}\left(\omega\right)=\frac{\varepsilon_0}{2\pi}\int_{-\infty}^{+\infty}{\chi}\left(\omega-\omega',\omega'\right){\tilde{\_E}}\left(\omega'\right){\rm d}\omega'\,, 
\label{eq:SXNW}
\end{equation}
in which ${\chi}(\omega,\omega')$ is the Fourier transform of the response function. As seen and discussed before, there are two angular frequency variables: one is due to the dispersion property (the delay time between the response and excitation), and the other is due to the temporal modulation of the material, which in this case would be caused by the time-varying resonance frequency. 

Equation~(\ref{eq:SXNW}) is general and holds for an arbitrary temporal modulation function. Since we consider a periodic time modulation here, we apply the Floquet theorem. Hence, in the frequency domain ($\omega$), the electric field from Eq.~\eqref{eq:BLOCHT} is re-written as 
\begin{equation}
\tilde{\_E}(\omega)=2\pi\sum_\ell\_E_\ell\delta(\omega-\Omega_\ell)e^{-jkz}\,,
\end{equation}
where $\Omega_\ell=\omega_{\rm{F}}+\ell\omega_{\rm m}$. By having the electric field as the summation of Dirac delta distributions, which all correspond to one single value of the phase constant $k$, and by using Eq.~\eqref{eq:SXNW}, we deduce the polarization density in terms of the susceptibility:
\begin{equation}
\tilde{\_P}(\omega)={\varepsilon_0}\sum_\ell\_E_\ell{{\chi}}\left(\omega-\Omega_\ell,\Omega_\ell\right)\,.
\label{eq:ssezb}
\end{equation} 
Now, it may be clear that our next step is to substitute the above equation into Eq.~\eqref{eq: 2}, and by this way, we achieve an equation that provides an expression for the susceptibility. Therefore, we infer that 
\begin{equation}
\begin{split}
&\left(\omega_{\rm r0}^2-\omega^2+j\gamma\omega\right)\sum_\ell\_E_\ell{\chi}\left(\omega-\Omega_\ell,\Omega_\ell\right)+\frac{m_{\rm r}\omega_{\rm r0}^2}{2}\sum_\ell\_E_\ell{\chi}\left(\omega-\Omega_\ell-\omega_{\rm m},\Omega_\ell\right)+\cr
&\frac{m_{\rm r}\omega_{\rm r0}^2}{2}\sum_\ell\_E_\ell{\chi}\left(\omega-\Omega_\ell+\omega_{\rm m},\Omega_\ell\right)=\omega_{\rm p}^2\sum_\ell\_E_\ell\delta(\omega-\Omega_\ell)\,.
\end{split} 
\end{equation} 
To solve this equation, we multiply the right- and left-hand sides by a function in the form of a Dirac delta distribution: $\delta(\omega-\Omega_{\ell'})$ in which $\Omega_{\ell'}=\omega_{\rm{F}}+\ell'\omega_{\rm m}$, and, subsequently, we integrate over all possible angular frequencies $\omega$. As a consequence, concerning each integer value for $\ell$ and $\ell'$, we express eventually that 
\begin{equation}
\begin{split}
&\left(\omega_{\rm r0}^2-\Omega_{\ell'}^2+j\gamma\Omega_{\ell'}\right){\chi}\left(\Omega_{\ell'}-\Omega_\ell,\Omega_\ell\right)+\frac{m_{\rm r}\omega_{\rm r0}^2}{2}{\chi}\left(\Omega_{\ell'}-\Omega_\ell-\omega_{\rm m},\Omega_\ell\right)+\cr
&\frac{m_{\rm r}\omega_{\rm r0}^2}{2}{\chi}\left(\Omega_{\ell'}-\Omega_\ell+\omega_{\rm m},\Omega_\ell\right)=\omega_{\rm p}^2\delta_{\ell'\ell}\,,
\end{split} 
\label{eq: 4}
\end{equation}
where $\delta_{\ell'\ell}$ is the Kronecker delta. As mentioned, the above equation has been written for each integer value $\ell$. Thus, we can fix this value ($\ell$) and accordingly repeat the above equation enough times ($2N+1$ times) by changing the integer value of $\ell'$ ($-N<\ell'<+N$). From this perspective, Eq.~\eqref{eq:ssezb} indicated that a square matrix $\overline{\overline{\chi}}$ is formed, which connects the polarization density at each Floquet angular frequency to the electric field. The columns of this matrix correspond to each value of $\ell$, and the rows correspond to each value of $\ell'$. Ideally, this matrix is infinite in size (i.e., $N\rightarrow\infty$). However, we can consider lower and upper limitations for $\ell$ and $\ell'$. This is because, at each Floquet angular frequency, only a few values are important and play a role if the modulation strength is not strong. In the above, fixing $\ell$ and changing $\ell'$, in fact, gives one column of the matrix $\overline{\overline{\chi}}$. In the next step, we change the integer value $\ell$ and again try to make the other column of the matrix. After performing this analysis for enough values of $\ell$, the whole square matrix is constructed. Having this matrix, using Eq.~\eqref{eq:ssezb}, knowing that $\tilde{\_D}(\omega)=\varepsilon_0\tilde{\_E}(\omega)+\tilde{\_P}(\omega)$, and, eventually, applying Maxwell's equations, one can deduce the band structure as explained in Section 2.1 of Ref.~\cite{ptitcyn2023floquet}.

% \red{Include formulation (with some but not all derivations) of eigenvalue equation for the two cases: when resonance frequency is modulated and when plasma frequency is modulated. something like 5-8 pages.}

% \red{Point out that more effects of dispersion will be discussed in Sec. 6.3}.

\subsubsection{Estimating the size of momentum bandgaps }\label{est}

% \red{Xuchen, this section should consist of 3 paragraphs. Please restructure the text accordingly. In the first one, you discuss the weak-modulation approximation. You can start by citing eq.\ref{eig} and then explaining eq \ref{eq: matrix}. Also, add a formula showing the bandgap size (eq. 1 in our preprint ``Unleashing infinite momentum bandgap using resonant material systems''). But do not discuss yet your 'recipe' for estimating bandgap size. You can show the figure (see my comment below). }

% \red{This is the second paragraph.}
The size of the momentum bandgap is crucial because it determines the range of light momenta (wavenumber) in an incident pulse that can be amplified in a PTC. One key factor for the size of the momentum bandgap is the relative modulation depth of the material parameter, such as permittivity, $m_\varepsilon$. By increasing the modulation depth, the size of the momentum bandgap can be expanded. However, this is not the only factor. Engineering the material dispersion is an alternative means to widen the bandgap~\cite{wang2023unleashing}, being especially important in those parts of the frequency spectrum where reaching a strong material modulation depth is very challenging~\cite{hayran2022ℏomega}.
% (see more detailed discussions in Section~\ref{sec: power constraints}). 

In order to find the size of the bandgap, one needs to plot the photonic band diagram. Nevertheless, calculating the band structure of a PTC with arbitrary general material dispersion, although possible, could be a difficult task, requiring a numerical solution of the matrix-type wave equation (\ref{eqeqeqRDSF}). There exists a simple yet powerful approach to estimating the size of the bandgap based on a closed-form analytical solution~\cite{martinez2018parametric,asadchy2022parametric,Koutserimpas2022parametric}. The approach provides a good qualitative and quantitative description of the bands in the vicinity of the momentum bandgap and is based on the assumption of the small modulation amplitude.    
% Predicting the size of the possible momentum bandgap easily is, therefore, an important task.   
% Generally, the size of the momentum bandgap can be accurately estimated once the band structure has been calculated. This can be done, e.g., by solving the matrix equations expressed in Eq.~(\ref{eqeqeqRDSF}). In the band structure, by drawing a horizontal line of $\omega_{\rm F}=\omega_{\rm m}/2$, and finding the distance of the two intersection points with the dispersion curves, one can measure the bandgap size. 
% This approach is general and applicable for arbitrary modulation depth. For strong modulation depth, higher-order harmonics are obvious, and therefore, the matrix size would be large. For this reason, this method should be based on numerical tools and can not reveal the physics of what factors determine the momentum bandgap size. \red{this paragraph is odd}
When the modulation depth is small (e.g., $m_{\rm p}\ll1$ for material with plasma frequency modulated), the higher-order harmonics excited inside the bandgap of the PTC are very weak and can be neglected, simplifying the description only to two dominant harmonics, $0$-th and $-1$-st. This assumption significantly simplifies the mathematics and makes it possible to solve the band structure analytically, providing more physical insights into the bandgap formation.

\begin{figure}[tb]
\centerline{\includegraphics[width= 0.7\columnwidth]{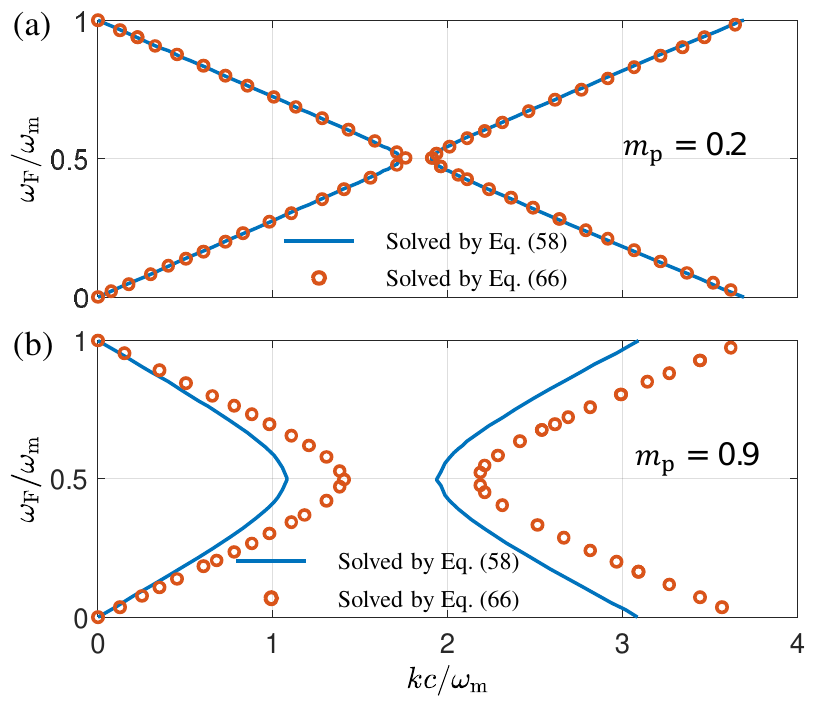}}
\caption{Comparison of the band structures of a dispersive PTC in the proximity of the momentum bandgap calculated using rigorous Eq.~(\ref{eqeqeqRDSF}) (the truncated matrix size in the calculation is  $9\times9$) and using Eq.~(\ref{eq: matrix}) based on the weak-modulation approximation (the matrix has $2\times2$ size). The modulation strength equals to (a) $m_{\rm p}=0.2$ and  (b) $m_{\rm p}=0.9$.
 The permittivity modulation function is $\hat{\varepsilon}_{\rm{T}}(\omega, t)=1+\omega^2_{\rm{p}}(t)/(\omega_{\rm r0}^2-\omega^2+j \gamma \omega)$, where $\omega_{\rm p}^2(t)=\omega_{\rm p0}^2[1+m_{\rm p}\cos(\omega_{\rm m}t)]$, $\omega_{\rm p0}=3.5\omega_{\rm r0}$, $\omega_{\rm m}=0.2\omega_{\rm r0}$, and $\gamma=0$.
}
\label{Fig:exact_approx}
\end{figure}

To exemplify such an analysis, we consider in the following a PTC with a sinusoidally modulated plasma frequency with modulation amplitude $m_{\rm p}=0.2$. 
The band structure of this PTC obtained by solving Eq.~(\ref{eqeqeqRDSF}) is plotted in Fig.~\ref{Fig:exact_approx}(a) (see blue solid line). 
The size of the bandgap is given by the distance between the two intersection points of the horizontal line representing the Floquet frequency $\omega_{\rm F}=\omega_{\rm m}/2$ with the photonic bands. 
Note that here, we have a rather small modulation depth. Therefore, the higher-order harmonics can be neglected. The main harmonics are excited at $\omega_0=\omega_{\rm m}/2$ and $\omega_{-1}=-\omega_{\rm m}/2$, respectively. By considering only these two modes, the matrix in Eq.~(\ref{eqeqeqRDSF}) can be reduced to a $2\times2$ matrix resulting in the wave equation in the form
% In this case, it is easy to solve the dispersion relation and the size of the momentum bandgap analytically so that it becomes possible to obtain more analytical insights into the bandgap formation. 
% The eigenequation with this reduced number of modes can be written as
\begin{equation}
\begin{bmatrix}
[1+\chi(\omega_{\rm F}-\omega_{\rm m})] (\omegaf-\omega_{\rm m})^2 - k^2c^2
 & m_{\rm p} \chi(\omega_{\rm F})(\omegaf-\omega_{\rm m})^2/2 \\
m_{\rm p} \chi(\omega_{\rm F}-\omega_{\rm m})\omega_{\rm F}^2/2 & [1+\chi(\omega_{\rm F})] \omega_{\rm F}^2 - k^2c^2
\end{bmatrix} \cdot
\begin{bmatrix}
  E_{-1}\\
  E_{0}  
\end{bmatrix}=0\, .
\label{eq: matrix}
\end{equation} 
The band structure predicted by Eq.~(\ref{eq: matrix}) is displayed in Fig.~\ref{Fig:exact_approx}(a) by red circles. It is seen that the two band structures calculated using Eq.~(\ref{eqeqeqRDSF}) and Eq.~(\ref{eq: matrix}) almost overlap, showing the high accuracy of the method based on the weak-modulation approximation. 
From the eigenvalue problem (\ref{eq: matrix}), we can obtain the bandgap size. This can be done by substituting $\omega_{\rm F}=\omega_{\rm m}/2$   and requiring that the determinant of the matrix vanishes. Then, we can obtain two positive solutions of $k$, 
\begin{equation}
k_\pm = {\omega_{\rm{m}}\over 2c} \sqrt{1+\chi(\omega_{\rm{m}}/2) [1 \pm  m_{\rm p}/2 ]}\,,
\label{eq:NKWX1}
\end{equation} 
corresponding to the edges of the dispersion relation that form the bandgap. The bandgap size is then given by $|k_+-k_-|$. 
Interestingly, the described approach for estimating the bandgap size can provide a reasonably good qualitative description even for strong modulations. As one can see from Fig.~\ref{Fig:exact_approx}(b), even when $m_{\rm p}=0.9$,  the approximate approach predicts a shifted but qualitatively similar band structure near the bandgap. The bandgap size has a small deviation compared to that calculated using the exact solution. 

% \red{And this is third paragraph. Only this paragraph should contain the new stuff that we have in our preprint. We may hide this paragraph for the first submission in case if our preprint is not published yet not to compromise it. Or we can shorten it later so that we do not reveal much. But for now, please structure it in the following way. First, talk about the recipe, cite fig. 7b. Second, talk briefly about findings in our paper \cite{wang2023unleashing}. Highlight that frequency dispersion is responsible for the improvement of the bandgap size. }
As one can see from (\ref{eq:NKWX1}), the two edges of the momentum bandgap, $k_+$ and $k_-$, correspond to the wavevectors of two stationary (time-invariant) media calculated at $\omega=\omega_{\rm m }/2$. One medium is characterized by the susceptibility  $\chi(\omega)(1+m_{\rm p}/2)$, while the other  is characterized by    $\chi(\omega)(1-m_{\rm p}/2)$~\cite{wang2023unleashing}. This simple but powerful observation points us to the fact that the bandgap size can be predicted even without solving equation (\ref{eq: matrix}). Instead, it can be done by simply plotting the dispersion relations of two auxiliary time-invariant media. 

\begin{figure*}[t!]
\centerline{\includegraphics[width= 1\columnwidth]{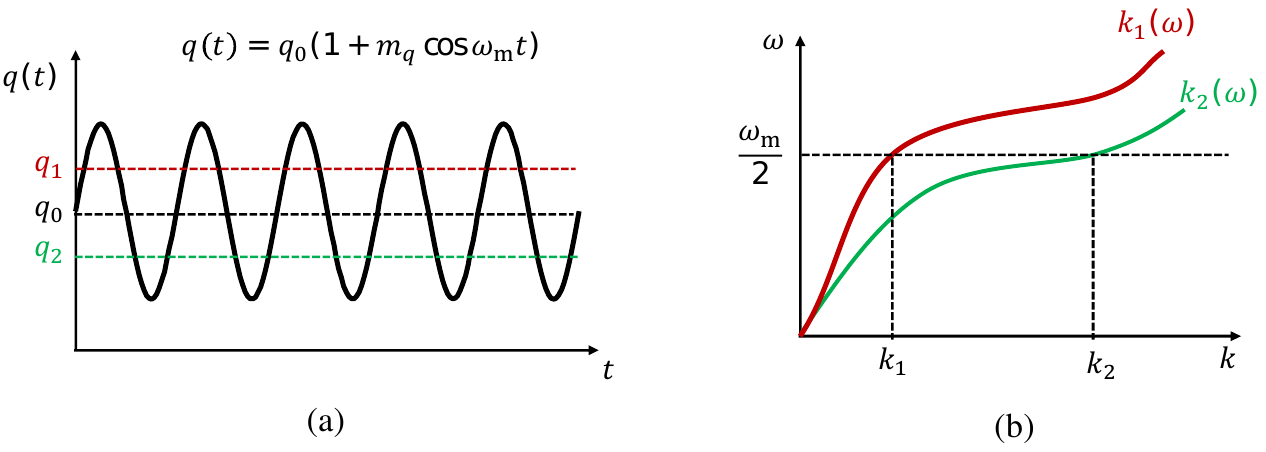}}
\caption{A simple rule for estimating the size of the momentum bandgap in a PTC. (a) A sinusoidal modulation form is applied to quantity $q(t)$.   (b) Conceptual dispersion relation curves $k_1(\omega)$ and $k_2(\omega)$ of two auxiliary time-invariant materials with $q(t) = q_1$ and $q(t) = q_2$, respectively. The size of the bandgap of the PTC whose modulation function is shown in (a) equals the distance $|k_2-k_1|$ between two crossing points shown in (b). 
}
\label{Fig:general rule}
\end{figure*}

The above-mentioned rule is not limited to PTCs where the time-varying parameter is the plasma frequency~\cite{wang2023unleashing}. Generally speaking, it can be used for PTCs with other time-varying physical quantity $q(t) = q_0(1 + m_{  q}\cos(\omega_{\rm m} t))$, 
where $m_{q}$ is the modulation depth and $q$ could be permittivity, permeability, etc. (see Fig.~\ref{Fig:general rule}(a)).
To apply the rule, we first plot the dispersion curves of the two time-invariant auxiliary materials described by \( q_1=(1 +\frac{m_{ q}}{2})q_0\) and \( q_2=(1 - \frac{m_{  q}}{2})q_0\). Their conceptual dispersion relations  \(k_1(\omega)\) and \(k_2(\omega)\) are sketched  in Fig.~\ref{Fig:general rule}(b). 
Next, we draw a horizontal line corresponding to \(\omega = \omega_{\rm m}/2\) and find the two points where it intersects with the dispersion curves of the two auxiliary materials. 
The intersection points with the curves are $[ k_1(\omega_{\rm m}/2), \omega_{\rm m}/2 ] $ and $[ k_2(\omega_{\rm m}/2), \omega_{\rm m}/2 ] $. The horizontal coordinates of these two points precisely define the edges of the momentum bandgap for a PTC modulated according to \(q(t)\). Thus, the bandgap width equals to $|k_2(\omega_{\rm m}/2)-k_1(\omega_{\rm m}/2)|$. 

This rule is very useful, especially in scenarios where the PTC is made of a material with complicated frequency dispersion. Since it is visual and does not imply determining the band structure of the PTC, it can be used not only to estimate the bandgap size of a given PTC (analysis purpose) but also to predict what material dispersion one would need to obtain the bandgap with desired characteristics (synthesis purpose). For example, in \cite{wang2023unleashing},   it was found that a highly dispersive material, e.g., a resonant metamaterial, can drastically enhance the size of the momentum bandgap for a given modulation depth. 
% Even a small change in material property in resonant structures leads to a large shift in resonant frequency. This results in a clear separation of the dispersion curves of Material 1 and 2 along the \(\omega\)-axis. Now, it is possible to achieve a configuration for a suitably chosen material where only one dispersion curve is intersected for a specific modulation frequency. This indicates that the momentum bandgap has only one edge. This, in turn, implies an infinite momentum bandgap with the other edge extending to infinity.

Altogether, this discussion finalizes the consideration of material dispersion in the context of PTCs. It remains to emphasize that material dispersion can always be considered in the analysis by suitable modifications of the theoretical framework. Material dispersion is particularly important when describing realistic materials, so when realistic predictions on observable quantities shall be made. However, assuming non-dispersive material is also of scientific value. The simplification allows us to obtain a better glimpse into basic effects, which is particularly useful when exploring fundamental phenomena. Therefore, the reader will find in this tutorial and in the general literature discussions that assume a non-dispersive but time-varying permittivity. Such assumptions are acceptable provided that the potential limitations are recognized.

After discussing the material dispersion aspect of a realistic PTC, we consider two other aspects that characterize a more realistic system in the following sections. First, we consider spatially finite PTCs. Second, we consider temporally finite PTCs.
%%%%%%%%%%%%%%%%%%%%%%%%%%%%%%%%%%%%%%%%%%%%%%%%%%%%%%%%%%%%%%%%%%%%%%%%%
%%%%%%%%%%%%%%%%%%%%%%%%%%%%%%%%%%%%%%%%%%%%%%%%%%%%%%%%%%%%%%%%%%%%%%%%%
%%%%%%%%%%%%%%%%%%%%%%%%%%%%%%%%%%%%%%%%%%%%%%%%%%%%%%%%%%%%%%%%%%%%%%%%%

\subsection{Spatially-finite PTCs }\label{sfptcs}
% Here, we will discuss the effects of periodic time-modulation in those structures which are finite in at least one spatial dimension. These include time-varying spheres and time-varying material slabs.\\

% References: \cite{panagiotidis2022inelastic}, \cite{wang2023metasurface}, \cite{sadafi2023dynamic}, \cite{asadchy2022parametric}, \cite{holberg1966parametric}. Cited all

While analyzing the eigenmodes of PTCs with infinite spatial extent is fundamentally important for understanding their physics, in real applications, the PTCs always have spatial boundaries. In optical applications, the size of the experimental sample can be tens of wavelengths or even larger; thus, it can be treated as a nearly infinite-sized structure. However, on many other occasions, including those at microwave frequencies, large-size PTCs may not be feasible. Therefore, analyzing how the finite spatial extent affects the properties of PTCs is crucial. 

For finite-sized PTCs, spatial boundaries exist between the PTCs and the background medium. 
Under external excitation, the scattering from the PTCs can be analyzed using a conventional mode-matching method. By listing all the eigenmodes inside and outside the PTCs, and then applying spatial boundary conditions, the amplitudes of all the eigenmodes inside and outside the PTCs can be uniquely solved. This approach enables the calculation of reflection, transmission, and the fields inside the PTCs~\cite{zurita2009reflection}. 

\begin{figure}[t]
\centerline{\includegraphics[width= 1\columnwidth]{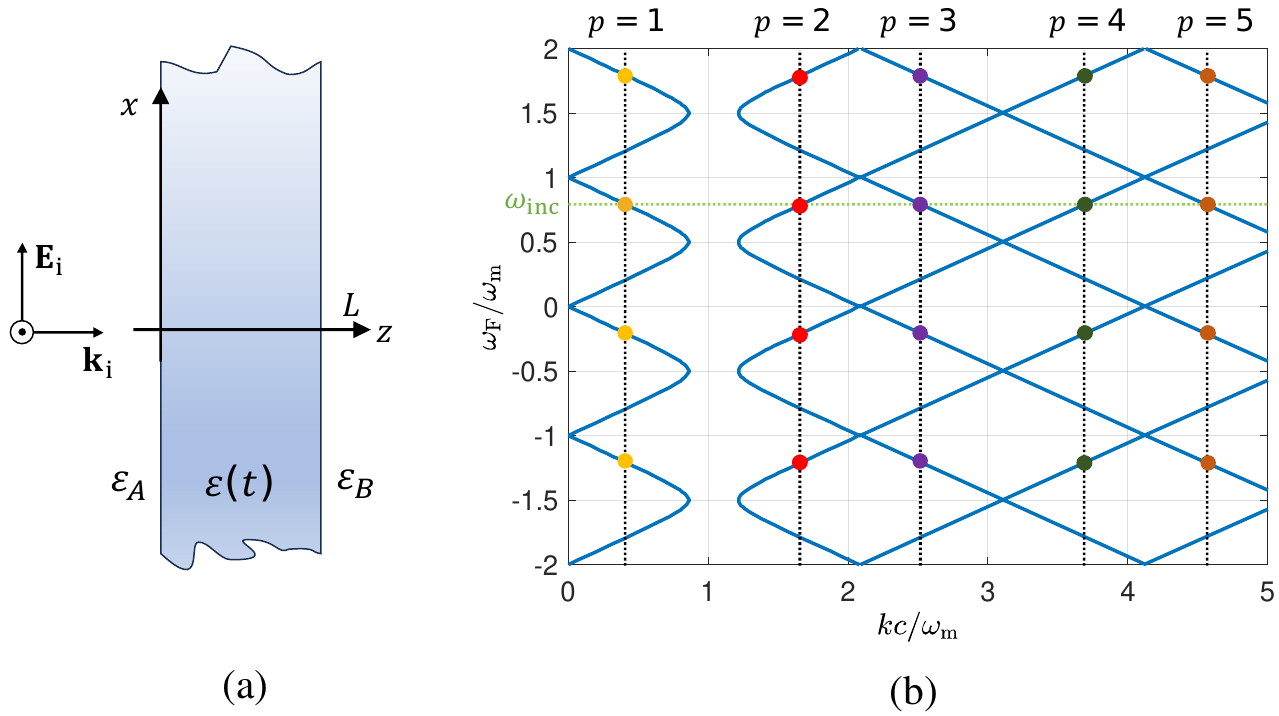}}
\caption{(a) A dielectric slab characterized by a time-varying permittivity $\varepsilon(t)=\varepsilon_{\mathrm{av}}(1+m_\varepsilon \cos(\omega_{\rm m}t))$. The slab has a thickness of $L$ and is surrounded by semi-infinite media characterized by $\varepsilon_A$ and $\varepsilon_B$, respectively. In the specific example, we consider $\varepsilon_{\mathrm{av}} = 5$ and $m_\varepsilon = 0.6$. (b) Band structure of the time-varying material. The points of the same color denote frequency harmonics of a single eigenmode of the PTC. When the PTC is illuminated at a spatial boundary by a plane wave with frequency $\omega_{\rm inc}$, harmonics of different frequencies and wavenumbers denoted by the colorful dots in the figure are excited. 
}
\label{Fig:finite_slab}
\end{figure}

Next, we consider an example to briefly explain the procedures for solving the reflection and transmission from a finite-size PTC.  
The time-varying slab of length $L$ is shown in Fig.~\ref{Fig:finite_slab}(a). The permittivity inside the slab is modulated periodically in time as $\varepsilon(t)$. Outside the slab, the permittivity is $\varepsilon_A$ for the range $z < 0$ and $\varepsilon_B$ for $z > L$. In this case, two spatial boundaries exist at $z = 0$ and $z = L$. 
The external excitation is an $x$-polarized plane wave illuminating the PTC along the $z$-direction. 
% Before solving the spatial boundary conditions, we must express the fields inside and outside each media by summing all possible modes that propagate in them.
Figure~\ref{Fig:finite_slab}(b) shows the band structure of the infinite PTC. An incident plane wave with frequency $\omega_{\mathrm{inc}}$ excites inside the PTC multiple eigenmodes with different wavenumber $k$ but the same eigenfrequency $\omega_{\rm inc}$. Each of these eigenmodes consists of an infinite number of frequency harmonics  $\omega_{\mathrm{i}} + n\omega_{\mathrm{m}}$ denoted as points with the same color in the figure.  
% The excited momenta can be determined by finding the wavenumbers in the band structure corresponding to the incident frequency. The modes are marked by yellow circles, representing the intersection of the light green dashed line with the band structure. This process is similar to the case when an incident field illuminates conventional spatial photonic crystals. 
% Then, due to time modulation, these modes with different momenta couple to multiple frequencies, $\omega_{\mathrm{i}} + n\omega_{\mathrm{m}}$, spaced by $\omega_{\mathrm{m}}$. 
Thus, all the frequency harmonics marked with points in Fig.~\ref{Fig:finite_slab}(b) are excited inside the PTCs, although the external excitation had a unique wavenumber $k$. 
% This is distinct from an eigenmode analysis, where an infinite medium and initial plane wave fields are assumed before modulation. In the finite structure case, although external excitation has a unique $k$, different wavenumbers in the time-varying medium are induced due to the existence of spatial boundaries.

We label each eigenmode with index $p$  according to its wavenumber. 
% Considering that all modes, marked by circles of different colors in Fig.~\ref{Fig:finite_slab}(b), are excited in the time-varying media, we group the modes according to their wavenumber using index $p$. 
The first eigenmode with $p=1$ contains $n$ frequency harmonics, with frequencies $\omega_{\mathrm{i}} + n\omega_{\mathrm{m}}$. Using Eq.~(\ref{eq:MATRIXFVEC}), we can solve the amplitude relations among all the harmonics of a given eigenmode $E_{p=1, n}(\omega_{\rm inc})$. A similar process applies to other eigenmodes. It should be noted that negative $k$ values, corresponding to backward waves traveling in the $-z$ direction, are also excited but are not shown in the band structure. 

Although each eigenmode has a fixed relation of harmonic amplitudes, different eigenmodes can have different amplitude weights. For the $p$-th eigenmode, we denote the weights for forward and backward waves as $C_p$ and $D_p$, respectively. The field inside the PTCs slab is a superposition of all the possible harmonics,
\begin{equation}
    \mathbf{E}_{\mathrm{sl}}(z, t) = \sum_{p=1}^{\infty} \sum_{n=-\infty}^{\infty} \left [C_p e^{-jk_p(\omega_{\mathrm{inc}}) z} + D_p e^{jk_p(\omega_{\mathrm{inc}}) z} \right] E_{p,n} (\omega_{\mathrm{inc}}) e^{j\omega_n t} \mathbf{a}_x\,.   
    \label{generalzurita}
\end{equation}

Outside the time-varying slab, the reflected field $\mathbf{E}_{\mathrm{r}}(z, t)$ and transmitted field $\mathbf{E}_{\mathrm{t}}(z, t)$ consist of infinite frequency harmonics, i.e., $\omega_{\rm inc}+n\omega_{\rm m}$. These harmonics must respect the dispersion relation of the stationary background media, $k_n^{ {A, B}}(\omega_{\mathrm{inc}}) = \sqrt{\varepsilon_{A,B}}(\omega_{\mathrm{inc}} + n\omega_{\mathrm{m}})/c$. The fields are a superposition of these frequency harmonics, with unknown harmonic amplitudes to be solved later.

After defining the fields inside and outside the PTCs, we can apply two spatial boundary conditions at $z = 0$ and $z = L$ to compute all the unknown coefficients:

\begin{subequations}
   \begin{equation}
    [\mathbf{E}, \mathbf{H}]_{\mathrm{i}}(0, t) + [\mathbf{E}, \mathbf{H}]_{\mathrm{r}}(0, t) = [\mathbf{E}, \mathbf{H}]_{\mathrm{sl}}(0, t)\,,
   \end{equation} 
   \begin{equation}
    [\mathbf{E}, \mathbf{H}]_{\mathrm{sl}}(L, t) = [\mathbf{E}, \mathbf{H}]_{\mathrm{t}}(L, t)\,.
   \end{equation} \label{eq: interface conditon}
\end{subequations}

This analysis reveals that a slab, when illuminated by a plane harmonic wave $\omega_{\mathrm{i}}$, effectively becomes a polychromatic light source, radiating at frequencies $\omega_{\mathrm{i}} + n\omega_{\mathrm{m}}$. We have thus established the general equations necessary to determine the reflection and transmission coefficients for the generated harmonics.

Using the method of eigenmode expansion combined with conditions of continuity of fields at the spatial boundaries, one can solve the scattering or eigenfields problem of PTCs with different shapes. Typical examples include finite planar slabs \cite{holberg1966parametric,martinez2016temporal,martinez2018parametric,zurita2010resonances,valdez2024parametric,globosits2024photonic} as discussed before, and sphere particles \cite{asadchy2022parametric}. In the latter scenario, vector spherical harmonics are employed for the spatial component of the wavefunctions instead of plane waves, as previously outlined. Nevertheless, aside from this distinction, the overall approach remains unchanged.

As discussed earlier, spatially infinite PTCs exhibit momentum bandgaps, with their widths being linearly proportional to the modulation depth when the modulation is weak. This raises the pertinent question of the conditions under which spatially-finite PTCs support bandgaps and the factors determining their widths.
% While for infinite, unbounded media, under time modulation, the momentum bandgap opens and parametric amplification happens, for a finite media, the question is if the parametric amplification always happens? 
It was demonstrated in \cite{martinez2018parametric,asadchy2022parametric} that while parametric amplification always occurs, it is typically finite. The exponential growth, that is, parametric oscillations, only happen at specific conditions for the modulation depth, and radiation and dissipation losses. 
In particular, for low or moderate modulation depths, finite PTCs exhibit exponential amplification only when $\omega_{\rm m}/2$ is close to the resonance frequency $\omega_{\rm r}$ of some mode inside the same material without temporal modulations. This translates into the requirement that $\Delta \omega = |\omega_{\rm m}/2- \omega_{\rm r}| \ll \omega_{\rm m}/2$. The resonance mode can be, for example, a Fabry-Perot mode~\cite{martinez2018parametric} or Mie resonance mode~\cite{asadchy2022parametric}. 
The qualitative description of this process can be given by the temporal coupled mode theory (see more details in Section~\ref{PAPTCs}). 
When a finite PTC with low modulation depth is illuminated by incidence at a frequency close to $\omega_{\rm m}/2$, one can use the weak-modulation approximation and describe the system by merely two coupled quasi-normal modes. They both oscillate at frequency $\omega_{\rm m}/2$ and are described by  temporal envelopes $a_1(t)$ and $a_2(t)$. The coupled-mode equations for this case then read~\cite{asadchy2022parametric}
\begin{subequations}
\begin{equation}
      \frac{\mathrm{d}}{\mathrm{d}t}a_1(t)=[-j\Delta \omega-\gamma_{\rm tot}]a_1(t)-j\eta a_2^*(t)\,,  
\end{equation}
\begin{equation}
      \frac{\mathrm{d}}{\mathrm{d}t}a_2^*(t)=[j\Delta \omega-\gamma_{\rm tot}]a_2^*(t)+j\eta^* a_1(t)\,, 
\end{equation}\label{eq: coupled}
\end{subequations}
where $\eta$ is the coupling parameter that is linearly proportional to the modulation depth $m_\varepsilon$, $\gamma_{\rm tot}$ is the total decaying rate caused by possible radiation and/or dissipation loss in the system, and  '$*$' denotes the complex conjugate operation. The condition of parametric amplification ($a_{1,2}(t)$ grow exponentially) can be solved from Eqs.~(\ref{eq: coupled}). The condition is strongly related to the modulation depth. The threshold modulation depth that induces the parametric amplification of such finite-sized PTCs is expressed as,
\begin{equation}
    m_\varepsilon^{\rm thr} \propto \gamma_{\rm tot}+\frac{{\Delta \omega}^2}{2\gamma_{\rm tot}}\,. \label{eq: thr}
 \end{equation} 
Equation~(\ref{eq: thr}) demonstrates that a higher decay rate of the mode necessitates a greater modulation depth to achieve parametric oscillation. Furthermore, a larger deviation of the half-modulation frequency from the resonance frequency also requires an increased modulation depth.

The above theory can be applied to various different spatially-finite PTC geometries. One example is the time-varying dielectric slab discussed in \cite{holberg1966parametric}. 
It was found in \cite{holberg1966parametric} that if exponentially increasing waves are to be produced, the modulation depth must exceed a certain critical value. This critical value is dependent on the thickness of the slab and on the relation of the permittivities of the slab and the surrounding medium. For the case that the modulation depth is less than the critical value, the transmission and reflection coefficients can be calculated by the mode expansion method as displayed in Eq.~(\ref{generalzurita}) combined with interface conditions in Eq.~(\ref{eq: interface conditon}).

\begin{figure}[tb]
\centerline{\includegraphics[width= 1\columnwidth]{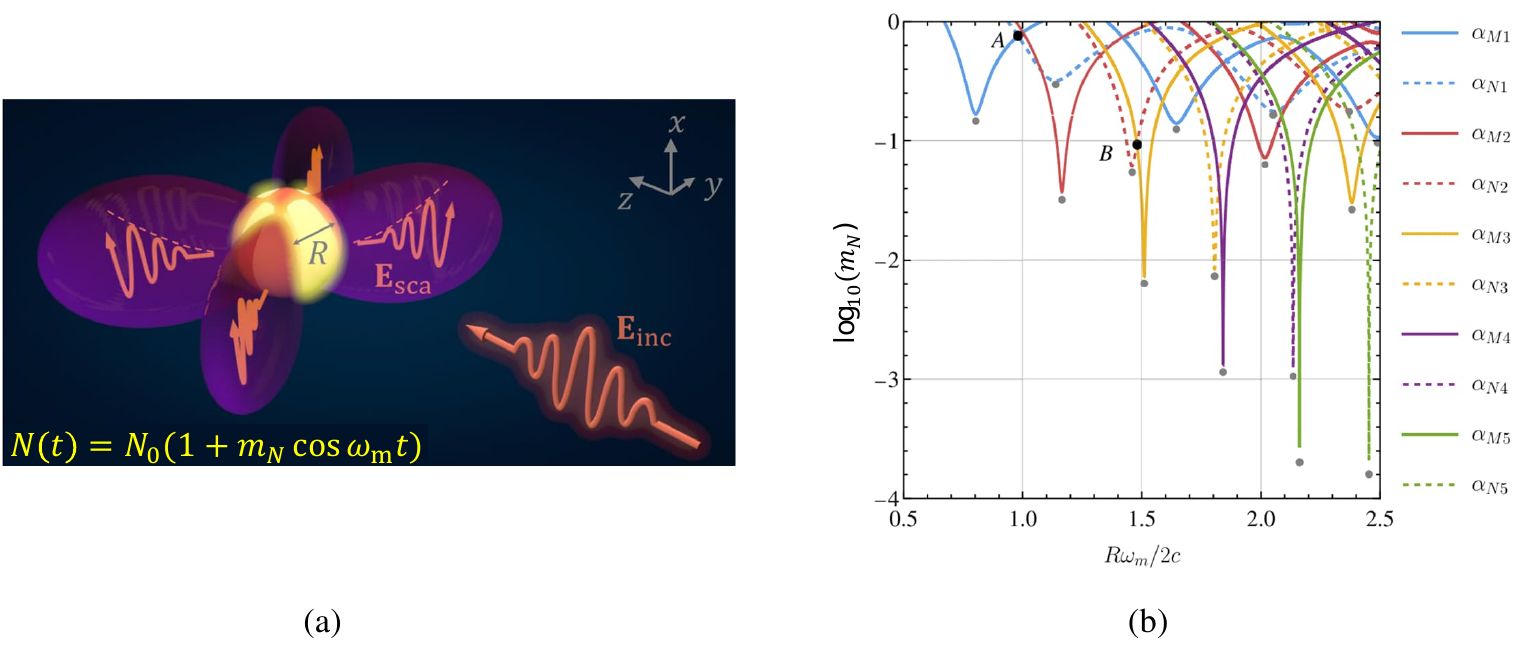}}
\caption{(a) A spherical particle with a time-modulated bulk carrier
density $N(t)$ illuminated by incident light. Temporal modulation leads to parametric Mie resonances with simultaneous scattered-field amplification and the possibility of far-field pattern manipulation. 
(b) Threshold values of the modulation depth $m_N$ 
that provide parametric oscillations at
fixed frequency $\omega_{\rm m}/2$  for different multipolar modes in the time-modulated
sphere versus its radius $R$.  It can be seen that the higher-order modes, as sustained at higher radii, require a lower modulation depth to observe parametric oscillations due to their higher quality factors (lower radiation losses). Figures were reprinted from~\cite{asadchy2022parametric}. 
}
\label{Fig:viktar figures}
\end{figure}

Another example, a sphere with time-varying charge carrier density $N(t)$, is shown in Fig.~\ref{Fig:viktar figures}(a). Figure~\ref{Fig:viktar figures}(b) shows the minimum modulation depth that can provide exponential  amplification (parametric oscillations) for different resonant modes of the time-varying dielectric sphere. When the sphere radius $R$ increases, higher order electric (denoted as $\alpha_N$) and magnetic (denoted as $\alpha_M$) modes can provide parametric oscillations. Those higher-order multipolar modes possess higher quality factors (lower $\gamma_{\rm tot}$) and, therefore, they have lower thresholds of modulation amplitudes for achieving parametric oscillations.

% A more general qualitative understanding of the conditions for parametric amplification is that, the incoming power (pump and incident) should exceed the loss including radiation and dissipation losses of the system. In the condition (\ref{eq: thr}), improving the modulation depth is an effective way to increase the external energy. Beyond the threshold value, the incoming power is larger than the losses, therefore, parametric amplification happens. 
% On the other hand, if the modulation depth is a fixed value, the parametric amplification condition can be determined by the structural parameters. 
% Adjusting the geometric parameters which creates a higher quality factor system can more easily induce parametric amplification.  

% the parametric amplification condition of a finite time-varying dielectric slab.   Below this thickness, the wave does not experience exponential growth, while beyond it, the wave enters an unstable region, experiencing parametric amplfication with exponentially growing amplitude. 

Note that the exponential field growth indicates an unstable system. In other words, the threshold value of modulation depth in \cite{asadchy2022parametric} and slab thickness in \cite{holberg1966parametric} separate stable and nonstable regimes. But even when we are above the threshold (unstable), the growth is always limited in practice. In particular, we quickly reach the nonlinear regime, and then the system becomes detuned from the optimal conditions. For example, in recent experimental work on metasurface-based PTCs \cite{wang2023metasurface}, a finite-sized metasurface was modulated, and parametric oscillations occurred, but finite gain and stable performance were observed.

It should be finally noted that term 'exponentially growing' indicates instability. In this case, calculating the scattering amplitudes using the mode expansion method is nonphysical, as these quantities are defined in the frequency domain, assuming the system is stable. 
The stability issue is also discussed in~\cite{salehi2023parametric}.

\subsection{Temporally-finite PTCs \label{tfptcs}}

%\red{If there is only 1-2 papers on this topic, let us combine this section with the previous one. Otherwise, keep it separate.}
%In this subsection, we discuss the temporal slabs whose permittivity variation is restricted to a finite temporal domain. Further, the %$temporal analog of the Bragg grating is discussed.
%References: \cite{koufidis2023temporal}, \cite{ramaccia2021temporal}.
By definition, PTCs imply a system whose material properties are modulated periodically in time. Such periodic modulation extends for all times $t$, i.e., $-\infty<t<\infty$. However, to probe the emerging features of the PTCs, we can only consider a finite number of cycles of such periodic time modulation. Theoretically, this requires solving the Maxwell equations with temporal interface conditions.

\begin{figure*}[tb]
\centerline{\includegraphics[width= 0.8\columnwidth,trim=6 6 6 6,clip]{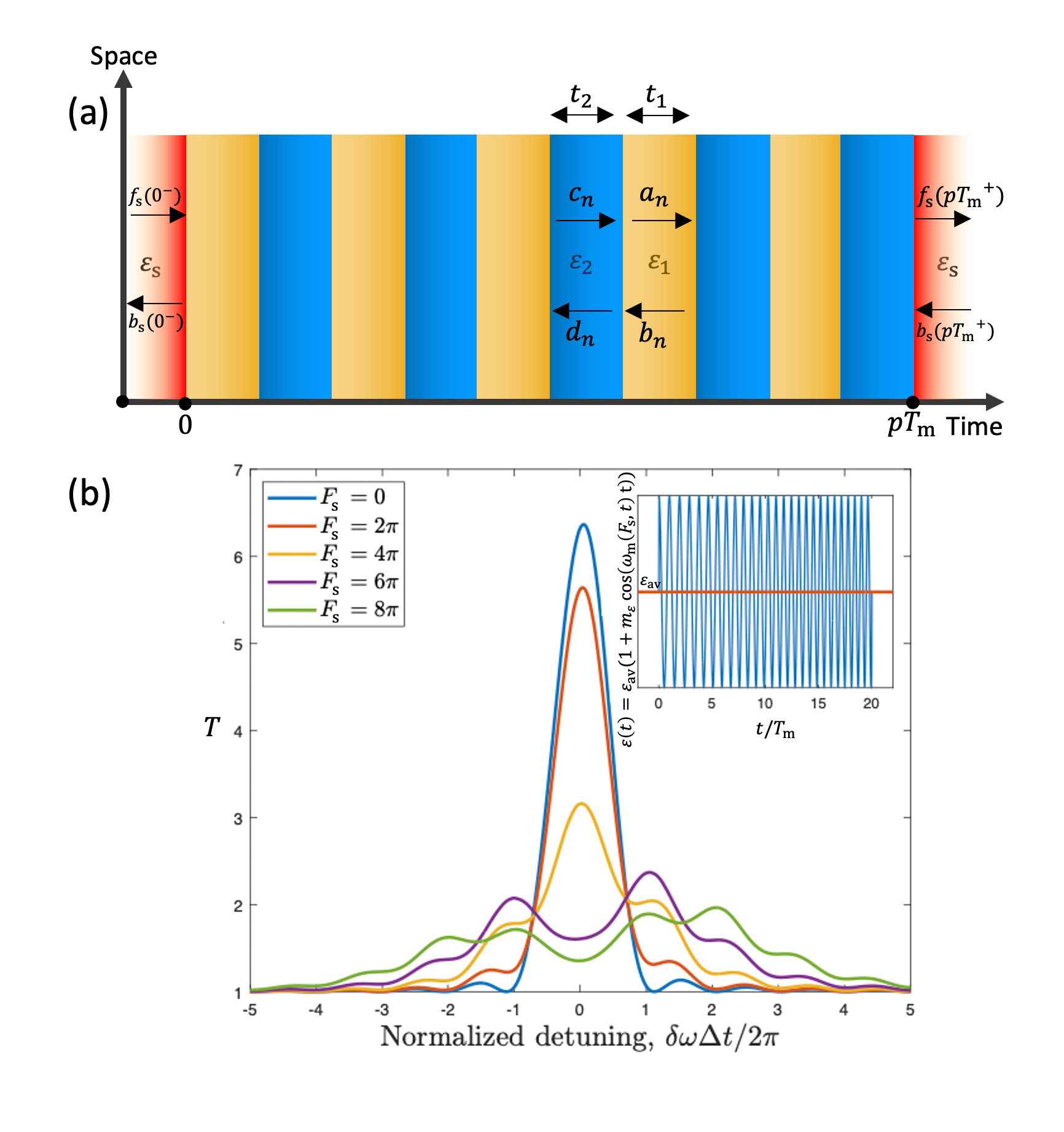}}
\caption{
(a) A temporal slab consisting of a stepwise temporal modulation.
(b) Transmittance $T$ of the temporal slab (in the inset) as a function of the detuning parameter $\delta\omega$ for various values of chirping coefficient $F_\mathrm{s}$. Here, $\varepsilon_\mathrm{av}=1$.
Figure (b) was reprinted from~\cite{koufidis2023temporal}.}
\label{fig:temporal_slabs}
\end{figure*}

In Fig.~\ref{fig:temporal_slabs}(a), a temporally finite PTC, i.e., a temporal slab, is shown. Such a temporal slab consists of $p$ cycles of a stepwise temporal modulation (see Fig.~\ref{fig:temporal_slabs}(a)). During one period of the modulation, the permittivity of the medium takes the values $\varepsilon_1$ (for time-interval $t_1$) and $\varepsilon_2$ (for time-interval $t_2$) (see Fig.~\ref{fig:temporal_slabs}(a)). The slab exists only for a finite time, i.e., $0<t<pT_\mathrm{m}$. Furthermore, we assume the permittivity of the medium for times $t<0$ and $t>pT_\mathrm{m}$ to be $\varepsilon_\mathrm{s}$. To solve the scattering problem for such a temporal slab, the transfer matrix method~\cite{ramaccia2021temporal} can be used as outlined here in Subsection~\ref{sec: transfer matrix}. Such an approach is helpful, as knowing the temporal matching matrix $\=J_{\varepsilon_\mathrm{a}\rightarrow\varepsilon_\mathrm{b}}$ of the interface at which the permittivity jumps from some $\varepsilon_\mathrm{a}$ to some $\varepsilon_\mathrm{b}$ and the ABCD-matrix of the temporal modulation (see Eq.~\eqref{eq: abcd}) is sufficient to construct the effective transfer-matrix $\=T$ of the scattering structure shown in Fig.~\ref{fig:temporal_slabs}(a). Note that we can compute the temporal matching matrix $\=J_{\varepsilon_\mathrm{a}\rightarrow\varepsilon_\mathrm{b}}$ using the continuity of the fields $D_x,B_y$ (see Eqs.~\eqref{eq: Dx_a}--\eqref{eq: by_a} and Eqs.~\eqref{eq: Dx_c}--\eqref{eq: by_c}) at the temporal interface $\varepsilon_\mathrm{a}\rightarrow\varepsilon_\mathrm{b}$ as 
\begin{equation}
\=J_{\varepsilon_\mathrm{a}\rightarrow\varepsilon_\mathrm{b}}=\begin{bmatrix}
       1+\frac{\omega_\mathrm{a}}{\omega_\mathrm{b}}&1-\frac{\omega_\mathrm{a}}{\omega_\mathrm{b}}\\
       1-\frac{\omega_\mathrm{a}}{\omega_\mathrm{b}}&1+\frac{\omega_\mathrm{a}}{\omega_\mathrm{b}}\\
   \end{bmatrix}\,,\label{eq:j_mat}
\end{equation}
where $\frac{\omega_\mathrm{a}}{\omega_\mathrm{b}}=\sqrt{\frac{\varepsilon_\mathrm{b}}{\varepsilon_\mathrm{a}}}$ (see Eq.~\eqref{eqdisp}). Here, Eq.~\eqref{eq:j_mat} is similar to \cite[Eq.~3]{ramaccia2021temporal}. Finally, we can write the effective transfer-matrix $\=T$ of the temporal slab shown in Fig.~\ref{fig:temporal_slabs}(a) as \cite[Eq.~11]{ramaccia2021temporal}

%In Fig.~\ref{fig:temporal_slabs}(a), a general scattering structure (temporal slab) made by cascading the layers of materials with refractive indices $n_m$ for time durations $\Delta t_m$ is shown. Note that the temporal modulation in the considered scattering structure exists only for a finite time, i.e., $t_1<t<t_{M+1}$ (see Fig.~\ref{fig:temporal_slabs}(a)). To solve the scattering problem for such a temporal slab, the transfer matrix method~\cite{ramaccia2021temporal} can be used as outlined here in Sec.~\ref{sec: transfer matrix}. Such an approach is helpful, as knowing the temporal matching matrix $\mathbf{M}_m$ of the $m^\mathrm{th}$ interface (occurring between $t=t_{m-1}$ and $t_m$) and the temporal delay matrix $\mathbf{D}_m$ during time $\Delta t_m$ is sufficient to construct the effective transfer-matrix $\mathbf{T}_M$ of the scattering structure shown in Fig.~\ref{fig:temporal_slabs}(a) as \cite[Eq.~11]{ramaccia2021temporal}

%
\begin{equation}
\=T=\=J_{\varepsilon_2\rightarrow\varepsilon_\mathrm{s}}\cdot{\=J}^{(-1)}_{\varepsilon_2\rightarrow\varepsilon_\mathrm{1}}\cdot{\=M}^{(-p)}\cdot\=J_{\varepsilon_\mathrm{s}\rightarrow\varepsilon_1}\,.\label{eq:tr_mat}
\end{equation}
Here, the transfer matrix $\={T}$ connects the forward and backward propagating fields $f_\mathrm{s}(pT_\mathrm{m}^+),b_\mathrm{s}(pT_\mathrm{m}^+)$ at time $t=pT_\mathrm{m}^+$ to the forward and backward propagating fields $f_\mathrm{s}(0^-),b_\mathrm{s}(0^-)$ at time $t=0^-$ as (see Fig.~\ref{fig:temporal_slabs}(a))

\begin{equation}
\begin{bmatrix} f_\mathrm{s}(pT_\mathrm{m}^+)  \\ b_\mathrm{s}(pT_\mathrm{m}^+) \end{bmatrix}
 =
 \={T}
  \cdot\begin{bmatrix}
   f_\mathrm{s}(0^-)  \\ b_\mathrm{s}(0^-)
   \end{bmatrix}\,.\label{eq:tr_mat2}
\end{equation}

Having calculated the transfer matrix $\={T}$, one can easily calculate the optical observables, i.e., transmittance ($T$), reflectance ($R$), and absorbance ($A$) of the underlying scattering structure \cite{ramaccia2021temporal}. In particular, the transmittance and reflectance for a forward propagating incident field with amplitude $f_\mathrm{s}(0^-)$ at the time $t=0^-$ is given by $T=\left|\={T}_{11}\right|^2$ and $R=\left|\={T}_{21}\right|^2$, respectively. These expressions are obtained from Eq.~\eqref{eq:tr_mat2} by considering that $b_\mathrm{s}(0^-)=0$, i.e., the absence of the backward wave at $t<0$ due to causality~\cite{galiffi2022photonics}. Furthermore, the transmittance and reflectance for a backward propagating incident field with amplitude $b_\mathrm{s}(0^-)$ at the time $t=0^-$ is given by $T=\left|\={T}_{22}\right|^2$ and $R=\left|\={T}_{12}\right|^2$, respectively (here due to causality, $f_\mathrm{s}(0^-)=0$). Moreover, the absorbance $A$ can be computed as $A=1-R-T$. 
Note that in Fig.~\ref{fig:temporal_slabs}(a), a temporal stepwise system is considered. However, one can use the transfer matrix method for an arbitrary temporal modulation, as shown in \cite{Chegnizadeh2018general}. Furthermore, the transfer matrix method can also be used to homogenize the finite PTCs shown in Fig.~\ref{fig:temporal_slabs}(a) using an eigenmode-based approach as discussed in \cite{garg2024two}. Moreover, the transfer matrix method is also useful to study the photon squeezing in the time-varying media \cite{echave2024photon}.

As an application of the finite PTCs, the authors of~\cite{koufidis2023temporal} consider the transfer matrices of more complex temporal slabs using the M\"obius transformation method. In particular, they investigate the reflectance and transmittance of temporally finite slabs when the modulation frequency $\omega_\mathrm{m}$ and/or the modulation strength $m_\varepsilon$ becomes time-dependent, that is $\varepsilon(t)=\varepsilon_\mathrm{av}[1+m_\varepsilon(t)\cos \omega_\mathrm{m}(t) t]$. Note that the former phenomenon is known as chirping and the latter is known as apodization of the temporal modulation. In the following, we will review the effects of a linear chirp on the temporal modulation. Note that a linearly chirped modulation is important to consider, as the chirping of light pulses occurs inevitably in dispersive systems due to the group velocity dispersion \cite{diels1996ultrashot}. Therefore, in the systems that involve temporal modulation of $\varepsilon$ using optical pumps, such chirping effect becomes crucial. Furthermore, as outlined later, a linearly chirped modulation leads to an increase in the amplification bandwidth of the temporal slabs \cite{koufidis2023temporal}. Such a change in the bandwidth can be, in principle, used to control the shape of the light pulses scattered off the temporal slabs. 

As shown in the inset of Fig.~\ref{fig:temporal_slabs}(b), a temporal slab with a temporally perturbed modulation frequency $\omega_\mathrm{m}(F_{s},t)$ is considered. Note that the modulation frequency $\omega_\mathrm{m}(F_{s},t)$ depends on $F_s$ as $\omega_\mathrm{m}(F_{s},t)=\omega_\mathrm{m0}+2F_\mathrm{s}t/\Delta t^2$. Here, $\omega_\mathrm{m0}$ is a constant modulation frequency over which the linear chirping is applied, $F_\mathrm{s}$ is the chirping coefficient, and $\Delta t$ is the total duration of the considered temporal slab. Here, the total duration of the slab is taken as $\Delta t=20T_\mathrm{m}$. Having chirped the temporal modulation, the transmittance of the temporal slab is calculated for different values of the chirping coefficient $F_\mathrm{s}$. 
Such transmittance as a function of the detuning parameter $\delta\omega$ is shown in Fig.~\ref{fig:temporal_slabs}(b). Note that $\delta\omega$ corresponds to the detuning of the incident frequency $\omega_\mathrm{inc}$ from $\frac{\omega_\mathrm{m}(F_\mathrm{s},t)}{2}$, i.e., $\delta\omega=\omega_\mathrm{inc}-\frac{\omega_\mathrm{m}(F_\mathrm{s},t)}{2}$. From Fig.~\ref{fig:temporal_slabs}(b), we observe that as $F_\mathrm{s}$ increases (chirping increases), the maximum value of the transmittance $T$ decreases. However, the $k$-bandwidth over which $T$ is substantially higher than unity increases. Therefore, we conclude that the linear chirping of the temporal modulation increases the amplification bandwidth at the cost of the maximum value of the amplification of the incident fields.

As a summary, we have been elaborating in the last two sections on aspects that matter for realistic systems, i.e., their finiteness. The assumption of an infinite spatial and temporal extent cannot hold up in reality, and we must consider the finiteness in both dimensions. In the following section, we would like to lift one more assumption made up to this point, which does not always hold in practical realizations of PTCs. It concerns the isotropy of the considered material. 

\subsection{Effects of anisotropy in PTCs \label{Anisotropicmedium}}

Up to now, only isotropic materials have been considered. 
% Isotropic materials are gases, liquids, glasses, and all amorphous or polycrystalline materials with a grain size much smaller than the wavelength. 
Isotropic materials are characterized by a scalar susceptibility or permittivity. In contrast, in this section, we discuss PTCs made from anisotropic materials. Examples of anisotropic materials include various types of crystals.
%References: \cite{li2023stationary}.
The anisotropy implies that the material properties are described by tensorial quantities. The induced polarization density does not need to have the same orientation as the electric field that induces it. Moreover, the isofrequency surfaces of the dispersion relation in anisotropic materials usually have ellipsoidal shapes. As such, the length of the wavenumber depends on the direction of propagation. 
% Finally, whereas the eigenmodes in an isotropic medium are elliptically polarized plane waves, the eigenmodes in an anisotropic medium are linearly polarized. 

\begin{figure*}[h]
\centerline{\includegraphics[width= 1\columnwidth]{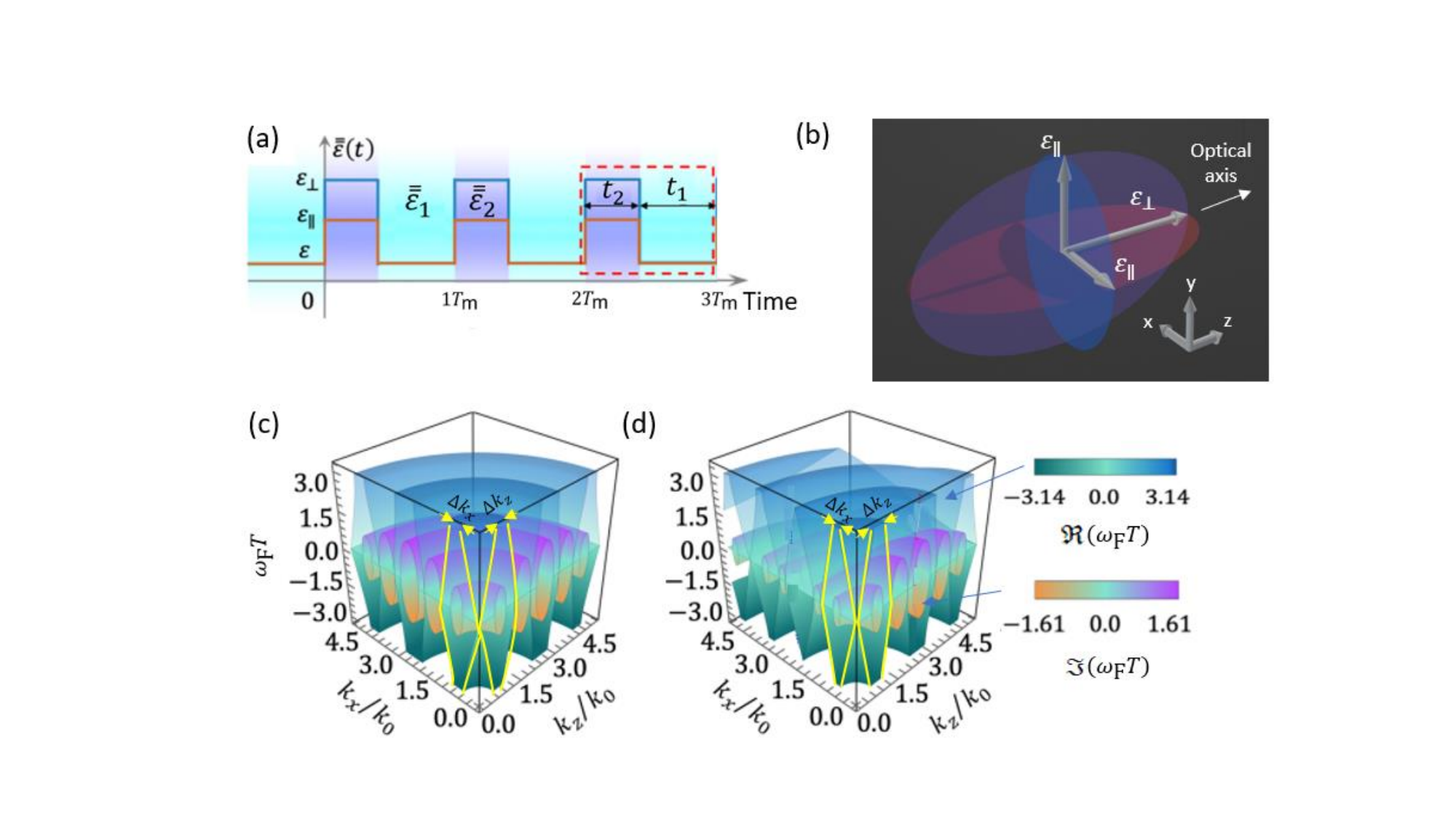}}
\caption{(a) Permittivity modulation in an anisotropic PTC. The permittivity oscillates between that of an uniaxial anisotropic medium and an isotropic medium where $\varepsilon=1$, $\varepsilon_{\parallel}=4$, $\varepsilon_{\perp}=25$, $t_1=t_2=T_\mathrm{m}2$, and $k_0=2\pi/T_\mathrm{m}c_0$.
(b) The permittivity ellipsoid of the PTC during the times when it is anisotropic and described by permittivity tensor $ \overline{\overline{\varepsilon}}_2$. The arrows depict the principal axes of the material permittivity. The incidence $xz$-plane is shown in pink. (c) The band structure for the ordinary light. The yellow lines depict the first and second bands, while the yellow arrows denote the bandgaps. The bandgaps $\Delta k_x $ and $\Delta k_z $ have the same widths. (d) The band structure for the extraordinary light. The bands depend on the propagation direction. In particular, the lowest-order bandgaps in the $k_x$ and $k_z$ directions are different. Figures (a), (c), and (d) were reprinted from~\cite{li2023stationary}. 
}
\label{Fig:APTC}
\end{figure*}

% Anisotropic photonic time crystals (APTCs) greatly extend beyond isotropic media by considering a periodic temporal modulation of anisotropic material properties. 
Anisotropic PTCs combine the anisotropy and periodic temporal modulations, resulting in more complex and exotic light-matter interaction phenomena, including light emission manipulation and control of radiative energy distribution in space.
In the following, we concentrate on the most fundamental configuration of anisotropic PTCs explored in~\cite{li2023stationary}, which is constructed by alternating in time between two types of lossless, nonmagnetic media periodically. The first medium is isotropic with scalar permittivity $\varepsilon$, while the second one is an uniaxial crystal with permittivity tensor $ \=\varepsilon_{2}$, as shown in Fig.~\ref{Fig:APTC}(a). For simplicity, material dispersion is disregarded, and the principal axes of the uniaxial crystal are aligned to the coordinate system, which can be seen in Fig.~\ref{Fig:APTC}(b). 
In this coordinate system, the permittivities of the PTC at the two temporal states are given by $\= \varepsilon_{1} = \varepsilon {\=I}$ and $ \=\varepsilon_{2} = \text{diag}(\varepsilon_{\perp}, \varepsilon_{\perp}, \varepsilon_{\parallel})$, where $\=I$ is the identity matrix.  
% In contrast, the isotropic crystal is characterized by \(\= \varepsilon_{1} = \varepsilon {I} \), with $I$ being the identity matrix. 

It is well known that in uniaxial anisotropic media, a light beam that propagates not along the optical axis can be split into two rays with orthogonal polarization. The first ray has the polarization orthogonal to the optical axis. It is called ordinary because it ``sees'' the same material permittivity $\varepsilon_{\parallel}$ independent of its incident direction. The second ray is referred to as extraordinary because it is direction-dependent, that is, depending on the propagation direction, it experiences a material permittivity within the range $[\varepsilon_{\parallel}; \varepsilon_{\perp}]$. Likewise, the polarization degeneracy of the eigenmodes in an anisotropic PTC is lifted. For a given incidence plane, there are two distinct photonic band structures: for ordinary and extraordinary light. Following Ref.~\cite{li2023stationary}, let us choose the incidence plane to be parallel to the $xz$-plane. From Fig.~\ref{Fig:APTC}(b), we can see that the ordinary light has transverse-electric (TE) or $s$-polarization. On the other hand, the extraordinary light has transverse-magnetic (TM) or $p$-polarization.

Since the light of orthogonal polarizations does not mix, the photonic band structure can be 
calculated using the transfer matrix method, analogously to how it was described in Subsection~\ref{sec: transfer matrix}. Performing the necessary derivations, one can obtain two independent eigenvalue equations similar to (\ref{eq:band diagram}), one for the ordinary and another for the extraordinary light. 
% from isotropic media to obtain the transfer matrix in an uniaxial crystal and reach an equation similar to equation \ref{eq:band diagram} for an anisotropic PTC. 
% It is worth mentioning that the transfer matrix would be block diagonal having ordinary and extraordinary blocks which are decoupled from each other. Then, the Flouqet frequency of ordinary waves is independent of extraordinary waves. 
Solving these equations yields two band structures of the anisotropic PTC. 
Figures~\ref{Fig:APTC}(c) and (d) show the band structures for a specific PTC configuration with a setting described in the figure caption. One can see from Fig.~\ref{Fig:APTC}(c) that
for the ordinary light, the band structure does not depend on the direction of propagation since it has radial symmetry in the $k_x -k_z$ plane. In other words, this band structure is similar to an isotropic PTC. Indeed, the lowest-order bandgaps in both the $k_x$ and $k_z$ directions, marked with yellow color, have the same widths ($\Delta k_x = \Delta k_z$). Note that in the figure, both real and imaginary parts of the eigenfrequency contours are plotted using different color maps. 
In contrast, the band structure for the extraordinary light is not symmetric in the $k_x-k_z$ plane, as shown in Fig.~\ref{Fig:APTC}(d). Here, the lowest-order bandgaps  $\Delta k_x$ and $\Delta k_z$ are different. 
Thus, light amplification now depends strongly on its propagation direction. This regime of anisotropic PTCs provides exciting opportunities to achieve direction-dependent light amplification.   

Furthermore, anisotropic PTCs possess another unique feature. It is well known that in a time-invariant medium, at the rest frame, a static (position- and time-invariant) charge cannot generate electromagnetic radiation. However, this is not true anymore if the stationary charge is positioned inside an anisotropic PTC~\cite{li2023stationary}. When the medium switches between the isotropic and anisotropic states, the static charge induces propagating electromagnetic waves. It appears that this phenomenon of DC-to-AC conversion is unique to anisotropic PTCs and was not achieved previously in the isotropic counterparts. A more comprehensive discussion on the phenomenon of electromagnetic radiation from a charge located inside a PTC is given in Section~\ref{radfreeelectron}.

In addition to~\cite{li2023stationary}, we note that the concept of PTC has also been extended to {\it{biaxial}} anisotropic materials. It has been shown that specific non-uniform plane waves can experience broad momentum band gaps even with small modulation depths in such materials\cite{dong2024non}. 

Recently, the study of wave propagation in time-varying anisotropic media has been linked to the concept of twistronics in condensed matter physics, serving as its temporal counterpart~\cite{ptitcyn2024temporal}. Condensed-matter twistronics involves the study of twisted two-dimensional material bilayers and the impact of the twisting angle on their electrical properties.
The photonic research outlined in~\cite{ptitcyn2024temporal} investigates light propagation through a spatially unbounded anisotropic medium experiencing a temporal jump in the relative permittivity tensor. This jump results in the creation of a new anisotropic medium which is a rotated version of the original. It was discovered that such temporal jumps lead to frequency conversion, the extent of which significantly depends on the propagation direction of the initial wave, the rotation angle, and the initial values of the material parameters.

\subsection{Defects in PTCs \label{defectPTC}}

%In this subsection, we will study the impact of defects in PTCs. We will describe how such defects can be harnessed to manipulate the light-matter %interactions inside the MBGs.\\
%References: \cite{sadhukhan2023defect}, \cite{sadhukhan2022bandgap}, \cite{apffel2023temporal}.
The defects in photonic crystals have given rise to various interesting applications such as enhanced light emission \cite{noda2000trapping}, photonic crystal based beam splitters \cite{Bayindir2000photonic}, and ultra-sensitive sensors \cite{aly2023ultra}. Similarly, the defects in PTCs may also lead to many intriguing physical effects. Their consideration is a further endeavor when studying more realistic PTCs.

\begin{figure*}[h]
\centerline{\includegraphics[width= 1\columnwidth,trim=4 4 2 4,clip]{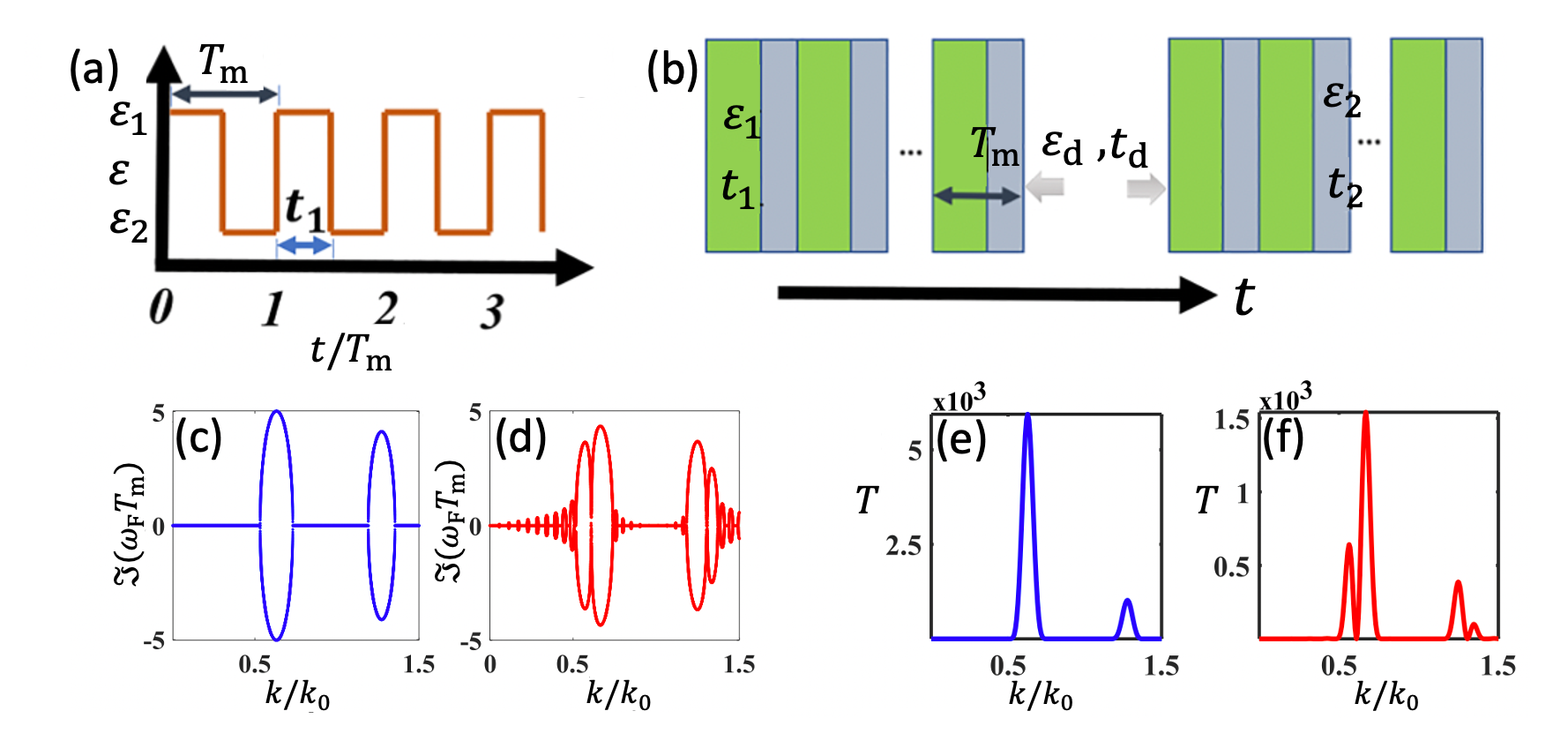}}
\caption{
(a) A PTC consisting of a time-periodic stepwise permittivity modulation.
(b) A PTC with a temporal defect. Here, the periodic nature of the temporal modulation of the PTC is broken for the time duration $t_\mathrm{d}$ due to a temporal defect with permittivity $\varepsilon_\mathrm{d}$. (c), (d) Imaginary part of the eigenfrequency $\omega$ as a function of the eigenwavenumber $k$ (band structure) in the absence and presence of the defect, respectively. (e), (f) Transmittance $T$ of the PTC in the absence and presence of the defect, respectively. Note that $k_0= 2\pi/(cT_\mathrm{m})$. Figures were reprinted from~\cite{sadhukhan2023defect}.
}
\label{fig:defect_PTC}
\end{figure*}

In \cite{sadhukhan2023defect}, the impact of temporal defects during the periodic temporal modulation of PTCs has been studied. In Fig.~\ref{fig:defect_PTC}(a), a PTC with a time-periodic permittivity $\varepsilon$ has been shown without any defects. Note that the permittivity $\varepsilon$ fluctuates between $\varepsilon_1=3$ for time interval $t_1=1$~fs and $\varepsilon_2=1$ (see Fig.~\ref{fig:defect_PTC}(a)). To introduce a temporal defect, the time-periodic modulation is perturbed for the time interval $t_\mathrm{d}$ during which the permittivity takes a different value as compared to the values during the periodic modulation (see Fig.~\ref{fig:defect_PTC}(b)). Such a value of the permittivity during the time interval $t_\mathrm{d}$ is denoted by $\varepsilon_\mathrm{d}$. After the time-interval $t_\mathrm{d}$, the permittivity modulation of the PTC returns to its earlier periodic state as shown in Fig.~\ref{fig:defect_PTC}(b). In the following, we assume, $\varepsilon_\mathrm{d}=1$ and $t_\mathrm{d}=1$~fs.

The photonic band structures of the PTCs characterize its optical response. Therefore, it is important to note the changes introduced in the band structures of the PTCs in the presence of temporal defects. In Figs.~\ref{fig:defect_PTC}(c)--(d), the imaginary part of the eigenfrequency $\omega$ as a function of the eigenwavenumber $k$ of the considered PTC is shown in the absence and presence of the temporal defect, respectively. From Fig.~\ref{fig:defect_PTC}(c), we observe the existence of non-zero imaginary parts of the eigenfrequency $\Im(\omega_\mathrm{F})$ in two different spectral regions ($k$ regions). Therefore, there exist two momentum bandgaps in the considered spectral range. However, from Fig.~\ref{fig:defect_PTC}(d), in the presence of the defect, we observe that within the bandgaps, there exist certain $k$ values for which $\Im(\omega_\mathrm{F})$ goes to zero, leading to the splitting of each bandgap. Such splitting is evident as each main lobe in Fig.~\ref{fig:defect_PTC}(c) splits into two main lobes in Fig.~\ref{fig:defect_PTC}(d). Furthermore, we notice that besides the main lobes in Fig.~\ref{fig:defect_PTC}(d), we observe certain side lobes of $\Im(\omega_\mathrm{F})$ outside the momentum bandgap. Note that despite the fact that the time-modulation is not strictly periodic anymore, it is still possible to calculate the bandstructure of the defective PTC. Such a calculation can be easily done by considering the impact of defect in terms of a defect matrix while calculating the bandstructure of the PTC using the transfer matrix approach \cite[Eq.~2]{sadhukhan2023defect}.

As a next step, the impact of the aforementioned features due to defects on the transmissivity $T$ of the PTC is investigated. As mentioned in Section~\ref{tfptcs}, we need a temporally-finite PTC to calculate such transmissivity. Therefore, in the following, we assume the defect-free PTC to have ten temporal unit cells. Further, the defective PTC is taken such that the resulting scattering structure has five temporal unit cells on both sides of the temporal defect (see Fig.~\ref{fig:defect_PTC}(b)). The transmissivity of such a finite defective PTC is shown in Figs.~\ref{fig:defect_PTC}(e)--(f). In Fig.~\ref{fig:defect_PTC}(e), we observe two maxima of the transmittance $T$ of the PTC in the absence of the defect. They occur due to the presence of two momentum bandgaps in the considered spectral range (see Figs.~\ref{fig:defect_PTC}(c)--(d)). However, in the presence of the defect, a minima of $T$ exists sandwiched between the maxima. This can be explained by the vanishing values of $\Im(\omega_\mathrm{F})$ at the spectral locations of the minima of $T$ (see Fig.~\ref{fig:defect_PTC}(d)). 

Such temporal defects endow PTCs with additional degrees of freedom. Therefore, the amplification within the momentum bandgaps can be manipulated more effectively by exploiting the defects.

\subsection{Effects of disorder \label{disorderPTC}}
\begin{figure*}[h]
\centerline{\includegraphics[width= 0.9\columnwidth,trim=4 4 2 4,clip]{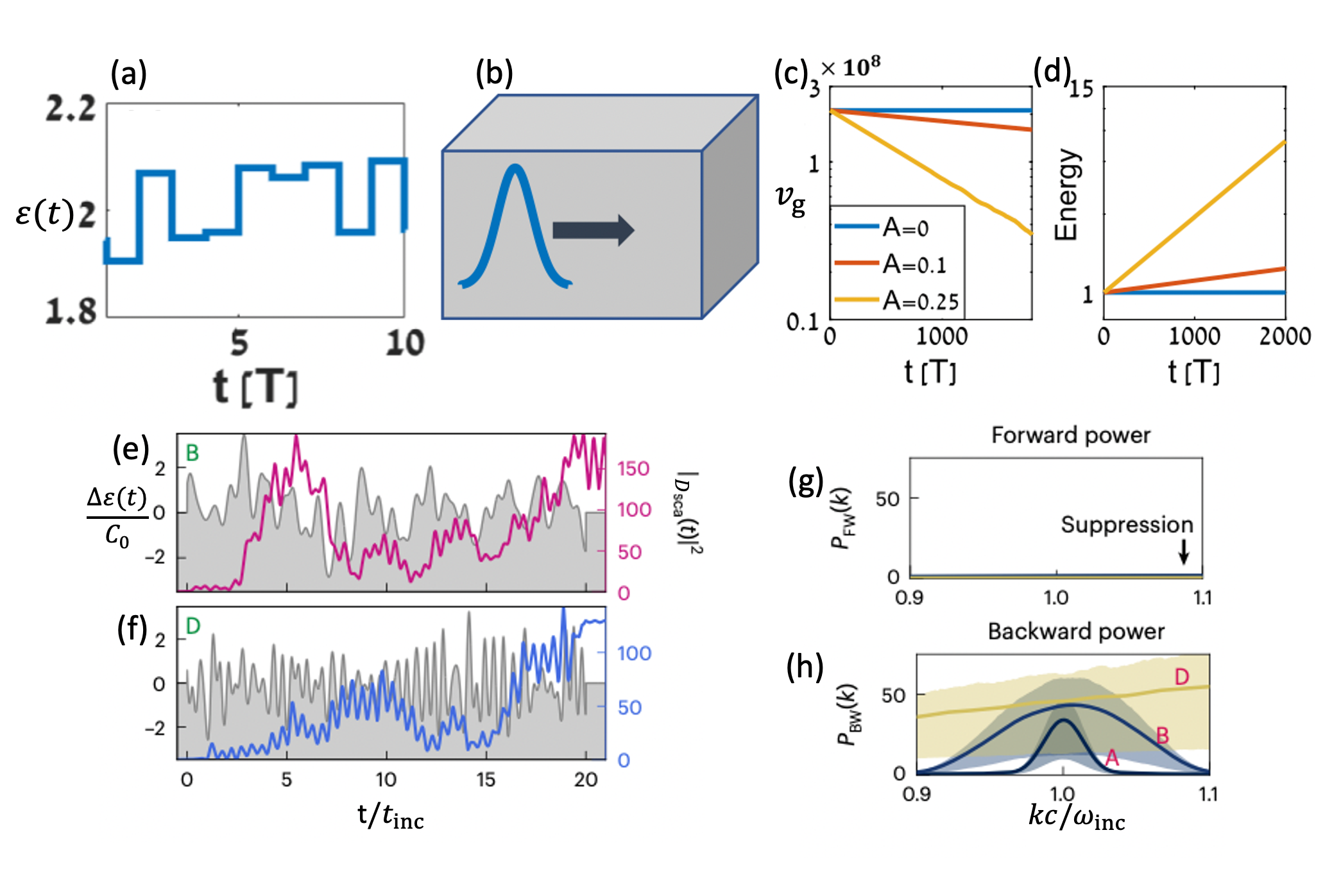}}
\caption{
(a) Permittivity $\varepsilon$ of a temporally disordered system as a function of time $t$.
(b) Propagation of a Gaussian pulse in a spatially homogeneous medium with temporally disordered permittivity shown in (a). (c) Group velocity $v_\mathrm{g}$ and (d) energy of the Gaussian pulse shown in (b) as a function of time. (e), (f) Realized permittivity disorder $\Delta\varepsilon$ (gray areas) and the corresponding scattering intensities $|D_\mathrm{sca}|^2$ (solid lines) of the systems that suppress backward and forward scattering, respectively. (g), (h) Forward and backward scattered powers ($P_\mathrm{FW}$, $P_\mathrm{BW}$) a function of wave momenta $k$, respectively, of a device designed to suppress forward scattering. Here, cases A, B, and D represent perfectly ordered, intermediate, and near-Poisson temporal modulations, respectively. Here, $t_\mathrm{inc}=2\pi/\omega_\mathrm{inc}$, $D_\mathrm{sca}(t)$ is the scattered electric displacement field, and $C_0$ is a normalization constant that depends on the correlation of $\varepsilon(t)$. Figures~(a)--(d) were reprinted from~\cite{sharabi2021disordered}. Figures~(e)--(h) were reprinted from~\cite{Kim2023unidirectional}.
}
\label{fig:disorder}
\end{figure*}

%In this subsection, we will study the effects of disorder in PTCs.\\
%References: \cite{sharabi2021disordered}, \cite{Kim2023unidirectional}, 
%\cite{apffel2021time}
The propagation of electromagnetic waves in temporally disordered media is another crucial aspect with regard to realistic PTCs. Such disordered PTCs correspond to those photonic systems that are spatially homogeneous, but their material properties change randomly as a function of time. In \cite{sharabi2021disordered}, the interaction of such disordered PTCs with an incident pulse was studied. Figure~\ref{fig:disorder}(a) shows the permittivity $\varepsilon(t)$ of the disordered PTC under consideration. The permittivity of such a disordered PTC consists of equal time segments of duration $T$. During each such segment, the permittivity can be written as $\varepsilon= 2+A \times U[-1,1]$. Here, $A$ represents the disorder magnitude, and $U$ is a uniform distribution. Next, the propagation of a Gaussian pulse is studied through such a disordered system [see Fig.~\ref{fig:disorder}(b)]. Note that we assume the full width at half maximum (FWHM) of the pulse to be $200~T$. Further, the central frequency of the pulse is taken as $2\pi/(5~T)$. In the following, the group velocity $v_\mathrm{g}$ and pulse energy of the Gaussian pulse are examined for different disorder amplitudes $A$. Figures~\ref{fig:disorder}(c) and (d) show the variation of the group velocity and energy of the pulse, respectively, as a function of the propagation time $t$. From Fig.~\ref{fig:disorder}(c), it can be seen that the group velocity of the pulse goes down exponentially as a function of the propagation time. Further, Fig.~\ref{fig:disorder}(d) shows an exponential growth of the pulse energy in the disordered systems with the propagation time. As discussed in \cite{sharabi2021disordered}, a further investigation of the temporally disordered systems reveals a strong dependence of the emerging effects on the band structure of the PTC. Moreover, effects analogous to the Anderson localization are also reported in the disordered PTCs \cite{sharabi2021disordered}.

Additionally, some interesting applications of the disordered media are discussed in \cite{Kim2023unidirectional}. Specifically, the temporal disorder was utilized to tailor light scattering from spatially homogeneous time-varying structures. The time-varying permittivity of the considered disordered system can be written as $\varepsilon(t)=\varepsilon_\mathrm{av}(1+\Delta\varepsilon(t))$. Here, $\Delta\varepsilon(t)$ incorporates the temporal disorder. First, \cite{Kim2023unidirectional} discusses how temporally disordered systems can be used to attain unidirectional scattering. Figures~\ref{fig:disorder}(e) and (f) show the permittivity profiles of the engineered temporal systems that exhibit negligible backward and forward scattering for an incident frequency $\omega_\mathrm{inc}$, respectively. Note that the incident frequency $\omega_\mathrm{inc}$ is such that,  $\omega_\mathrm{inc}=kc/\sqrt{\varepsilon_\mathrm{av}}$. Here, $k$ is the magnitude of the incident wavevector. Next, a transition from ordered temporal modulation (PTC regime) to uncorrelated temporal disorder has been studied. Such an investigation reveals that the uncorrelated disorder increases the momentum bandwidth over which non-zero backward scattering can be attained while maintaining negligible forward scattering [see Figs.~\ref{fig:disorder}(g)--(h)]. Note that such a negligible forward scattering is maintained by inversely designing the temporal disorder.

\subsection{Nonlinear PTCs \label{NLPTC}}
%So far, in the paper, only linear PTCs are discussed. In this subsection, we will depart from linearity and discuss the nonlinear effects in PTCs. In particular, gap solitons will be discussed in detail.\\
%References: \cite{pan2022superluminal2}
%cite{pan2022superluminal} -- I think it is the same as their journal paper. If yes, keep only the journal paper reference. 

\begin{figure*}[h]
\centerline{\includegraphics[width= 1\columnwidth,trim=4 4 2 6,clip]{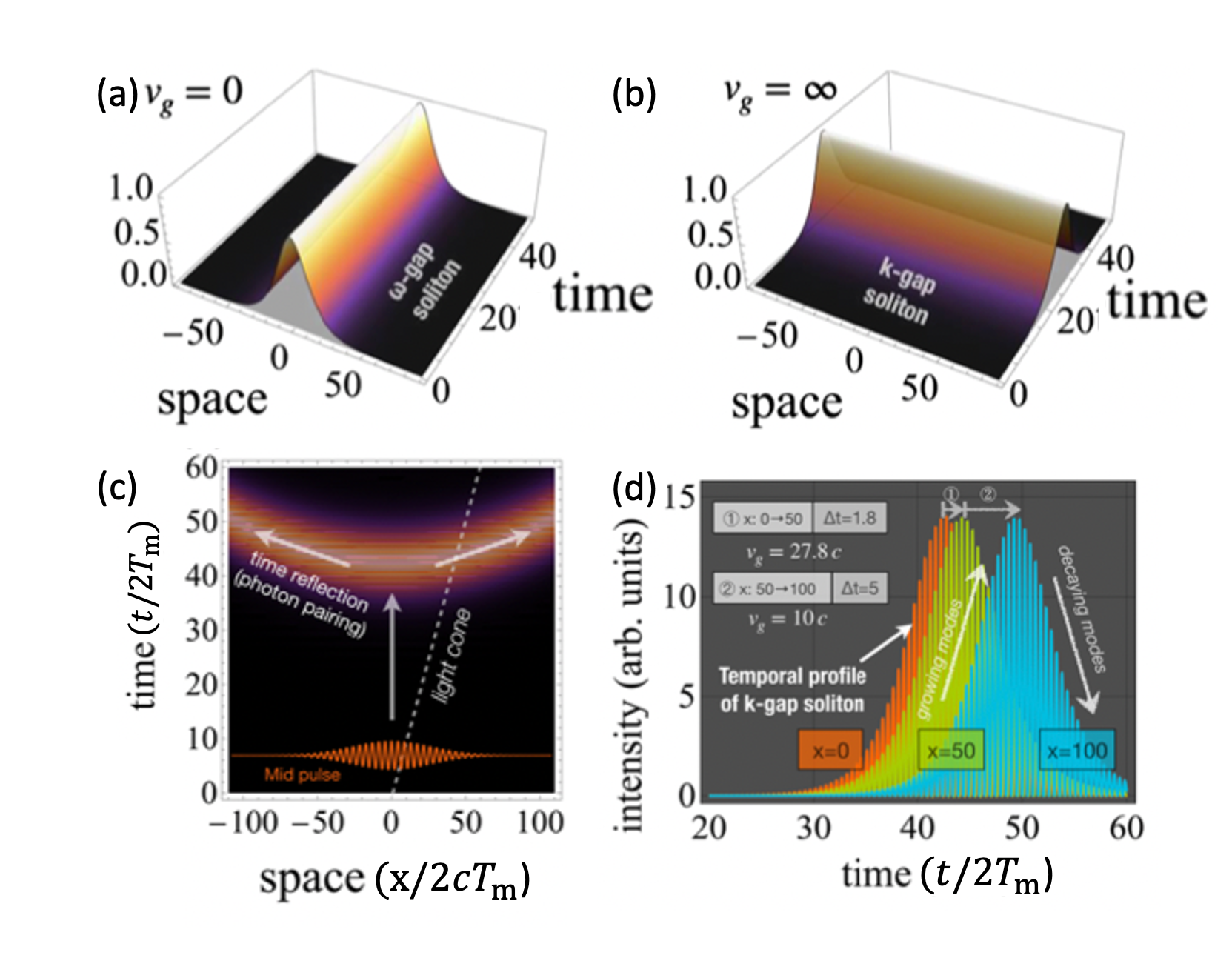}}
\caption{
(a), (b) Soliton solutions inside the $\omega$-gap and $k$-gap, respectively. Here, the vertical axis denotes the intensity of the soliton wave pulse. (c) Spatiotemporal dynamics of $k$-gap solitons generated by an input Gaussian pulse (shown by solid red curve). (d) The temporal profiles of the solitons generated by the input pulse in (b) at different spatial locations. Figure was reprinted from~\cite{pan2023superluminal}.}
\label{fig:solitons_PTC}
\end{figure*}

Finally, we discuss in this chapter the propagation of electromagnetic fields in the PTCs formed inside nonlinear media (see also a relevant discussion for photonic space-time crystals in Section~\ref{NLSTPC}).
The permittivity of such nonlinear PTCs has a time-periodic linear part and a stationary nonlinear contribution term. One example of a nonlinear PTC supporting Kerr nonlinearity has the form $\varepsilon(t,|\mathbf{E}|^2)= \varepsilon_1(t)+\chi^{(3)}|\mathbf{E}|^2$ \cite{pan2023superluminal}. Here, $\varepsilon_1(t)$ is the time-periodic linear permittivity given by $\varepsilon_1(t)=\varepsilon_\mathrm{av}[1+m_\varepsilon\mathrm{cos}(\omega_\mathrm{m}t)]$, $\chi^{(3)}$ is the third-order nonlinear susceptibility, and $\mathbf{E}$ is the electric field of light. Such a nonlinear PTC has been studied in detail in \cite{pan2023superluminal}. In particular, such PTCs are shown to support solitonic solutions inside their momentum bandgap. Solitons refer to those wave pulses that maintain their shapes while traveling through a medium despite the dispersion of the medium \cite{boyd2008nonlinear}. Note that the solitons found inside the momentum bandgap ($k$-gap) of the PTCs differ from the solitons found inside the energy bandgaps ($\omega$-gap) of the spatial photonic crystals (see Figs.~\ref{fig:solitons_PTC}(a)--(b)). First, the solitons existing inside the $\omega$-gaps are finite wave packets in space, but they have a plane wave dependence in time $t$ (see Fig.~\ref{fig:solitons_PTC}(a)). Further, their group velocity, $v_\mathrm{g}=0$ due to the fact that the edge states of $\omega$-gap satisfy $\frac{\partial \omega}{\partial k}=0$. 

On the contrary, the $k$-gap solitons are finite wave packets in time, but they have a plane wave dependence in space (see Fig.~\ref{fig:solitons_PTC}(b)). Further, their group velocity, $v_\mathrm{g}=\infty$ due to the fact the edge states of the $k$-gap satisfy $\frac{\partial k}{\partial \omega}=0$. Such infinite group velocity indicates a superluminal behavior of the $k$-gap solitons as they travel faster than light inside the nonlinear PTCs.
Figure~\ref{fig:solitons_PTC}(c) shows the generation of such a $k$-gap soliton. To produce the results in the figure, the units are chosen such that the speed of light in vacuum $c=1$, the modulation frequency is $\omega_\mathrm{m}=4\pi$,  the modulation strength $m_\varepsilon=0.12$, and the third-order nonlinear susceptibility is $\chi^{(3)}=\varepsilon_0\varepsilon_\mathrm{av}^3/300$. Further, in Fig.~\ref{fig:solitons_PTC}(d), the temporal profiles of the gap soliton excited by the input pulse in Fig.~\ref{fig:solitons_PTC}(c) for different spatial locations are plotted. The intensity of the solitons in Fig.~\ref{fig:solitons_PTC}(d) first grows as a function of time. Such growth is due to the exponentially growing modes inside the $k$-gap. However, after attaining a certain peak, the intensity decays with time. Such decay is driven by the nonlinearities of the system that transfer the power from the growing modes to the decaying modes of the $k$-gap. Furthermore, as noted in \cite{pan2023superluminal}, the infinite group velocity of the solitons does not violate Einstein's causality principle. This is because, in active media, the information velocity is defined by the velocity of the leading edge of the wave packet. Therefore, the group velocity, which quantifies the motion of the center of the wave packet, does not relate to the velocity at which information travels. Of course, as shown numerically, the velocity of the leading edge of the wave packet does not exceed the speed of light \cite{pan2023superluminal}. Finally, recently \cite{kiselev2024symmetry} showed that the interplay between a time-varying permittivity and nonlinearity induces broken spatial and time translation symmetries in the PTCs.

\section{Relations to other parametric systems: Similarity and distinction}\label{secRelations}
% In this section, we will discuss those concepts in electrical engineering and optics that closely resemble PTCs in terms of name or emerging effects. In particular, we will review the underlying similarities and differences.

In the rapidly evolving field of photonics, distinguishing PTCs from related phenomena is essential for a comprehensive understanding. This discussion clarifies the nuances that differentiate PTCs, such as their unique temporal and frequency characteristics, from similar-sounding or -appearing phenomena in other fields of electrical engineering, optics, and condensed matter physics. By elucidating these distinctions and parallels, we facilitate deeper insights into the behavior and applications of PTCs, contributing to a more accurate scientific discourse. Such an analysis not only aids in theoretical comprehension but also guides practical advancements, ensuring that researchers and practitioners accurately apply these concepts to innovate in different areas.

Generally, PTCs are material systems that involve parametric amplification effects. In particular, light with a wavenumber located inside the momentum bandgap of a PTC is parametrically amplified. In a general sense, parametric amplification is called ``parametric" because it involves the modulation of a system's parameter (such as capacitance, inductance, or permittivity) to achieve signal amplification. This process does not add energy directly to the signal, but rather, through the periodic variation of the system parameter, energy is transferred from an external source to a signal that is amplified.
Parametric amplification can occur in various physical systems, such as electrical circuits, nonlinear optical materials, transmission lines, mechanical parametric pendula, etc. In the following sections of this chapter, we discuss and compare parametric amplification effects in different systems: 
electrical circuits in Section~\ref{tvcapacitor}, nonlinear materials in Section~\ref{OPA}, and PTCs in Section~\ref{PAPTCs}.

Finally, we compare in Section~\ref{TCPTC} PTCs to the recently discovered ``time crystals'' in condensed matter physics. 

%%%%%%%%%%%%%%%%%%%%%%%%%%%%%%%%%%%%%%%%%%%%%%%%%%%%%%%%%%%%%%%%%%%%%%%%%%%%%%%%%%%%%%%%%%%%
%%%%%%%%%%%%%%%%%%%%%%%%%%%%%%%%%%%%%%%%%%%%%%%%%%%%%%%%%%%%%%%%%%%%%%%%%%%%%%%%%%%%%%%%%%%%

%Based on our Sec. S7 in our Sci.Adv.
%Explain phase insensitivity. 

{\subsection{Parametric amplification in electrical circuits }\label{tvcapacitor}}

Parametric amplification, a cornerstone in modern circuit design, exploits the principle of parameter modulation to amplify signals. This mechanism, distinct from direct energy transfer seen in traditional amplifiers, modulates a system parameter (like capacitance or inductance) periodically at a specific frequency. This modulation can transfer energy from the power supply to the signal, resulting in amplification. That concept of parametric amplification, first discovered in 1892~\cite{fitzgerald1902driving}, has found profound applications ranging from telecommunications to quantum computing. In radio and microwave engineering, it enhances signal strength while preserving phase information, which is essential for high-fidelity communications. 
% Similarly, in quantum systems, parametric amplification is pivotal for signal processing at minimal energy levels, which is crucial for maintaining quantum coherence. The broad applicability of parametric effects, from enhancing weak signals in nano-electromechanical systems to enabling sensitive measurements in scientific instrumentation, underscores their significance in advancing technology and deepening our understanding of complex physical phenomena. The fact that a system parameter is periodically changed to amplify a signal for a parametric amplification and a PTC prompts the following comparison, and we try to clarify similarities and differences.
% \subsubsection{Time-varying capacitance}

\begin{figure}[htb]
\centerline{\includegraphics[width= 1\columnwidth]{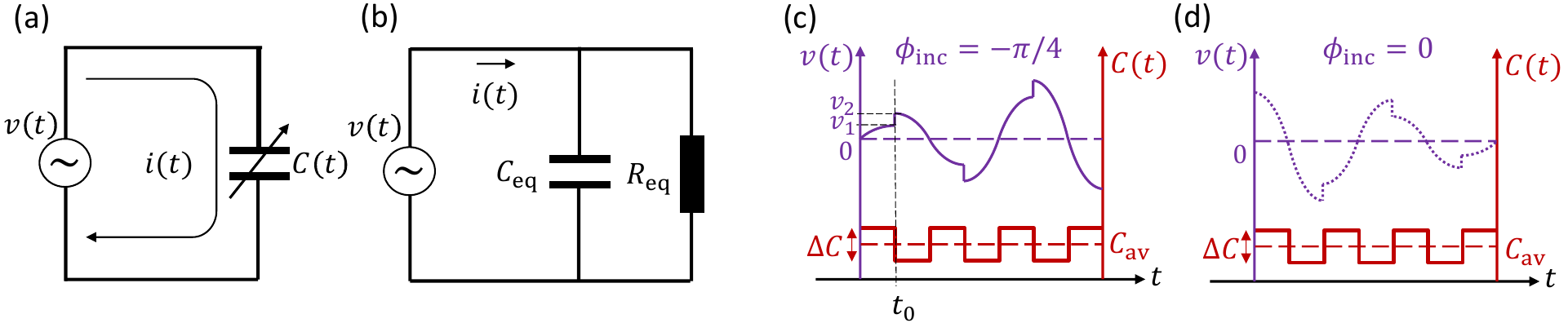}}
\caption{(a) Electrical circuit with a time-varying capacitance. (b) Circuit equivalent to that in (a) where the time-varying capacitance is replaced by a parallel connection of a time-invariant capacitance $C_{\rm eq}$ and a resistance $R_{\rm eq}$. The resistance can have positive and negative values depending on the phase $\phi_{\rm inc}$. (c) The time evolution of the voltage in the circuit shown in (a) for the case of a stepwise capacitance modulation. Phase $\phi_{\rm inc}$ is chosen such that there is parametric amplification in the circuit. (d) Same as (c) but where the phase is shifted by $\pi/4$, which leads to a parametric de-amplification.
}
\label{circuit}
\end{figure}

Let us consider the parametric amplification phenomenon in the most simple electric circuit consisting of a time-varying capacitance $C(t)$ connected to a voltage source $v(t)$, as shown in Fig.~\ref{circuit}(a). This source is time-harmonic and described by voltage $v(t)=V_0\cos(\omega_{\rm inc}t+\phi_{\rm inc})$, where $\phi_{\rm inc}$ is an arbitrary phase. 
First, let us consider for simplicity a scenario where the capacitance is modulated in time in a stepwise manner (see the red curve in Fig.~\ref{circuit}(c)). 
Such time dependency could be mechanically induced by altering the spacing between a capacitor's plates. When these plates are moved apart or closer together, the capacitance decreases or increases correspondingly. Consider a scenario where at the moment $t=t_0$, when the voltage across the capacitor reaches its peak value $v_1$, the plates are instantly separated. This action necessitates work against the attractive force existing between the charged plates. 
Consequently, the capacitance shifts from $C_1=C_{\rm av}+ \Delta C/2$ to $C_2=C_{\rm av}- \Delta C/2$. It is convenient to introduce the relative capacitance change $\delta_C = C_1/C_2>1$. It is critical to note that during this transition, the electric charge on the capacitor remains unchanged~\cite[p.~389]{haus2012electromagnetic}. As a result, the voltage across the capacitor must rise to $v_2= \delta_C v_1$.

Accordingly, the energy stored in the capacitor, expressed as $C(t) v(t)^2 /2$, increases by $\delta_C$. This increase in energy originates from the mechanical work done in separating the capacitor plates. Next, at the time moment when the voltage across the capacitor is zero, the plates are pushed together. But this time, no energy is stored or released, as the charge at the capacitor plates is zero. Likewise, when the voltage reaches its maximum negative value, the plates are moved apart once again, leading to another increase in the capacitor's energy (refer to Fig.~\ref{circuit}(c)). Thus, the energy in the circuit escalates throughout one complete oscillation cycle, resulting in parametric amplification (here, the modulated ``parameter'' is the capacitance $C$). The energy growth in the circuit is exponential. 
One can notice that to have an energy growth without any energy decrease within each cycle of the voltage (i.e., for the time-interval $\Delta t=2\pi/\omega_{\rm inc}$), it is important to modulate the capacitance at twice the frequency, i.e., $2\omega_{\rm inc}$. Importantly, the modulation of the capacitance must be synchronized with the voltage in the circuit to obtain the maximum parametric amplification. 

One can follow the same logic and consider the case when the voltage oscillation is shifted by a $\pi/4$ phase with respect to the voltage signal in the above example. Then, the power in the circuit would decrease twice per cycle (i.e., for $\Delta t=2\pi/\omega_{\rm inc}$) and never increase, as shown in Fig.~\ref{circuit}(d). This regime is called parametric de-amplification~\cite{yurke1989observation}. 

Next, let us conduct a more detailed analysis of parametric amplification and de-amplification in the circuit. To simplify the analysis, instead of a stepwise capacitance modulation, we assume a time-harmonic modulation of the form $C(t)=C_{\rm av} (1+m_{\rm C} \cos(\omega_{\rm{m}}t)) $, where we assume upfront that $\omega_{\rm{m}}=2\omega_{\rm{inc}}$. Note that we choose the modulation phase to be zero because only the relative phase between the modulation function and the voltage is important and is described by $\phi_{\rm inc}$ in $v(t)=V_0\cos(\omega_{\rm inc}t+\phi_{\rm inc})$. 
% Here, we supposed that the source voltage does not have a phase for simplicity. 
% On the other hand, the time-varying capacitance is expressed as $C(t)=C_0(1+\delta\cos(\omega_{\rm{m}}t+\phi_0))$, in which $C_0$ is the static capacitance, $\delta$ denotes the modulation strength, $\omega_{\rm{m}}$ represents the modulation angular frequency, and $\phi_0$ is the corresponding phase. In fact, $\phi_0$ can be interpreted as the phase difference between the voltage over the capacitance and the temporal function of the modulation. 
The alternating electric current flowing through the capacitance can be calculated using  $i(t)=\frac{{\rm d}[C(t) v(t)]}{{\rm d}t}$. By substituting inside this expression the capacitance and voltage functions and dropping the term oscillating at $3\omega_{\rm inc}$ (practically, it is accomplished by adding a frequency filter), we arrive at the electric current consisting of two terms oscillating at $\omega_{\rm inc}$:

\begin{equation}
i(t)=-C_{\rm av} V_0 \omega_{\rm inc} \sin(\omega_{\rm inc}t+\phi_{\rm inc})
- \frac{m_{\rm C} C_{\rm av} V_0}{2} \omega_{\rm inc} \sin(\omega_{\rm inc}t -\phi_{\rm inc})= i_1(t)+i_2(t). 
\l{currentnew}
\end{equation} 
% By doing this, we explicitly observe that the electric current has two harmonics, and to have $\omega_0$ as one of those harmonics, the modulation angular frequency should be $\omega_{\rm{m}}=2\omega_0$. Accordingly, the generated harmonics are $\omega_0$ and $3\omega_0$. If the modulation strength is small enough, we can assume that the third harmonic generation is neglected. Therefore, the electric current reduces to $i(t)=i_{\rm{s}}(t)+i_{\rm{A}}(t)=-\omega_0C_0V_0\sin(\omega_0t)-[(\omega_0C_0V_0\delta)/2]\sin(\omega_0t+\phi)$. 
The first term $i_1(t)$ is the electric current associated with a conventional linear and time-invariant capacitance $C_{\rm eq}$. It is easy to see that the time-averaged dissipated power due to this current term is always zero. That is the case because $v(t)$ and $i_1(t)$ have a $90^\circ$ phase shift, i.e., $ \langle P(t) \rangle_t = \langle v(t) i_1(t) \rangle_t=0$.

The second term $i_2(t)$ is drastically different because it is a consequence of the temporal modulation. For determining the contribution of this second term, the phase $\phi_{\rm inc}$ plays a significant role. The power dissipated at the capacitor for this term is given by 

\begin{equation}
\langle P(t) \rangle_t = \frac{m_{\rm C} C_{\rm av} V_0^2 \omega_{\rm inc}}{4} \sin 2\phi_{\rm inc}.
\l{powernew}
\end{equation} 

Thus, the current term $i_2(t)$ for certain values of $\phi_{\rm inc}$ leads to nonzero dissipated or accumulated power in the time-varying capacitor like it had some equivalent resistance $R_{\rm eq}$. 
Therefore, we can view the circuit in Fig.~\ref{circuit}(a) as equivalent to a parallel connection of some static capacitance $C_{\rm eq}$ and the static resistance $R_{\rm eq}$, as shown in Fig.~\ref{circuit}(b). The resistance can be determined from $ \langle P(t) \rangle_t = \frac{V_0^2}{2 R_{\rm eq}}$.
One can see that depending on the value of $\phi_{\rm inc}$, the resistance 
% Let us focus on the particular values of $\phi_0=\pi/2$ and $3\pi/2$. The discussion on other phase values is not in the scope of this tutorial, but it has been mentioned, for example, in Ref.~\eqref{PrasadFuPRAppliedLE}. Concerning these two values, we see that the second term equals $i_{\rm{A}}(t)=\pm(\omega_0C_0V_0\delta/2)\cos(\omega_0t)$, which is in phase with the voltage source ($\phi_0=3\pi/2$). Alternatively, it is 180 degrees out of phase when $\phi_0=\pi/2$. In this scenario, hence, $i_{\rm{A}}(t)$ must correspond to a positive or negative resistance, where the latter one is identical with 
\begin{equation}
R_{\rm{eq}}= \frac{2}{\omega_{\rm inc} m_{\rm C} C_{\rm av}  \sin 2\phi_{\rm inc}}
\l{resipa}
\end{equation} 
can have different signs. The equivalent time-invariant capacitance is given by $C_{\rm eq} = C_{\rm av} + \frac{m_{\rm C} C_{\rm av} }{2} \cos 2 \phi_{\rm inc}$. Note that, in general, $C_{\rm eq} \neq  C_{\rm av}$. Next, let us consider three special cases. 
First, when  $\phi_{\rm inc}= -\pi /4 +\pi p$ ($p \in \mathds{Z}$), the equivalent resistance is negative $R_{\rm{eq}}= -\frac{2}{\omega_{\rm inc} m_{\rm C} C_{\rm av}}$ and $C_{\rm eq} = C_{\rm av}$, resulting in parametric amplification in the circuit. On the other hand, when $\phi_{\rm inc}= \pi /4 +\pi p$, $R_{\rm{eq}} =\frac{2}{\omega_{\rm inc} m_{\rm C} C_{\rm av}} >0$ and $C_{\rm eq} = C_{\rm av}$, there is a parametric de-amplification in the circuit. Finally, when $\phi_{\rm inc} = \pi p/2$, we obtain that the equivalent parameters are given by $R_{\rm{eq}} \rightarrow \infty$ and $C_{\rm eq} = C_{\rm av} + (-1)^p \frac{m_{\rm C} C_{\rm av} }{2}$. This means that instead of the equivalent resistor, we have an open circuit in Fig.~\ref{circuit}(b), and no power accumulates or dissipates in the circuit. In this case, $\langle P(t) \rangle_t=0$. 

The parametric amplification occurs when the modulation frequency is exactly twice the signal frequency $\omega_{\rm{m}}=2\omega_{\rm{inc}}$. For that reason, the process is referred to as \textit{degenerate}~\cite[Sec.~11.4]{haus2012electromagnetic}. In the context of the considered circuit, this terminology means that $i_1(t)$ and $i_2(t)$ oscillate at the same signal frequency $\omega_{\rm inc}$, i.e., the two current modes degenerate regarding their frequency. Degenerate parametric amplification in electrical circuits is a process that is strongly sensitive to the phase difference of the modulation function and the signal~\cite[Ch.~11]{yamamoto2004fundamentals}, as one can see from Eq.~\r{resipa}. In fact, this statement is very general and applies to other degenerate parametric physical systems. For example, as we show in Section~\ref{OPA}, degenerate parametric amplification in nonlinear optics strongly depends on the relative phase difference between the signal and pump photons. Further, amplification inside the momentum bandgap in PTCs is typically studied in the case when the signal frequency $ \omega_{\rm{F}}$ is located in the first Brillouin zone, that is, in the degenerate regime when $\omega_{\rm{m}}=2\omega_{\rm{F}}$. Therefore, strictly speaking, PTCs are also phase-sensitive. However, as discussed in Section~\ref{PAPTCs}, this phase sensitivity additionally depends on the spatial distribution of the signal wave inside the PTC. In fact, in most scenarios of degenerate PTCs, the phase sensitivity vanishes unless the crystal is engineered in a specific way. This makes parametric amplification in PTCs different compared to that in other systems. A qualitative explanation of this difference was made in the Supplementary Information of Ref.~\cite{wang2023metasurface}, where a PTC was viewed as a transmission line comprising a cascade of circuits similar to that in Fig.~\ref{circuit}(a). Although amplification/de-amplification in each circuit is phase-sensitive, in total, the energy of the standing mode (since it is inside the momentum bandgap) inside the transmission line grows exponentially. A more rigorous explanation of the phase-sensitivity properties of PTCs is given in Section~\ref{PAPTCs}.

Phase sensitivity of parametric amplification often leads to difficulties in practical setups, as one needs to control the modulation and signal phases carefully. In such scenarios, engineers opt for non-degenerate amplification, that is, when  $\omega_{\rm{m}} \neq 2\omega_{\rm{inc}}$. It should be noted that non-degenerate amplification is qualitatively different from the degenerate one~\cite[p.~388]{haus2012electromagnetic} and is a phase-insensitive process.
 % because in the former case the signal (at $\omega_{\rm inc}$) and idler (at $\omega_{\rm m}-\omega_{\rm inc}$) share the same frequency and thus determine jointly the input signal excitation
On the other hand, the phase sensitivity of the degenerate parametric amplification also has very important applications, such as for generating squeezed states of light. 
These states have reduced noise in one quadrature of the electromagnetic field at the expense of increased noise in the other, allowing for measurements that surpass the quantum noise limit. This is invaluable in fields where noise reduction can significantly enhance performance, such as in quantum metrology and gravitational wave detection.

{\subsection{Parametric amplification in nonlinear optics}\label{OPA}}

In the previous section, we considered parametric amplification in a single lumped circuit element. However, more related to PTCs are distributed (bulk) systems where the modulation is in the traveling-wave form. Among others, they include transmission lines and nonlinear optical materials, as was pointed out in the Introduction. In this section, we overview coupled-mode equations for optical parametric amplification (difference-frequency generation process) in materials with a second-order nonlinearity. We start with the general (non-degenerate) case and conclude with the special case of degenerate parametric amplification. In the following Section~\ref{PAPTCs}, we derive coupled-mode equations for light propagation inside the momentum bandgap of a PTC and compare them with those in the present section.

\begin{figure}[tb]
\centerline{\includegraphics[width= 1\columnwidth]{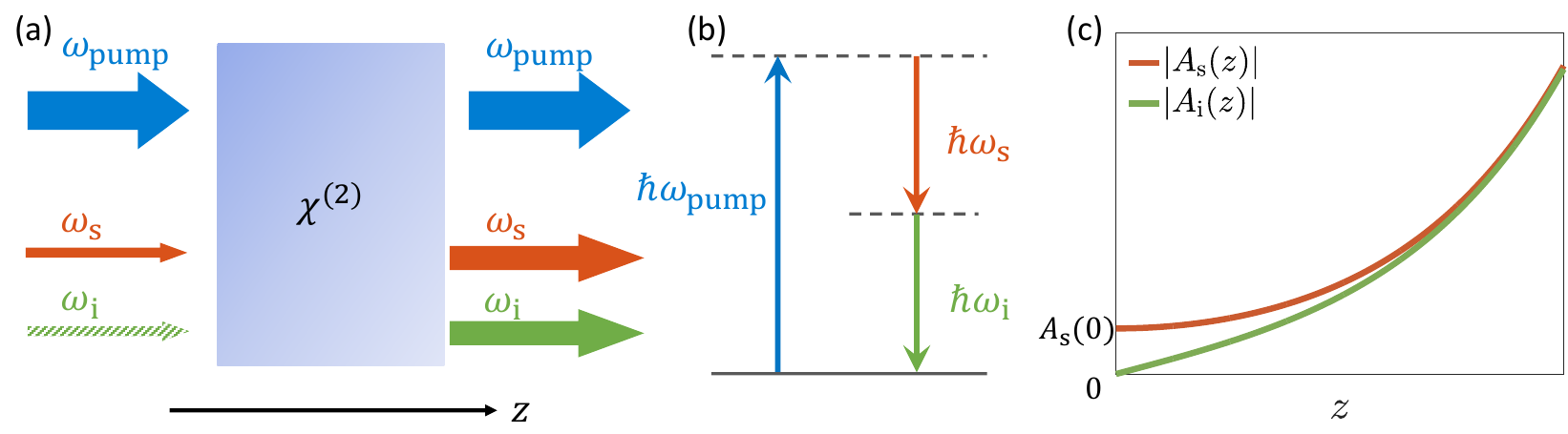}}
\caption{(a) Difference-frequency generation process resulting in optical parametric amplification along the spatial coordinate $z$. Typically, the idler frequency is absent at the input. (b) Photon description of the interaction of three optical waves in the process shown in (a). (c) Spatial evolution of the envelope amplitudes of the signal and idler waves in the assumption of perfect phase-matching $\Delta k=0$ and an undepleted pump. The initial conditions are given by $A_{\rm s}(0) \neq 0$ and  $A_{\rm i}^*(0) = 0$.
}
\label{Fig:nonlinear}
\end{figure}

Let us consider the difference-frequency generation process shown in Fig.~\ref{Fig:nonlinear}(a). There, a pump wave at the frequency $\omega_{\rm pump}$ and a signal wave at the frequency $\omega_{\rm s}$ interact in a lossless optical medium with a $\chi^{(2)}$ nonlinearity. Out of that interaction, they produce an output idler wave at frequency $\omega_{\rm i} = \omega_{\rm pump} - \omega_{\rm s}$. For simplicity, we assume that the pump wave is strong, and we neglect back-action from the generated signal and idler onto the pump. This implies that it is undepleted by the nonlinear interaction so that we can treat its amplitude as constant over distance $z$. The photon description of the interaction of the three optical waves in the considered process is shown in Fig.~\ref{Fig:nonlinear}(b). The spatial dependence of the amplitude of the pump wave propagating along the $z$-direction is given by 
$a_{\rm pump}(z) = A_{\rm pump} \exp(-j k_{\rm pump} z)$, where $A_{\rm pump}$ is the constant pump amplitude and $k_{\rm pump}$ is the wavenumber of the pump wave. For the signal and idler waves $a_{\rm s}(z) = A_{\rm s}(z) \exp(-j k_{\rm s} z)$ and $a_{\rm i}(z) = A_{\rm i}(z) \exp(-j k_{\rm i} z)$ where the amplitudes are a slowly varying function in space compared to the fastly oscillating exponential function. The well-known coupled-mode equations expressing how the amplitudes of the three waves are related to one another inside the nonlinear material along the $z$ direction read as~\cite[Sect.~2.8]{boyd2008nonlinear}
\begin{equation}
\begin{array}{l}
\displaystyle \frac{{\rm d} A_{\rm s}}{{\rm d} z} = -j \, \eta_{\rm OPA} \, A_{\rm i}^* e^{-j\Delta k z}\,, \vspace{2mm} \\
\displaystyle \frac{{\rm d} A^*_{\rm i}}{{\rm d} z} = j \, \eta^*_{\rm OPA} \, \frac{\omega_{\rm i}n_{\rm s}}{\omega_{\rm s}n_{\rm i}} A_{\rm s} e^{j\Delta k z}\,,
\end{array}
\l{eq:param1}
\end{equation}
where $\Delta k= k_{\rm pump} - k_{\rm s} -k_{\rm i}$, the coupling coefficient is $\eta_{\rm OPA} = \omega_{\rm s} \chi^{(2)} A_{\rm pump}/(c n_{\rm s})$, $n_{\rm s}$, and the $n_{\rm i}$ denote the refractive indices of the medium for the signal and idler waves. Subscript ``OPA" refers to the optical parametric amplification process. In the expression for the coupling coefficient, $\chi^{(2)}$ is the second-order nonlinear susceptibility. 
From the first equation in Eq.~\r{eq:param1}, one can see that the increase in signal amplitude $A_{\rm s}$ along $z$ is proportional to the amplitude of the idler wave $A_{\rm i}^*$ and the pump wave amplitude $A_{\rm pump}$ (hidden inside coefficient $\eta_{\rm OPA}$). 
Note that the idler amplitude appears complex-conjugated because the product of the two time dependencies $\exp(j\omega_{\rm pump} t)$ of the pump amplitude and $\exp(-j\omega_{\rm i} t)$ of the complex conjugate of the idler amplitude results in the time dependence $\exp[j\omega_{\rm pump} t -j \omega_{\rm i}t] = \exp(j\omega_{\rm s} t)$ of the signal wave.  
Therefore, from Eq.~\r{eq:param1}, the presence of a field at frequency $\omega_{\rm i}$ stimulates the downward transition from $\omega_{\rm pump}$ that leads to the generation of the $\omega_{\rm s}$ field. Likewise, the $\omega_{\rm s}$ wave stimulates the generation of the $\omega_{\rm i}$ wave. Hence, the generation of the signal wave reinforces the generation of the idler wave and vice versa, leading to the exponential growth of each wave.

The optical parametric amplification process requires nearly perfect phase matching $\Delta k \approx 0$ since otherwise, the coupling between the signal and idler waves gets out of phase rapidly as light travels along the nonlinear material. The requirement of phase matching
\begin{equation}
k_{\rm pump} = k_{\rm s} +k_{\rm i}
\l{eq:phasematchreq}
\end{equation}
in practice usually translates into the necessity to engineer the birefringence of the nonlinear material, but a detailed consideration would go beyond the scope of this tutorial.  

Assuming the typical initial conditions $A_{\rm s}(0) \neq 0$ and $A_{\rm i}^*(0) = 0$, the solution of the differential equations~\r{eq:param1} reads~\cite[Sect.~2.8]{boyd2008nonlinear}
\begin{equation}
\begin{array}{rcl}
\displaystyle A_{\rm s}(z) & = & A_{\rm s}(0) \cosh \left[   |\eta_{\rm OPA}| \sqrt{\frac{\omega_{\rm i} n_{\rm s}}{\omega_{\rm s} n_{\rm i}}} z \right]\,, \vspace{2mm} \\
\displaystyle A_{\rm i}^*(z) & = &  \frac{j}{\eta_{\rm OPA}} A_{\rm s}(0) \sinh \left[   |\eta_{\rm OPA}| \sqrt{\frac{\omega_{\rm i} n_{\rm s}}{\omega_{\rm s} n_{\rm i}}} z \right]\,.
\end{array}
\l{eq:param2}
\end{equation}
The nature of this solution is shown in Fig.~\ref{Fig:nonlinear}(c). Note that the signal and the idler fields experience monotonic growth and that asymptotically, each field experiences exponential growth. Importantly, this exponential growth is independent of the phase difference of the pump and signal waves (the phase of $A_{\rm pump}(z)$ does not affect the value of $|\eta_{\rm OPA}|$ appearing in Eq.~\r{eq:param2} which determines the mode amplitude). 

In the special case of degenerate parametric amplification, i.e., when $\omega_{\rm s}=\omega_{\rm i}$, coupled-mode equations~\r{eq:param1} simplify to
\begin{equation}
\begin{array}{l}
\displaystyle \frac{{\rm d} A_{\rm s}}{{\rm d} z} = -j \, \eta_{\rm OPA} \, A_{\rm s}^*\,, \vspace{2mm} \\
\displaystyle \frac{{\rm d} A^*_{\rm s}}{{\rm d} z} = j \, \eta^*_{\rm OPA}   A_{\rm s}\,,
\end{array}
\l{eq:param3}
\end{equation}
where perfect phase matching was assumed ($\Delta k=0$). Nevertheless, the degenerate scenario is qualitatively very different. Now, the excitations $A_{\rm s}(0)$ and $A_{\rm s}^*(0)$ lie in the same frequency band and thus determine jointly the input signal excitation. In other words, one cannot, in practice, satisfy the previously assumed initial conditions $A_{\rm s}(0) \neq 0$ and  $A_{\rm s}^*(0) = 0$. Therefore, it is convenient to rewrite two decoupled mode envelopes as~\cite[Sec.~11.4]{haus2012electromagnetic} 
\begin{equation}
\begin{array}{rcl}
A_{\rm OPA}^{(1)}(z) &=& \frac{1}{2} \left[ A_{\rm s}(z) e^{-j{\rm \psi_{\rm OPA}}/2} + A_{\rm s}^*(z) e^{j{\rm \psi_{\rm OPA}}/2} \right]\,, \vspace{2mm} \\
A_{\rm OPA}^{(2)}(z) &=& -\frac{1}{2} \left[ A_{\rm s}(z) e^{-j{\rm \psi_{\rm OPA}}/2} - A_{\rm s}^*(z) e^{j{\rm \psi_{\rm OPA}}/2} \right]\,,
\end{array}
\l{eq:param4}
\end{equation}
where we defined the phase $\psi_{\rm OPA}$ according to $j\eta_{\rm OPA} = |\eta_{\rm OPA}| e^{j \psi_{\rm OPA}}$ as related to the phase of the pump wave.

Then, the solution of Eq.~\r{eq:param3} in the basis of the decoupled mode envelopes Eq.~\r{eq:param4} reads as
\begin{equation}
\left[ \begin{array}{c}
A_{\rm{OPA}}^{(1)}(z) \\
A_{\rm{OPA}}^{(2)}(z) 
\end{array} \right]
= 
\left[ \begin{array}{cc}
e^{-|\eta_{\rm{OPA}}| z} & 0 \\
0 & e^{|\eta_{\rm{OPA}}| z} 
\end{array} \right]
\cdot
\left[ \begin{array}{c}
A_{\rm{OPA}}^{(1)}(0) \\
A_{\rm{OPA}}^{(2)}(0) 
\end{array} \right]\,.
\l{param5}
\end{equation}
These equations predict an exponential spatial growth of $A_{\rm OPA}^{(2)}$ and an exponential decay of $A_{\rm OPA}^{(1)}$. The two decoupled modes are $90^\circ $ out of phase.  
In contrast to Eq.~\r{eq:param2}, now the phase of the pump wave $A_{\rm pump}$ affects the light propagation inside the nonlinear material, providing the possibility for both parametric amplification and de-amplification. Indeed, this phase affects $\psi_{\rm OPA}$ and, therefore, also the values of the amplitudes $A_{\rm{OPA}}^{(1)}(0)$ and $A_{\rm{OPA}}^{(2)}(0)$, as seen from Eq.~\r{eq:param4}. 

Thus, degenerate parametric amplification in materials with $\chi^{(2)}$ nonlinearity is phase-sensitive, similarly to that in electrical circuits discussed in Section~\ref{tvcapacitor}. In the next section, we analyze the coupled-mode equations for the dominant modes inside the momentum bandgap of a PTC and discuss the differences and similarities with the considered case.

\subsection{Parametric amplification in PTCs }\label{PAPTCs}

In this section, we derive the temporal coupled-mode theory for light inside an infinite (both in space and time) PTC with negligible frequency dispersion, in particular, when its wavenumber resides inside the momentum bandgap. The derivations are similar to those published in \cite[Suppl. Mat.]{asadchy2022parametric},\cite{khurgin2023photonic}. 
In contrast to frequency-domain analysis, where eigenmodes within the momentum bandgap are described by complex frequencies, here we will analyze temporal evolution in the time domain. Although both approaches yield the same solution for wave propagation inside the PTC, the time-domain analysis allows for a clearer comparison of wave processes in PTCs and nonlinear optics.

\begin{figure}[tb]
\centerline{\includegraphics[width= 1\columnwidth]{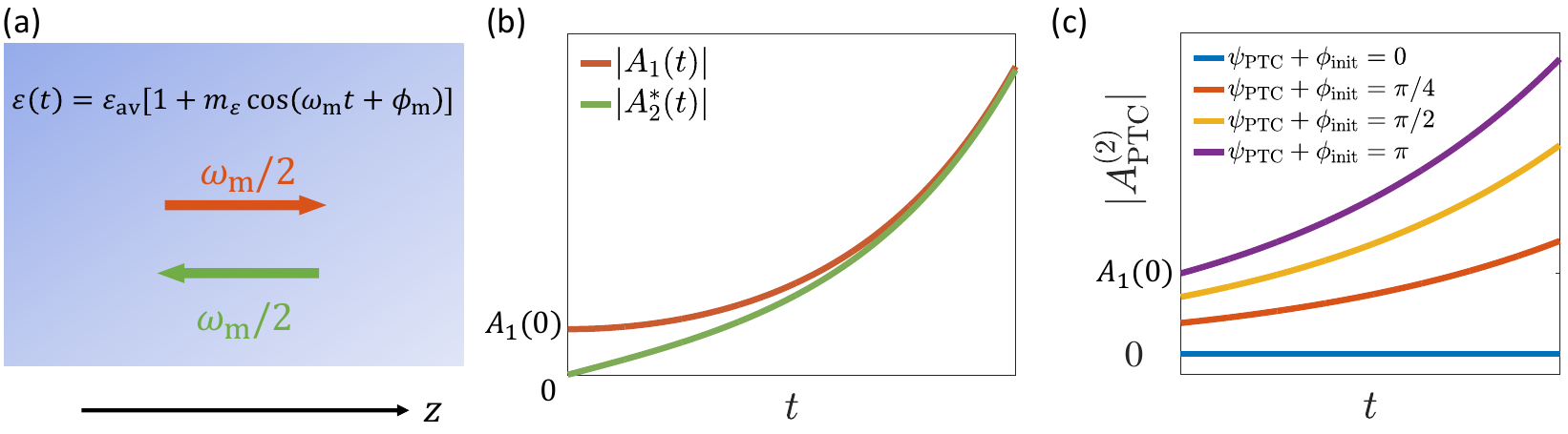}}
\caption{(a) Two dominant modes inside a momentum bandgap of a PTC. They correspond to plane waves at frequency $\omega_{\rm m}/2$ propagating in the opposite directions. 
(b) Temporal evolution of the envelope amplitudes of the two modes shown in (a). The initial conditions are given by $A_1(0) \neq 0$ and  $A^*_2(0) = 0$ (fields before the temporal modulation started corresponded to a single propagating wave). (c) Phase sensitivity analysis of a PTC with initial conditions in the form $A_2^*(0) =  A_1(0) e^{j \phi_{\rm init}}$ (fields before the temporal modulation started corresponded to a standing wave).
}
\label{PTCCMT}
\end{figure}

We start with the assumption that inside the bandgap, there are two dominant modes, the $0$-th and the $-1$-st, oscillating at frequency $\omega_{\rm m}/2$ and $-\omega_{\rm m}/2$, respectively (see illustration in Fig.~\ref{PTCCMT}(a)). 
As discussed in Section~\ref{momband}, this approximation typically works well (see also Fig.~\ref{Fig:standingwave}(a)). Moreover, the proposed theory can be further generalized to cases where the number of dominant modes is four or more. Therefore, from~(\ref{eq:BLOCHT}), we write the real-valued electric field of an eigenmode in the following form:
\begin{equation} 
% \begin{array}{cc}
 \ve{E} (z,t) =   A_1(t) \, e^{j \w_{\rm m}/2 t} \, e^{-j k z}\_a_x +
 A_2^*(t) \, e^{-j \w_{\rm m}/2 t} \, e^{-j k z} \_a_x + 
 {\rm c.c.}\,, 
 % \\ \vspace{2mm}
 % =  A_1(t) \, e^{-i \w_{\rm m}/2 t} \, e^{i k z}\_a_x +
 % A_2(t) \, e^{-i \w_{\rm m}/2 t} \, e^{-i k z} \_a_x + 
 % {\rm c.c.}
 % \end{array}
      \l{S18}
\end{equation}
where ``c.c.'' stands for the complex conjugated terms, $k$ denotes the real-valued wavenumber of the mode that is assumed to be inside the momentum bandgap, and $A_1(t)$ and $A_2^*(t)$ are the slowly varying (compared to $\exp (j\w_{\rm m}/2 t)$) unknown temporal mode envelopes which define the temporal evolution of the modes inside the momentum bandgap. Note that in Eq.~\r{S18} we write the $A_2^*(t)$-mode as complex conjugated to follow the same notations as in Section~\ref{OPA} for an easier comparison. 
We aim to determine $A_1(t)$ and $A_2^*(t)$. 

As in the previous chapters and sections, we assume here that the permittivity is modulated according to $\varepsilon(t) = \varepsilon_{\rm av} (1+ m_\varepsilon  \cos (\omega_{\rm m} t+\phi_{\rm m}) )= \varepsilon_{\rm av} (1+ m_\varepsilon ( e^{j \omega_{\rm m}t +j \phi_{\rm m}} +e^{-j \omega_{\rm m}t -j \phi_{\rm m}}) /2)$ with some arbitrary modulation phase $\phi_{\rm m}$. 
By substituting the modulation function and the ansatz~Eq.~\r{S18} into the wave equation~(\ref{Eq: governing}) and dropping the higher-order harmonics oscillating at $3\omega_{\rm m}/2$, we obtain~\cite[Suppl. Mat.]{asadchy2022parametric}
\begin{equation} 
\begin{array}{cc}\displaystyle
- \frac{k^2 c^2 A_1(t)}{\varepsilon_{\rm av}}    = 
 \aattone +2j \frac{\w_{\rm m}}{2} \atone - \frac{\w_{\rm m}^2}{4} A_1(t)
    \vspace{0.3cm} \\ \displaystyle
+\frac{ m_\varepsilon e^{j\phi_{\rm m}}}{2}   \left[ \aattconjtwo +2j \frac{\w_{\rm m}}{2} \atconjtwo - \frac{\w_{\rm m}^2}{4} A_2^*(t)\right]\,, 
\end{array}
      \l{S192}
\end{equation}
%%%%%
\begin{equation} 
\begin{array}{cc}\displaystyle
- \frac{k^2 c^2 A_2(t)}{\varepsilon_{\rm av}}    = 
 \aatttwo +2j \frac{\w_{\rm m}}{2} \attwo - \frac{\w_{\rm m}^2}{4} A_2(t)
    \vspace{0.3cm} \\ \displaystyle
+\frac{ m_\varepsilon e^{j\phi_{\rm m}}}{2}   \left[ \aattconjone +2j \frac{\w_{\rm m}}{2} \atconjone - \frac{\w_{\rm m}^2}{4} A_1^*(t)\right]\,.
\end{array}
      \l{S192b}
\end{equation}
The complex conjugate terms in~Eq.~\r{S18} yield the same equations as~Eq.~\r{S192} and Eq.~\r{S192b}.
It should be noted that from a single wave equation, we ended up with two equations, i.e. Eq.~\r{S192} and Eq.~\r{S192b}, because even in the parametric degenerate case, photons at frequencies $\omega_{\rm m}/2$ and $-\omega_{\rm m}/2$ should be distinguished. This leads to one coupled equation for each kind of photon.  

Now, we can use the fact that the same medium without temporal modulation, that is, when $\Delta \varepsilon  \rightarrow 0$  and $\varepsilon(t) = \varepsilon_{\rm av}$, has the dispersion relation of the conventional form
\begin{equation} 
k^2 c^2 = \varepsilon_{\rm av}  \omega_{\rm st}^2\,,
      \l{Snew}
\end{equation}
which can be observed, e.g., from Fig.~\ref{fig:bandstructure_harmonic}. Here, we add a small imaginary part to the eigenfrequency in the stationary medium $\omega_{\rm st}= \frac{\omega_{\rm m}}{2} +j g $, where $g\ll \frac{\omega_{\rm m}}{2}$, to consider a possible small absorption in the medium. Note that $k$ is assumed to be real-valued. By substituting  $\frac{\omega_{\rm m}}{2}=\omega_{\rm st}-j g $ inside Eq.~\r{S192b} and using Eq.~\r{Snew}, we obtain~\cite[Supp. Inf.]{asadchy2022parametric}
\begin{equation} 
\begin{array}{cc}\displaystyle
 \aattone +2j (\omega_{\rm st}-j g) \atone + g (2j \omega_{\rm st} +g ) A_1(t)
    \vspace{0.3cm} \\ \displaystyle
+\frac{ m_\varepsilon e^{j\phi_{\rm m}}}{2}   \left[ \aattconjtwo +2j (\omega_{\rm st}-j g) \atconjtwo - (\omega_{\rm st}-j g)^2  A_2^*(t)\right] = 0\,, 
\end{array}
      \l{S19244}
\end{equation}
%%%%%
\begin{equation} 
\begin{array}{cc}\displaystyle
 \aatttwo +2j (\omega_{\rm st}-j g) \attwo + g (2j \omega_{\rm st} +g ) A_2(t)
     \vspace{0.3cm} \\ \displaystyle
+\frac{ m_\varepsilon e^{j\phi_{\rm m}}}{2}   \left[ \aattconjone +2j (\omega_{\rm st}-j g) \atconjone - (\omega_{\rm st}-j g)^2  A_1^*(t)\right] = 0\,.
\end{array}
      \l{S19244b}
\end{equation}

Next, by using $g \ll |\omega_{\rm st}|$ and subsequently imposing the slowly varying envelope approximation, which requires that $\frac{{\rm d} A_{1,2}(t)}{{\rm d} t} \ll \w_{\rm st} A_{1,2}(t) $ and $\frac{{\rm d}^2 A_{1,2}(t)}{{\rm d} t^2} \ll \w_{\rm st} \frac{{\rm d} A_{1,2}(t)}{{\rm d} t} $, we obtain 
\begin{equation} 
\begin{array}{cc}\displaystyle
 \frac{{\rm d} }{{\rm d} t} A_1(t) = - g  A_1(t)   -j\eta_{\rm PTC} A_2^*(t)\,,
 \vspace{0.3cm} \\ \displaystyle
\frac{{\rm d} }{{\rm d} t} A_2^*(t) =   - g   A_2^*(t)  +j\eta_{\rm PTC}^* A_1(t)\,,
 \end{array}
      \l{S24}
\end{equation}
where
\begin{equation}  
\eta_{\rm PTC} = \frac{m_\varepsilon \omega_{\rm st}}{4 } e^{j \phi_{\rm m}}.
      \l{etaeq}
\end{equation}

We can see that coupled-mode equations of a PTC inside the bandgap Eq.~\r{S24} closely resemble those of a degenerate parametric amplification in nonlinear optics  Eq.~\r{eq:param3} (especially in the case of lossless medium when $g=0$). In the present case, the roles of signal and idler photons are played by the $0$-th and $-1$-st harmonics, respectively. Instead of evolution in space along the $z$-axis, the modes' envelopes now evolve in time.

It should be noted that Eqns.~\r{S24} were derived assuming that the two dominant modes in the PTC have the same frequencies, corresponding to the degenerate case (see \r{S18}). In the non-degenerate case, the coupled-mode equations will have an exponential factor similar to that in \r{eq:param1} with the argument proportional to the frequency difference of the two dominant modes~\cite{khurgin2023photonic}.
This implies that PTCs also have a phase-matching condition. However, in contrast to Eq.~\r{eq:phasematchreq}, here the phase mismatch occurs in time rather than space. Nevertheless, satisfying the phase-matching condition in PTCs is much simpler in practice since one only needs to ensure that the signal frequency equals exactly half the modulation frequency (no need to use the effect of birefringence).

Next, making replacement $j\eta_{\rm PTC} = |\eta_{\rm PTC}| e^{j\psi_{\rm PTC}}$, we solve system~Eq.~\r{S24} regarding $A_1(t)$ and $A_2^*(t)$:
\begin{equation}
\left(\begin{array}{c}
A_1(t)   \vspace*{.4cm}\\
A_2^*(t)    \vspace*{.4cm} 
\end{array}\right)
=
e^{-g t}
\left(\begin{array}{cc}
\cosh |\eta_{\rm PTC}|  t     &      -  e^{j\psi_{\rm PTC}} \sinh |\eta_{\rm PTC}| t \vspace*{.4cm}\\
  -   e^{-j\psi_{\rm PTC}} \sinh |\eta_{\rm PTC}| t           &       \cosh |\eta_{\rm PTC}|  t    \vspace*{.4cm}
\end{array}\right) 
\cdot 
\left(\begin{array}{c}
A_1(0)   \vspace*{.4cm}\\
A_2^*(0)    \vspace*{.4cm} 
\end{array}\right)\,.
\l{S25}
\end{equation}
Here, $A_1(0)$ and $A_2^*(0)$ are the initial conditions for the mode envelope amplitudes. The zero within the function arguments denotes an arbitrarily chosen reference point in time, highlighting the relative nature of the time axis that lacks an absolute scale.
In contrast to Eq.~\r{eq:param3}, where the signal and idler waves in the degenerate scenario were inevitably coupled in the initial excitation, in the PTC case, modes $A_1(t)$ and  $A_2^*(t)$ are oppositely propagating plane waves along the $z$-axis, which could be easily decoupled and excited separately.
Specifically, it is feasible to configure a PTC such that, prior to introducing temporal modulations, a single plane wave propagates in the medium with minimized reflections at any spatial interface. This setup corresponds to a temporally finite PTC, as discussed in Section~\ref{tfptcs}. In this setup, initial conditions like $A_1(0) \neq 0$ and $A^*_2(0)= 0$ are naturally attainable.
% In fact, it is possible to prepare a PTC setup where before the temporal modulations appear, there is a single plane wave propagating in the medium (reflections at possible spatial interfaces are minimized). Such a scenario corresponds to a temporally finite PTC (see related discussion in Section~\ref{tfptcs}). In this setup,  initial conditions of the form $A_1(0) \neq 0$ and  $A^*_2(0)= 0$ can be naturally achieved.
Then, the solution is given by
\begin{equation}
\begin{array}{rcl}
\displaystyle A_1(t) & = & A_1(0) e^{-g t} \cosh     |\eta_{\rm PTC}|   t  \,, \vspace{2mm} \\
\displaystyle A_2^*(t) & = & -   A_1(0)e^{-g t}   e^{-j \psi_{\rm PTC}} \sinh    |\eta_{\rm PTC}| t\,.
\end{array}
\l{eq:param2fd}
\end{equation}
Figure~\ref{PTCCMT}(b) depicts this solution. It is clear that similar to Eq.~\r{eq:param2}, both $A_1(t)$ and $A_2^*(t)$ modes experience monotonic exponential growth assuming that $g$ is sufficiently small (compare to Fig.~\ref{Fig:nonlinear}(c)). However, this exponential growth is independent of the modulation phase $\phi_{\rm m}$. Therefore, PTCs provide a phase-insensitive parametric amplification under the chosen initial conditions, even in the degenerate case. This is in stark contrast to the optical parametric amplification considered in Section~\ref{OPA}. One can also observe from Eq.~\r{eq:param2fd} that after a sufficiently long time duration ($|\eta_{\rm PTC}|  t \gg 1$), the modes become related as  $A_2^*(t) =  A_1(t) e^{-j \psi_{\rm PTC}+j\pi}$. This fact will be discussed further at the end of the section.

Nevertheless, one can also identify initial conditions that provide phase-sensitive amplification inside the bandgap of a PTC. 
To see this, let us diagonalize the matrix in~Eq.~\r{S25}. By doing this and replacing mode envelopes according to
\begin{equation} 
\begin{array}{cc}\displaystyle
A_{\rm PTC}^{(1)}(t) = \frac{1}{2} \left[ A_1(t) e^{-j{\rm \psi_{\rm PTC}}/2} + A_2^*(t) e^{j{\rm \psi_{\rm PTC}}/2} \right]\,, \vspace{2mm} \\ \displaystyle
A_{\rm PTC}^{(2)}(t) = -\frac{1}{2} \left[ A_1(t) e^{-j{\rm \psi_{\rm PTC}}/2} - A_2^*(t) e^{j{\rm \psi_{\rm PTC}}/2} \right]\,,
 \end{array}
      \l{S26}
\end{equation}
we obtain the final solution for the modified mode envelopes:
\begin{equation}
\left[\begin{array}{c}
A_{\rm{PTC}}^{(1)}(t) \\
A_{\rm{PTC}}^{(2)}(t) 
\end{array} \right]
=  e^{-g t}
\left[ \begin{array}{cc}
e^{-|\eta_{\rm{PTC}}| t} & 0 \\
0 & e^{|\eta_{\rm{PTC}}| t} 
\end{array} \right]
\cdot
\left[ \begin{array}{c}
A_{\rm{PTC}}^{(1)}(0) \\
A_{\rm{PTC}}^{(2)}(0) 
\end{array} \right]\,.
\l{S27}
\end{equation}
One can note a striking equivalence of Eq.~\r{S27} with Eq.~\r{param5} when $g=0$. 

If one wants to find the solution where a PTC regardless of phase $\phi_{\rm m}$ always exhibits parametric de-amplification (even when $g=0$), one needs to satisfy $A_{\rm{PTC}}^{(2)}(0)=0$ and $A_{\rm{PTC}}^{(1)}(0) \neq 0$.
Then, from Eq.~\r{S26}, one finds that the initial condition must be of the form
$A_2^*(0) = e^{-j \psi_{\rm PTC}} A_1(0)$, which implies that prior to the temporal modulations, there was a standing-wave excitation in the medium. 

To find the mode envelope evolution for an arbitrary initial standing-wave excitation, let us substitute in Eq.~\r{S27} initial conditions in the form $A_2^*(0) =  A_1(0) e^{j \phi_{\rm init}}$ with some arbitrary phase $\phi_{\rm init}$. The solution then reads
\begin{equation} 
\begin{array}{cc}\displaystyle
A_{\rm PTC}^{(1)}(t) = \frac{1}{2} e^{-g t - |\eta_{\rm PTC}| t} 
\left[  e^{-j{\psi_{\rm PTC}}/2} + e^{j \phi_{\rm init}} e^{j{\psi_{\rm PTC}}/2} \right] A_1(0)\,, \vspace{2mm} \\
\displaystyle
A_{\rm PTC}^{(2)}(t) = -\frac{1}{2} e^{-g t + |\eta_{\rm PTC}| t} 
\left[  e^{-j{\psi_{\rm PTC}}/2} - e^{j \phi_{\rm init}} e^{j{\psi_{\rm PTC}}/2} \right] A_1(0)\,.
 \end{array}
      \l{S178}
\end{equation}
While the first mode is always decaying, the second (dominant) mode can be exponentially growing given that $g<|\eta_{\rm PTC}|$ and the expression in the square brackets is non-zero. Figure~\ref{PTCCMT}(c) depicts the time evolution of the dominant mode in the case when $g=0$. We can see that the phase $\psi_{\rm PTC}$ (which is a function of the modulation phase $\phi_{\rm m}$ according to  Eq.~\r{etaeq}) strongly affects the parametric amplification in the PTC.  
Thus, parametric amplification in a PTC with the initial excitation corresponding to a standing wave is strongly phase-sensitive and is similar to that in the degenerate optical parametric amplifier described by Eq.~\r{param5}. Interestingly, this fact of the phase sensitivity was exploited in \cite{hayran2023beyond}. In practice, standing-wave initial excitation can be naturally achieved when the PTC is surrounded at least from one side by a highly reflective surface, such as metal backing.

It is interesting to look back at the solution given by Eq.~\r{eq:param2fd} for the initial condition when $A_1(0) \neq 0$ and  $A^*_2(0) = 0$. As was mentioned before, after the transient time duration $t_{\rm trans}$ such that $ |\eta_{\rm PTC}| t_{\rm trans} \gg 1$, the modes inside the PTC form a standing wave described by $A_2^*(t) =  A_1(t) e^{-j \psi_{\rm PTC}+j\pi}$. In the present notations of a standing wave
$A_2^*(t_{\rm trans}) =  A_1(t_{\rm trans}) e^{j \phi_{\rm init}}$, this state is described by $\psi_{\rm PTC} +  \phi_{\rm init} = \pi$. It corresponds to the fastest growing (purple) line in Fig.~\ref{PTCCMT}(c).

Thus, in contrast to the degenerate difference-frequency generation in nonlinear optics, we see that for PTCs, parametric amplification (measured after a sufficient number of wave cycles) depends on the modulation phase only in a special case of initial excitation, that is, when a pure standing wave was inside the material prior to the temporal modulations. When the initial excitation corresponds to a single plane wave, amplification in PTCs is phase-insensitive.
This distinction in phase sensitivity compared to that in nonlinear optics arises because, in the latter scenario, signal and idler photons at the input stage are indistinguishable. Conversely, in the context of PTCs, one can readily generate an input that comprises exclusively one mode (single propagating wave).
Also, as mentioned, the phase-matching condition in PTCs is much easier to satisfy in practice than that in the optical parametric amplification process. Finally, while in the former process, the amplification occurs in space, in the latter process, it occurs in time. This could be an advantage of PTCs in some situations, as they can provide high amplification even in compact geometries.  

Finally, it should be noted that in a recent work~\cite{khurgin2023photonic}, an important comparison was made between the PTC regime and the process of backward optical parametric amplification for the case of transverse pumping~\cite{ding_transversely_1995,lanco2006semiconductor}. It was pointed out that although the geometry and the eigenmodes of the medium are similar in both cases, the two parametric processes have qualitative differences due to the different initial conditions. In particular, the backward optical parametric amplification process has a finite growth of light energy (due to the oscillatory nature of light propagation), while there is an exponential growth of the modes in PTCs.

\subsection{Comparison of time crystals and photonic time crystals\label{TCPTC}}
PTCs are fundamentally different from time crystals in condensed matter physics. In this subsection, we will first explain the basic concepts of time crystals and then discuss their differences from PTCs. 
% \\
% References: \cite{giergiel2019topological}, \cite{wang2022observation}
% \red{to viktar: please add the topological works 140, and 142 to the text. }

Time crystals, first conceived by Frank Wilczek in 2012, are to some extent dual to spatial crystals. While spatial time crystals have a periodic pattern in space, and they break the continuous spatial translational symmetry due to the periodic repeating patterns, in time crystals, the particles oscillate periodically in time in a harmonic manner, which breaks continuous time translational symmetry \cite{wilczek2012quantum}. Time crystals represent a new phase of matter, which is of fundamental importance to material science. As they bear similarity to the PTCs considered in this tutorial, we would like to elaborate on them shortly to distinguish them clearly. 

The originally proposed time crystals by Frank Wilczek were supposed to oscillate at their ground state in thermal equilibrium in a closed system, without energy exchange with the external surroundings.
Such oscillation is spontaneous and is caused by the interactions of the particles themselves. In time crystals, the Hamiltonian does not change in time, while the particles periodically oscillate in time and persist forever. 
However, it was later demonstrated by Watanabe and Oshikawa that such a self-sustained motion quantum system is fundamentally forbidden \cite{watanabe2015absence} since it violates the laws of thermodynamics and quantum mechanics.
In 2017, Yao, {\it et al.} pointed out that in a non-equilibrium open system pumped by a periodic drive, the persistent oscillation of particles is possible \cite{yao2017discrete}.
In other words, if energy exchange with the external environment (open system) is allowed, the idea of a time crystal can still be realized.
These are called discrete time crystals because they exhibit discrete time translational symmetry. 
External periodic forces should drive discrete time crystals. This force keeps the system out of equilibrium. However, the system's response to this driving force exhibits a periodicity that is \textit{different} (usually an integer multiple) from that of the driving force itself. 
This incoherence of the particle oscillation and the external pump indicates that the oscillation of particles is mainly caused by the interactions of particles themselves rather than directly driven by the external pump. 
It is a kind of semi-spontaneous oscillation.
The most important feature of a discrete time crystal is that it is robust to the interaction strength among particles and the imperfection in the driving pulses \cite{yao2017discrete,choi2016observation}, showing the characteristics of a new phase of matter.  

Discrete time crystals with external periodic driving are different from the originally proposed time crystal by Wilczek, which is a spontaneous process without any dependence on external driving. Although Wilczek's time crystals are impossible to realize, it is possible to step closer to them if the external driving is present (to respect the no-go theorem and provide energy to the system) but is independent of time, for example, driven by a DC pump. This is called a continuous time crystal because the pump is not periodically repeating but continuous and constant in time. With such a time-invariant pump, the particles oscillate periodically, which is purely determined by the interactions among particles. The oscillation is independent of the pump, which only compensates for the intrinsic loss of the material during oscillation. 
A continuous time crystal has been experimentally observed in many quantum systems (see, e.g., \cite{greilich2024robust}).

\begin{figure}[t]
\centerline{\includegraphics[width= 1\columnwidth]{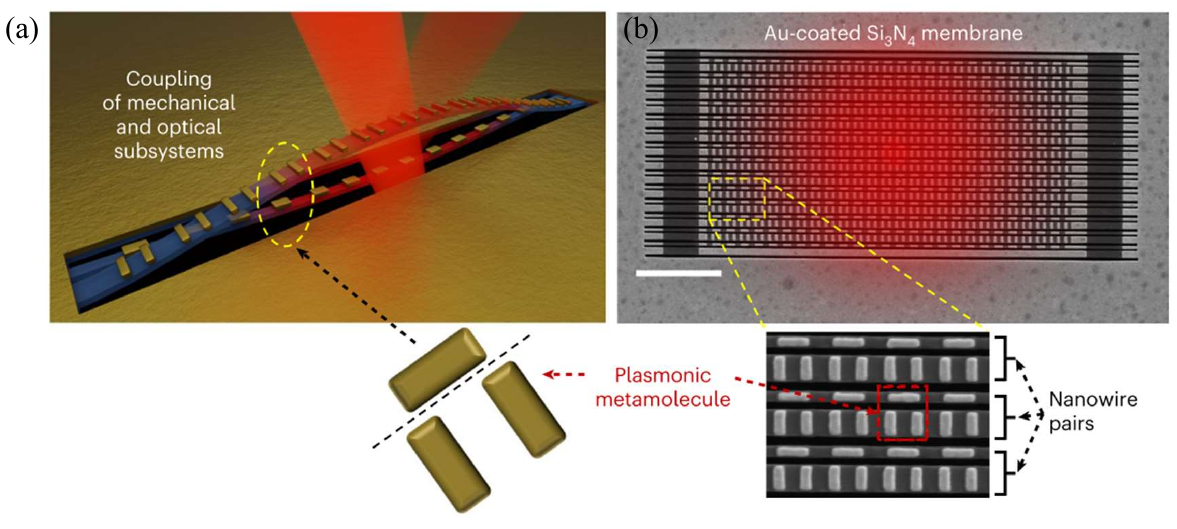}}
\caption{(a) Artistic representation of the fundamental unit of a classical continuous-time crystal, synthesized within a photonic metamaterial.
(b) Scanning electron microscope imagery capturing the entire two-dimensional array of metamolecules. When illuminated with coherent light (depicted schematically by the overlaid laser spot), the system transitions into a state characterized by persistent, synchronized oscillations of nanowires. Figures were reprinted from~\cite{liu2023photonic}.
}
\label{fig: zheludov}
\end{figure}

A recent work demonstrated that continuous time crystals can form in a classical mechanic-photonic coupled system \cite{liu2023photonic}.
There, the researchers created a two-dimensional array of plasmonic metamolecules (see Fig.~\ref{fig: zheludov}). These are tiny, gold-coated resonant structures placed on mechanically flexible nanowires cut from a semiconductor membrane. 
The nanowires were illuminated with laser light at a frequency that matched the plasmonic resonance of the metamolecules and many orders of magnitude higher than the frequency of mechanical nanowire oscillations, effectively representing a continuous pump. The continuous beam of laser light heats the nanowires, causing them to start oscillating due to thermal expansion. These oscillations are random at first but become quickly synchronized across the array of nanowires due to the effect of spontaneous synchronization. This synchronization is a key feature, as it represents the ordered, repeating pattern in time that represents a time crystal. As the nanowires oscillate, they change probe light transmission through the system. The synchronized oscillations can be observed directly by measuring these changes in light reflectivity. 
% Such a system was called in~\cite{liu2023photonic} as ``photonic metamaterial analogue of a continuous time crystal''. 

It is important to note that such a system emulates a continuous time crystal because the oscillations induced in the metamaterial are at a much lower frequency than the pump light itself. This allows the system to effectively "average out" the pump's periodicity over the timescale of the oscillations, making the pump appear continuous. As a result, the metamaterial transitions to a state showing persistent and synchronized oscillations, which are the defining features of a continuous time crystal.
This innovative approach extends the concept of time crystals to photonic systems, where the interaction between light and matter can create new phases of matter with unique temporal properties.

The combination of both spatial and temporal translational symmetry breaking defines a space-time crystal, exhibiting periodicity in both space and time. This phenomenon has been recently observed experimentally in a magnon platform \cite{trager2021real}. In this magnonic system, the oscillations of each particle in space have different phases, leading to a periodic modulation in both dimensions. This modulation results in the formation of a magnonic band structure due to the back folding of modes at the Brillouin zone boundaries of the space-time crystal. The study also demonstrates interactions between magnons and the space-time crystal, leading to lattice scattering and the generation of ultrashort spin waves that cannot be described by classical dispersion relations for linear spin wave excitation.

Unlike condensed-matter ``time-crystals'', a PTC is a medium with time-varying macroscopic optical properties. To achieve these time-varying periodic optical properties, an external periodic pump is required, similar to discrete time crystals. However, from a microscopic view, in discrete time crystals, particles generally oscillate at a frequency different from that of the external pump. The interactions among the atoms themselves mainly determine this oscillation. In contrast, in PTCs, the material's particles are directly driven by an external pump and oscillate in synchronization with it. 
% Thus, the photonic time crystals represents an independent research direction that studies the interactions between light and matter in time-varying media, while  time crystal focuses on spontaneous time-translational symmetry breaking and its robustness, which are properties of a phase of matter.

%%%%%%%%%%%%%%%%%%%%%%%%%%%%%%%%%%%%%%%%%%%%%%%%%%%%%%%%%%%%%%%%%%%%%%
%%%%%%%%%%%%%%%%%%%%%%%%%%%%%%%%%%%%%%%%%%%%%%%%%%%%%%%%%%%%%%%%%%%%%%
%%%%%%%%%%%%%%%%%%%%%%%%%%%%%%%%%%%%%%%%%%%%%%%%%%%%%%%%%%%%%%%%%%%%%%

\section{Material platforms to realize PTCs } \label{secMaterials}
The experimental realization of PTCs is challenging, but recent advances have shown promising potential for the fabrication of such devices. We will review different types of material platforms that are being used for the realization of PTCs and the observation of their wave phenomena. We start this section by discussing initially a material platform operating at radio frequencies where PTCs can be implemented using transmission lines. The possibility of reliably changing their properties on time scales comparable to the oscillation period of the electromagnetic fields renders them excellent candidates to study many of the fundamental effects that were just discussed on experimental grounds. Then, we discuss material platforms to observe comparable effects at optical frequencies. Next, we concentrate on metasurfaces made from time-varying photonic materials, so, essentially, thin films only. They are easier to realize and interrogate experimentally, which allows for a better understanding of the properties of PTCs. 
% Finally, we discuss in two dedicated subsections the possibility of observing effects discussed in PTCs in a different setting or even in non-photonic material platforms.
In the last section, we highlight that not just electromagnetic fields can propagate in systems with time-varying material properties. Other waves, such as acoustic, elastic, or water waves, can also be used to study waves in time-varying media. Finally, more exotic material systems supporting synthetic dimensions can be envisioned, which we will discuss here briefly. This should comprehensively overview the material systems used to study PTCs.

\subsection{Transmission lines}

Modulating the material parameters, such as the permittivity, remains challenging in many frequency ranges. As discussed in Subsection~\ref{Sec:TimeVaryingResonanceFrequency}, the change in resonance frequency would require a change in the properties of the atoms that make up the materials. It is easy to appreciate that this is rather demanding. However, when instead of an actual atom, a meta-atom is considered, and if the required modulation frequency of the material properties is fairly small and accessible by electronic means, we can get excellent control over a system that offers us the desired effects. 

Indeed, at radio frequencies, it is possible to use time-varying transmission lines to emulate a time-varying medium \cite{Reyes-Ayona2015Observation,reyes2016electromagnetic,moussa2023observation, kazemi2019exceptional}. That is possible because the governing equations that express the voltage along the circuit can be mapped to an ordinary wave equation in optics. Moreover, modulating a transmission line is reasonably easy by considering a time-dependent capacitance as part of the transmission line. Voltage-controlled varactor diodes provide such a time-dependent capacitance. With that, all ingredients are at hand to study PTCs.
\begin{figure}[t]
\centerline{\includegraphics[width= 0.5\columnwidth]{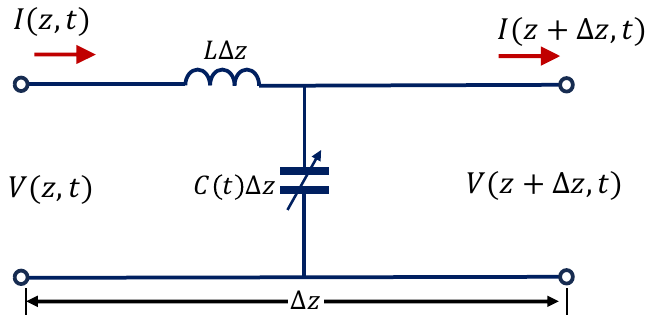}}
\caption{Transmission line with lumped elements extended along the $z$-direction. The distributed capacitance is modulated in time. The transmission line is electromagnetically equivalent to a material with time-varying permittivity. In other words, voltage waves propagate along the transmission line in the same manner as electromagnetic waves propagate in bulk material with time-varying permittivity. 
Here, $\Delta z$ is the length of an infinitesimal section of the transmission line. 
}
\label{Fig:distributed circuit}
\end{figure}

Figure~\ref{Fig:distributed circuit} shows a typical transmission line that supports the propagation of voltage-current waves. The transmission line is composed of distributed inductances (in series) $L$ with the unit of F/m and capacitance (shunt) $C$ with the unit of H/m~\cite[Sec.~2.1]{pozar_microwave_2012}. The series inductance $L$ represents the total self-inductance of the two conductors of the transmission line, and the shunt capacitance $C$ is the capacitance due to the close proximity of the two conductors. By multiplying both quantities by $\Delta z$ (infinitesimal length along the transmission line), we would obtain conventional (lumped) capacitance and inductance. 
% Other arrangements can also be imagined, but that would be a basic one. 
For simplicity, we assume the system is lossless, and the capacitance is modulated in time as a function of $C(t)$. 

According to Kirchhoff’s voltage and current laws, the circuit displayed in Fig.~\ref{Fig:distributed circuit} must satisfy the following relations,
\begin{subequations}
  \begin{equation}
    V(z, t)-L\Delta z\frac{\partial I(z, t)}{\partial t}-V(z+\Delta z, t)=0\,,
\end{equation}  
and
\begin{equation}
    I(z, t)- \Delta z \frac{\partial }{\partial t} \left[ C(t) V(z+ \Delta z, t)\right]-I(z+\Delta z, t)=0\,.
\end{equation} \label{eq: kirchhoff}
\end{subequations}
Taking the limit of $\Delta z\rightarrow0$, Eqs.~(\ref{eq: kirchhoff}) can be rearranged in differential forms, and becomes,
\begin{subequations}
  \begin{equation}
     \frac{\partial V(z, t)}{\partial z} =-L  \frac{\partial I(z, t)}{\partial t}
  \end{equation}
   and
   \begin{equation}
     \frac{\partial I(z, t)}{\partial z} =-  \frac{\partial [C(t)V(z, t)]}{\partial t}\,.
  \end{equation}\label{eq: vI}  
\end{subequations}
The above equations have exactly the same form as the Faraday and Ampere laws in Maxwell's equations when written for a wave propagating in a bulk medium with absolute permeability $\mu$ and permittivity $\varepsilon(t)$ along the $z$-direction with a given polarization: 
\begin{subequations}
\begin{equation}
     \frac{\partial E(z, t)}{\partial z} =-\mu  \frac{\partial H(z, t)}{\partial t}
\end{equation}
\begin{equation}
     \frac{\partial H(z, t)}{\partial z} =- \frac{\partial [\varepsilon(t) E(z, t)]}{\partial t}\,.
\end{equation}
    \end{subequations}
Thus, the governing electromagnetic equations for a lossless transmission line with time-varying distributed capacitance correspond to those for bulk dielectric material with time-varying permittivity. The equivalence of the circuit components and the material parameters suggests that
\begin{equation}
    C(t) \sim \varepsilon(t)\,, \quad L \sim \mu\,. 
\end{equation}

\begin{figure}[t]
\centerline{\includegraphics[width= 1\columnwidth]{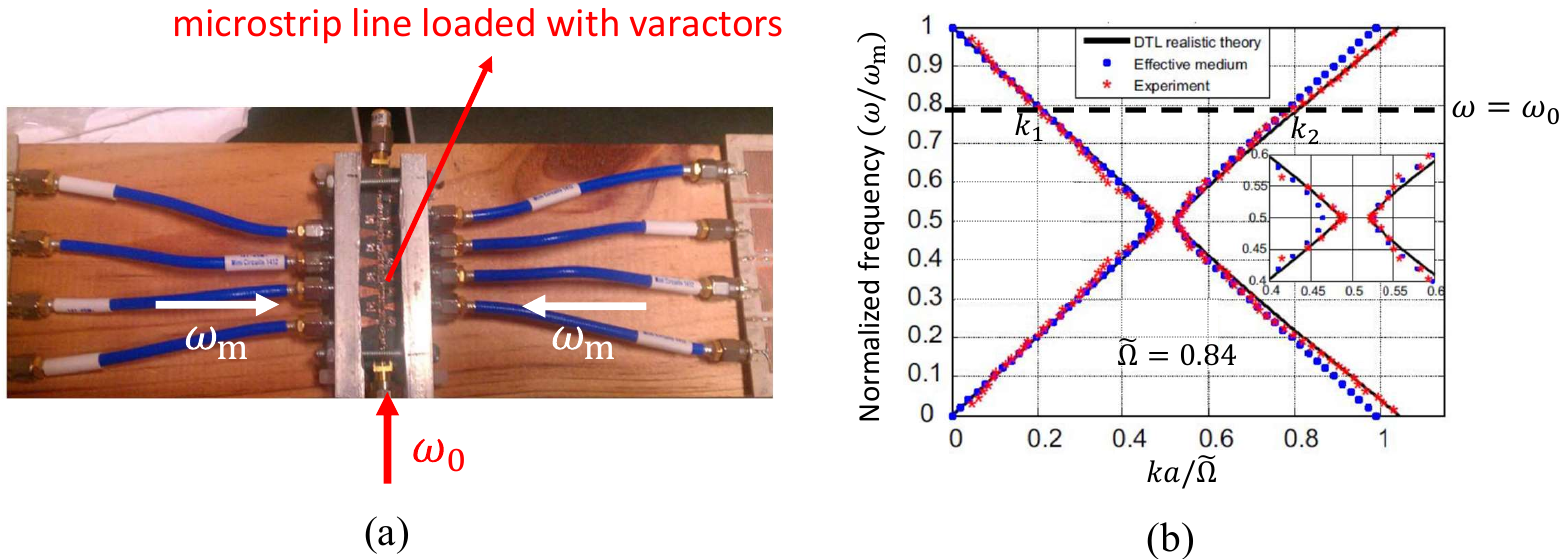}}
\caption{Experimental demonstration of PTCs using a microstrip line loaded with varactors. (a) Photograph of an experiment prototype. (b) An experimentally determined band structure. The band structure is determined by measuring the allowed wavenumbers $k_1$, $k_2$ when altering the input signal frequency $\omega_0$. Here, $a$ is the spatial period of the transmission line. Figure (a) was reprinted from~\cite{reyes2016electromagnetic}. Figure (b) was reprinted from~\cite{Reyes-Ayona2015Observation}.
}
\label{Fig:transmission_line}
\end{figure}

Indeed, this equivalence allows us to study some of the wave phenomena of bulk PTCs using one-dimensional transmission lines with time-varying parameters.  
For example, the momentum bandgap of a PTC was experimentally observed using transmission lines in \cite{Reyes-Ayona2015Observation, reyes2016electromagnetic}. In those works, the transmission line was formed by a microstrip line whose lumped capacitances were realized with varactors (tunable capacitors). The setup comprised a finite-sized PTC consisting of eight unit cells, and the varactors were controlled with the same signal to ensure the in-phase time variation of all components (see Fig.~\ref{Fig:transmission_line}(a)). That should guarantee the spatial homogeneity of the PTC. 
In the reported experiment, a voltage signal wave was launched along the transmission line whose frequency $\omega=\omega_0$ could be varied. Then, the phases of the voltage in each unit cell are measured to determine the allowed eigen wavenumbers $k_1$ and $k_2$ that belong to the two lowest bands of the band diagram (see Fig.~\ref{Fig:transmission_line}(b)).
% of the wave that propagates along the transmission line when the varactors are modulated in their capacitance at a specific modulation frequency and modulation amplitude. 
By varying the signal frequency (for a given modulation frequency), the dispersion relation was experimentally extracted, leading to the functional dependency between $\omega$ and $k$.
% The experimental system and selected results are shown in Fig.~\ref{Fig:transmission_line}.
The band structure contains a momentum bandgap in the momentum space. Whereas these experiments convincingly demonstrate the core aspect of PTCs, the possibility of wave amplification inside the momentum bandgap has not been reported in these earlier works. 

% \subsection{Power constraints in PTCs and possible solutions (Viktar)} \label{sec: power constraints}
% Here, we will discuss the power constraints in opening wide MBGs in PTCs. In addition, we will also discuss possible solutions to overcome these limitations.\\
% References: \cite{hayran2022ℏomega}.

% \subsection{Epsilon-near-zero (ENZ) materials (Mahdi and Sajjad)}
\subsection{Optical materials}\label{optmatsec}
% \red{Still references to cite:\\
% References: \cite{kinsey2015epsilon}, \cite{li2021refractive}, \cite{clerici2017controlling}.
% %\cite{saha2022engineering}.}
% }
%Epsilon-near-zero materials have been the most promising candidates for realizing PTCs. We will discuss the relevant published work in this subsection.\\
%References: \cite{kinsey2015epsilon}, \cite{alam2016large}, \cite{li2021refractive}, \cite{zhou2020broadband}, \cite{lustig2021towards}, \cite{lustig2023time}, \cite{caspani2016enhanced}, \cite{clerici2017controlling}, \cite{saha2020broadband}, \cite{saha2022engineering}, \cite{lobet2023new}.

The creation of PTCs operating at optical frequencies is very challenging since the modulation must be at optical speeds and have sufficiently strong amplitude~\cite{saha2023photonic}. Indeed,  the relative width of the momentum bandgaps in a PTC is proportional to the relative modulation strength, as was discussed in Section~\ref{est}. 
Various mechanisms can conceivably modify the refractive index in bulk materials. However, the majority, including electro-optic, acousto-optic, and thermo-optic mechanisms, are constrained by their operational speeds~\cite{williamson2020integrated}. Consequently, these methods are deemed unsuitable for realizing PTCs at optical frequencies. 
% Nonetheless, electro-optic and acousto-optic mechanisms find utility in applications such as time-modulated optical isolators, where the required modulation frequencies are significantly lower than the operational frequency, sufficient only for coupling different modes. Both electro-optic and acousto-optic modulations are limited to refractive index changes of a fraction of a percent, with electro-optic modulators achieving speeds up to hundreds of GHz.

Optical modulation has emerged as the most prominent method in this context.
The most natural way to change the refractive index at optical speeds is probably to exploit the Kerr-type nonlinearity. The Kerr nonlinearity is a third-order nonlinear effect occurring in many centrosymmetric materials. 
In the Kerr effect, a strong pump beam changes the refractive index of a nonlinear medium through which it propagates. By changing the pump intensity in time (e.g., by using pulsed pump excitations), one can obtain modulation of the refractive index of the material.  
When a weak probe light with electric field $E(\omega')$ propagates through the nonlinear medium illuminated by a pump beam with amplitude $E(\omega)$, we obtain the nonlinear polarization   $P^{\rm NL}(\omega) = 6\varepsilon_0 \chi^{(3)} \left( \omega' = \omega' + \omega - \omega \right) \left| E(\omega) \right|^2 E(\omega')$ \cite[Sec.~4.1]{boyd2008nonlinear}, where $\chi^{(3)}$ is the third-order nonlinear susceptibility. 
In this context, a cross-coupling between the pump and probe is observed, and the refractive index of the medium experienced by the probe light can be expressed as $n = n_0 + 2n_2 \left| E(\omega) \right|^2$(notice that  $n$ here denotes the refractive index and should not be confused with the  harmonic number), where $n_0$ is the linear and $n_2 = \frac{3 \chi^{(3)}}{2n_0}$ is the nonlinear refractive index.  The optical Kerr effect is \textit{nearly instantaneous} in the sense that the response of the material's refractive index to changes in pump intensity occurs extremely rapidly (on the femtosecond level and faster). Thus, to induce a PTC in the nonlinear material, one needs to be able to change the intensity of the pump beam quickly in time so that it leads to the periodic modulation of the refractive index. 
However, there are two challenges in practical setups with nonlinear materials. The first one is related to the fact that most materials have small nonlinear susceptibilities. Indeed, the maximum observed change of the relative modulation strength of the refractive index is in the range of 1\%, even at power densities as high as 1 ${\rm TW}/{\rm cm}^2$~\cite{borchers2012saturation}. With such modulation depths, the momentum bandgap becomes negligible. The second challenge is to change the pump intensity at optical speed in a periodic manner. Although ultrashort pulse excitation is nowadays technologically accessible, making a train of many such pulses with a high duty cycle (high repetition rate) is not feasible. Current technology allows obtaining repetition rates in trains of pulses up to 10~GHz~\cite{bartels200910}. 

Therefore, so far, the experimental developments have been limited to\textit{ single time interfaces} (requiring a single ultrashort pump pulse) in a special type of materials supporting large modulation strengths, so-called epsilon-near-zero (ENZ) materials~\cite{fomra2024nonlinear}. Since permittivity and refractive index are related as   $n=\sqrt{\varepsilon}$, differentiating $n$ with respect to $\varepsilon$, we can find that  $\Delta n = \frac{\Delta \varepsilon}{2\sqrt{\varepsilon}}$. From that expression, we see that since $\varepsilon$ approaches zero in ENZ materials, small variations in $\Delta \varepsilon$ can induce significant changes in $\Delta n$~\cite{alam2016large}.
The ENZ regime occurs close to the plasma frequency of a material whose dielectric function is described by a Drude model. However, to use this effect, one needs to ensure that the imaginary part of the material permittivity is sufficiently small.  Typical materials discussed in the literature for that purpose are transparent conductive oxides (TCOs)~\cite{jaffray2024spatio} such as indium tin oxide (ITO) and aluminum-doped zinc oxide (AZO) and others ~\cite{kinsey2015epsilon,alam2016large,caspani2016enhanced,clerici2017controlling,saha2020broadband,li2021refractive}. They have the benefit that the ENZ-region can be suitably tuned, but it is roughly at telecommunication wavelengths, which is an asset for potential applications. In the ENZ wavelength region, the imaginary part of the permittivity in TCOs is in the range of 0.2--0.4~\cite{kinsey2015epsilon,alam2016large,caspani2016enhanced}.
The nonlinear refractive index of ITO at its ENZ wavelength of 1240~nm reaches $n_2= 0.11~\text{cm}^2/\text{GW}$~\cite{alam2016large}, while for AZO at 1390~nm $n_2= 3.07 \times 10^{-4}~\text{cm}^2/\text{GW}$ \cite{caspani2016enhanced}. 
% Another TCO example is cadmium oxide (CdO). In \cite{saha2020broadband}, a significant reflection change of up to 135\% in yttrium-doped CdO has been reported in a broadband range from near-infrared to mid-infrared.
%sajjad I have a problem here adding the paper about molybdenum, they reported  based on the permittivity change but other references reported based on the effective nonlinear refractive index
A comprehensive overview of nonlinear refractive indices of different materials can be found in~\cite{lobet2023new}.
%\begin{table}[ht]
%\centering
%\begin{tabular}{ |p{1.5cm}||p{1.5cm}|p{1.5cm}|p{1.5cm}|p{1.5cm}| }
 %\hline
% \multicolumn{5}{|c|}{The nonlinear refractive index in ENZ material\(^*\)} \\
% \hline
% Material & $n_{2\text{(eff)}}$ (cm$^2$/GW) & Probe (nm) & Pump (nm) & Pulse width (fs) \\
 %\hline
%ITO & 0.11 & 1247 & 1250 & Z scan \\
% AZO & 3.07e-4 & 758 & 1350 & R/T \\
% Si & 5.19e-18 & 830 & 1550 & 90 \\
% Al & DZ & DZA & 012 & 66 \\
% Am & AS & ASM & 016 & 4 \\
% An & AD & AND & 020 & 43 \\
% An& AO & AGO & 024 & 34 \\
% \hline
%\end{tabular}
%\caption{Effective refractive index change in ENZ materials.  * This table is adapted from\cite{lobet2023new}.}
%\end{table}

It should be noted that the nonlinearities, represented by $n_2$, in transparent conducting oxides (TCOs) originate differently compared to those in conventional Kerr nonlinear materials. This distinction arises from the non-parabolic dispersion of the conduction band, which leads to a change in the average effective mass of the electron sea due to intraband absorption~\cite{secondo2020absorptive,khurgin2021fast}.
Significant changes in refractive index $\Delta n/n$, on the order of 100\%, have indeed been observed in these materials~\cite{saha2020broadband}. In early experimental studies, rapid changes (on the order of a few hundred femtoseconds) in the refractive index were observed~\cite{shaltout2016doppler,zhou2020broadband,bruno_broad_2020}. However, these changes were still two to three orders of magnitude slower than necessary for achieving PTCs in the optical (infrared or visible) regime. As a result, this type of nonlinearity was initially termed ``slow"~\cite{khurgin2021fast}. More recent theoretical work has suggested that refractive index changes in transparent conducting oxides (TCOs) could be nearly instantaneous\cite{un_electronicbased_2023,pendry_avalanche_2024}. Subsequent experimental studies have confirmed such ultrafast modulations, showing excitation and relaxation times of the refractive index in ITO in the order of 5--10~fs and 10--20~fs, respectively~\cite{lustig2023time,tirole_doubleslit_2023}, limited only by the pump-pulse duration. These findings represent a significant step towards the implementation of PTCs at optical frequencies.
% For example, in Ref.~\cite{lustig2021towards}, an AZO sample was modulated with an 800 nm pump beam, and the reflection and transmission for a 1300~nm probe were measured. Since the pump was much shorter than the probe pulse duration, a significant (approaching unity) refractive index change was detected, making the first step toward implementing PTCs. More recently, modulation of an ITO sample with a single-cycle pump pulse was demonstrated~\cite{lustig2023time}.
% Furthermore, broadband frequency translation has also been reported using similar time-interfaces in ENZ materials~\cite{zhou2020broadband}.  

Periodic optical temporal modulation of ENZ materials using the nonlinearity mechanism described above (arising from the non-parabolic dispersion) remains impractical due to unrealistically high power requirements for modulating TCOs at optical speeds. It was theoretically predicted that to obtain a moderately wide momentum bandgap, one would need pump power density on the order of tens of TW/${\rm cm}^3$ even for PTCs operating in the THz frequency range~\cite{hayran2022ℏomega}. Such intense pump power, compounded by the nonzero dissipation inherent to the materials, would invariably result in the materials' degradation, precluding the observation of PTCs.

Furthermore, TCOs are also capable of supporting another type of nonlinearities where temporal modulations are generated entirely through optical means via the third-order susceptibility, $\chi^{(3)}$, in the four-wave mixing process~\cite{khurgin2021fast}. Specifically, in degenerate four-wave mixing, two counter-propagating pump beams interact with a signal wave to generate a third-order nonlinear polarization~\cite{shen1986basic}. During this process, the two pump waves combine to create a uniform intensity pattern oscillating at the frequency $2\omega$. This oscillation modulates the material's refractive index at the same frequency, thereby offering the potential to form a PTC. However, such effects are typically overshadowed by other more dominant nonlinear phenomena, such as the ``slow dynamic grating"~\cite{eichler1977laser,khurgin2021fast,hayran2022ℏomega}.

To summarize, the realization of optical PTCs remains a significant challenge, necessitating further experimental and theoretical exploration. We wish to highlight several innovative approaches that have recently been proposed to facilitate the observation of PTCs within the optical domain. The first approach involves utilizing a second-order nonlinear process, where a standing pump wave at the second harmonic frequency $2\omega$ interacts with counter-propagating waves at the fundamental frequency~\cite{ding_transversely_1995, lanco2006semiconductor, hayran2022ℏomega}. This mechanism is instantaneous and remains unaffected by the spurious ``dynamic-grating'' effects. Additionally, the pump wave is confined within a resonant cavity, enhancing the modulation depth of the material. The second strategy focuses on inducing intrinsic or structural resonances within the time-modulated material, strongly enhancing the system's quality factor~\cite{wang2023unleashing}. This method significantly reduces the required modulation depth (2 orders of magnitude), enabling the implementation of PTCs that possess considerable momentum bandgaps in low-loss materials characterized by Kerr-type nonlinearity. Furthermore, recently, in Ref.~\cite{michael2024photonic} it was proposed that the PTC behavior can be mediated by phonon squeezing. Finally, in recent works, it was suggested that the large momentum bandgap can be obtained in low-loss systems supporting biaxial anisotropy~\cite{dong2024non} or longitudinal phonon modes~\cite{zhang2024longitudinal}.

The discussion presented above is related to PTCs with time-modulated permittivity. However, the concept of PTCs extends beyond just variations of permittivity over time. Some theoretical studies have explored the modulation of other material parameters. For instance, varying the permeability \cite{martinez2016temporal}, and magneto-optical coefficients \cite{he2023faraday}. In such scenarios, the modulation of these parameters also leads to a momentum bandgap, paving the way for more complex and exotic effects. Despite this theoretical advancement, practically implementing modulation of parameters other than permittivity remains a significant challenge, particularly within optical frequency ranges.

\subsection{Two-dimensional platforms}
%\red{Sec. S1--S3 in Sci.Adv.}
%Metasurfaces offer a practical power-efficient platform to realize PTCs. The advances in that direction will be discussed.\\

\begin{figure}[t]
\centerline{\includegraphics[width= 1\columnwidth]{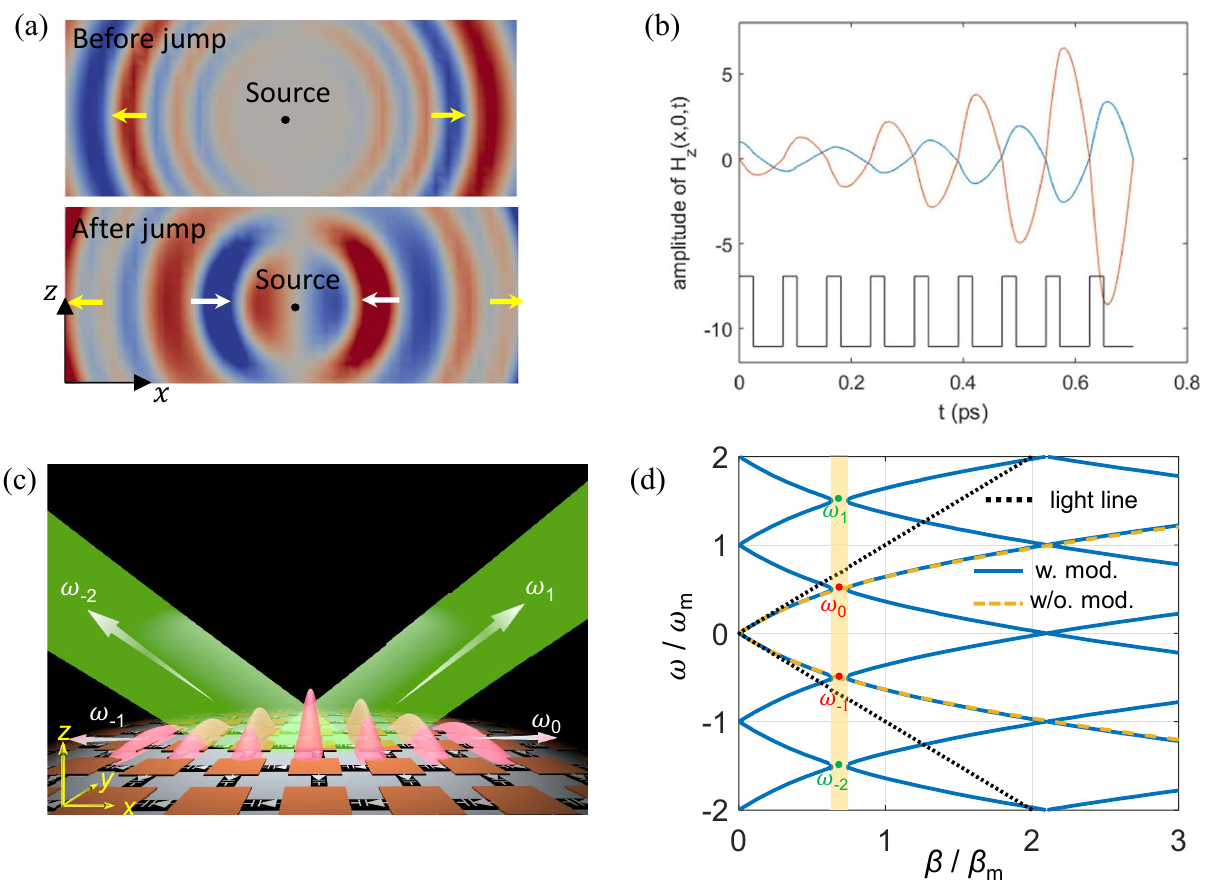}}
\caption{(a) Propagation of a surface plasmon polariton on a graphene sheet ($xz$-plane) before (top) and after (bottom) a temporal jump of the surface conductivity of graphene. The arrows represent the propagating directions. (b) Field growth of a graphene plasmon polariton as a function of time when applying a periodic stepwise modulation of surface conductivity. The field is recorded at one spatial point on the surface. The blue and red curves correspond to real and imaginary parts of the $z$-component of the magnetic field. (c) Metasurface realization of a PTC. Due to the periodic temporal modulation, both surface modes and propagating modes become exponentially growing inside the momentum bandgap. 
(d) Band structure of the metasurface in (c). The orange dashed line represents the dispersion curve of a stationary (non-modulated) capacitive surface. Figures (a) and (b) were reprinted from~\cite{wilson2018temporal}. Figures (c) and (d) were reprinted from~\cite{wang2023metasurface}.
}
\label{Fig:metasurfaces}
\end{figure}

In the previous sections, we have discussed PTCs made of time-varying volumetric (three-dimensional) materials or one-dimensional transmission lines that emulate them. These were the original types of PTCs proposed. However, the concept of PTCs  can be also applied to two-dimensional systems. In a two-dimensional (2D) material system, the modulation is applied to surface properties, such as surface conductivity or impedance.
% , instead of modulating the volumetric material parameters as seen in PTCs based on bulk media.
% Whereas we discussed previously that the realization of PTCs as a bulk material continues to be challenging, an alternative approach is to realize them as a thin layer of a material characterized by temporally varying material parameters. In analogy to conventional metamaterials with spatially varying material properties that are experimentally easier to implement as thin films, we will call these PTCs made from thin films here as metasurfaces. In the extreme case of an ultrathin film, the actual material essentially only constitutes a modification to the surface.
In contrast to bulk media, which support plane-wave propagation in all three spatial dimensions, the 2D PTCs have negligible thickness and sustain eigenmodes that propagate along the surface. This leads to a restricted dimensionality of the problem since the surface waves can propagate only along two directions (in specific settings, propagation in only one direction is permissible). This dimensional reduction significantly simplifies the implementation of PTCs, as modulation is required only on the surface, exempting the need to modulate along the thickness direction.

The periodic fast temporal modulation of material properties in a 2D material system, probably, for the first time were considered in Ref.~\cite{wilson2018temporal}. The authors studied a suspended graphene layer as shown in Fig.~\ref{Fig:metasurfaces}(a), where the layer is located in the $xz$ plane. Without a temporal modulation, the graphene sheet supports TM-polarized surface plasmon polaritons as its eigenmode. The top panel of Fig.~\ref{Fig:metasurfaces}(a) shows that under a dipole source excitation, a TM-polarized surface plasmon polaritons is generated and propagates in the outward direction as indicated by the yellow arrows. First, the authors considered a simple case of a temporal jump, that is, when the surface conductivity was instantaneously switched to another value at a specific moment. Therefore, time-reflected and time-transmitted waves were generated, as indicated by white and yellow arrows in the bottom panel of Fig.~\ref{Fig:metasurfaces}(a). This effect is similar to the temporal jump of bulk media, as discussed in Section~\ref{sec: transfer matrix}, but happens in a two-dimensional system here. The amplitudes of temporal reflection and transmission can be derived using temporal boundary conditions~\cite{wilson2018temporal}. 
Next, the authors of Ref.~\cite{wilson2018temporal} considered a temporal cascade of such temporal jumps resulting in a periodic step-wise modulation of the surface conductivity. The modulation frequency was twice the frequency of the surface plasmon polaritons. It was discovered that such modulation leads to constructive interference between time-reflected and time-transmitted waves, resulting in an exponentially growing standing-wave pattern, as depicted in Fig.~\ref{Fig:metasurfaces}(b). Similar results have been observed more recently in~\cite{kiselev_light_2023, shirokova2023surface,kimgraphene}.

At the time of those initial investigations, the amplification was not attributed to the momentum bandgap as the band structure of such time-varying surfaces has not been investigated in depth. Only recently, the band structure of a time-varying material surface was calculated \cite{wang2023metasurface}. An artistic illustration of the experimental setup with an actual implementation of that surface is shown in Fig.~\ref{Fig:metasurfaces}(c). In that work, it was assumed that the surface is capacitive, and the surface capacitance varies periodically in time as a function of $C(t)=C_{\rm av}[1+m_{\rm c}\cos(\omega_{\rm m}t)]$, where $C_{\rm av}$ is the average surface capacitance. 
The band structure for the TE-polarized surface wave supported by the time-varying surface is shown in Fig.~\ref{Fig:metasurfaces}(d). It was revealed that a momentum bandgap is open (see yellow shaded region). Inside the momentum bandgap, the eigenfrequencies ($\cdots, \omega_{-2}, \omega_{-1}, \omega_{0}, \omega_{1}, \cdots$) are complex. Some modes ($\omega_{-1}, \omega_{0}$) correspond to surface modes (below the dotted light line in the figure), while others correspond to free-space propagating modes (above the light line). This implies that modulating a surface boundary can provide amplification of the surface and free-space propagating modes simultaneously.
This observation suggests that the reduction in the dimensionality of PTCs does not compromise their characteristic momentum bandgap and the consequent amplification effects. Instead, it underscores an advantage over bulk media, limited to amplifying only the propagating plane waves.

Due to the dimensionality reduction, 2D PTCs are much easier to realize than 3D PTCs. This eliminates the need for uniform modulation of the material parameters into the third dimension (thickness direction), a requirement for bulk media.
For this reason, the authors in \cite{wang2023metasurface} implemented the time-varying homogeneous surface in the microwave region using a metasurface. The capacitive metasurface was composed of periodic metallic patches with effective capacitance of $C_{\rm av}$ (see Fig.~\ref{Fig:metasurfaces}(c)). The metallic patches were connected with varactors. By applying uniform periodical voltage signals on the varactors, the effective capacitance of the whole metasurface can be modulated in time. The experimental results observed near 30 dB amplification of surface wave in the center of the momentum bandgap.

\subsection{Mimicking PTCs with other material platforms}

% \red{Also cite these papers about time reflection in synthetic dimensions and with cold atoms. PTC could be realised there too:
% \begin{itemize}
%     \item K. Sacha, Reflection and refraction at a time boundary
% \item Quantum time reflection and refraction of
% ultracold atoms
% \item Time reflection and refraction in synthetic frequency dimension
% \end{itemize}
% }

% References: \cite{dong2013experimental}, \cite{apffel2021time}. 
% \red{Check also works of Romain Fleury}

Electromagnetic waves exhibit similarities with various types of waves, such as mechanical, phonon, and sound waves. This similarity allows the concept of PTCs, characterized by the periodic repetition of electromagnetic properties in time, to be extended to other wave physics domains. For instance, a material with time-modulated acoustic properties may be considered an acoustic time crystal. This section explores a range of proposed materials that align with the broader concept of ``wave time crystals". 
The study of wave time crystals within various physics domains serves a dual purpose. First, alternative material platforms may offer more practicality, thereby facilitating testing fundamental concepts shared by all wave time crystals. Second, such advancements could pave the way for novel technological applications in diverse technological fields.

\begin{figure}[t]
\centerline{\includegraphics[width= 1\columnwidth]{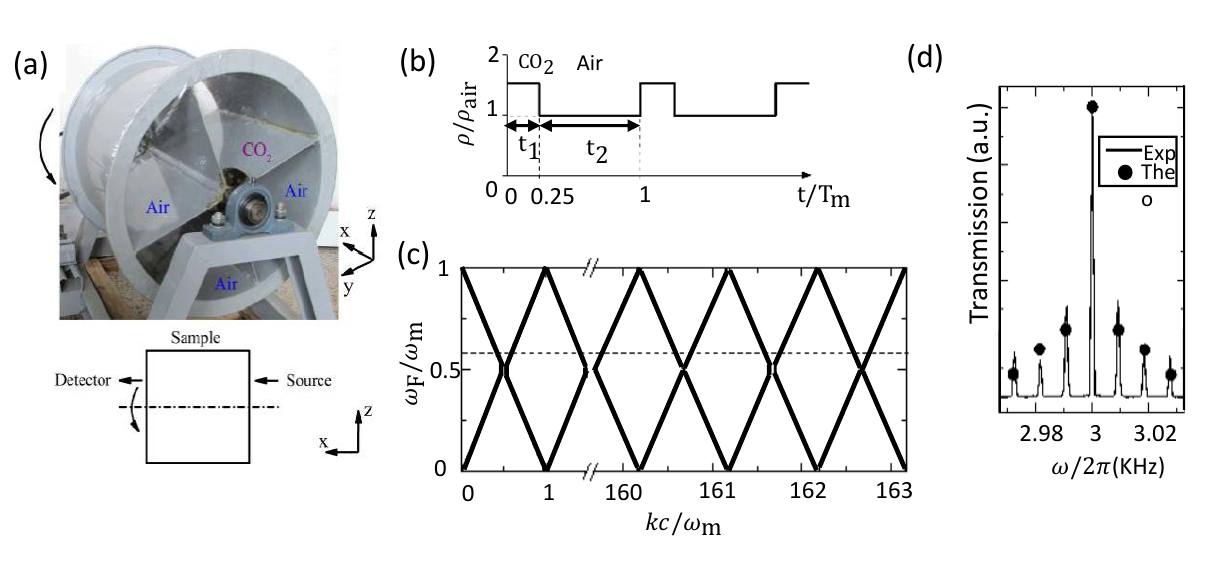}}
\caption{(a) The experimental setup of an acoustic time crystal consisting of a rotating cylinder with four chambers through which acoustic waves propagate along the $x$-direction. (b) The theoretical temporal variation of the mass density of the acoustic time crystal. In fact, a stepwise modulation in practice is hard to obtain in the explored setup, but it continues to be a good approximation. (c) The band structure of the acoustic time crystal. Note the discontinuity of the x-axis. The spectral domain shows the experimentally relevant bands because the modulation frequency is rather small compared to the operational frequency where the response of the acoustic time crystal is probed. (d) The transmission spectrum of the acoustic time crystal when illuminated by waves with a frequency of 3.0 kHz. Figures were reprinted from~\cite{dong2013experimental}. 
}
\label{Fig:npf1}
\end{figure}

One of the early realizations of the wave time crystal has been proposed in \cite{dong2013experimental} using an acoustic system. Figure~\ref{Fig:npf1}(a) depicts the experimental setup, consisting of a rotating cylinder with four chambers. One chamber is filled with ${\rm CO}_2$, while the remaining three contain air. 
An acoustic source placed on the right side of the cylinder emits waves along the $+x$-direction, and a detector is located on the left side. Rotating the cylinder alters the mass density of the medium through which the wave propagates over time. Figure~\ref{Fig:npf1}(b) shows the temporal variation of the mass density that the incident acoustic waves experience, varying between the densities of $\rho_{\rm air}$ and  
$\rho_{\rm CO_2}$, which $\rho_{\rm CO_2}$ is 1.53 times than $\rho_{\rm air}$. The acoustic band structure for this time-varying acoustic medium is presented in Fig.~\ref{Fig:npf1}(c), illustrating the first Brillouin zone. Since in the experiment the modulation frequency $\omega_{\rm m}$ was limited to the value of $\omega_{\rm m}/2\pi=9.3$~Hz, being much smaller than the incident acoustic frequency $\omega_{\rm m} \ll \omega_{\rm inc} $, the dominant excited harmonics had very large values of the normalized wavenumber $kc/\omega_{\rm m}$. In other words, the dominant harmonics belonged to very high-order bands shown in Fig.~\ref{Fig:npf1}(c) (note that the diagram is discontinued at $kc/\omega_{\rm m}=1.5$ for clarity). At such high-order bands,  momentum bandgap does not open, as is seen in Fig.~\ref{Fig:bandstructrue_3cases}.
% It comprises 326 bands, which can be attributed to the significantly lower rotating frequency of the cylinder (modulation frequency) compared to the frequency of the incident wave, denoted as 
% There are two bandgaps one at the start and another at the end of the zone.
The authors of Ref.~\cite{dong2013experimental} measured experimentally the transmission spectra of such an acoustic time crystal. For an incident frequency of $\omega_{\rm inc}/2\pi=3$~kHz, the transmission spectrum is depicted in Fig.~\ref{Fig:npf1}(d). One can see the main peak, corresponding to the incident wave frequency, surrounded by additional side harmonics separated from one another by $\omega_{\rm m}/2\pi$ distance.  
It should be mentioned that the proposed acoustic time crystal, despite its straightforward configuration and absence of any nonlinear effects, possesses a substantial disadvantage of limited feasible modulation frequencies due to the mechanical rotating parts.

\begin{figure}[t]
\centerline{\includegraphics[width= 1\columnwidth]{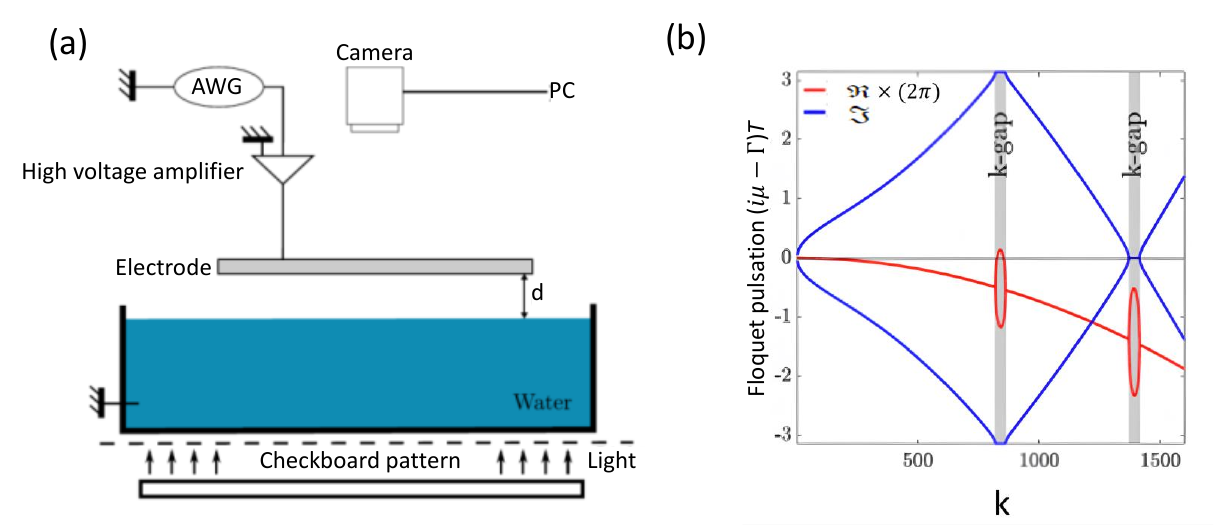}}
\caption{(a) The experimental setup to study time crystals in water waves. The time crystals for water waves are enabled by the modulation of a potential above the water surface to excite a Faraday wave. A Faraday wave refers to standing waves that appear on the surface of a liquid in a container undergoing vertical oscillations. The vertical oscillation is induced in this experimental setup by an electrode placed above the water surfaces and on which a sequence of voltage pulses is applied. The sequence of these pulses takes the form of a comb of Dirac delta distribution. The applied electric field exerts an attractive force on the water surface that modifies the wave speed, forming a time crystal for these specific water waves. The water waves are imaged by a camera above the setup that takes images of a checkerboard underneath the water tank. From the deformation of the checkerboard, the water waves can be reconstructed. (b) The band structure of the system exemplarily for $d = 5$~mm and an amplitude of 8~kV and a frequency of 60~Hz for the Dirac comb excitation. Real (red) and imaginary (blue) parts of the dispersion relation can be seen. 
Let $\Gamma = 2\nu k^2$, where $\nu$ represents the kinematic viscosity of the liquid. The eigenvalue is expressed in the form $\exp\left(i\mu - \Gamma\right)$, where $\mu$ is a complex number.
 Clearly visible momentum gaps form. Figure was reprinted from~\cite{apffel2021time}.
}

\label{Fig:npf4}
\end{figure}

Another platform for observing physics of wave time crystals is based on water waves. A possible experimental setup, proposed in \cite{apffel2021time} to study the physics of time-varying medium using water waves, can be seen in Fig \ref{Fig:npf4}(a). The setup consists of a grounded water tank that is conductive. An electrode is placed at a distance $d$ above the water surface. The electrode is connected to a high-voltage amplifier. A Faraday instability, which manifests as parametric water waves, can be observed by modulating the gravitational acceleration. The gravitational potential is periodically changed by applying a sequence of voltage pulses to the electrode, which exerts an electrostatic force on the water surface. This system couples hydrodynamics and electrostatics. Therefore, instead of modulating the gravitational acceleration through physical displacement, we can use electrostatic modulation to achieve this effect\cite{Ward_Matsumoto_Narayanan_2019}. A comb-like modulation of potential in time is applied to the electrode on the top of the water to implement the time crystal. A light source illuminates a checkerboard pattern below the water tank, and with a camera above the setup, the evolution of Faraday waves can be captured. A band structure can be extracted as the potential is modulated in time in the system and uniform in space. The band structure with the specific parameters of $d = 5$~mm and the amplitude of the Dirac comb potential equals 8 kV is shown in Fig.~\ref{Fig:npf4}(b). There are two bandgaps in the band structure.

In Refs.~\cite{chong2023modulation,PhysRevE.107.034211}, phononic time crystals were analyzed both theoretically and experimentally. These studies focused on a one-dimensional phononic lattice characterized by time-periodic elastic properties. This lattice was composed of magnetic masses (cylindrical bar magnets) that repel each other, with their grounding stiffness modulated by electrical coils that are driven by periodically varying electrical signals. For weak excitations (in the linear regime), the authors observed the generation of momentum bandgaps. Furthermore, it was discovered that such phononic time crystals can support what are termed wavenumber breathers~\cite{chong2023modulation}. Unlike classical breathers, which are spatially localized and time-periodic solutions to nonlinear lattice differential equations found across photonics, phononics, and electrical systems~\cite{flach2008discrete}, wavenumber breathers are localized in time and exhibit periodicity in space. This positions them as the dual counterparts to the classical breathers.
%\begin{figure}[t]
%\centerline{\includegraphics[width= 1\columnwidth]{two_ladder.PNG}}
%\caption{(a) Two-legged ladder model
%(b) The band diagram for the model
%}

%\label{Fig:two_ladder}
%\end{figure}
PTCs have also been investigated in elastic and electromechanical waveguides
\cite{trainiti2019time,xia2021experimental}. For instance, when the stiffness is modulated in time, the complex frequency has been demonstrated. In such a platform, the modulation is implemented experimentally by an array of piezoelectric patches shunted through a negative electrical capacitance controlled by a switching circuit \cite{trainiti2019time}.

So far, in this section, we discussed the possible implementation of wave time crystals based on water, elastic, and acoustic waves, in which the modulation frequency was rather small. Let us investigate other platforms to find possible realizations of PTCs at optical frequencies. In that context, the concept of synthetic dimension is a powerful tool for analyzing a higher-dimensional system using a lower-dimensional system. For instance, arranging and coupling certain states of a system makes it possible to form a one-dimensional lattice in the synthetic space~\cite{yuan2018synthetic}. One example of such a system is a ring resonator supporting resonant modes at a discrete set of frequencies (equally spaced from one another). By time-modulating the permittivity of the ring, it is possible to couple the neighboring resonance modes, mimicking the tight-binding model in the synthetic dimension. 
Moreover, one can configure a two-dimensional synthetic lattice with a different arrangement and coupling~\cite{dutt2020single,ozawa2019topological}. 
% This approach enables the simulation of the system characteristic of 1D or 2D lattices, which represent effective dimensions rather than physical ones, using a system that contains only modes and couples them together\cite{}. 

Recent proposals suggest leveraging synthetic dimensions, specifically the one-dimensional Su-Schrieffer-Heeger (SSH) lattice model, to explore phenomena such as time reflection and time refraction~\cite{dong2023quantum} -- phenomena essential for constructing PTCs.
% The idea of time reflection and refraction in synthetic dimension has been proposed in \cite{dong2023quantum}, considering the Su-Schrieffer-Heeger (SSH) one-dimensional lattice.
Such a lattice contains two energy bands, and an abrupt change in the lattice leads to a new band structure after the time boundary. The modes in that modified band structure contain information on the propagation characteristics of the time-reflected and time-refracted wavepackets. In \cite{dong2023quantum}, the SSH model was implemented as a momentum-space lattice. Here, the sign of the group velocity can be changed by altering the sign of the hopping parameters, effectively changing the sign of the refractive index. The experimental realization has been based on a Bose-Einstein condensate of a cloud of ultracold atoms that forms the momentum states as a synthetic dimension. These states can be coupled through a two-photon Bragg transition, using two laser beams with opposite momentum and one containing multiple frequencies. By adjusting the intensity and phase of the laser beams, it is possible to control the coupling coefficient and implement variable coupling, which enables the observation of time reflection and time reflection \cite{dong2023quantum}. Notably, the coupling between each state can be controlled independently in this platform.

% Synthetic dimensions can be formed with a photonic platform as well. For example, different modes in a ring resonator can be coupled by applying a time modulation of the permittivity to form a synthetic frequency dimension. The coupling between frequency modes can be adjusted by the modulation \cite{yuan2018synthetic}. 
Time reflection and refraction were also recently observed in photonic platforms supporting synthetic dimensions~\cite{long2023time}. Importantly, it was found that modulation at microwave frequencies (as low as 0.72~MHz) was sufficient to observe time-boundary effects for optical waves in the synthetic frequency dimension. In that work, a two-leg ladder model comprising two energy bands was considered. Such a system can be described by a dedicated model that features two coupled sites, characterized by a coupling coefficient. Additionally, the model accommodates a self-coupling at each site. Exciting the system in one particular mode and applying an abrupt change to the coupling coefficient within sites causes a new band structure. The new band structure sustains modes into which the incident wavepacket can be time-reflected and refracted. This model can be physically realized with two ring resonators with frequency modes representing each site of the ladder lattice. As mentioned before, each resonator is made from time-varying permittivity that couples the modes. The coupling between sites happens with a directional coupler between two resonators. By applying a variation in the amplitude of modulation in the ring resonator, the time boundary occurs, and the reflected and refracted waves are observable. 
% It should be noted that the primary frequency of the resonator operates in the optical regime, while the modulation occurs in the microwave regime.
In conclusion, the two platforms mentioned above allow us to obtain strong time reflections. Therefore, they can be also used for creating a PTC using synthetic dimensions.

\section{Potential applications of PTCs }\label{secApplications}
The flow of the tutorial has been chosen in such a way that we introduced the basics and some advanced properties of PTCs first. Then, we discussed specifically material systems and spectral domains that allow us to implement PTCs. In the following section, we discuss various applications of PTCs. It is important to note that the field of PTCs is now predominantly in the realm of fundamental research, with applications of PTCs still being relatively limited. Nevertheless, these opportunities stimulate future research efforts, and we should keep them in mind. 
It should also be noted that the strict requirements for rapid temporal modulation have hindered the optical realization of PTCs. Consequently, at the same time, research has been expanding in the direction of photonic space-time crystals, where these stringent requirements are to some extent relaxed. The applications of space-time crystals will be discussed in Section~\ref{secSpacetime}. 

% \red{In the end, let us combine these subsections into a single continuous section.}

%%%%%%%%%%%%%%%%%%%%%%%%%%%%%%%%%%%%%%%%%%%%%%%%%%%%%%%%%%%%%%%%%%%%%%%%%%%%%%%%%%%%%%%
%%%%%%%%%%%%%%%%%%%%%%%%%%%%%%%%%%%%%%%%%%%%%%%%%%%%%%%%%%%%%%%%%%%%%%%%%%%%%%%%%%%%%%%
%%%%%%%%%%%%%%%%%%%%%%%%%%%%%%%%%%%%%%%%%%%%%%%%%%%%%%%%%%%%%%%%%%%%%%%%%%%%%%%%%%%%%%%

\subsection{Realizing thresholdless lasers using PTCs}
% \red{Note that for the emission from an atom inside a PTC we would need to discuss density of states near the bandgap. I advise to follow section IID in Halevi's 2009 paper.}

% \red{Cite also this new work: Electrodynamic Modeling of Threshold-free Lasing in Photonic time crystals
% }

% \red{There are nice presentations online in youtube from Segev. We can also cite our recent work with dipole at the metasurface PTC. Also please check carefully the comment by Bumki Min to the Science paper \cite{park_comment_2022}.}
% In this subsection, we review the recently proposed thresholdless laser using PTCs \cite{xu2023thresholdless}. 

One of the crucial subjects worth studying concerning any periodic structure or, more general, photonic material is the investigation of the radiation of a source placed inside or close to it. The photonic materials may significantly affect the local density of states that change, in turn, the radiation properties of a source compared to the case that the source is in a vacuum. Besides metamaterials, this subject has attracted particular attention in the photonic crystal community. In fact, for spatial photonic crystals, this study resulted in influential novel discoveries. It was shown that if a three-dimensional periodic dielectric structure has an electromagnetic bandgap that overlaps with the transition frequency of an emitter placed into the photonic material, spontaneous emission from the source can be entirely suppressed~\cite{yablonovitch_inhibited_1987a}. There will be simply no electromagnetic mode in space to which an excited emitter could release its energy. The idea is quite general and can also be extended to other types of photonic materials. For example, an emitter whose specific transition is given by different multipolar contributions can be placed and oriented in close proximity to an individual plasmonic nanoantenna so that its emission is suppressed. In this case, the emission along the different multipolar contributions can be brought into a destructive interference \cite{rusak2019enhancement}. Modifying the spontaneous emission of emitters thanks to a structured photonic environment led to many applications, e.g., in the context of solar cells, light-emitting devices, or displays \cite{callahan2012solar,abebe2018rigorous}. 

Hence, it will be reasonable to scrutinize the problem for PTCs: How is the emission of electromagnetic waves (or light) by a source embedded in a PTC modified? For instance, one can think about the point dipole radiation~\cite{lyubarov2022amplified}. Importantly, it is intriguing to see how the radiation depends on the excitation frequency, which determines the oscillation of the dipole moment. This is a valid question due to the existence of momentum bands and gaps in a PTC. Here, we discuss the problem initially classically and afterwards quantum-optically, as done in \cite{lyubarov2022amplified}. We will work out that radiation is exponentially amplified if the emission is linked to the momentum bandgap.

After having discussed the details of the dispersion relation in depth in the previous chapters, we should be familiar with the fact that the imaginary part of the eigenfrequency, which expresses the growth in time, is strongest at the frequency corresponding to half the modulation frequency. Therefore, the longer the evolution of the field, the stronger that frequency dominates over all the others.  Of course, there are also exponentially decaying fields that vanish after enough time has elapsed. But from a classical perspective, we shall appreciate that as soon as there is a finite projection of some incident field onto the mode that will exponentially grow in time, the amplitude of that mode will dominate the field eventually. It is worth mentioning that the presence of an exponentially growing and decaying solution is in contrast to spatial photonic crystals in which the excitation within the bandgap results in two decaying waves in space (evanescent waves).

However, what happens if we excite with a frequency associated with the passband? Nevertheless, in this scenario, we must keep in mind that we switch on the source at a specific moment. The frequency spectrum launched into the system will not correspond to a single frequency, as that would require a time-harmonic source that oscillates forever and will oscillate forever. Any possible source will consist of a time-harmonic signal modulated with some envelope function. Then, the spectrum launched into the system corresponds to the Fourier transform of that envelope function convoluted with the Dirac-delta distribution at the signal frequency.

To be precise, we consider a point dipole that is turned on at a specific moment: $\_J(\_r,t)=\_J_0\delta^3(\_r)\exp(j\omega_0t)\theta(t)$. Here, $\delta^3(\_r)$ represents the three-dimensional Dirac delta distribution, and $\theta(t)$ is a Heaviside step function describing that the electric current density is generated at $t=0$. Such a dipole excites modes with all possible frequencies, including the exponentially growing gap modes. Of course, predominantly, it emits radiation at the carrier frequency, but the multiplication with the Heaviside step function also initiates other frequencies. The emission from such a source leads to fields amplified constantly over time, drawing energy from the modulated material parameters.

\begin{figure}[t]
\centerline{\includegraphics[width= 1\columnwidth]{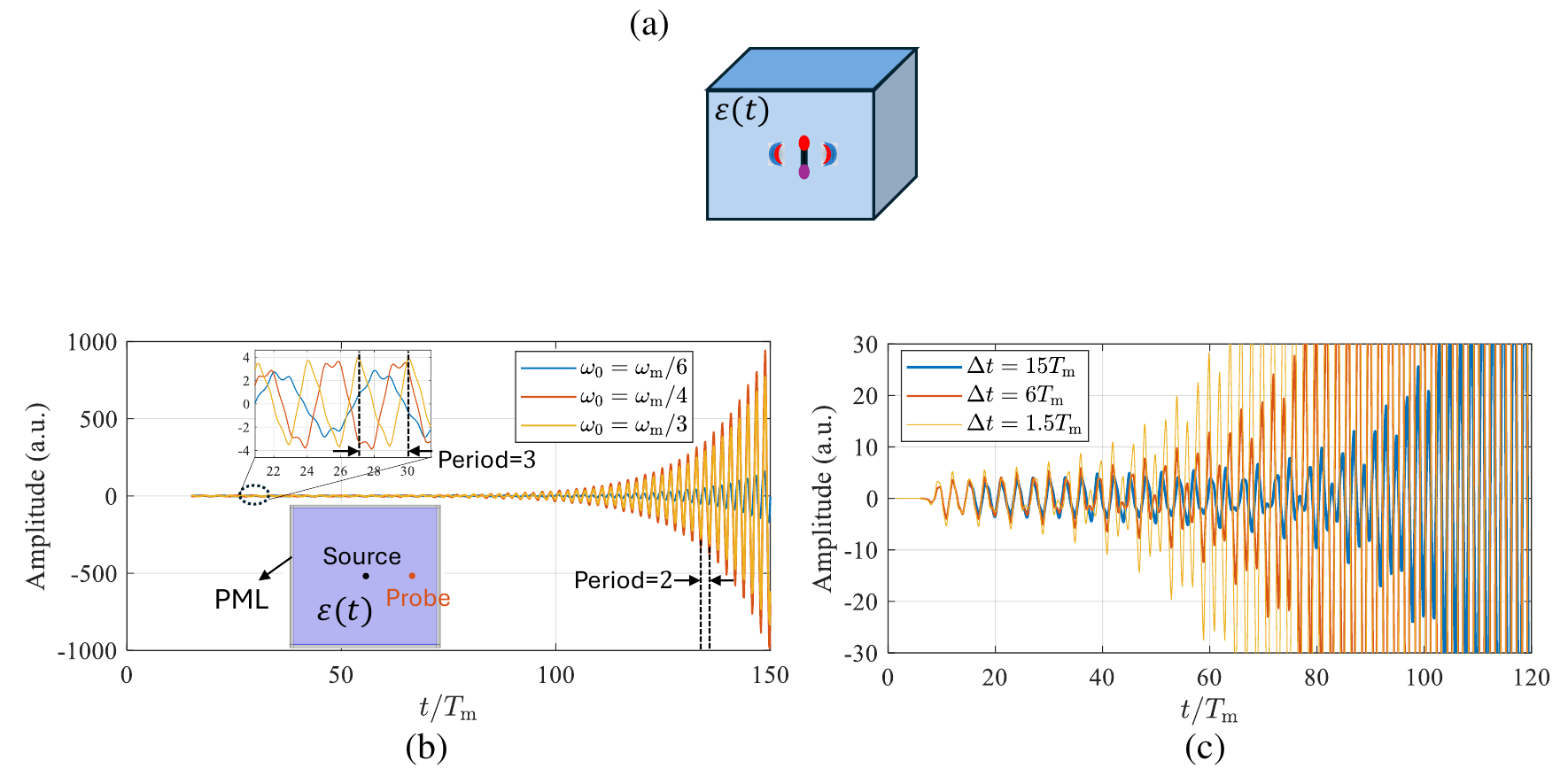}}
\caption{(a) Schematic view and (b), (c) simulation results for a point dipole that is embedded in a PTC. In the simulation, the relative permittivity of PTC is set as $\varepsilon(t)=3(1-0.2\cos(\omega_{\rm m}t))$. 
(b)~Evolution of the radiated field in time for three different excitation frequencies. Here, $\omega_0$ denotes the excitation frequency, and $\omega_{\rm{m}}$ represents the modulation frequency. The simulation domain is shown in the inset picture (the dipole source is located in the center, and the probing point is at the right side of the source).~(c)~Evolution of the radiated field in time for different time intervals $\Delta t$ between the excitation and modulation. Here, the incident frequency is $\omega_0=\omega_{\rm m}/3$. The COMSOL Multiphysics software has been used to achieve these numerical results.}
\label{Fig:TDRPTCPD}
\end{figure}

To validate the above statements, we use COMSOL Multiphysics software to simulate the emission from a point dipole located in a time-varying material. Figure~\ref{Fig:TDRPTCPD}(a) depicts the geometry of the problem. The permittivity is modulated harmonically in time. 
% so that  $\varepsilon(0) =\varepsilon(T_{\rm m}) = \varepsilon_1$ and $\varepsilon(T_{\rm m}/2) = \varepsilon_2$. 
The dipole  is considered to emit radiation at one of the three different carrier frequencies. These frequencies are denoted as $\omega_0=\omega_{\rm{m}}/6$, $\omega_{\rm{m}}/4$, and $\omega_{\rm{m}}/3$, where $\omega_0$ and $\omega_{\rm{m}}$ are the carrier frequency and modulation frequency, respectively. Notice that the dipole source is turned on at $t=0$ in the simulation. Figure~\ref{Fig:TDRPTCPD}(b) shows the field evolution as a function of time for $t>15T_{\rm m}$ where $T_{\rm m}$ is the reference time defined as $T_{\rm m}=2\pi/\omega_{\rm m}$. The time variation is switched on $15T_{\rm m}$ after the source has been switched on. There are a few aspects we can notice. Initially and after the time modulation kicks in, at a very low amplitude, the fields of the dipole source oscillate at the frequency corresponding to the carrier frequency of the source (see the inset of Fig.~\ref{Fig:TDRPTCPD}(b)). This makes sense because the source predominantly emits fields of such a frequency. Outside the momentum bandgap, the eigenmodes of the medium correspond to time-harmonic fields.

However, with time passing, we note that an exponentially growing field emerged. The frequency of that ever-increasing field corresponds to half the modulation frequency $\omega_{\rm m}/2$. Independent of the chosen carrier frequency $\omega_0$, this established field oscillation will always have the same frequency. This makes perfect sense in light of the permissible eigenmodes in such a PTC, as it corresponds to the eigenmode with the strongest exponential growth in time. At one moment, it simply dominates all other frequency components. We can think of it as the onset of lasing. It is crucial to consider the dispersion relation in our analysis. The field inside the PTC can always be written as a superposition of the eigenmodes. After launching the source and switching on the time modulation of the material parameter, we have a temporal interface between a stationary medium and the PTC. After switching on the source and after switching on the time modulation where temporal refraction occurs, we excite nearly all possible modes. The exact amplitude depends on the details of the source (in our case, it is mostly the consideration of a Heaviside step function used to switch on the source), but it will be finite at nearly all frequencies. The fields propagate outward as time progresses and are absorbed in the Perfectly Matched Layer (PML) region surrounding the computational domain of interest. Adding PML is solely for establishing a radiation boundary and is not related to the physics of PTCs. Still, even though the initial amplitude will be exponentially suppressed after some time, the amplitude of all the frequencies of the eigenmodes inside the momentum bandgap will grow exponentially once the modulation is switched on. Eventually, the field at the frequency half the modulation frequency is strongest, i.e., the field at $\omega_{\rm{m}}/2$. This is precisely what we see in the simulation results.

In this simulation described, the corresponding host medium is modulated at a later moment with respect to the source excitation moment. Hence, it is intriguing to see how the time difference between the excitation and modulation moments also affects the amplification phenomenon. The expectation suggests that the longer we wait, the stronger all the frequencies are damped (because of the dissipation in the PMLs) that do not correspond to the carrier frequency. Therefore, the later we switch on the time modulation, the longer we need to wait until the exponentially growing field corresponding to the momentum bandgap compares in amplitude to the time-harmonic carrier signal that is neither damped nor amplified. Thus, we fix the excitation frequency to $\omega_0= \omega_{\rm m}/3 $ and simulate the field where we switch on the time modulation of the material properties after some time interval since the dipole source was turned on, which is indicated by $\Delta t$ in Fig.~\ref{Fig:TDRPTCPD}(c). As expected, it is clearly seen that as $\Delta t$ becomes larger, the amplification becomes noticeably later. The exponentially damped spurious frequencies need more time to get amplified and reach the value where they are dominant.

The classical picture provided above about the radiation of a classical point dipole can be extended to the emission of light by atoms that are in excited states by using the principles of quantum electrodynamics. However, the problem is not straightforward. When we have a static medium, the excited atom will decay to the ground state which is known as spontaneous emission. On the other hand, for the PTC, it is challenging to analyze this type of emission due to the generated photons by the PTC regardless of the frequency of the atomic transition. If we neglect the impact of the gap modes, the spontaneous emission rate is expressed as~\cite{lyubarov2022amplified}
\begin{equation}
\gamma=\frac{V}{\pi\hbar^2}\sum_m k_m^2\left| V_f^m\right|^2\left\vert\frac{\partial\omega}{\partial k}\right\vert_{k=k_m}^{-1}\,,
\label{eq:SERPTCWRO}
\end{equation}
in which $V_f^m$ is the coupling constant between the initial and the final Floquet eigenstates through the interaction Hamiltonian, and $k_m$ denotes the wave number of the mode corresponding to the $m$th harmonic of the atomic transition. This equation shows that when we are very close to the band edge of the PTC, the spontaneous emission rate becomes very small (ideally zero). This is because, at the band edge, the slope of the dispersion curve is basically vertical ($\partial\omega/\partial k\rightarrow\infty$). This issue has interesting consequences. Since the exited atom does not decay into the lower state, it will always remain in the excited state. It is important to remark that this theoretical conclusion has been criticized by Ref.~\cite{park_comment_2022}. It claims that the vertical slope of the dispersion relation at the band edge does not guarantee zero spontaneous emission rate in non-Hermitian systems. In fact, if the analysis (Eq.~\eqref{eq:SERPTCWRO}) properly considers the non-Hermiticity of the PTC, the spontaneous emission rate is non-zero at the band edge~\cite{park_spontaneous_2024}.

The above results show the application of PTCs in lasing, where we can provide a resonator by locating mirrors on both sides of the PTC. The length of the cavity must be large enough compared to the desired wavelength. The existence of a saturation mechanism gives rise finally to a stable monochromatic emission. Importantly, such emission can be engineered because we can control the modulation system, giving us the possibility to have a tunable laser. We would like to mention that in Ref.~\cite{xu2023thresholdless}, a model based on a four-level system has been used to demonstrate this possibility of having a (threshold-free) lasing operation.

%%%%%%%%%%%%%%%%%%%%%%%%%%%%%%%%%%%%%%%%%%%%%%%%%%%%%%%%%%%%%%%%%%%%%%%%%%%%%%%
%%%%%%%%%%%%%%%%%%%%%%%%%%%%%%%%%%%%%%%%%%%%%%%%%%%%%%%%%%%%%%%%%%%%%%%%%%%%%%%
%%%%%%%%%%%%%%%%%%%%%%%%%%%%%%%%%%%%%%%%%%%%%%%%%%%%%%%%%%%%%%%%%%%%%%%%%%%%%%%

\subsection{Enhancing the emission rate of radiation by free electrons }\label{radfreeelectron}
% Please cite these 2 works as well:
%References: \cite{dikopoltsev2020free},\cite{li2023stationary}. 

\begin{figure}[t]
\centerline{\includegraphics[width= 1\columnwidth]{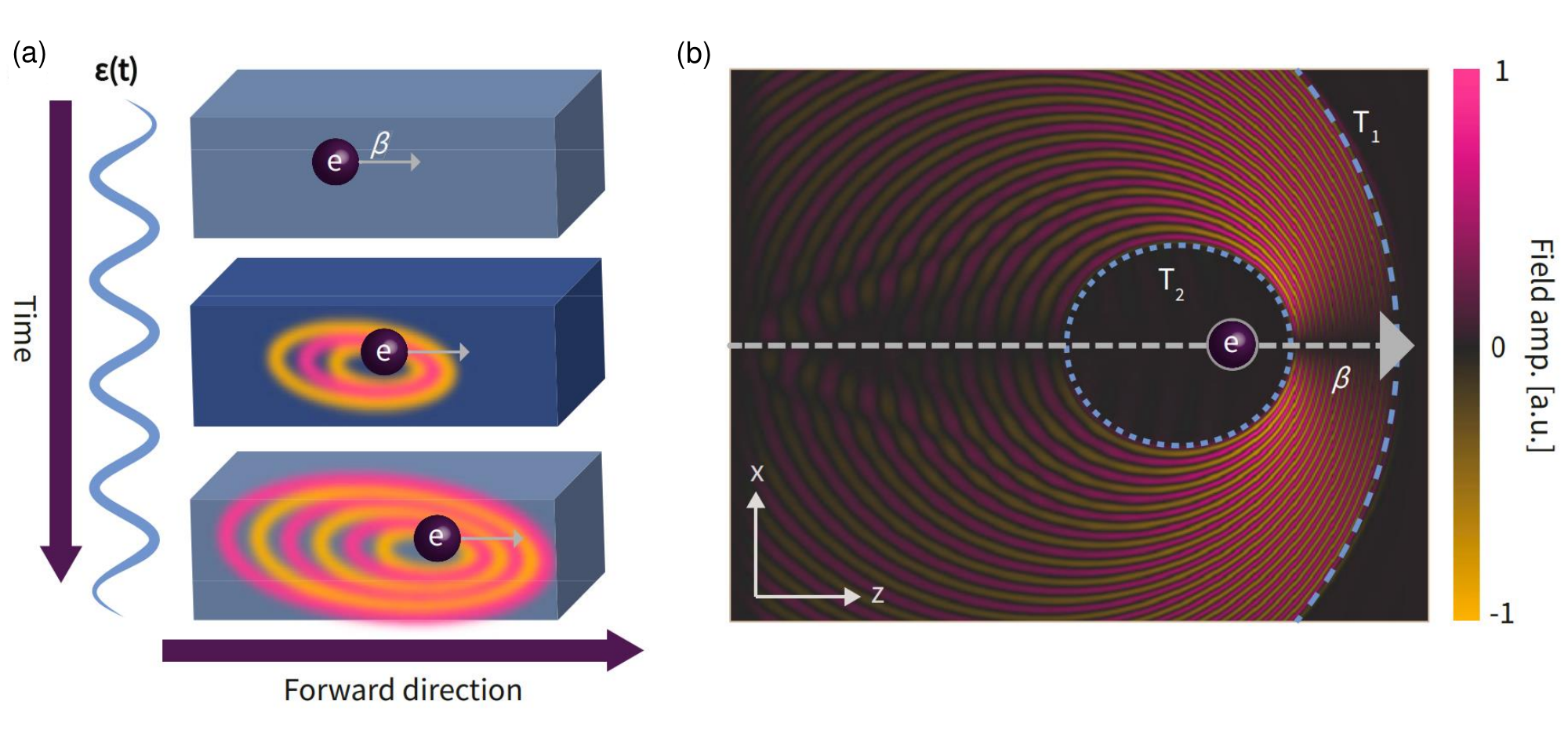}}
\caption{The radiation of a free electron with a finite velocity, which is located in a PTC. (a)~Schematic view of the evolution of radiation and electron position as time moves forward. The permittivity of the medium is periodically changing in time. (b)~Finite-difference time-domain (FDTD) simulation result for the amplitude of the magnetic field of the electromagnetic radiation. Here, the free electron is moving in a PTC of sinusoidal modulation starting at $t=T_1$ and ending at $t=T_2$. The simulation was done for the scenario in which the electron velocity is below the Cherenkov threshold. Figures were reprinted
from~\cite{dikopoltsev2022light}.}
\label{Fig:ERPIPTCSY}
\end{figure}

In the previous section, we described radiation from mainly point dipoles and explained the corresponding mechanism. However, the radiating source embedded in a PTC or close to a PTC can differ. This subject becomes important if the PTC can dramatically affect the radiation phenomenon regarding those sources such that it removes fundamental limitations that exist when the same source is placed in conventional static materials. Accordingly, besides point dipoles, the interaction of free electrons with PTCs and investigation of the radiation from these free electrons has also attracted attention~\cite{dikopoltsev2022light,dikopoltsev2020free}. Within a bulk unbounded static medium, if a free electron moves with a constant velocity, it emits in the form of Cherenkov radiation. This radiation mechanism requires that the constant velocity of the electron is greater than the phase velocity of the electromagnetic wave propagating in the static medium. Therefore, the question is: Does the PTC eliminate such a condition? Indeed, it is shown that free electrons moving in a spatially homogeneous PTC radiate spontaneously even if the constant velocity is below the Cherenkov threshold as illustrated in Fig.~\ref{Fig:ERPIPTCSY}~\cite{dikopoltsev2022light}. This is an interesting result. Of course, another benefit is that similar to the radiation from the point dipole explained above, if the wavevector is located within the momentum gap, we expect that the radiation is enhanced exponentially (for that, the energy is provided by the modulation). 

To describe the physics, we can model the moving electron classically and define an electric current density expressed as $\_J(\_r,t)=\delta(\_r_\perp)\delta(z-\beta ct)\_a_z$, in which $\_r_\perp$ refers to the coordinates transverse to $z$, and  $v=\beta c $ determines the constant velocity of the electron. This expression assumes that the electron is moving in the $z$ direction. On the other hand, based on the theory of PTCs explained above, each Floquet frequency is accompanied by an ideally infinite number of harmonics. Thus, according to this fact about harmonics and the expression for the electric current density, we can conclude that the optimum scenario for radiation from a free electron is when $k_z\beta c=\omegaf+m\omega_{\rm{m}}$. Here, $k_z$ is the $z$-component of the wavevector (recall that $\omegaf$ is the Floquet frequency, $\omega_{\rm{m}}$ is the modulation frequency, and $m$ represents the order of the harmonic). This equality is a phase-matching condition, which plays a significant role. It means that the temporal modulation provides the electron with the energy to interact with frequencies that are higher or lower than the Floquet frequency of the radiated modes. This has important consequences. As mentioned, concerning a static medium, the electron emits if its constant velocity is larger than the phase velocity of modes in the medium. However, now, regarding a PTC, the situation is different. Due to this phase-matching condition, we can have interaction with lower harmonics ($m<0$) that allow the free electron with a lower speed to emit (although we are in the regime where Cherenkov radiation cannot exist). For example, in \cite{dikopoltsev2022light}, the phase-matching condition is met for the order $m=-1$, and, therefore, the radiation occurs (i.e., $k_z\beta c=\omegaf-\omega_{\rm{m}}$). Temporal modulation generates harmonics that have small eigenfrequencies for a given wavenumber $k_z$. Therefore, these harmonics have reduced phase velocity $\omega/k_z$, enabling slow electrons to radiate. Above the Cherenkov threshold, one can expect that the electron emits ordinary Cherenkov radiation that happens in a time-invariant medium as well. However, besides, it can also radiate to higher harmonics with $m>0$. 

Hence, in conclusion, there are two different regimes of radiation: The ``subluminal" regime, where only lower harmonics contribute to the process, and the ``superluminal" regime, in which all the harmonics can be present in addition to the fundamental harmonic ($m=0$). In the end, we would like to mention two points. First, if a free electron in a PTC starts to radiate (within the subluminal or superluminal regime), the velocity cannot be constant. Due to the radiation, the electron speed must immediately change in time (which means that the acceleration is not zero). From this point of view, the same scenario may happen as the flash point dipole (described in the previous section). In other words, such a source may excite all the modes including the gap modes, which results in the exponentially growing fields in any case. However, a more rigorous and sophisticated theory is needed to validate this statement. Second, the electron with the constant velocity should not necessarily be embedded in a PTC in order to radiate. It can be in the vacuum and travel close to such a time-varying system~\cite{gao2023free,SPRFETG}.  

In the above, we focused on the radiation from a moving electron. Based on electrostatics, we know that an electron or any charge with zero velocity, which is placed in a static medium, cannot radiate. Such a stationary charge generates only static electric fields in the medium. How about locating this stationary charge in a PTC? For a static isotropic medium, the corresponding electric flux density is not a function of the relative permittivity of the medium. Thus, definitely, in a conventional isotropic PTC, the charge does not emit as well. The electric flux density is still time-independent. However, in the case of a static anisotropic medium unlike isotropic one, we see that the electric flux density depends on the components of the permittivity tensor. Hence, one can think now about what happens if those components become a function of time. It is shown that if the stationary charge is placed in an anisotropic PTC, the electric flux density becomes time-dependent, and the time derivative of the electric flux density does not become zero. In consequence, as demonstrated in~\cite{li2023stationary}, for such a stationary charge that is within an anisotropic PTC, we have a radiation phenomenon.

\subsection{Controlling  the spectral flow of light\label{sec: spectral}}
% Here, we discuss how the spectral properties of light can be manipulated using PTCs.\\
% References: \cite{hayran2021controlling}.

\begin{figure}[tb]
\centerline{\includegraphics[width= 1\columnwidth]{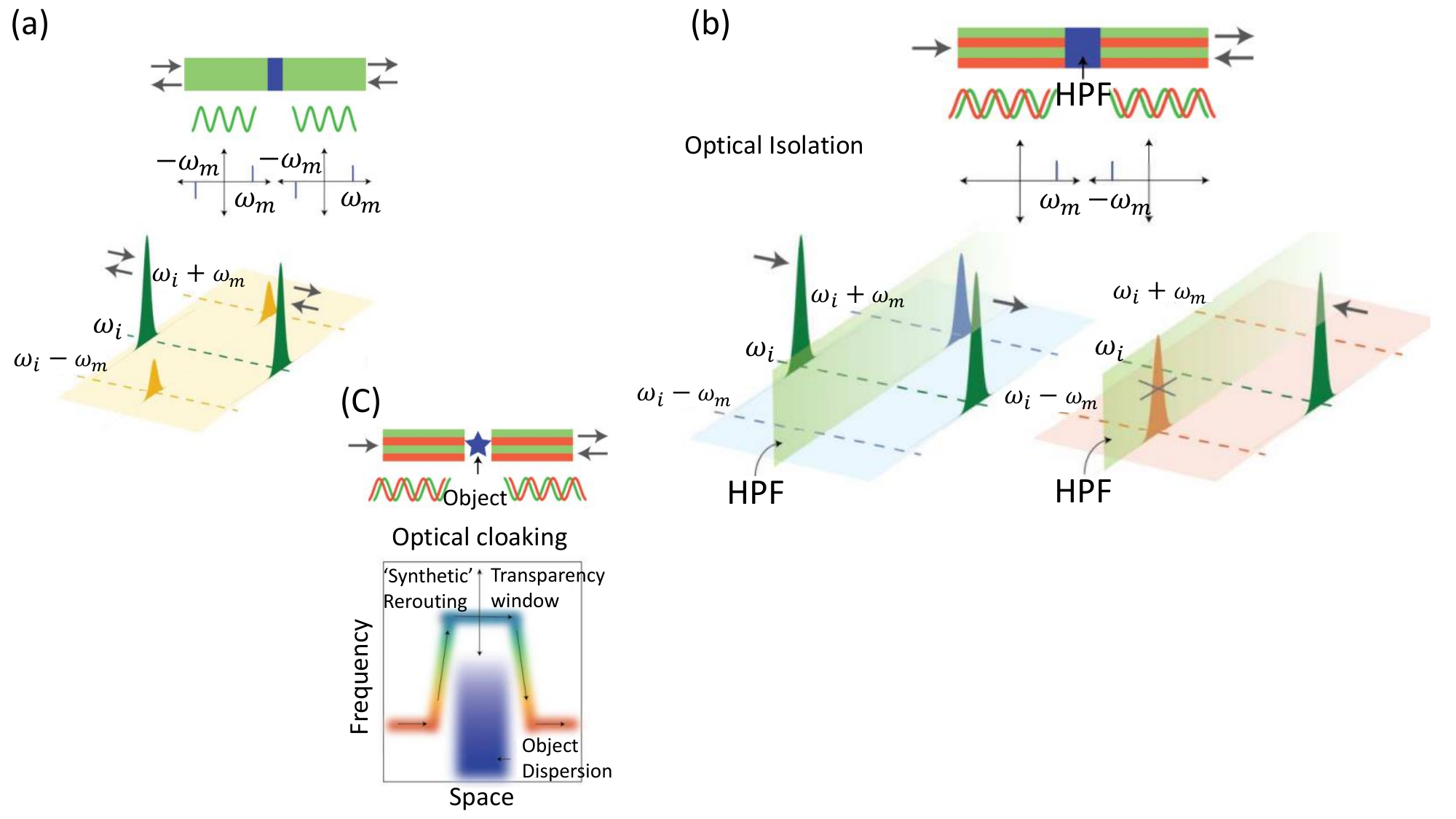}}
\caption{(a) A reciprocal tandem of two PTC slabs (shown in green) with real-valued periodically time-varying permittivities (Hermitian modulation). The two PTC slabs are separated by a high-pass filter (shown as a blue box). Light transmission for illuminations from the left and from the right is the same. Due to the Hermitian modulation function, incident signal at $\omega_{\rm i}$ experiences symmetric up-conversion and down-conversion to $\omega_{\rm i} + \omega_{\rm m}$ and $\omega_{\rm i} - \omega_{\rm m}$, respectively.  (b) Same as in (a) but both PTC slabs have non-Hermitian modulation (both real and imaginary parts of permittivity are periodically modulated at different regions of the slabs depicted in green and orange). Light can be transmitted only from left to right, while it is fully blocked by the filter when propagating in the opposite direction.  (c) The asymmetric frequency conversion can be used for ``frequency'' rerouting of incident waves providing the opportunity for direction-dependent optical cloaking. The object (shown in blue) has a small transparency window for blue wavelengths. By up-converting incident light to these wavelengths using the PTC, one can obtain high light transmission through the object. The transmitted light is then down-converted to the original incident frequency. Figure was reprinted
from~\cite{hayran2021controlling}.
}
\label{TCA2}
\end{figure}

PTCs are interesting not only from the perspective of light amplification inside their momentum bandgaps but also from the perspective of spectral transformations they can provide for a given signal. As was demonstrated in Fig.~\ref{fig:bandstructure_harmonic}, the spectrum of excited harmonics in a typical PTC is nearly symmetric with respect to the fundamental frequency ($n=0$). Nevertheless, this symmetry can be broken in PTCs whose real and imaginary parts of permittivity are periodically modulated in time with a specific phase difference~\cite{hayran2021controlling}.
These PTCs with non-Hermitian modulation functions can provide control over the scattering response in the frequency space, e.g., fully up-converting or down-converting in frequency incident signals.  Such ``one-way'' frequency conversion can be used for creating optical magnetless isolators. Indeed, consider a tandem of two PTCs with non-Hermitian modulation designed such that the left one up-converts and the right one down-converts incident signal of a given frequency $\omega_{\rm i}$ (see the geometry in the upper part of Fig.~\ref{TCA2}(b)). The two PTCs are separated by a high-pass filter (HPF) transmitting only $\omega_{\rm i}+\omega_{\rm m}$ harmonic, shown as a blue box in the figure. In this scenario, the light can propagate through the tandem of the PTCs only from left to right, as shown in the bottom of Fig.~\ref{TCA2}(b). A similar nonreciprocal effect due to one-way frequency conversion was also reported previously in other optical systems~\cite{koutserimpas2018nonreciprocal,li2014photonic}. On the other hand, if the PTCs with Hermitian modulation (only the real part of permittivity is modulated) form a tandem, it will be reciprocal, as shown in Fig.~\ref{TCA2}(a). 

Moreover, it was suggested in~\cite{hayran2021controlling} that the same PTCs with non-Hermitian modulation functions can be used for the optical cloaking of objects. As is shown in Fig.~\ref{TCA2}(c), if an object is made out of a dispersive material, it can be hidden for a given illumination by shifting the incidence spectrum into a frequency domain for which the object is transparent. Then, the light will weakly interact with the object. After shifting the frequency back to the original value, the incident field continues to propagate as if there would have been no object. This ``frequency rerouting'' could possibly allow to render large objects invisible, overcoming conventional cloaking limitations. However, it is important to acknowledge that this cloak would not work for the light coming from the opposite direction.

\subsection{Advanced optical absorbers beyond the Rozanov bound}

In this section, we discuss PTCs with specifically designed losses and their role in achieving ultra-wideband absorption that exceeds the Rozanov limit. 
In 2000, Rozanov \cite{rozanov2000ultimate} established that the absorption bandwidth of a lossy material slab is related to its thickness. If a time-invariant material slab has a reflection spectrum denoted as $\Gamma(\lambda)$, then the integral of $\ln{|\Gamma(\lambda)|}$ across the entire wavelength spectrum is bounded above by a value related to the slab's thickness $d$ and slab permeability $\mu_{\rm s}$. This limit is expressed as follows:
\begin{equation}
\left|\int_0^{+\infty} \ln{|\Gamma(\lambda)|} , \mathrm{d} \lambda\right| \leq 2\pi^2 \mu_s d\,. \label{eq: rozonov}
\end{equation}
% Here, the material is assumed to be non-magnetic, hence, $\mu_s=1$. 
The left side of Eq.~(\ref{eq: rozonov}) is called the Rozanov integral $I_R$.

The Rozanov bound in Eq.~(\ref{eq: rozonov}) is derived under the premise that materials are linear and time-invariant. 
Therefore, a natural possible strategy to surpass the Rozanov limit is to introduce material temporal modulations. For example, in \cite{li2021temporal,PhysRevApplied.17.014017, PhysRevApplied.17.044003}, the authors use temporal switching of a lossy material, to go beyond the Rozanov limit. However, temporal switching requires precise synchronization of the switching events, which is challenging to achieve in practice.

\begin{figure}[t!]
\centerline{\includegraphics[width= 1\columnwidth]{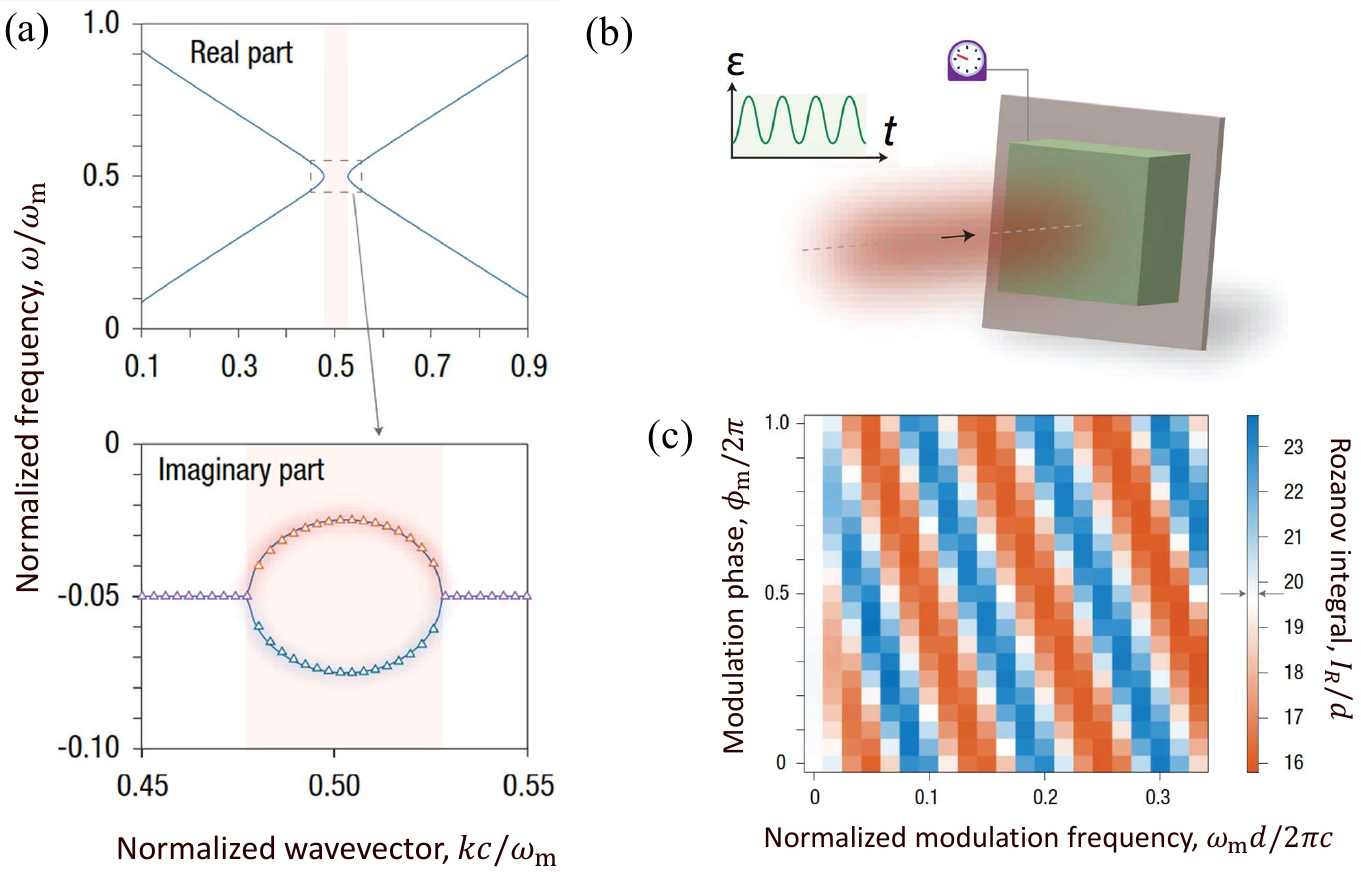}}
\caption{(a) Band structure of a lossy PTCs, calculated from the Floquet theorem. The modulation depth here is $m_{\rm p}=1$. The momentum bandgap in the band structure induced by the periodic temporal modulation (upper panel) results in the imaginary part of the eigenfrequency splitting into two distinct values within the gap (lower panel). Colored triangular markers denote results obtained using the coupled mode theory. (b) The structure of the advanced absorber is made of a time-varying dielectric slab and a metal ground plane. The incoming pulse has a finite size in the space and time domain. (c)
Rozanov integral values, $I_R$, for a range of modulation parameters. The modulation depth here is $m_{\rm p}=0.4$. Figure was reprinted from~\cite{hayran2023beyond}.}
\label{fig: Rozanov}
\end{figure}
A promising solution, as proposed in \cite{hayran2023beyond}, is to apply continuous (periodic) modulation on the dissipative material. Here, the material is dispersive following the Lorentz dispersion relation. The plasma frequency is modulated according to $\omega^2_{\rm p}(t) = \omega_{\rm p0}^2[1+m_{\rm p} \sin(\omega_{\rm m}t+\phi_{\rm m})]$, which effectively modulates the permittivity continuously.
This continuous modulation results in the formation of a PTC. The band structure of such a lossy PTC can be analyzed using the Floquet theorem, as presented in Subsection~\ref{sec: planewave expansion}. With the proper loss factor, a momentum bandgap is formed in the band structure, depicted in Fig.~\ref{fig: Rozanov}(a). Unlike lossless PTCs, the two complex eigenfrequencies within this bandgap indicate decaying modes due to the negative imaginary part of the eigenfrequencies ($e^{-i\omega t}$ convention is used in \cite{hayran2023beyond}), as illustrated in the lower panel of Fig.~\ref{fig: Rozanov}(a).
To overcome the Rozanov limit, one needs to predominantly excite the lossier eigenmode (shown by the line with blue triangles in the figure). For that, the authors of \cite{hayran2023beyond} position a time-varying dielectric slab on a metal plate, as shown in Fig.~\ref{fig: Rozanov}(b). As it was discussed in Section~\ref{PAPTCs}, then the PTC becomes phase-sensitive to the excitation, and by carefully adjusting the modulation phase, it is possible to selectively excite one of its two eigenmodes. Since the mode with larger attenuation is excited and dominant, the absorptance in the material can be higher than in its time-invariant counterpart. 
% When an incident broadband pulse impinges on the slab, it reflect off the metal plate and form a standing wave pattern within the slab. This ensures uniform phase distribution of the fields inside the slab. The absorption behavior of the PTC is determined by the phase of this standing wave in relation to the temporal modulation—either enhancing or reducing  absorption. By adjusting the modualtion phase, the absorber's performance can exceed the Rozanov limit.
Figure~\ref{fig: Rozanov}(c) displays the Rozanov integral $I_R$ as a function of modulation parameters (phase $\phi_{\rm m}$ and frequency $\omega_{\rm m}$). One can see that with appropriate phase selection, for any modulation frequency, performance beyond the Rozanov limit is achievable. Thus, PTCs can be used to create ultra-thin broadband absorbers of electromagnetic radiation. Such absorbers are of great importance for many applications, especially at GHz and mm-wave regimes. 

Recently, it was proposed that PTCs under proper configuration can be used for creating customizable multi-band absorbers~\cite{dong2023tunable}. Such absorbers are capable of absorbing light at a predefined number of frequency bands, forming a frequency comb. 
It was demonstrated, in particular, that by changing the incidence angle or the temporal modulation function, it is possible to control the spectral spacing between the neighboring bands in the comb. This functionality can find applications for the control of thermal emission, direction-selective filters, and switches.

\subsection{Enhanced resolution imaging}

Recently, it has been also suggested that PTCs can also find applications in the field of imaging for resolution enhancement~\cite{manzoor2020enhanced}.  
% The resolution of an imaging system is critically influenced by the imaging resolution.
In order to obtain an image with resolution not bounded by the diffraction limit, one needs to preserve the information about its high spatial frequencies (wavenumbers)~\cite{pendry2000negative}. Since the light components with high spatial frequencies reside below the light line of the medium where light propagates, they get quickly attenuated in space and cannot reach the image sensor. A perfect lens based on a material slab with a negative refractive index was proposed to solve this problem and restore high-resolution details of an image through amplification of evanescent modes inside the slab~\cite{pendry2000negative}. 
However, materials with negative refractive index inevitably possess dissipation loss, which diminishes the perfect lens effect.  Due to the inherent amplification nature of PTCs inside the momentum bandgaps, in Ref.~\cite{manzoor2020enhanced}, it was suggested to exploit PTCs for enhancing high spatial frequencies of the 2D Fourier expansion of an image. Moreover, it was found that by additionally applying aperiodic perturbations to the temporal modulations of the PTC, the image resolution can be further enhanced.

% \red{Please check where to add this paper ~\cite{carminati_universal_2021}. It is important one. We need to describe it too.}

\section{Introducing spatial periodicity in PTCs: Spatiotemporal photonic crystals \label{secSpacetime}} 

This last bigger chapter of this tutorial shall widen its scope, and we concentrate here on photonic crystals with optical properties periodic in space and time. As such, we will call them here spatiotemporal photonic crystals (ST-PCs). 
To a certain extent, this can be considered as a generalization of concepts previously considered as independent. Initially, we consider a specific kind of ST-PCs where the modulation is in the form of a traveling wave. Then, we summarize the main contributions towards a generalization of this concept. We elaborate in this context particularly on the aspect of how a suitably structured spatial photonic crystal can enhance the effects associated with a temporal modulation. That is important because it shows that even though we may experimentally achieve especially at optical frequencies only a small to modest modulation of the material properties in time, these effects can be enhanced by suitably structuring the time-varying material. We finalize this section with discussions on topological aspects of ST-PCs and properties of nonlinear ST-PCs.

\subsection{Traveling-wave modulation}
% \red{Remember also to cite Pendy's PRL, Mario Silverinha, and Shanhui's works.}
% In this subsection, ST-PCs based on traveling wave modulation of permittivity are considered.\\
% References: \cite{deck2019uniform}, \cite{biancalana2007dynamics}, \cite{taravati2020space}, \cite{deck-leger2023orthogonal}. Also \cite{chamnara2017optical}. \red{There are much more references! Also have a look the refs in our tutorial paper on nonreciprocity. }

In ST-PCs, the material properties are not only locally periodic in time but also periodic in space. The most simple ST-PCs have a traveling wave modulation form. 
It assumes that the material properties are modulated as a traveling wave in a general form, which is equivalent to a moving media. As we will see in this section, PTCs represent a special case of a broader class of artificial materials, that is, ST-PCs \cite{biancalana2007dynamics}. 

Under the assumption that the traveling wave modulation is propagating into the $+z$-direction, such modulation can be written in a Fourier series as
\begin{equation}
    \varepsilon (z, t)=\sum_p a_p e^{-jp(k_{\rm m} z-\omega_{\rm m} t)}\,, \label{eq: permittivity}
\end{equation}
where $k_{\rm m}$ and $\omega_{\rm m}$ are modulation wavenumber and frequency respectively, and $\omega_{\rm m}/k_{\rm m}=v_{\rm m}$ is the speed of modulation wave. When an $x$-polarized plane wave characterized by the frequency $\omega$ and wavevector $k$, in the form of $A(z, t)=A_0e^{-j(k z-\omega t)}$ ($A_0$ is the wave amplitude), travels through such a modulated media with modulation $\varepsilon (z, t)$, due to the frequency mixing effect, the electric field inside the media should contain infinite numbers of harmonics. This can be mathematically explained by multiplying $\varepsilon (z, t)$ and $A(z, t)$ and generating harmonics in the form of
\begin{equation}
    \mathbf{E}(z, t)= \sum _n  E_n e^{jn(\omega_{\rm m}t-k_{\rm m}z)}e^{j(\omega t-k z)}\mathbf{a}_x= \sum _n  E_n e^{j(\omega_n t-k_n z)}\mathbf{a}_x\,,  \label{eq: E}
\end{equation}
where the generated harmonics have an equal order in space and time, $k_n =k+n  k_{\rm m}$ and $\omega_n=\omega+n  \omega_{\rm m}$. Note that Eq.~(\ref{eq: E}) is an application of Floquet's theorem for periodicity in both space and time.
The electric field of the eigenwave in such a media must respect the wave equation
\begin{equation}
    \nabla \times \nabla \times \mathbf{E}(z, t)+\mu_0 \frac{\partial^2}{\partial t^2}[\varepsilon(z, t) \mathbf{E}(z, t)]=0\,. \label{eq: wave equation}
\end{equation}
Substituting \eqref{eq: E} and \eqref{eq: permittivity} into \eqref{eq: wave equation}, the band structure can be calculated by using mode-matching analysis \cite{chamnara2017optical}.

The traveling wave that modulates the material properties propagates at a speed $v_{\rm m}=\frac{\omega_{\rm m}}{k_{\rm m}}$. Therefore, we can classify the possible scenarios for the dispersion relation into three types, depending on the relation between the modulation speed $v_{\rm m}$ and the speed of light in the same medium in the absence of temporal modulations $v_{\rm ph}=c/\sqrt{a_0}$, where $a_0$ is the 0-th Fourier coefficient in (\ref{eq: permittivity}) which is the temporal average of the dielectric function\cite{deck2019uniform,galiffi2019broadband, pendry2021gain}. These scenarios are typically referred to as sub-luminal ($v_{\rm m} < v_{\rm ph}$), luminal ($v_{\rm m} = v_{\rm ph}$), and super-luminal ($v_{\rm m} > v_{\rm ph}$).
\begin{figure}[t!]
\centerline{\includegraphics[width= 0.6\columnwidth]{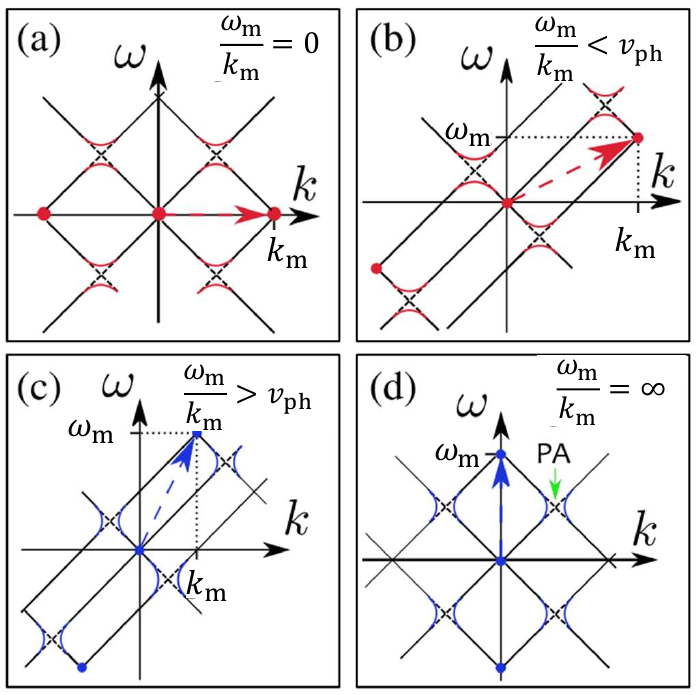}}
\caption{(a) Dispersion relation of a photonic spatial crystal where $\omega_{\rm m}/k_{\rm m}=0$ with $\omega_{\rm m}=0$ and $k_{\rm m}\neq0$. (b) Same dispersion relation but now for a spatiotemporal photonic crystal with a traveling-wave modulation. Here, the modulation wave propagates at speeds lower than the phase velocity in the time-invariant medium, i.e., $\omega_{\rm m}/k_{\rm m}< v_{\rm ph}$. (c) Same as in (b) but for the super-luminal scenario ($\omega_{\rm m}/k_{\rm m}>v_{\rm ph}$). (d) Dispersion relation of a PTC with $\omega_{\rm m}/k_{\rm m}=\infty$. The bandgaps between red and blue bands correspond to energy and momentum bandgaps, respectively. Here, ``PA'' means parametric amplification.
Figure was reprinted from~\cite{galiffi2019broadband}.
}
\label{Figtravellingwave}
\end{figure}
% . One where  is smaller than
% , i.e., $v_{\rm m}<c$, one where the modulation speed is larger than the speed of light, i.e., $v_{\rm m}>c$, and one where it is equal to the speed of light, i.e., $v_{\rm m}=c$. 
% We can start the discussion in the limit of a vanishing modulation frequency ($\omega_{\rm m} \rightarrow 0$) and a non-zero spatial modulation frequency ($k_{\rm m}\neq0$), resulting in vanishing speed $v_{\rm m}$. Then, the material under consideration becomes a traditional (spatial) photonic crystal. The band structure of such material under weak spatial modulation represents merely a folded dispersion curve\footnote{ In the context of photonic crystals, a folded dispersion curve represents the dispersion relation of a material when its periodic structure causes the original dispersion relation to ``fold'' into a smaller region, typically within the first Brillouin zone \cite[p.~146]{joannopoulos_photonic_2008}. } of a stationary material (see the red curves in Fig.~\ref{Figtravellingwave}(a)). The band structure is periodic with respect to $k$ with period $k_{\rm m}$, as depicted by the reciprocal vector of ($k_{\rm m}, 0$) in the figure. At the band crossings, the energy (frequency) bandgaps appear.
We can start the discussion in the limit of a vanishing modulation frequency ($\omega_{\rm m} \rightarrow 0$) and a non-zero spatial modulation frequency ($k_{\rm m}\neq0$), resulting in vanishing speed $v_{\rm m}$. Then, the material under consideration becomes a traditional (spatial) photonic crystal. The band structure of such material under weak spatial modulation represents merely a folded dispersion curve, which represents the dispersion relation of a material when its periodic structure causes the original dispersion relation to ``fold" into a smaller region, typically within the first Brillouin zone~\cite[p.~146]{joannopoulos_photonic_2008}. This folded dispersion curve is seen in the red curves in Fig.~\ref{Figtravellingwave}(a). The band structure is periodic with respect to $k$ with a period $k_{\rm m}$, as depicted by the reciprocal vector of ($k_{\rm m}, 0$) in the figure. At the band crossings, the energy (frequency) bandgaps appear.

\begin{figure}[t]
\centerline{\includegraphics[width= 0.8\columnwidth]{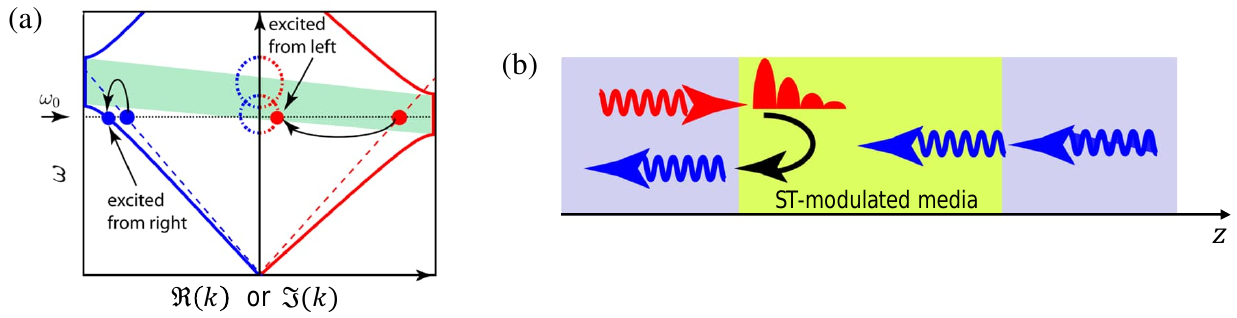}}
\caption{ (a) Asymmetric energy bandgaps in a sub-luminal traveling-wave modulated ST-PC. The vertical axis represents the frequency and the horizontal axis represents either real (solid lines) or imaginary (dotted circles) part of the propagation constant $k$. (b) Nonreciprocal wave propagating through the same crystal of finite size. Figures were reprinted from~\cite{chamnara2017optical}.
}
\label{Fig. nonreciprocity}
\end{figure}

When the speed of the modulation wave increases but still being smaller than $v_{\rm ph}$, the band structure under weak modulation approximation is formed by folding the dispersion curve of a stationary material into a new \textit{tilted} Brillouin zone~\cite{biancalana2007dynamics} with a reciprocal vector ($k_{\rm m}$, $\omega_{\rm m}$) (see the red arrow in Fig.~\ref{Figtravellingwave}(b)). A tilted Brillouin zone occurs when there is both spatial and temporal modulation in the material. This causes the original Brillouin zone (which is defined for purely spatial modulation) to tilt due to the additional temporal component.
Likewise, at the crossings, energy bandgaps are formed. One can see that in this case, the band structure is asymmetric with respect to the frequency and momentum axes. 
% Clearly, at least for a given direction, there is a frequency gap within which no propagating solution to Maxwell's equations is found. Inside the bandgap, the eigenwavevector is a complex value, resulting in the exponential decrease of amplitude. 
One of the most important physical effects induced by the asymmetric band structure is nonreciprocity~\cite{yu2009complete}. The asymmetric band structure of a typical traveling-wave modulated media is shown in Fig.~\ref{Fig. nonreciprocity}(a). When the excitation frequency is $\omega_0$ (indicated by the black dashed line in the figure), it corresponds to a real wavenumber in the negative $k$ domain and a complex wavenumber in the positive $k$ domain. The real eigen wavenumber means that the wave can propagate through the media without attenuation, while the complex eigen wavenumber means that the wave exponentially decays in space. Therefore, as shown in Fig.~\ref{Fig. nonreciprocity}(b), the device works as a wave isolator. For an incident wave coming from the right side (corresponding to negative $k$), the wave passes through the media. For incidence from the left side (corresponding to positive $k$), the wave is strongly attenuated. 
Similar nonreciprocal effects have been reported in traveling-wave modulated metasurfaces platform \cite{wang2020theory, guo2019nonreciprocal, cardin2020surface, hadad2015space, taravati2022microwave}. 

When $v_{\rm m}> v_{\rm ph}$, the modulation is superluminal. The conceptual band structure for this scenario is shown in Fig.~\ref{Figtravellingwave}(c). Similar to the subluminal modulation, the band structure is asymmetric, which means that such modulation can also induce nonreciprocity. However, in contrast to the subluminal case, for superluminal modulation, the crossing of bands generates a momentum bandgap. 
% It is known that waves inside the momentum bandgap can be amplified. 
Therefore, superluminal modulation can induce nonreciprocal amplification \cite{lee2021parametric}.

Regardless of whether the modulation is subluminal or superluminal, it was discovered in \cite{huidobro2019fresnel} that a space-time modulated metamaterial with both permittivity and permeability modulation can always generate the Fresnel drag effect, typically observed in moving media. This finding suggests that space-time traveling wave modulation can be represented by effective bianisotropic parameters with nonreciprocal magnetoelectric coupling, which can, in turn, be mapped to a moving homogeneous medium. A more complete study of moving interface between two materials were considered in a recent paper \cite{li2024wave}.

% Moreover, a plane wave traveling inside the luminally modulated media can be compressed into a train of pulses, as shown in   Fig.~\ref{Fig:luminal}(d). The reason is explained in Fig.~\ref{Fig:luminal}(b). The field in Fig.~\ref{Fig:luminal}(b) is recorded at the output face of the ST-PC ($x=d$). The modulation imposes exponential growth to the signal at time moments where $\omega_{\rm m}t=\pi/2$ and exponentially suppresses it at $\omega_{\rm m}t=3\pi/2$. This can be understood from Fig.~\ref{Fig:luminal}(c): those field amplitudes that sit at $-\pi/2<\omega_{\rm m}t<\pi/2$  experience a lower permittivity, and hence a higher phase velocity. Therefore, the wave is compressed and amplified. On the contrary,  the field amplitudes that sit at $\pi/2<\omega_{\rm m}t<3\pi/2$ experience a higher permittivity and thus a lower phase velocity, resulting in field attenuation. 

% Coming back to Fig.~\ref{Figtravellingwave}, when $k_{\rm m}=0$, the modulation velocity is infinite, this corresponds to a PTC where the material property is uniformly modulated in time without spatial periodicity. 

%\clearpage 

\begin{figure}[h]
\centerline{\includegraphics[width= 0.9\columnwidth]{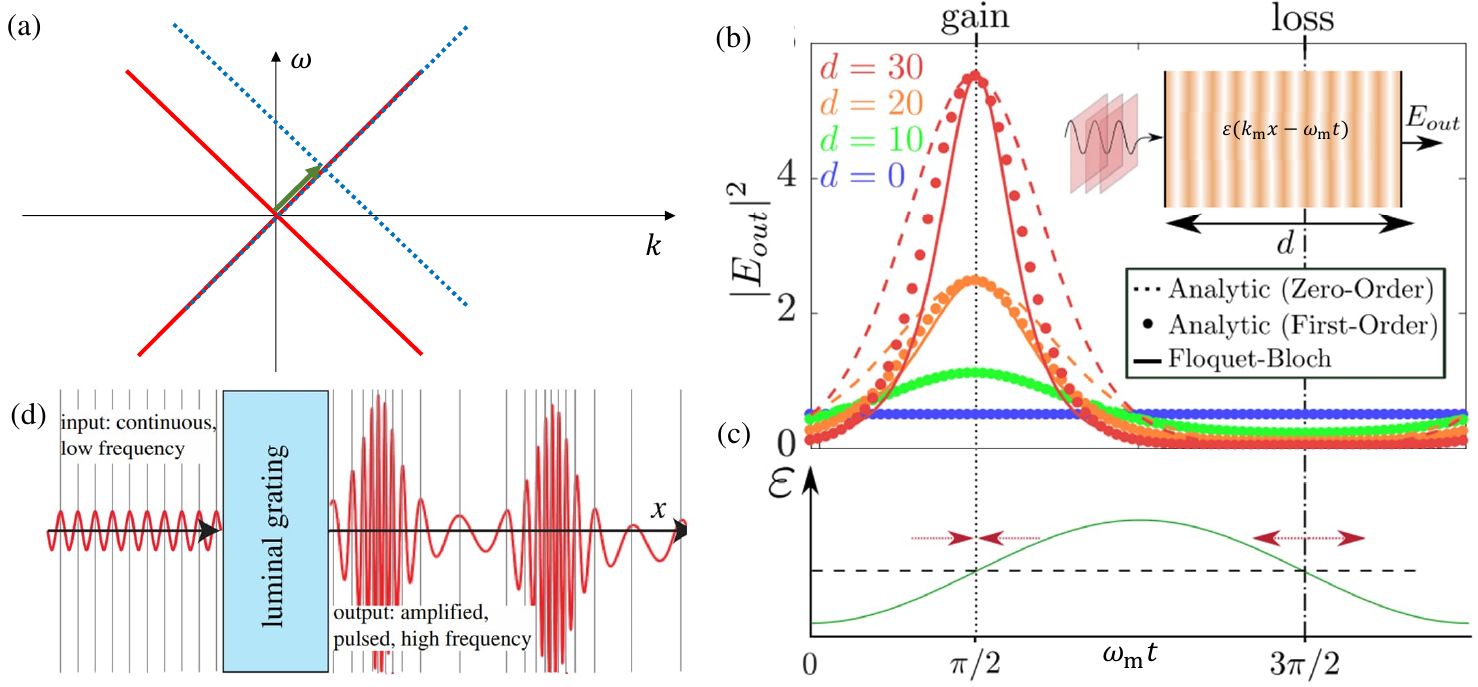}}
\caption{ (a) Band degeneracy for the case of luminal traveling-wave modulation. (b) Field amplitude at the right boundary of the ST-PC at different time moments. (c) Permittivity variation at different time moments. (d) Luminal grating when illuminated with a continuous wave at low frequency. The output is a compressed pulse train.
% Figures were adapted from~\cite{galiffi2019broadband} and \cite{pendry2021gain}. 
Figures (b) and (c) were reprinted from \cite{galiffi2019broadband}. Figure (d) was reprinted from \cite{pendry2021gain2}.
}
\label{Fig:luminal}
\end{figure}

Between the subluminal and superluminal regimes, there is a special scenario that the modulation phase velocity is equal to the speed of light, which is called luminal, i.e., $v_{\rm m}= v_{\rm ph}$. This is an exceptional case where exotic wave effects can happen \cite{galiffi2019broadband}. As shown in Fig.~\ref{Fig:luminal}(a), the reciprocal vector (green arrow)aligns with the dispersion curve of a stationary material. The band structure of a space-time varying media is formed by folding the dispersion curve of the stationary material by reciprocal vectors. The crossing of the bands results in that all the forward-travelling states are degenerate in a broadband region and therefore strongly coupled. Therefore, luminal modulation can induce broadband nonreciprocal amplification \cite{galiffi2019broadband}. 
Moreover, a plane wave propagating through a luminally modulated medium can be transformed into a pulse train, as illustrated in Fig.~\ref{Fig:luminal}(d). This phenomenon is elucidated by the dynamics presented in Fig.~\ref{Fig:luminal}(b), where the field captured at the ST-PC's output face ($x=d$) exhibits modulation-induced variations. Specifically, at instances when $\omega_{\rm m}t=\pi/2$, the signal undergoes exponential growth, while at $\omega_{\rm m}t=3\pi/2$, it experiences exponential suppression. This effect stems from the observations in Fig.~\ref{Fig:luminal}(c), indicating that field amplitudes within the range $-\pi/2<\omega_{\rm m}t<\pi/2$ encounter a reduced permittivity and, consequently, a heightened phase velocity, leading to wave compression and amplification. Conversely, amplitudes within the range $\pi/2<\omega_{\rm m}t<3\pi/2$ are subjected to increased permittivity, resulting in decreased phase velocity and field attenuation.

Revisiting Fig.~\ref{Figtravellingwave}, in the scenario where $k_{\rm m}=0$, the modulation velocity becomes infinite. This situation means that in a PTC, the material's properties change uniformly over time without any repeating patterns in space. Thus, we can see that PTCs represent a special scenario of a broader notion of ST-PCs.

\subsection{Arbitrary spatiotemporal modulation }
The traveling-wave modulation, as discussed in the previous section, is a special type of space-time modulation function. In general, the modulation function of an ST-PC does not necessarily need to be in the traveling-wave form. Indeed, the modulation can be an arbitrary periodic function with a temporal period $T_{\rm m}$ and spatial period $P_{\rm m}$, respectively. The scattering property of such generalized space-time periodic media was studied, e.g., in \cite{taravati2019generalized}.
Next, we aim to derive the eigenvalue problem of such generalized space-time modulation, following the steps in \cite{park2021spatiotemporal}. 
For isotropic and non-dispersive materials, Maxwell's curl equations simplify to
\begin{equation}
\frac{\partial}{\partial x}
\begin{bmatrix}
\mathbf{E} \\
\mathbf{H}
\end{bmatrix}
= -\frac{\partial}{\partial t}
\begin{bmatrix}
0 & \mu(x, t) \\
\varepsilon(x, t) & 0
\end{bmatrix}
\begin{bmatrix}
\mathbf{E} \\
\mathbf{H} 
\end{bmatrix}\, . \label{eq: spacetime}
\end{equation}

Assuming the permittivity and permeability are modulated with identical spatial and temporal periods, the material parameter matrix can be expressed as a two-dimensional Fourier series:
\begin{equation}
\begin{bmatrix}
0 & \mu(x, t) \\
\varepsilon(x, t) & 0
\end{bmatrix}
= \sum_{m,n}  {\={U}}_{m,n} e^ { j(nk_{\rm m}x - m\omega_{\rm m}t) }\, , \label{eq: generalized modulation}
\end{equation}
where $k_{\mathrm{m}} = \frac{2\pi}{P_{\rm m}}$ and $\omega_{\mathrm{m}} = \frac{2\pi}{T_{\rm m}}$ represent the spatial and temporal angular modulation frequencies, respectively, and ${\={U}}_{m,n}$ are $2\times 2$ anti-diagonal matrix storing the Fourier coefficients of permittivity and permeability, with $n$ and $m$ being integers that range from $-\infty$ to $+\infty$.

Using the Bloch-Floquet theorem, the solutions of Eq.~(\ref{eq: spacetime}) can be written as 
\begin{equation}
\begin{bmatrix}
\mathbf{E} \\
\mathbf{H} 
\end{bmatrix}
= e^{j(k_{\rm B}x - \omega_{\rm F} t)} \sum_{m,n} \mathbf{\Psi}_{m,n}e^{j(n k_{\rm m} x - m\omega_{\rm m} t)}\, , \label{eq: generalized fild}
\end{equation}
where \( k_{\rm B} \) and \( \omega_{\rm F} \) are the Bloch wavenumber and Floquet frequency of the Bloch–Floquet mode, respectively, and  $\mathbf{\Psi}_{m,n}$  is $2\times 1$ column vector containing the harmonic amplitudes of electric and magnetic fields for harmonic index $(m, n)$.  Substituting Eqs.~(\ref{eq: generalized fild}) and (\ref{eq: generalized modulation}) into Eq.~(\ref{eq: spacetime}) gives
\begin{equation}
\sum_{m,n} (k_{\rm B} + nk_{\rm m}) \mathbf{\Psi}_{m,n} = \sum_{m,n} (\omega_{\rm F} + m\omega_{\rm m}) \sum_{p,q} {\={U}}_{m-p,n-q} \mathbf{\Psi}_{p,q}\, . \label{eq: summation}
\end{equation}
In principle, the ranges of $m, n, p, q$ can extend from $-\infty$ to $+\infty$. However, to derive a finite number of equations, these indices are truncated. Specifically, the spatial order is limited to $\{n, q\} \in [-Q, Q]$ and the temporal order to $\{m, p\} \in [-P, P]$, where $P$ and $Q$ are positive integers. Consequently, (\ref{eq: summation}) corresponds to a finite set of equations. For each field quantity (electric or magnetic field), the number of equations is $S = (2P+1)(2Q+1)$. These equation sets can be expressed through matrix operations. By defining convolution matrices $ \={C}_{\varepsilon,\mu} $ and a $2S$-dimensional column vector of field Fourier components $ \mathbf{\Phi} = ({\mathbf{E}_a}, {\mathbf{H}_a})^T $ (here, $\mathbf{E}_a$ and $\mathbf{H}_a$ are $S$-dimensional row vector storing the Fourier amplitudes of electric and magnetic waves), the matrix operation is formulated as follows:
\begin{equation}
    [\={I}_2 \otimes (k_{\rm B}\={I}_S+\={G})]
    \mathbf{\Phi}=[\={I}_2 \otimes (\omega_{\rm F}\={I}_S+\={W})]\cdot
    \begin{bmatrix} 
        0 & \={C}_{\mu}\\
       \={C}_{\varepsilon} &0 
    \end{bmatrix}\mathbf{ \Phi}\, . \label{eq: matrix generalized}
\end{equation}
Here, the symbol $\otimes$ represents the Kronecker product. The subscript of the unit matrix $\={I}$ represents its dimensionality.
Matrices $\={G}$ and $\={W}$ are square matrices of size $S$ which are defined as $\={G}=\={I}_{2P+1}\otimes {\rm diag
}(nk_{\rm m})$ and $\={W}= {\rm diag
}(m\omega_{\rm m}) \otimes \={I}_{2Q+1}$. Matrices $\={C}_{\varepsilon, \mu}$ are $S$-dimensional square matrices. 
After rearranging (\ref{eq: matrix generalized}), the eigenvalue problem can be established,
\begin{equation}
   \begin{bmatrix}
        0 & \={C}_{\mu}\\
       \={C}_{\varepsilon} &0 
    \end{bmatrix}^{-1} \cdot
     \begin{bmatrix}
        k_{\rm B}\={I}_S+\={G} & -\={W}\cdot\={C}_{\mu}\\
       -\={W} \cdot \={C}_{\varepsilon} & k_{\rm B}\={I}_S+\={G}
    \end{bmatrix}\mathbf{\Phi}=\omega_{\rm F} \mathbf{\Phi}\, .
\end{equation}

The above derivation originally obtained in~\cite{park2021spatiotemporal} is for the most general periodic space-time modulation. However, it is important to consider also a special scenario where the modulation function of the permittivity can be separated in both space and time variables, i.e.,
$\varepsilon(x,t)=\varepsilon_x(x) \varepsilon_t(t)$. The method to analytically determine the band structure under the separable modulation function of this form was presented in \cite{gonzalez2010mode,sharabi2022spatiotemporal}. 
\begin{figure}[h]
\centerline{\includegraphics[width= 0.55\columnwidth]{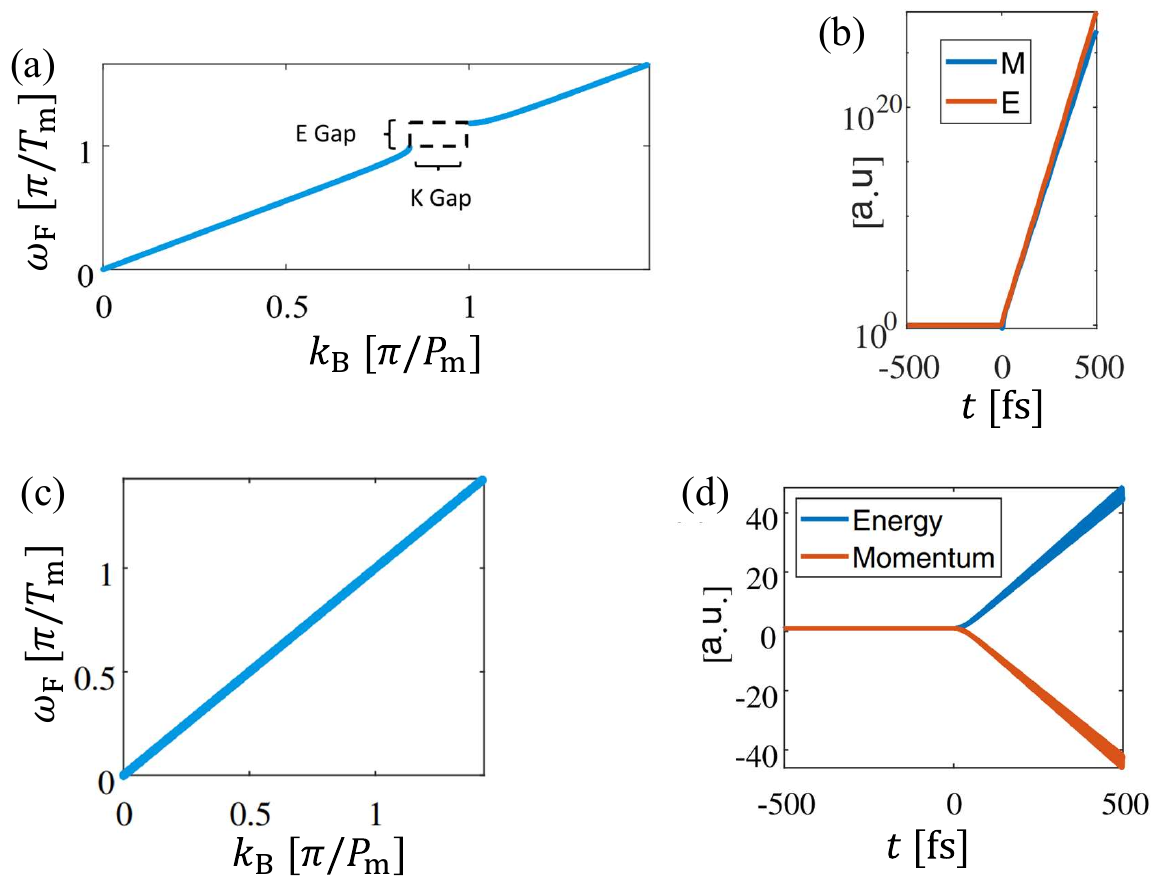}}
\caption{ (a) Mixed momentum-energy bandgap. The spatiotemporal permittivity modulation is  in the form of $\varepsilon(x, t)=\varepsilon_x(x)\varepsilon_t(t)$, where $\varepsilon_x(x)=(1+2B)/[1+B(1+\cos(k_{\rm m}x))]$ with $B=0.5$ and $\varepsilon_t(t)=1+A[1+\cos(\omega_{\rm m}t)]$ with $A=0.5$. (b) The exponential growth of momentum $M$ and energy $E$ of the eigenmode inside the mixed bandgap with the modulation starting at $t=0$. (c) Closed mixed bandgap. Here, the spatial and temporal variation functions must be inversely proportional to each other.  (d) Linear growth of momentum and energy in the closed mixed bandgap with the modulation starting at $t=0$. Figures were reprinted from~\cite{sharabi2022spatiotemporal}.
}
\label{FIg. mixed}
\end{figure}
With such modulation, it is possible to generate the so-called ``mixed'' bandgaps which show the features of both momentum and energy bandgaps \cite{sharabi2022spatiotemporal}. Figure~\ref{FIg. mixed}(a) shows that under proper space-time modulation, the energy and momentum bandgaps overlap in the frequency-momentum plane. In the overlapping region, the eigenmode features complex frequency and complex wavenumber. This results in unique propagation phenomena where exponential growth induced by temporal modulation and exponential decay caused by spatial modulation can counteract each other. When the temporal growing mode dominates, its momentum and energy exponentially grow, as shown in Fig.~\ref{FIg. mixed}(b).  Moreover, under very specific modulation parameters,  the counteracting forces of growing in time, and decaying in space are exactly matched, causing the bandgap to close, as shown in Fig.~\ref{FIg. mixed}(c). Inside this closed mixed bandgap, a pulse wave stops and expands its width in space, maintaining the constant amplitude. This results in the linear growth of its momentum and energy, as shown in Fig.~\ref{FIg. mixed}(d).

\subsection{Enhancing the size of a momentum bandgap using resonant ST-PCs \label{Sec:ExploitinResonantSTPC}}
\begin{figure*}[tbh]
\centerline{\includegraphics[width= 1\columnwidth,trim=4 4 4 1,clip]{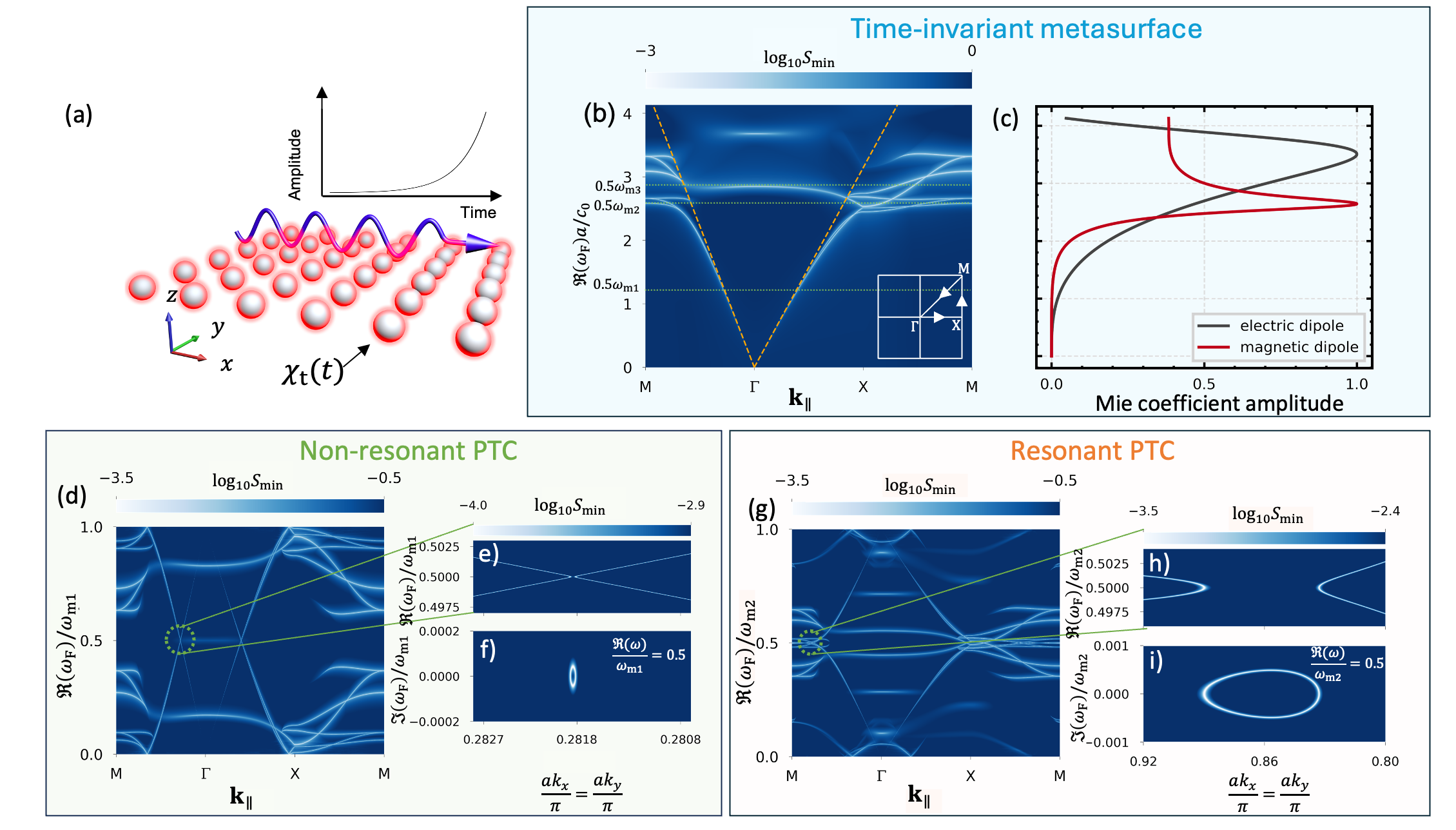}}
\caption{
(a) An ST-PC based on a metasurface consisting of dielectric spheres with time-varying material susceptibility. (b) The band structure of the time-invariant metasurface. (c) The dipolar Mie coefficients of an isolated static sphere. (d)--(f) The band structure of the non-resonant ST-PC with the modulation frequency $\omega_\mathrm{m1}$. 
(g)--(i) The band structure of the resonant ST-PC with the modulation frequency $\omega_\mathrm{m2}$. Note that $a$ represents the spatial period of the metasurface in (a). Further, $S_{\rm min}$ is a metric based on the T-matrix of the ST-PC. A local minimum of $S_\mathrm{min}$ determines the spectral location of an eigenmode of the ST-PC (see \cite{wang2023unleashing} for details).
Figure was reprinted from\cite{wang2023unleashing}.}
\label{fig:resonant_STPC}
\end{figure*}

%In this subsection, we will describe how the resonances of a spatially modulated photonic crystal assist in lowering the power %requirements to open large MBGs.\\ 
%References: \cite{garg2022modeling,wang2023unleashing}.
Another class of ST-PCs with a separable spatiotemporal modulation are those whose susceptibility is of the form $\chi(\mathbf{r},t)=\chi_{\rm r}(\mathbf{r})\chi_{\rm t}{(t)}$. Here, $\chi_{\rm r}(\mathbf{r})$ corresponds to the spatial part of the susceptibility encoding the spatial modulation of the system. Further, $\chi_{\rm t}{(t)}$ is the temporal part of the susceptibility encoding the temporal modulation of the system. Note that the modulation of susceptibility in such a form is qualitatively different from the modulation of permittivity in the same form discussed in the previous section. 
One example of a system having such a separable spatiotemporal profile of susceptibility is a metasurface made from a periodic arrangement of time-varying spheres embedded in vacuum (see Fig.~\ref{fig:resonant_STPC}(a)) \cite{garg2022modeling,wang2023unleashing}. Here, each sphere is assumed to have a time-dependent susceptibility $\chi_\mathrm{t}(t)= \chi_\mathrm{st}[1+M_\mathrm{s}\mathrm{cos}(\omega_\mathrm{m}t)]$. Here, $\chi_\mathrm{st}$ is the electrical susceptibility of the static spheres, $M_\mathrm{s}$ is the modulation depth, and $\omega_\mathrm{m}$ is the modulation frequency. Moreover, the susceptibility variation of the ST-PC due to the spatial arrangement of the spheres can be physically expressed in the form of a space-dependent susceptibility $\chi_{\rm r}(\mathbf{r})$. 

As discussed earlier, the existence of momentum bandgaps is one of the key features of PTCs \cite{zurita2009reflection}. However, the PTCs often require large modulation depths of the material parameters to show discernible bandgaps \cite{hayran2022ℏomega}. As was discussed in Section~\ref{optmatsec}, such a requirement of large modulation depth poses two major challenges. First, the modulation of material parameters with high modulation depths requires enormously high pump powers. Second, the intrinsic losses of the material may lead to its rapid thermal damage in the presence of such high-power pumps. To overcome these challenges, the structural resonances of ST-PCs can be harnessed. This is because they lower the required modulation depths drastically to attain large momentum bandgaps \cite{wang2023unleashing, khurgin2023energy}. 

In \cite{wang2023unleashing}, the metasurface-based design of the ST-PC shown in Fig.~\ref{fig:resonant_STPC}(a) was used to attain large momentum bandgaps for very low material modulation strengths. In the following, we discuss the proposed method in \cite{wang2023unleashing}. First, the band structure of a static metasurface (i.e., with $M_\mathrm{s}=0$) is plotted in Fig.~\ref{fig:resonant_STPC}(b). Here, the flat bands appear as prominent features. These flat bands occur due to the dipolar Mie resonances of the static spheres of the metasurface (see Fig.~\ref{fig:resonant_STPC}(c)). Next, the temporal modulation was switched on with the modulation depth as low as $M_s=0.01$. In the first (non-resonant) scenario, the modulation frequency was chosen as $\omega_\mathrm{m}=\omega_\mathrm{m1}$. Note that such a configuration of the metasurface corresponds to a non-resonant case because $\omega_\mathrm{m1}$ is spectrally far away from the locations of the flat bands (see Fig.~\ref{fig:resonant_STPC}(b)). The band structure for such a non-resonant ST-PC is plotted in Fig.~\ref{fig:resonant_STPC}(d). From Figs.~\ref{fig:resonant_STPC}(e)--(f), we observe a small momentum bandgap.

In the second (resonant) scenario, the modulation frequency was chosen as $\omega_\mathrm{m}=\omega_\mathrm{m2}$, while keeping the same modulation strength $M_s=0.01$.
% As a next step, the utility of the Mie resonances to widen the momentum bandgaps is demonstrated. Therefore, 
Note that, since $0.5\omega_\mathrm{m1}$ is at the spectral location of one of the flat bands in Fig.~\ref{fig:resonant_STPC}(b), such a configuration corresponds to a resonant case. In Fig.~\ref{fig:resonant_STPC}(g), the band structure of such a resonant ST-PC is plotted. From Figs.~\ref{fig:resonant_STPC}(h)--(i), a wide momentum bandgap is observed. As reported in~\cite{wang2023unleashing}, such structural resonances of the metasurface lead to the enhancement of the bandgap size by a factor of $350$.

%Topological effects have been recently studied in PTCs, we will review these advancements in this subsection. 
%References: \cite{lustig2018topological}, \cite{ma2019band}, \cite{serra2023rotating}, \cite{dong2023band}, \cite{ding2024non}, %\cite{ren2024observation}.

\begin{figure*}[h]
\centerline{\includegraphics[width= 0.97\columnwidth,trim=4 4 4 4,clip]{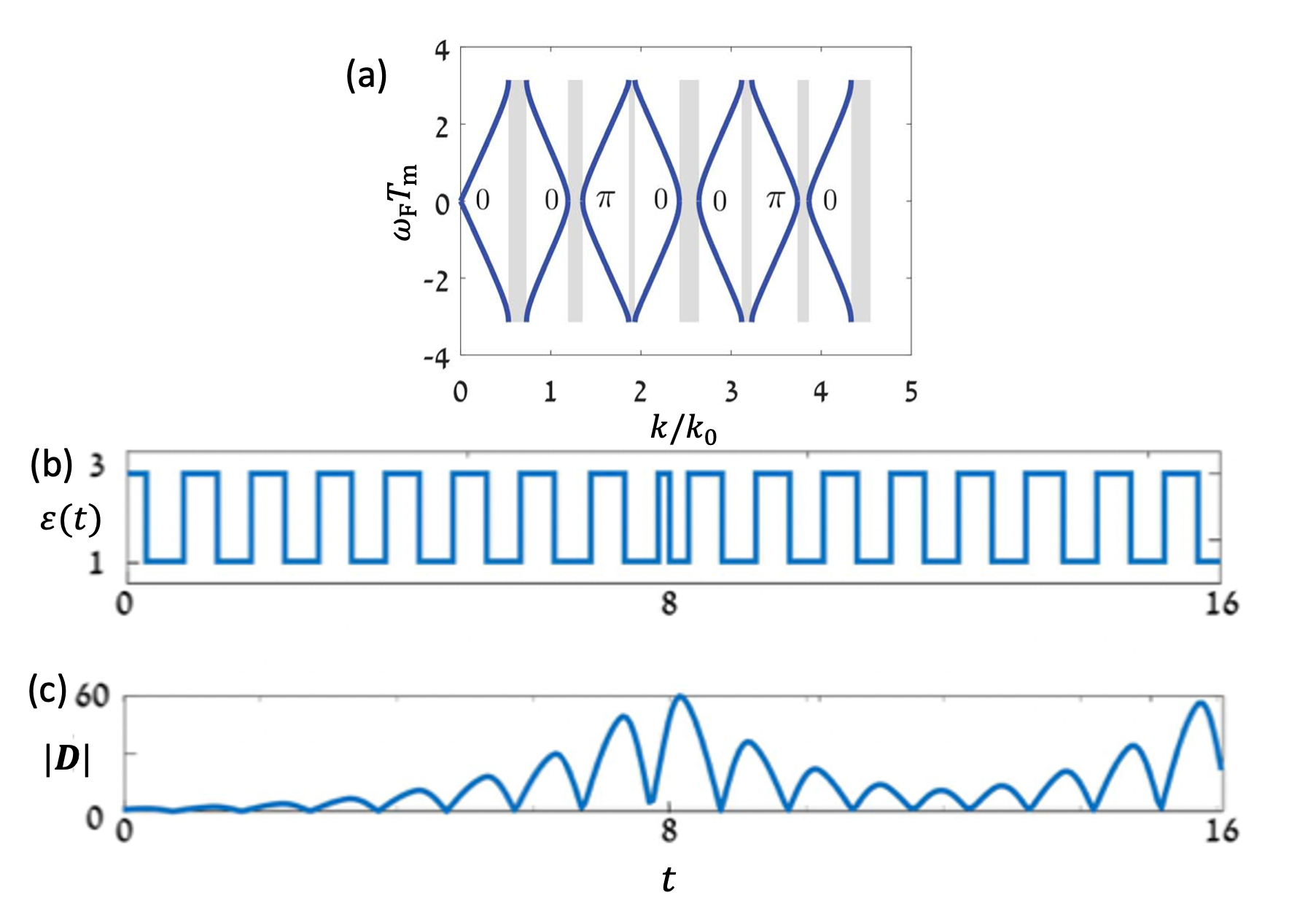}}
\caption{(a) The band structure of a PTC with step-wise modulation. Here, Zak phases $\theta^\mathrm{Zak}_p$ of the bands are shown beside the relevant bands. (b) The permittivity profile $\varepsilon(t)$ as a function of time $t$ of the composite system made from two PTC that occur one after another. (c) The variation of the displacement field amplitude $|\mathbf{D}|$ as a function of time $t$ of the topological edge state that occur as an eigenmode of the composite system shown in (b). Figure was reprinted from~\cite{lustig2018topological}.}
\label{fig:topological1}
\end{figure*}

\subsection{Topological aspects of PTCs \label{topology}}
The understanding of topological phases of photonic systems has given rise to various interesting phenomena in optics such as unidirectional light propagation \cite{wang2009observation} and unidirectional lasing \cite{bahari2017nonreciprocal}. An important characteristic of the topological phases of matter is that they are robust against defects and disorders in the underlying system \cite{Segev2021topological}. Therefore, studying the topology of physical systems is crucial for classifying them for stability against such defects and disorders. 

Recently, the topological phases of the PTCs have also been studied \cite{lustig2018topological,ma2019band,Long2024inverse}. Since the PTCs correspond to a 1D periodic system (with periodicity about temporal dimension), the topological invariant of interest in such a system is the Zak phase~\cite{zak1989berry}. The Zak phase of each band of a PTC is given by \cite[Eq.~5]{lustig2018topological}

\begin{figure}[h]
\centerline{\includegraphics[width= 0.5
\columnwidth,trim=4 4 4 1,clip]{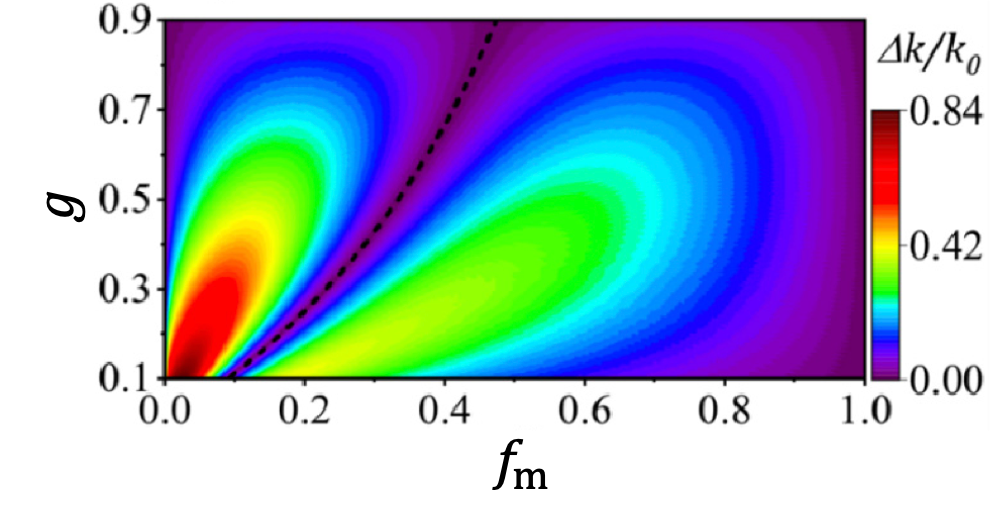}}
\caption{The variation of the second momentum bandgap size $\Delta k$ of the PTC as a function of the parameter $g$ and the modulation frequency $f_\mathrm{m}=\omega_\mathrm{m}/2\pi$. Here, the dotted black line corresponds to the condition of topological phase transition as the gap size vanishes, i.e., $\Delta k \rightarrow 0$.
Figure was reprinted from~\cite{ma2019band}.}
\label{fig:topological2}
\end{figure}

\begin{equation}
\theta^\mathrm{Zak}_p=\int_{-\pi/T_{\rm m}}^{\pi/T_{\rm m}}d\omegaf\left[j\int_0^{T_{\rm m}}dt\,\varepsilon(t)D^*_{p,\omegaf}(t)\frac{\partial{D_{p,\omegaf}(t)}}{\partial{\omegaf}}\right]\,. \label{eq:zak}
\end{equation}
Here, $\varepsilon(t)$ is the time-periodic permittivity of the PTC, $p$ is the band index (lowest band index being $p=0$), and $D_{p,\omegaf}(t)$ is the displacement field of the $p$-th band of the PTC at the Floquet frequency $\omegaf$. Note that the quantity $\theta^\mathrm{Zak}_p$ in Eq.~\ref{eq:zak} is the adaptation of the spatial Zak phase to the temporally periodic systems \cite{lustig2018topological}.  %Note that the quantities analogous to the Zak phase in the 2D and 3D spatially periodic systems are known as the Berry phase \cite{hassani2017berry}.
Using Eq.~\eqref{eq:zak}, we can calculate the Zak phases of the bands of the PTC with the same material modulation as shown in Fig.~\ref{Fig:photonic_space_time}(b). The Zak phases of the bands of the considered PTC are shown beside the corresponding bands in Fig.~\ref{fig:topological1}(a). 

The Zak phase $\theta^\mathrm{Zak}_p$ dictates the sign of the relative phase between the reflected and transmitted waves for an incident excitation that falls inside the momentum bandgap of the considered PTC~\cite{lustig2018topological}. For this purpose, consider a temporal slab made from a finite number of cycles of a stepwise modulation (see Fig.~\ref{fig:temporal_slabs}(a)). Further, let us assume an incident excitation to the temporal slab to be $E_{\rm inc}=E_0e^{i{(\omega_{\rm inc}t-k_\mathrm{inc}z)}}$. Furthermore, let us assume that $k_\mathrm{inc}$ falls within one of the momentum bandgaps of the PTC formed by the infinite periodic extension of the considered stepwise permittivity profile shown in Fig.~\ref{fig:temporal_slabs}(a). Further, from Eq.~\eqref{eq:tr_mat2}, we know that a plane wave incident to the temporal slab gives rise to reflected and transmitted plane waves. Let $r$ and $t$ to be the complex reflection and transmission coefficients of the reflected and transmitted plane waves, respectively. Moreover, let the phase difference between $r$ and $t$ to be $\phi_{s}$, i.e., $\angle r-\angle t=\phi_{s}$. Then the sign of $\phi_{s}$ is dictated by the Zak phase $\theta^\mathrm{Zak}_p$ as \cite{lustig2018topological}
\begin{equation}
\mathrm{sgn}(\phi_{s})=\eta\,(-1)^{l+s}\,\mathrm{exp}\left(j\sum_{p=1}^{s-1}\theta^\mathrm{Zak}_p\right)\,. \label{eq:zakgap}
\end{equation}
Here, $s$ is the gap number (lowest gap number being $1$), $\eta=\mathrm{sgn}\left(1-\varepsilon_1/\varepsilon_2\right)$, and $l$ is the number of bands below gap $s$ (see Fig.~\ref{fig:topological1}(a)).

Next, the authors of \cite{lustig2018topological} studied the topological edge state of the system that consists of two different PTCs with different topology occurring one after the other (see Fig.~\ref{fig:topological1}(b)). Note that the first PTC has $\varepsilon_1=3$ for time $t_1=0.5T_\mathrm{m}$ and $\varepsilon_2=1$ for time $t_2=0.5T_\mathrm{m}$. On the other hand, the second PTC has $\varepsilon_1=1$ for time $t_1=0.5T_\mathrm{m}$ and $\varepsilon_2=3$ for time $t_2=0.5T_\mathrm{m}$. Further, the interface of the two PTCs occurs at the time $t_\mathrm{edge}=8T$. Clearly, the two PTCs shown in Fig.~\ref{fig:topological1}(b) have the same bandgaps but different topologies. For such a system, \cite{lustig2018topological} report a topological edge state as an eigenstate. The corresponding edge state is shown in Fig.~\ref{fig:topological1}(c). From Fig.~\ref{fig:topological1}(c), we observe that for the topological edge state, the field $|\mathbf{D}|$ increases exponentially up to time $t=t_\mathrm{edge}$ followed by an exponential decay afterwards. Further, after the field amplitude has decayed it starts increasing again as a function of time. Note that such a topological edge state is robust with respect to the defects and impurities in the underlying time-varying system.

Furthermore, the authors of Ref.~\cite{ma2019band} discuss the conditions for the topological phase transitions in PTCs. Let us again consider a PTC made from the stepwise modulations of the permittivity as shown in Fig.~\ref{Fig:photonic_space_time}(a). Let us assume that $\varepsilon_2=4$ and $\varepsilon_{1}=g\varepsilon_{2}$, where $g$ characterizes the modulation strength of the PTC. The size $\Delta k$  of the second momentum bandgap of the PTC (lying between the bands $p=1$ and $p=2$) as a function of $g$ and the modulation frequency $f_\mathrm{m}=\omega_\mathrm{m}/2\pi$ is plotted in Fig.~\ref{fig:topological2} \cite{ma2019band}. From Fig.~\ref{fig:topological2}, we find that when $g=f_\mathrm{m}/(1-f_\mathrm{m})$ (dashed curve), the gap size $\Delta k$ is zero. Therefore, the topological phase transition occurs whenever the condition $g=f_\mathrm{m}/(1-f_\mathrm{m})$ is satisfied. This implies that the bands $p=1$ and $p=2$ have different topological phases $\theta^\mathrm{Zak}_p$ on either side of the dashed line. As discussed earlier, \cite{ma2019band} suggests that such a topological phase transition can be probed by measuring the phase difference $\phi_{s}$ between the complex reflection and transmission coefficients of the temporal slabs formed by truncating the modulation of the PTC in time (see Fig.~\ref{Fig:finite_slab}). In contrast to \cite{lustig2018topological,ma2019band}, the authors of \cite{dong2023band} study the topological phases (see Eq.~\eqref{eq:zak})  of the PTCs with continuous profiles of $\varepsilon(t)$ such as sinusoidal and exponential modulation. Moreover, \cite{Lin2024temporally} discusses the effects of temporal defects on the topological features of the PTCs.

\begin{figure*}[tb]
\centerline{\includegraphics[width= 0.8
\columnwidth,trim=4 4 4 4,clip]{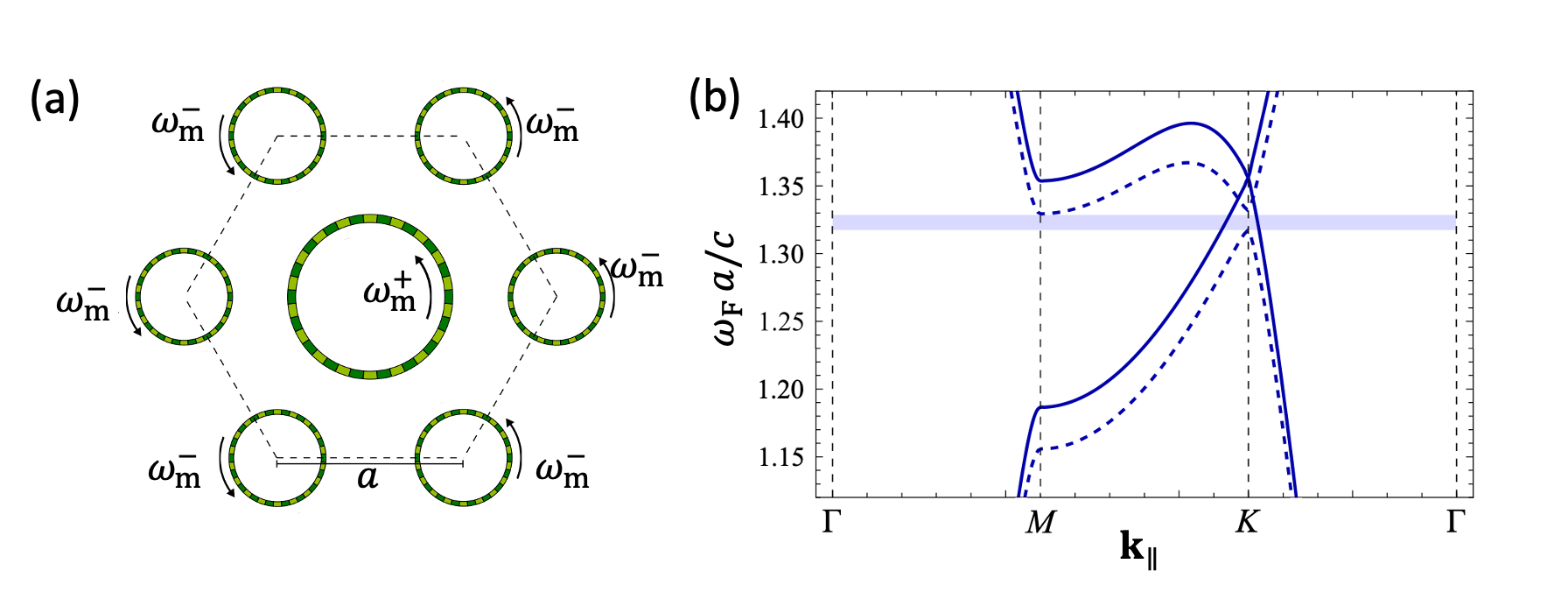}}
\caption{(a) The unit cell of an ST-PC formed by cylindrical ring resonators. b) The band structure of the ST-PC shown in (a) when the spatiotemporal modulation of the ring resonators is turned off, i.e., $\omega_\mathrm{m}^{\pm}=0$ (solid blue lines), and when the spatiotemporal modulation of the ring resonator at the center of the unit cell is turned on, i.e., $\omega_\mathrm{m}^{+}\neq0$, $\omega_\mathrm{m}^{-}=0$ (dotted blue lines). Figures (a) and (b) were reprinted from~\cite{serra2023rotating}.}
\label{fig:topological3}
\end{figure*}

Similar to PTCs, the topological effects in ST-PCs have also been studied. The authors of \cite{serra2023rotating} study the ST-PCs made from the inclusions that are subjected to rotating wave-modulation (see Fig.~\ref{fig:topological3}(a). The unit cell of the ST-PC consists of cylindrical ring resonators. Here, each resonator is subject to a spatiotemporal modulation of the permittivity and permeability given by $\varepsilon=\varepsilon(\phi-\omega_\mathrm{m} t)$ and $\mu=\mu(\phi-\omega_\mathrm{m} t)$, respectively. Here, $\phi$ is the azimuthal angle in the cylindrical coordinates and $\omega_\mathrm{m}$ is the angular frequency of the rotating spatiotemporal modulation. For such spatiotemporal crystals, the relevant topological invariant to study is the gap Chern number \cite{lu2014topological,serra2023rotating}. Note that, here, the term ``gap'' refers to the energy bandgaps arising in the ST-PCs. The gap Chern number for a specific bandgap of the considered ST-PCs can be defined as \cite{serra2023rotating}
\begin{equation}
C_\mathrm{gap}=\frac{1}{2\pi}\iint_\mathrm{BZ}F_\mathbf{k}\mathrm{d}^2\mathbf{k}\,. \label{eq:chern}
\end{equation}
Here, $\mathrm{BZ}$ refers to the spatial Brillouin zone of ST-PC, and $F_\mathbf{k}$ corresponds to the Berry curvature associated with the bandgap. Note that $F_\mathbf{k}$ can be computed using the Green's function of the ST-PC (see \cite{serra2023rotating} for more details). Next, to utilize the aforementioned gap Chern number, the band structure of the ST-PC is plotted (see Fig.~\ref{fig:topological3}(b)). In Fig.~\ref{fig:topological3}(b), the solid blue lines correspond to the case when the spatiotemporal modulations of the ring resonators are turned off (i.e., $\omega_\mathrm{m}^{\pm}=0$). On the other hand, the dotted blue lines correspond to the case when the spatiotemporal modulation of the ring resonators at the center of the unit cell of the ST-PC is turned on (i.e., $\omega_\mathrm{m}^{+}\neq 0, \omega_\mathrm{m}^{-}=0$). From Fig.~\ref{fig:topological3}(b), we observe that the spatiotemporal modulation opens an energy bandgap. The gap Chern number of the bandgap is then calculated using Eq.~\eqref{eq:chern}. The value of the gap Chern number turns out to be $C_\mathrm{gap}=-1$. Furthermore, it was shown in~\cite{serra2023rotating} that upon manipulating the relative signs of $\omega_\mathrm{m}^{+}$ and $\omega_\mathrm{m}^{-}$, one can engineer the sign of $C_\mathrm{gap}$. 

Further, in \cite{serra2023engineering}, the authors engineer non-trivial topological phases in ST-PCs. They show the emergence of scattering immune topological edge states in such ST-PCs.

\subsection{Nonlinear ST-PCs \label{NLSTPC}}
\begin{figure*}[h]
\centerline{\includegraphics[width= 0.8\columnwidth,trim=4 4 4 4,clip]{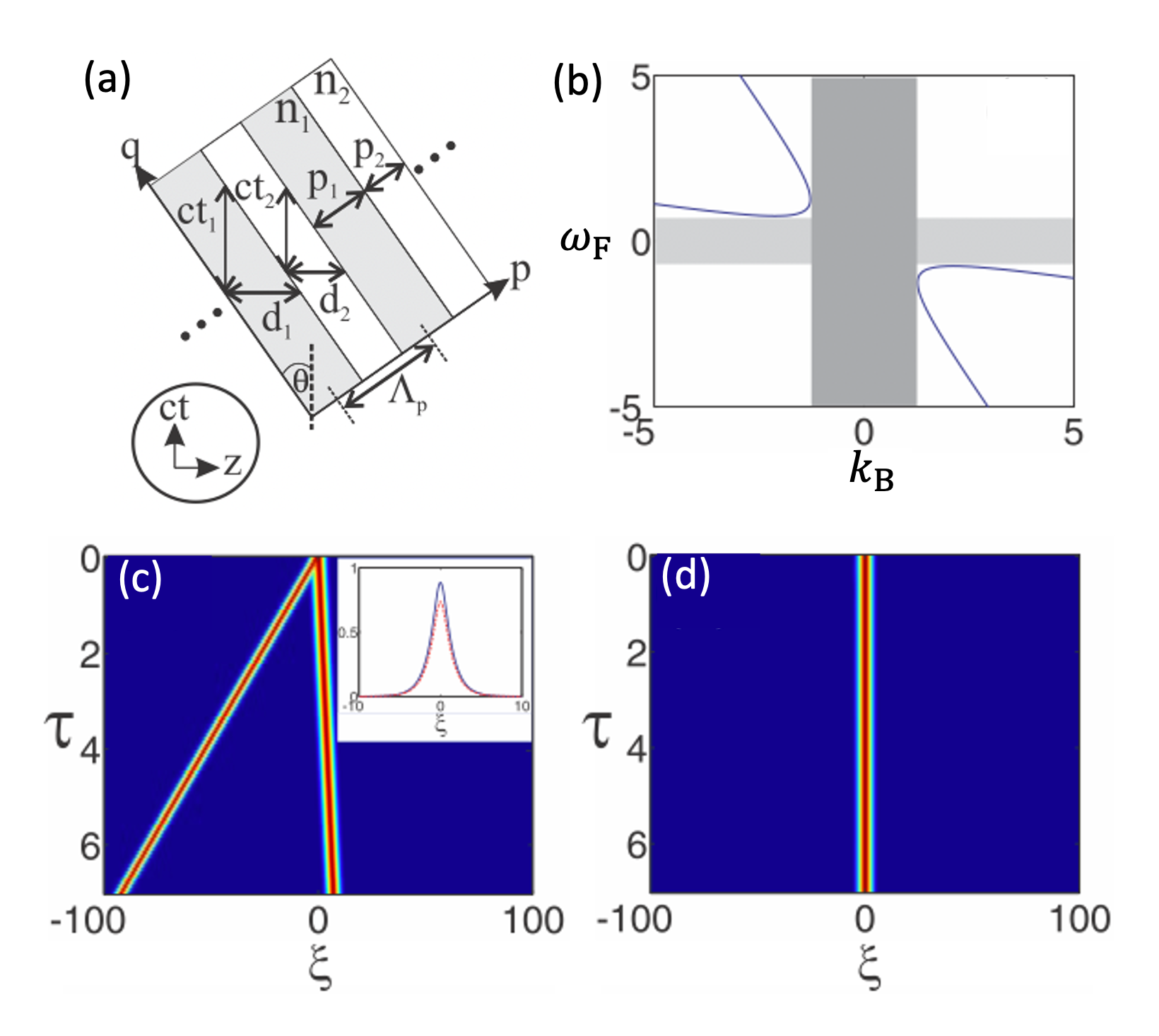}}
\caption{
(a) An ST-PC made from a nonlinear material. (b) The band structure of the nonlinear ST-PC. (c) Contour plot of the total intensity of an initial gap soliton as it propagates in the absence of the ST-PC (the inset shows the forward and backward intensity profiles of the input with blue and red curves, respectively. (d) The corresponding contour plot of the total intensity in the presence of the ST-PC. Here, $(\tau,\xi)$ quantifies the propagation length and localization of the gap soliton computed with respect to the rotated coordinate system $(p,q)$ shown in (a), respectively.
Figures were reprinted from~\cite{Biancalana2008gap2}.}
\label{fig:gapsolitons}
\end{figure*}

%In this subsection, nonlinear ST-PCs are described. In particular, emerging gap soliton solutions are discussed.\\
%References: \cite{biancalana2008gap}, \cite{Biancalana2008gap2}.
In addition to the linear ST-PC, the ST-PCs made from nonlinear media have also been studied. In particular, in Ref.~\cite{Biancalana2008gap2},  the emergence of gap soliton solutions of the underlying nonlinear wave equation satisfied by such nonlinear ST-PCs was demonstrated. The considered system is assumed to have a space-time-dependent linear refractive index $n(z,t)$ (see Fig.~\ref{fig:gapsolitons}(a)). Therefore, the linear part of the electric polarization is written as ${P}_\mathrm{L}=[n^2(z,t)-1]{E}$. Further, the system is assumed to have a third-order nonlinearity such that the nonlinear electric polarization is written as ${P}_\mathrm{\rm NL}=\chi_{\rm NL}E^3$. Here, $\chi_{\rm NL}$, is the third-order nonlinear susceptibility of the considered system. Such systems support mixed energy-momentum bandgaps (see Fig.~\ref{fig:gapsolitons}(b)). Note that the band structure shown in Fig.~\ref{fig:gapsolitons}(b) exhibits a bandgap with respect to both Floquet frequency $\omega_\mathrm{F}$ (energy bandgap) and Bloch wavenumber $k_\mathrm{B}$ (momentum bandgap). Moreover, the soliton solutions found inside such mixed bandgaps are termed spatiotemporal gap solitons.

The applicability of the developed method to find the spatiotemporal gap soliton solutions in \cite{Biancalana2008gap2} is shown in Figs.~\ref{fig:gapsolitons}(c)--(d). The authors of \cite{Biancalana2008gap2} used the rotated coordinate system $(p,q)$ to compute the soliton solutions (see Fig.~\ref{fig:gapsolitons}(a)). Note that in Figs.~\ref{fig:gapsolitons}(c)--(d), $\tau$ quantifies the propagation length of the soliton as it travels. On the other hand, $\xi$ quantifies the localization of the soliton solution (see the inset of Fig.~\ref{fig:gapsolitons}(c)). In Fig.~\ref{fig:gapsolitons}(c), the dynamics of the intensity of an initial gap soliton in the absence of the ST-PC are shown. Here, the inset of Fig.~\ref{fig:gapsolitons}(c) shows the input profile of the initial soliton. Note that the blue (red) curve denotes the forward (backward) propagating envelope of the electric fields of the soliton. From Fig.~\ref{fig:gapsolitons}(c), we observe that in the absence of the ST-PC, the forward and backward components of the initial soliton separate and do not interact as they propagate. Hence, expectedly, the initial soliton does not preserve its localization property. On the other hand, Fig.~\ref{fig:gapsolitons}(d) demonstrates such intensity dynamics for the case when the soliton is coupled to the ST-PC. From Fig.~\ref{fig:gapsolitons}(d), the undisturbed soliton dynamics is observed. Therefore, the initial soliton stays well localized as it propagates inside the ST-PC. Such undisturbed dynamics convincingly show the robustness of the proposed method to find the soliton solutions inside the mixed bandgaps of the ST-PCs \cite{Biancalana2008gap2}.

%

% \subsection{Optical Isolator Based on ST-PCs}
% Here, we will discuss how the non-reciprocity of a spatiotemporal crystal can be used to design an optical isolator.\\
% References: 

% \section{Unsorted papers}
% \red{Maybe papers by Francesco Monticone related to causality issue in time-varying systems. Also the group of Davide Ramaccia.  Check all other groups from Metamaterials who could have related works: maybe Paloma, Capolino, etc.}

\section{Future outlook }
%\red{If safe, mention current Sajjad's works}

After elaborating on all the work done in the context of PTCs and their foundational principles, we summarize in this short chapter aspects that we, as an entire community, shall address in the future to develop the topic of PTCs further. With the purpose of avoiding wrong impressions, we wish to fully acknowledge that much of the research has been driven so far by theoretical and computational studies. At the same time, the first experiments that address specific aspects of PTCs were reported, while many have yet to be done. Therefore, the outlook is written in a humble manner in that we do not wish to suggest that PTCs will change the world. Still, working on PTCs is fascinating and it is interesting to disclose new physical effects in these materials. 

First, we shall continue to explore the degrees of freedom PTCs offer us to control light propagation. We shall do so out of intellectual curiosity to shift the boundaries that define our current understanding of the world surrounding us. But of course, also with an eye on possible applications and devices that exploit these phenomena in their design. When looking back in time, we can rationalize how novel research themes in the context of nanophotonics were established. Most new trends emerged by writing down the constitutive relations and asking us how these constitutive relations can be modified to enlarge the space of possible effects. One line of future developments in the field of PTCs signifies the transition from isotropic to anisotropic materials, further towards bianisotropic~\cite{mirmoosaBTI24,OMEPaper,koufidis_light_2024} and eventually nonlocal~\cite{rizza_nonlocal_2022} and magnetic materials~\cite{shaposhnikov_probing_2024}. In a generalization, higher-order nonlocal materials can be explored. Also, assuming materials to be periodic in space gave significant momentum to many sub-fields of optics by considering them as photonic crystals. Adding a time variation to those material properties, generally to all, adds many novel opportunities. While most of these previously considered extensions concentrated on the spatial propagation characteristics of light propagation, a time variation adds control over the spectral composition of the field to the portfolio. Compared to nonlinear optical effects that provide similar capabilities, the advantage of a time modulation would be much more rational control of the spectral composition. The spectral content of the light is controlled deterministically by changing the waveform and the period of the time modulation on demand. Besides exploring these aspects, we would like to motivate the wider community to explore other possible extensions in the constitutive relations. 

When thinking about more practical aspects, the most urgent work that needs to be accomplished is the establishment of an ordinary PTC characterized by a time-dependent permittivity at optical frequencies. But generally, it would be great to explore more experimental systems that permit the observation of effects offered by PTCs. This holds for experimental platforms operating at different frequency domains and for experimental platforms that allow us to access wave phenomena outside that of electromagnetics and optics. Acoustics or fluid dynamics would be two examples. Also, unconventional material platforms could be explored. For example, scattering at high index spheres at extremely low frequencies was experimentally explored using voids filled with water, with water having a permittivity as high as 80 up to 10 GHz~\cite{jacobsen2021water}. The system invites exploration because voids can be created in a stretchable rubber material, similar to balloons filled with water. The time-variation of such systems might be feasible with some mechanical efforts. It would constitute an excellent implementation for some of the effects described in Section~\ref{Sec:ExploitinResonantSTPC}. 

However, the biggest challenge remains to establish reliable material platforms that allow us to observe the described effects at optical frequencies. We need to push the range of frequencies where PTCs were demonstrated toward the visible or at least toward the infrared domain, where many applications that would benefit from a deterministic control over the spectral control of light exist. The first steps along these lines were done by all-optical fast tuning of material properties, but many open questions remain. For example, it is not just the switching from one material property to another that matters, which permits the observation of time reflection and time refraction. It is the periodic modulation between two states of the material on time scales comparable to the oscillation period of light that needs to be accomplished. After the switching, in most cases, thermal or electronic processes are responsible for driving the material back to its original state. These are slow processes that need to be accelerated. Therefore, we need to find ways to change material properties on optical time scales between two states that, hopefully, differ substantially. And even if the change in the material properties is only modest, we can exploit our understanding of how to enhance the light-matter interaction thanks to a suitable structured spatial environment (see Section~\ref{Sec:ExploitinResonantSTPC}), by choosing suitable frequency domains, or a combination thereof. Spatiotemporal photonic crystals could be one solution~\cite{wang2023unleashing}. Also, operating in the epsilon-near-zero domain was already exploited~\cite{hayran2022ℏomega}. But other schemes continue to be uncharted, for example, operating in an integrated photonic system close to the cut-off frequency of a guided mode \cite{edwards2008experimental} or exploiting propagating surface plasmon polaritons~\cite{li2019resonance}. A part of these explorations should also concern the delineation from nonlinear optical effects that have already been discussed in the past~\cite{khurgin2023photonic}. While in some specific situations, the nonlinear polarization can be written so that it appears to be a time-varying material property, equating both effects under all circumstances would be misleading. Therefore, elaborating on the differences and exploring the unique aspects would be crucial.

Furthermore, it is reasonable to anticipate additional discoveries of new light-matter interaction phenomena in PTCs. In just the past five years, a wealth of significant fundamental studies on PTCs has surfaced. These studies encompass a range of phenomena, including among others the amplification of spontaneous emission from excited atoms, subluminal Cherenkov radiation, superluminal momentum-gap solitons, and temporal Anderson localization. Much like traditional photonic crystals have played a pivotal role in quantum optics, PTCs hold the potential to achieve similar or even greater significance due to their capacity for extreme light-matter interactions~\cite{quantum2023ETI}. However, realizing this potential necessitates PTCs to operate within the optical domain close to electronic transitions in solids. But, certainly, realizing PTCs in the infrared part of the spectrum or THz frequencies would be also exciting and important. There, vibrational and rotational excitations in molecules and phonons in solids do play a role.
 
With time passing and progress being made concerning the different material platforms with which we can realize PTCs operating at different frequencies, we can ask further questions. For example, how to actually design and later realize PTCs so that they offer predefined optical functionalities on demand, an emphasis should be here on spectral control and field amplification. This is the vast field of inverse design, where major developments are witnessed in nearly all fields of science. Most notably, the notion of differential programming attracted more and more attention. It would allow us to use gradient-based optimization to design functional devices easily. Therefore, when setting up computational tools to describe PTCs, this aspect should be considered from the beginning in implementing computational routines. Of course, also techniques from the field of machine learning and artificial intelligence shall be exploited. However, the benefit of such techniques compared to traditional approaches needs to be demonstrated. 

Moreover, emphasis shall be put on exploring possible novel effects of PTCs on theoretical, computational, and experimental grounds. A prime example could be in the context of synthetic dimensions~\cite{yuan2018synthetic}. Synthetic dimensions in photonics involve creating additional degrees of freedom for light propagation beyond the conventional three spatial dimensions. PTCs offer unique possibilities by exploiting the frequency as a synthetic dimension, where the temporal periodicity of PTCs allows us to address it in a highly efficient manner, possibly also driven by some of the inverse design techniques just described. This can enable studies of high-dimensional topological phenomena and complex light-matter interactions. Also, the temporal modulation in PTCs provides a means to dynamically control the properties of these synthetic dimensions so that topological properties can be adjusted. By exploiting the synthetic dimensions, PTCs can enhance light-matter interactions, enabling efficient photon-photon interactions and nonlinear processes that are meaningful for a future quantum information processing architecture. With that, PTCs can simulate higher-dimensional physical or quantum systems that are challenging to study in conventional settings, providing a platform for exploring new physics.

Finally, working towards more compact, ideally fully integrated PTCs would be fantastic. Currently, the experimental schemes are great to provide proof for the principles, but they are unlikely to constitute a base for future applications in, e.g., wireless communication systems. That domain could especially benefit from the possibility of transducing information among different frequencies and amplifying weak signals in unconventional manners. However, bulky experimental schemes are unlikely to be attractive, and fully integrated schemes would be necessary to make practical use of PTCs in the long run. However, before that level is reached, many more questions need to be answered, and we hope that some of the readers of this tutorial will be among those who deliver the answers.

\section{Concluding remarks}
In a short conclusion, we wish to wrap up our tutorial on PTCs. We started by appreciating that a PTC consists of a spatially homogeneous material whose properties periodically change in time. While there is no clear delineation, it is assumed that the oscillation period of the property compares to the oscillation period of the probe light. Or at least, being comparable in the order of magnitude allows us to observe many of the effects usually attributed to PTCs. While permittivity is a common property to modulate in these materials, other material properties can also be varied over time to achieve similar phenomena. When implemented in different physical systems, it would be a parameter describing the system that appears in the governing equations, which is on equal footing as the permittivity for optical phenomena. Motivated by experiments, we gave examples such as transmission lines, where the capacitance of the respective LC circuit was time-varying. Then, the description of how voltage waves propagate along the transmission line is the same as that of a wave equation in a homogeneous medium, and the time-varying capacitance emerges instead of the time-varying permittivity.
 
To explore the fundamental properties of PTCs, we outlined two computational techniques that can also be used for analytical explorations. On the one hand, we showed how to find elementary solutions to Maxwell’s equations in reciprocal space, both in space and time. On the other hand, an ABCD transfer-matrix technique can be used for similar purposes. Both techniques are beneficial for specific modulation profiles. The former is for sinusoidal modulations, while the latter is for modulations where the material property jumps between discrete values. Both methods make the same predictions, and it is instead a question of convenience which method to choose. The techniques can be used to explore key features of PTCs.

On the one hand, PTCs sustain momentum gaps for a sufficiently strong modulation. It implies that for a given frequency, no propagating wave exists for a specific range of momenta. Inside the momentum gaps, two inhomogeneous solutions exist to the wave equation. One is exponentially decaying in time, while the other is exponentially growing. Next, we have explored multiple aspects of PTCs that are relevant for realistic systems, leading also to a more complex description. Dispersion, the finiteness in space and time, an anisotropy, possibly nonlinearities, or deviations from the perfect time-harmonic modulation of the material properties were discussed. 

In the following chapters, we elaborated on the possibilities of implementing PTCs. First, we distinguished PTCs from other physical systems with similar effects, especially specific nonlinear processes. The different material platforms that were discussed provide access to PTCs in different spectral domains. Finally, we elaborated on the effects of light-matter interaction and possible applications of PTCs. We emphasized applications that exploit the critical properties of PTCs but also other aspects that look appealing.

Considering that many of the fundamental effects in PTCs were just explored, we expect many more applications to emerge shortly. However, we emphasize that the field of PTCs itself is relatively nascent, and many primary effects still await demonstration. But independent of these future achievements, we hope to have convinced the reader in the tutorial that the topic is fascinating and full of potential. The motivation to explore PTCs lies in their unique manipulation of light in time and, with that, also in the frequency domain, offering a novel playground for exploring fundamental physics and pushing the boundaries of optical technologies. Though in its infancy, this field holds the promise of revolutionary applications, ranging from advanced communication systems to groundbreaking quantum computing platforms.

The potential to control and manipulate light in intricate ways opens doors to uncharted territories in photonics. We envision that the continued research in PTCs will lead to the development of more efficient, faster, and compact photonic devices, which could transform the landscape of technology and industry. Furthermore, the interplay of PTCs with nonlinear optics and quantum phenomena presents a rich vein of research that could yield unprecedented insights into the nature of light-matter interactions.

In conclusion, the journey into the realm of PTCs is not just a pursuit of practical applications but a venture into the depths of scientific curiosity and innovation. It is a field where each discovery paves the way for new questions and deeper understanding, inviting researchers to continually push the frontiers of what is possible.

\begin{backmatter}
\bmsection{Funding}
P.G. and C.R. gratefully acknowledge financial support by the Deutsche Forschungsgemeinschaft (DFG, German Research
Foundation) through Project-ID No. 258734477 - SFB 1173.
P.G. and C.R. are part of the Max Planck School of Photonics, supported by the Bundesministerium für Bildung und Forschung, the Max Planck Society, and the Fraunhofer Society. P.G. acknowledges support from the Karlsruhe School of Optics and Photonics (KSOP). 
C.R. acknowledges support by the Deutsche Forschungsgemeinschaft (DFG, German Research Foundation) under Germany’s Excellence Strategy via the Excellence Cluster 3D Matter Made to Order (EXC-2082/1-390761711) and from the Carl Zeiss Foundation via the CZF-Focus@HEiKA Program.
X.W. and C.R. acknowledge support by the Helmholtz Association via the
Helmholtz program “Materials Systems Engineering” (MSE). 
M.M.A. and V.A. acknowledge the Academy of Finland (Project No. 356797), the Finnish Foundation for Technology Promotion, and Research Council of Finland Flagship Programme, Photonics Research and Innovation (PREIN), decision number 346529, Aalto University. 

\bmsection{Acknowledgments}
The authors would like to thank Mr.~Bahman Amrahi for the fruitful discussions about synthetic dimensions. The authors would like to thank all past and current co-workers in their groups that have contributed to the discussion and the understanding of photonic time crystals. Particularly, we would like to thank Grigorii Ptitcyn, Theodosios D. Karamanos, Aristeidis Lamprianidis, Sergei Tretyakov, Shanhui Fan, and Mohamed Mostafa.

\bmsection{Disclosures}
The authors declare no conflicts of interest.

\bmsection{Data availability} No data were generated or analyzed in the presented research.

\bigskip

% \bmsection{Supplemental document}
% See Supplement 1 for supporting content. 

\end{backmatter}

%%%%%%%%%%%%%%%%%%%%%%% References %%%%%%%%%%%%%%%%%%%%%%%%%
% \newpage
% \renewcommand{\notesname}{Endnotes}
% \theendnotes
%%%%%%%%%% If using BibTeX:
%\bibliographystyle{plainnat} 
\bibliography{references}
%\nocite{*}
%\printbibliography
%\bibliographystyle{plain}
%\bibliographystyle{plainnat}
% Biography section
\section*{Author Biographies}

\begin{biography}[{\includegraphics[width=1in,height=1.25in,clip,keepaspectratio]{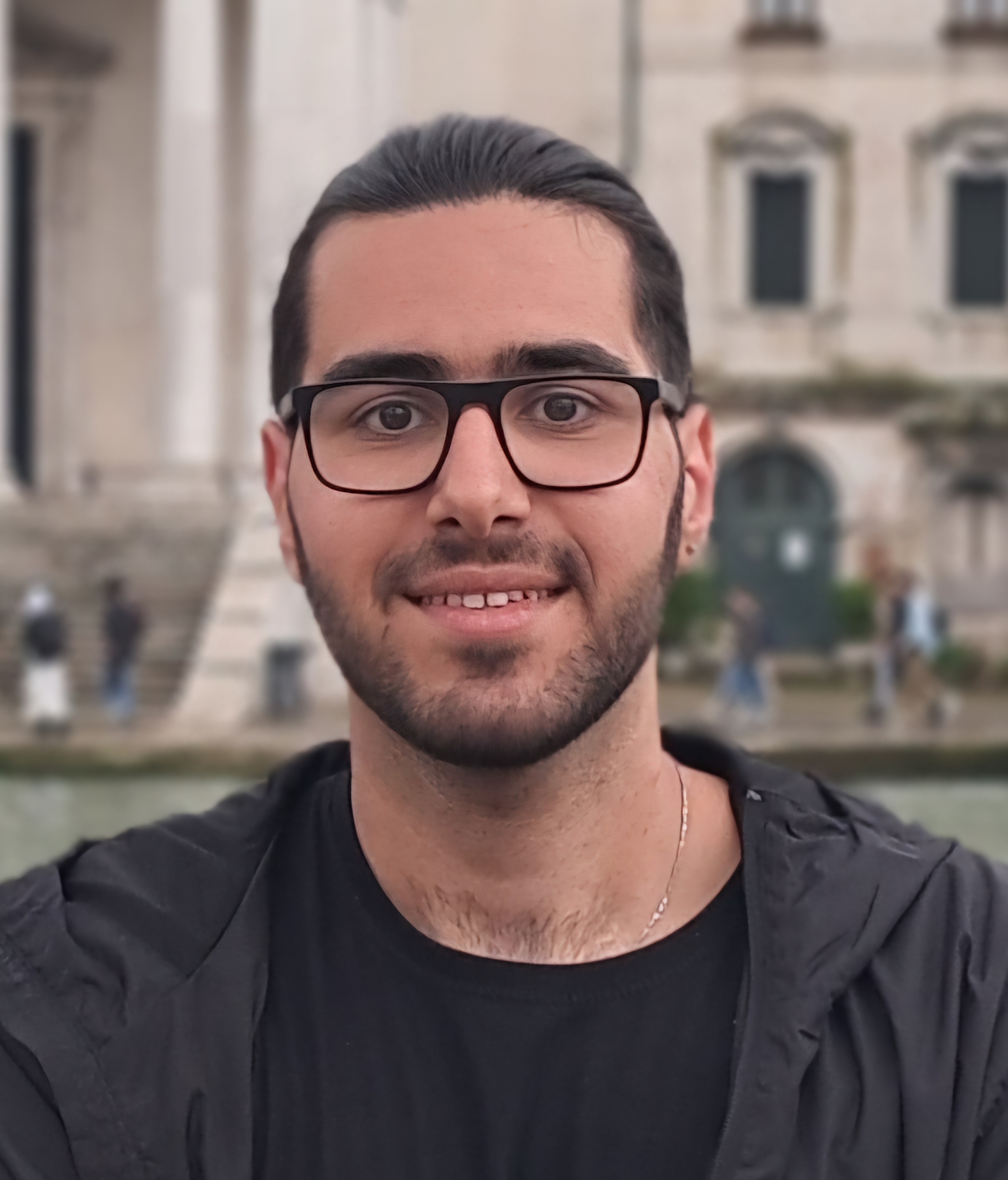}}]{Mohammad Mahdi Asgari}
    Mohammad Mahdi Asgari received his B.Sc. degree in Electrical Engineering (Telecommunications) from Babol Noshirvani University of Technology, Mazandaran, Iran in 2018, and his M.Sc. in Electrical Engineering (Field and Waves) from the Sharif University of Technology, Tehran, Iran in 2021. He has been a doctoral researcher in the Department of Electronics and Nanoengineering at Aalto University, Espoo, Finland, since 2022. His main research topics are metamaterials, time-varying systems, and inverse design.
\end{biography}

\begin{biography}[{\includegraphics[width=1in,height=1.25in,clip,keepaspectratio]{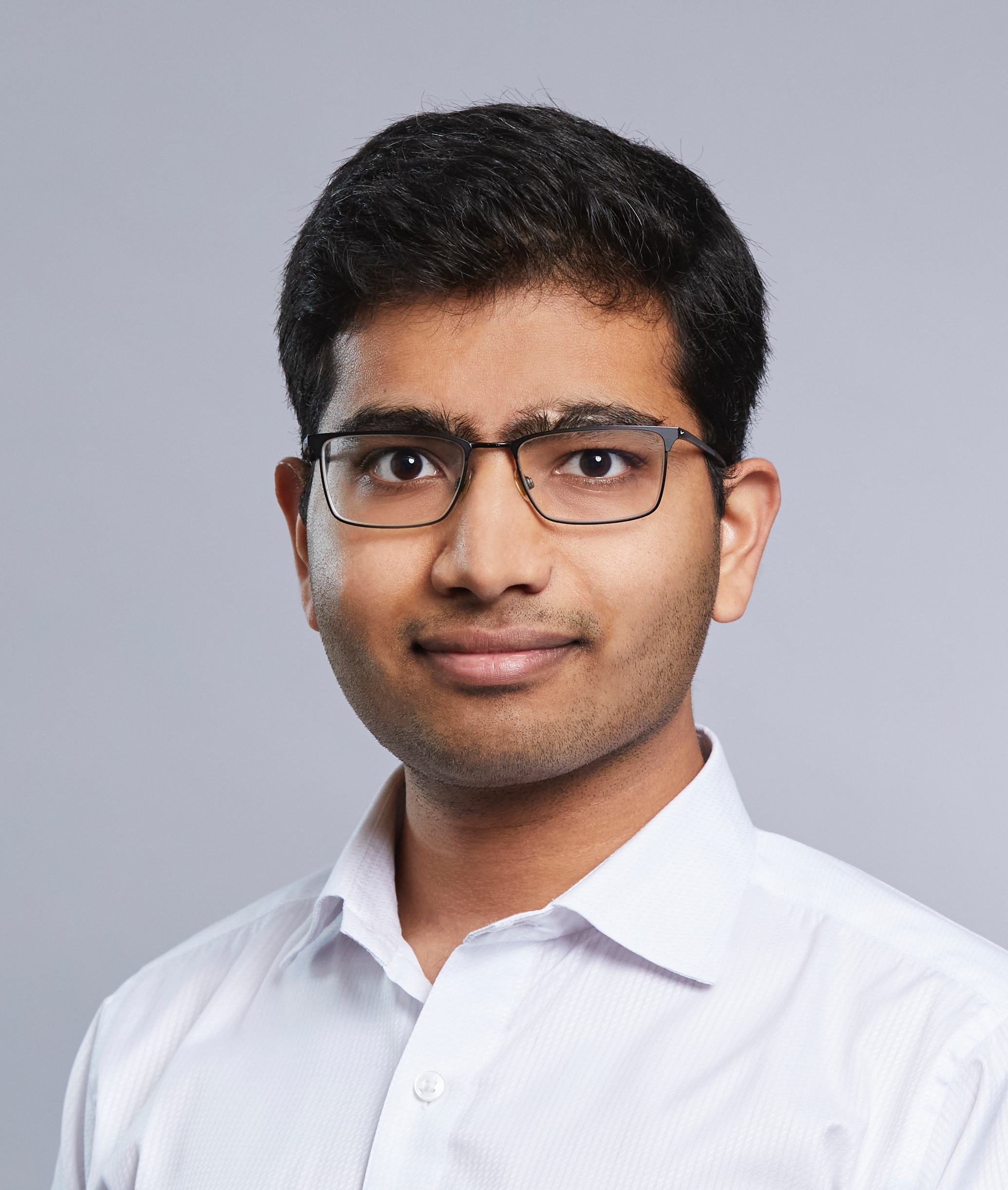}}]{Puneet Garg}
  Puneet Garg received his Bachelor’s degree in Physics (Honours) from St. Stephen’s College, India, in 2020. He then moved to Karlsruhe Institute of Technology (KIT), Germany, to pursue a Master’s degree in Optics and Photonics, which he received in 2022. Currently, he works as a doctoral researcher at the Institute of Theoretical Solid State Physics at KIT. His research interests include the theoretical and numerical investigation of time-varying metamaterials.
\end{biography}

\begin{biography}[{\includegraphics[width=1in,height=1.25in,clip,keepaspectratio]{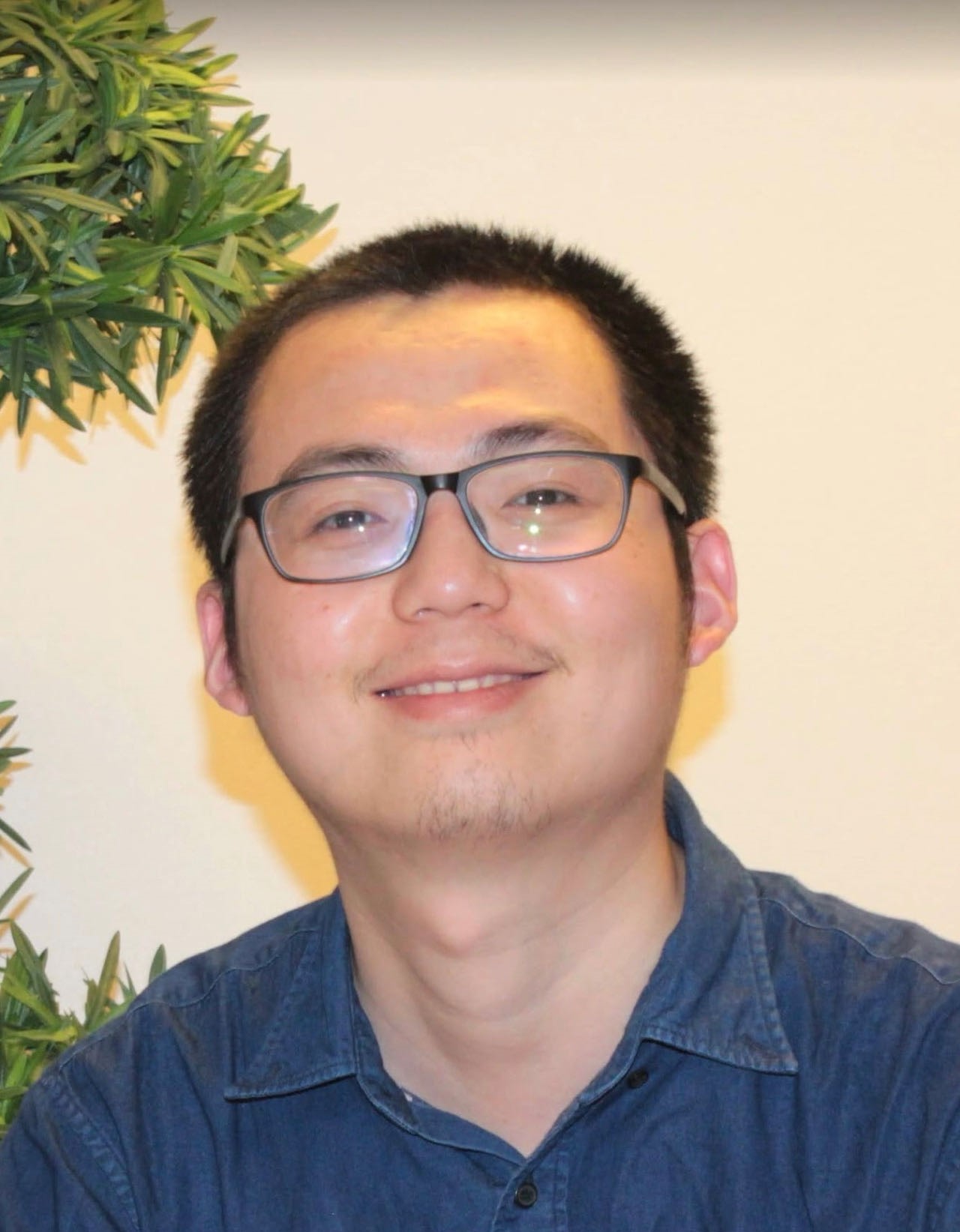}}]{Xuchen Wang}
    Xuchen Wang received a B.Sc. degree in optical information science and technology from Northwestern Polytechnical University, Xi’an, China, in 2011, a master’s degree from the Department of Optical Engineering, Zhejiang University, Hangzhou, China, in 2014, and a Ph.D. degree (Hons.) from the Department of Electronics and Nanoengineering, School of Electrical Engineering, Aalto University, Aalto, Finland, in 2020. He worked as a Radio Frequency Engineer at Huawei (Shanghai, China), and TP-Link (Shenzhen, China) from 2014 to 2016. He worked as a Post-Doctoral Researcher at Karlsruhe Institute of Technology, Germany from 2022 to 2023. He is currently working as a Professor in the Harbin Engineering University, China.
\end{biography}

\begin{biography}[{\includegraphics[width=1in,height=1.25in,clip,keepaspectratio]{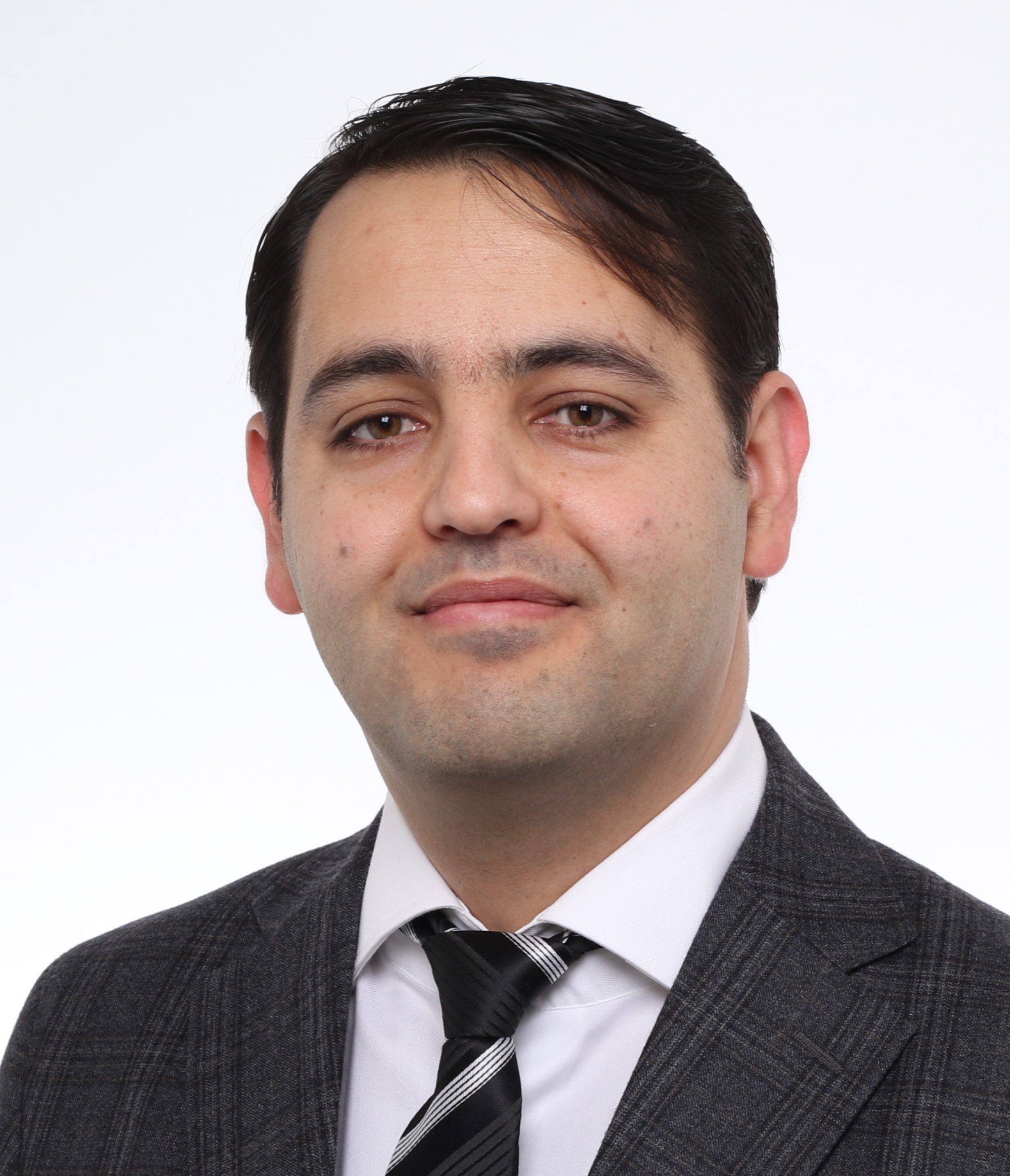}}]{Mohammad Sajjad Mirmoosa}
 Mohammad Sajjad Mirmoosa  received a B.Sc. degree from the Shahid Bahonar University of Kerman, Kerman, Iran, in 2011, and the M.Sc. and D.Sc. degrees from Aalto University, Aalto, Finland, in 2013 and 2017, respectively, all in Electrical Engineering. He is currently with the Department of Physics and Mathematics, University of Eastern Finland, as a Project Researcher. His main research interests include theories of classical and quantum electromagnetism. 
\end{biography}

\begin{biography}[{\includegraphics[width=1in,height=1.25in,clip,keepaspectratio]{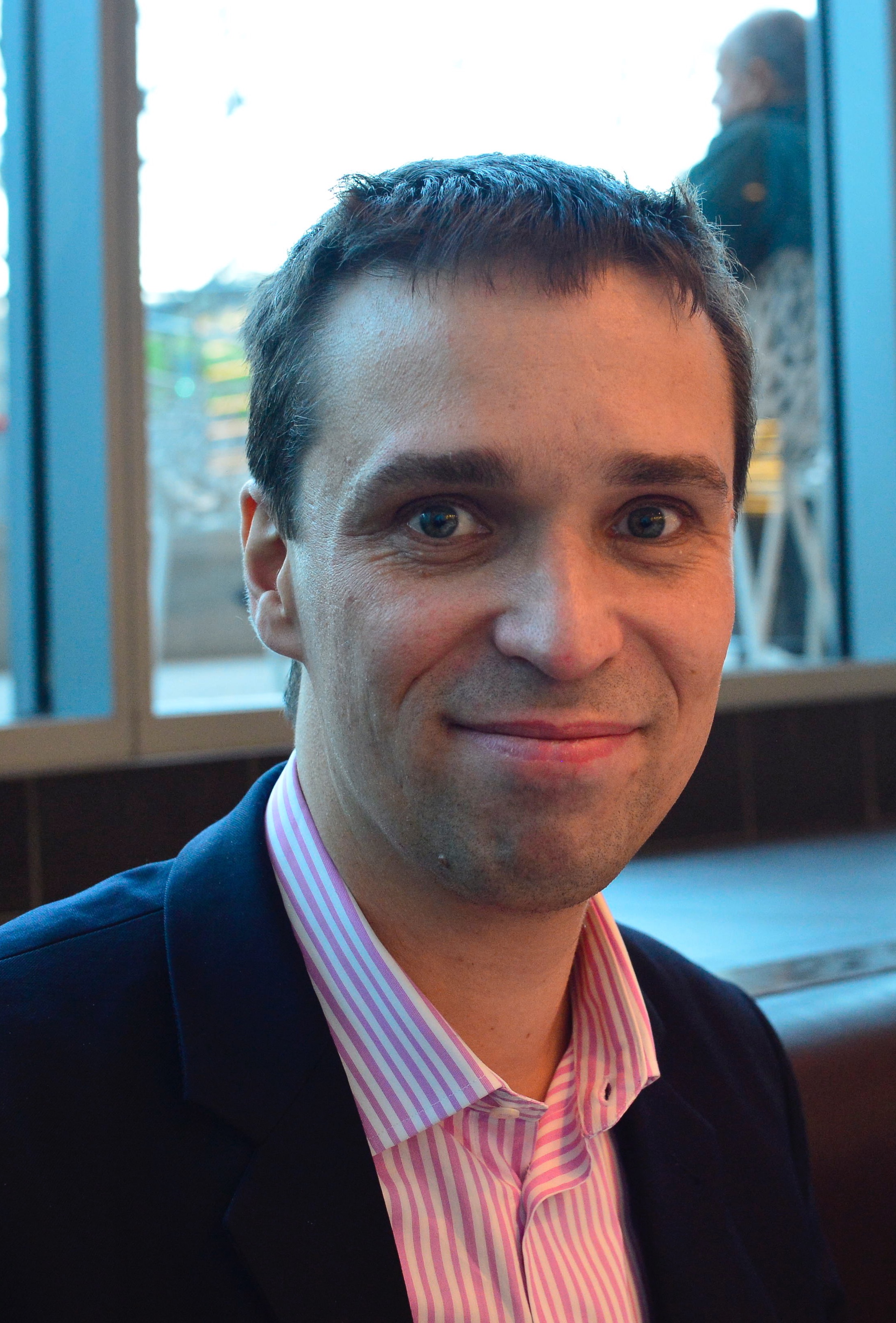}}]{Carsten Rockstuhl}
Carsten Rockstuhl received a Ph.D. from the University of Neucha\^tel, Neucha\^tel, Switzerland in 2004. After a Postdoc period at AIST in Tsukuba, Japan, he has been since 2005 with the Friedrich Schiller University of Jena, Jena, Germany. In 2013, he was appointed a full professor at the Karlsruhe Institute of Technology, Karlsruhe, Germany. He works on many aspects in the context of theoretical and computational nano-optics. He serves the community as an editor with multiple journals. Moreover, he is a member of the Karlsruhe School of Optics \& Photonics, where he currently acts as the dean of study, the Max Planck School of Photonics, and is a fellow of Optica.
\end{biography}

\begin{biography}[{\includegraphics[width=1in,height=1.25in,clip,keepaspectratio]{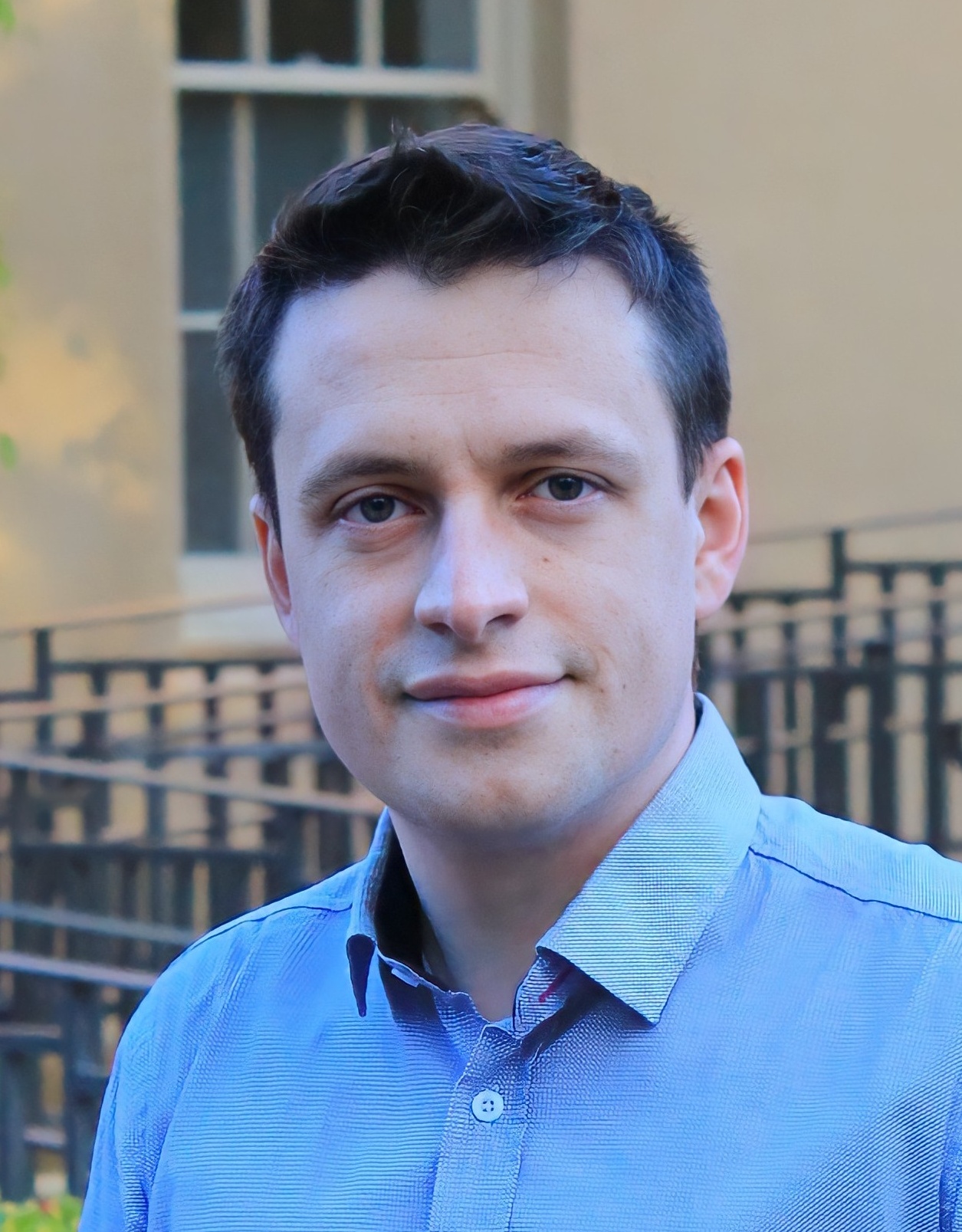}}]{Viktar Asadchy}
 Viktar Asadchy received his Diploma and M.Sc. degrees in Physics from Gomel State University, Belarus, in 2013 and 2014, respectively. In 2017, he obtained his D.Sc. degree in Electrical Engineering from Aalto University, Finland. From 2019 to 2022, he served as a Postdoctoral Fellow at the Department of Electrical Engineering, Stanford University, CA, USA. He is currently an Assistant Professor in the Department of Electronics and Nanoengineering at Aalto University, Finland. He is an Elected Associate Member of URSI. His primary research interests include metasurfaces, reconfigurable intelligent surfaces, metamaterials, photonic crystals, time-varying systems, and nanophotonics.

\end{biography}
\end{document}